%% file: jil_arxiv.tex
\documentclass[12pt]{article}
\usepackage{amsmath}
\usepackage{graphicx,psfrag,epsf,color,subfigure}
\usepackage{enumerate}
\usepackage{natbib}

\usepackage[unicode=true,pdfusetitle,
   bookmarks=true, 
   bookmarksnumbered=false, 
   bookmarksopen=false,
   breaklinks=false, 
   pdfborder={0 0 1}, 
   backref=false, 
   colorlinks=false]{hyperref}

\usepackage{amssymb}
\usepackage{multirow}
\usepackage{bm, color, comment}
\usepackage{array}
\usepackage{bm}
\usepackage{booktabs}
\usepackage{float}
\usepackage[font=small,labelfont=bf]{caption}
\setcitestyle{citesep={;}}
\usepackage[total={6.4in,8.6in}]{geometry}
\usepackage[utf8]{inputenc} 
\usepackage[T1]{fontenc}    
\usepackage{hyperref}       
\usepackage{url}            
\usepackage{booktabs}       
\usepackage{amsfonts}       
\usepackage{nicefrac}       
\usepackage{microtype}      
\usepackage{xcolor}         

\input{math_commands.tex}

\usepackage{algorithm2e} 

\usepackage{graphicx,psfrag,epsf,color}
\usepackage{amssymb}
\usepackage{epsfig, graphicx, amsmath}
\usepackage{subfigure}

\newcommand{\Mean}{{\mbox{E}}}

\newcommand{\diag}{{\mbox{diag}}}
\newcommand{\prob}{{\mbox{Pr}}}
\DeclareMathOperator*{\sargmax}{sarg\,max}

 \def\floor#1{\lfloor #1 \rfloor}

\newcommand\independent{\protect\mathpalette{\protect\independenT}{\perp}}
\def\independenT#1#2{\mathrel{\rlap{$#1#2$}\mkern2mu{#1#2}}}

\newcommand{\change}[1]{{\leavevmode\color{black}{#1}}}

\usepackage{authblk}

 \newcommand*\samethanks[1][\value{footnote}]{\footnotemark[#1]}
 
\begin{document}

\def\spacingset#1{\renewcommand{\baselinestretch}%
{#1}\small\normalsize} \spacingset{1}

\title{\bf Jump Interval-Learning for Individualized Decision Making}
\author[1]{Hengrui Cai  \thanks{Equal contribution.} \thanks{hengrc1@uci.edu} }
\author[2]{Chengchun Shi  \samethanks[1] \thanks{C.Shi7@lse.ac.uk} }
\author[3]{Rui Song\thanks{rsong@ncsu.edu}}
\author[3]{Wenbin Lu\thanks{wlu4@ncsu.edu}} 
\affil[1]{Department of Statistics, University of California Irvine}
\affil[2]{Department of Statistics, London School of Economics and Political Science}
\affil[3]{Department of Statistics, North Carolina State University}
 \date{}
 \maketitle 

\baselineskip=21pt

\begin{abstract}
An individualized decision rule (IDR) is a decision function that assigns each individual a given treatment based on his/her observed characteristics. Most of the existing works in the literature consider settings with binary or finitely many treatment options. In this paper, we focus on the continuous treatment setting and propose a jump interval-learning to develop an individualized interval-valued decision rule (I2DR) that maximizes the expected outcome. Unlike IDRs that recommend a single treatment, the proposed I2DR yields an interval of treatment options for each individual, making it more flexible to implement in practice. To derive an optimal I2DR, our jump interval-learning method estimates the conditional mean of the outcome given the treatment and the covariates via jump penalized regression, and derives the corresponding optimal I2DR based on the estimated outcome regression function. The regressor is allowed to be either linear for clear interpretation or deep neural network to model complex treatment-covariates interactions. To implement jump interval-learning, we develop a searching algorithm based on dynamic programming that efficiently computes the outcome regression function. Statistical properties of the resulting I2DR are established when the outcome regression function is either a piecewise or continuous function over the treatment space. We further develop a procedure to infer the mean outcome under the (estimated) optimal policy. Extensive simulations and a real data application to a warfarin study are conducted to demonstrate the empirical validity of the proposed I2DR. 
\end{abstract}
%

\noindent%
{\it Keywords:} Continuous treatment, Dynamic programming, Individualized interval-valued decision rule, Jump interval-learning, Precision medicine

\section{Introduction}

Individualized decision making is an increasingly attractive artificial intelligence paradigm that proposes to assign each individual a given treatment based on their observed characteristics. 
In particular, such a paradigm has been recently employed in precision medicine to tailor individualized treatment decision rule. 
Among all individualized decision rules (IDR), the one that maximizes the expected outcome is referred to as an optimal IDR. There is a huge literature on learning the optimal decision rule. Some popular methods include Q-learning \citep{watkins1992,chak2010,qian2011performance,song2015penalized}, A-learning \citep{robins2004optimal,murphy2003,shi2018high}, policy search methods \citep{zhang2012,zhang2013,wang2018quantile,nie2020learning}, outcome weighted learning \citep{zhao2012,zhao2015,zhu2017greedy,meng2020near}, concordance-assisted learning \cite{fan2017concordance,liang2017sparse}, decision list-based methods \citep{zhangyc2015,zhangyc2016}, and direct learning \citep{qi2020multi}. 
We note however, all these methods consider settings where the number of available treatment options is finite.

In this paper, we consider individualized decision making in continuous treatment settings. These studies occur in a number of real applications, including personalized dose finding \citep{chen2016personalized} and dynamic pricing \citep{den2020discontinuous}.  
For instance, in personalized dose finding, one wishes to derive a 
dose level or dose range for each patient. 
Due to patients' heterogeneity in response to doses, it is commonly assumed that there may not exist a unified best dose for all patients. Thus, one major interest in precision medicine is to develop an IDR that assigns each individual patient with a certain dose level or a specified range of doses 
based on their individual personal information, to optimize their health status. Similarly, in dynamic pricing, we aim to identify an IDR that assigns each product an optimal price according to their characteristics to maximize the overall profit.

{In contrast to} developing the optimal IDR under discrete treatment settings, individualized decision making with a continuous treatment domain has been less studied. Among those available,  
\cite{rich2014simulating} modeled the interactions
between the dose level and covariates to recommend  personalized dosing strategies.
\cite{laber2015tree} developed a tree-based method to derive the IDR by dividing patients into subgroups and assigning each subgroup the same dose level.
\cite{chen2016personalized} proposed an outcome weighted learning method to directly search the optimal IDR 
among a restricted class of IDRs. 
\cite{zhou2018parsimonious} proposed a dimension reduction framework to personalized dose finding that effectively reduces the dimensionality of baseline characteristics from high to a moderate scale. 
{\cite{kallus2018policy} and \cite{chernozhukov2019semi} evaluated and optimized IDRs for continuous treatments by replacing the indicator function in the doubly-robust approach with the kernel function, and by modeling the conditional mean outcome function (i.e., the value) through a semi-parametric form, respectively.}
\cite{zhu2020kernel} focused on the class of linear IDRs and proposed to compute an optimal linear IDR by maximizing a kernel-based value estimate. \cite{schulz2020doubly} proposed a doubly robust estimation method for personalized dose finding. The estimated optimal IDRs computed by these methods typically recommend one single treatment level for each individual, making it hard to implement in practice.

The focus of this paper is to develop an individualized interval-valued decision rule (I2DR) that returns a range of treatment levels based on individuals' baseline information. Compared to the IDRs recommended by the existing works, the proposed I2DR 
is 
more flexible to implement in practice. 
Take personalized dose finding as an illustration. First, interval-valued dose levels may be applied to patients of the same characteristics,  when arbitrary dose within the given dose interval could achieve the same efficacy. Studies of the pharmacokinetics of vancomycin conducted by \cite{rotschafer1982pharmacokinetics} suggested that adults with normal renal function should receive an initial dosage of 6.5 to 8 milligram of vancomycin per kilogram intravenously over 1 hour every 6 to 12 hours. 
In the review of warfarin dosing reported by \cite{kuruvilla2001review}, when the international normalized ratio (INR) approaches the target range or omit dose, they suggested to give 1-2.5 milligram vitamin K1 if a patient has a risk factor for bleeding, otherwise provide Vitamin K1 2-4 milligram orally. \change{Second, in cases where the available dose levels are limited, recommending a single dose is not practical. The proposed interval-valued dose rule gives more options. Based on the proposed interval, the decision maker can select the most appropriate dose by taking some other factors (e.g., patient affordability or side effects) into consideration. Third, a range of doses gives instructions for designing the medicine specification and helps to save cost on manufacturing dosage. Finally, many medical applications including treating chronic disease \citep{flack2020blood} and the radiation therapy for cancer \citep{scott2017genome} prefer optimal dose interval recommendation.} 

{Our contributions are summarized as follows. Scientifically, individualized decision making in a continuous treatment domain is a vital problem in many applications such as precision medicine and dynamic pricing. To the best of our knowledge, this is the first work on developing individualized interval-valued decision rules. Our proposal thus fills a crucial gap, extends the scope of existing methods that focus on recommending IDRs, and offers a useful tool for individualized decision making in a number of applications. 

{Methodologically, we propose a novel jump interval-learning (JIL) by integrating personalized decision making with multi-scale change point detection \citep[see][for a selective overview]{niu2016multiple}. Our proposal makes useful contributions to the two aforementioned areas simultaneously.}
 
	First, to implement personalized decision making, we propose a data-driven I2DR in a continuous treatment domain. Our proposal is motivated by the empirical finding that the expected outcome can be a piecewise function on the treatment domain in various applications.  Specifically, in dynamic pricing \citep{den2020discontinuous}, \change{the expected demand (outcome $Y$ of interest) for a product has jump discontinuities as a function of the charged price (action $A$) and baseline information such as income (covariates $X$). In other words, a small price change will lead to a considerably different demand given fixed covariates. In these applications, it is reasonable to impose
	a piecewise-function model for the outcome regression function.}
	We then leverage ideas from the change point detection literature and propose a jump-penalized regression to estimate the conditional mean of the expected outcome as a function of the treatment level and the baseline characteristics (outcome regression function). This partitions the entire treatment space into several subintervals. The proposed I2DR is a set of decision rules that assign each subject to one of these subintervals. In addition, we further develop a procedure to construct a confidence interval (CI) for the expected outcome under the proposed I2DR and the optimal IDR.

	
	{Second, we note that most works in the multi-scale change point detection literature either focused on models without covariates, or required the underlying truth to be piecewise constant \citep[see e.g.,][and the references therein]{boysen2009consistencies, Frick2014, Piotr2014}. Our work goes beyond those cited above in that we consider a more complicated (nonparametric) model with covariates, and allow the underlying outcome regression function to be either a piecewise or continuous function over the treatment space. 
To approximate the expected outcome as a function of baseline covariates, we propose a linear function model and a deep neural networks model. We refer to the two procedures as L-JIL and D-JIL, respectively. Here, the proposed L-JIL yields a set of linear decision rules that is easy to interpret. See the real data analysis in Section \ref{realdata} for details. 
On the contrary, the proposed D-JIL employs deep learning \citep{lecun2015deep} to model the complicated outcome-covariates relationships that often occur in high-dimensional settings. 
We remark that both procedures are developed by imposing a piecewise-function model to approximate the outcome-treatment relationship. Yet, they are valid when the expected outcome is a continuous function of the treatment level as well.}
Theoretically, we systematically study the statistical properties of the jump-penalized estimators with linear regression or deep neural networks. Our theoretical approaches can be applied to the analysis of general covariate-based change point models. The model could be either parametric or nonparametric. 
Specifically, we establish the almost sure convergence rates of our estimators. 
When the underlying outcome regression function is a piecewise function of the treatment,
we further derive the almost sure convergence rates of the estimated change point locations, and show that with probability $1$, the number of change points can be correctly estimated with sufficiently large sample size. These findings are nontrivial extensions of classical results derived for models without covariates. For instance, 
deriving the asymptotic behavior of change point estimators for these models typically relies on the tail inequalities for the partial sum process \citep[see e.g.,][]{Frick2014}. However, these technical tools are not directly applicable to our settings where deep learning is adopted to model the outcome regression function. Moreover, we expect our theories to also be of general interest 
to the line of work on developing theories for deep learning methods \citep[see e.g.,][]{imaizumi2019deep,schmidt2020nonparametric,farrell2021deep}. 

The rest of this paper is organized as follows. In Section \ref{sec:statsframe}, we introduce the statistical framework, define the notion of I2DR, and posit our working model assumptions. 
In Section \ref{secmethod}, we propose the jump interval-learning method and discuss its detailed implementation. 
Statistical properties of the proposed I2DR and the estimator for the mean outcome under the proposed I2DR are presented in Section \ref{sectheory}. We further develop a confidence interval for the expected outcome under the estimated I2DR.
Simulation studies are conducted in Section \ref{simul} to evaluate the finite sample performance of our proposed method. 
We apply our method to a real dataset from a warfarin study in Section \ref{realdata}. In Section \ref{secproofthm1}, we provide the technical proof for one of our main theorem, followed by a concluding discussion in Section \ref{secdis}. 
The rest of proofs are provided in the supplementary article.

\section{Statistical Framework}\label{sec:statsframe} This section is organized as follows. We first introduce the model setup in Section \ref{secnots}. The definition of I2DR is formally presented in Section \ref{secI2DR}. In Section \ref{secmodels}, we posit two working models assumptions for the expected outcome as a function the treatment level. 
We aim to develop a method that works under both working assumptions.

 \subsection{Model Setup}\label{secnots} 
We begin with some notations. 
Let $A$ denote the treatment level assigned to a randomly selected individual in the population from a compact interval. Without loss of generality, suppose $A$ belongs to $[0,1]$. Let $X \in \mathbb{X}$ be that individual's baseline covariates where the support $\mathbb{X}$ is a subset in $\mathbb{R}^p$. We assume the covariance matrix of $X$ is positive definite. Let $Y \in \mathbb{R}$ denote that individual's associated outcome, the larger the better by convention. Let $p(\bullet|x)$ denote the probability density function of $A$ given $X=x$. 
 {In addition, for any $a\in [0,1]$, define the potential outcome $Y^*(a)$ as the outcome of that individual that would have been observed if they were receiving treatment $a$. 
The observed data consists of the covariate-treatment-outcome triplets $\{(X_i,A_i,Y_i):i=1,\dots,n\}$ where $(X_i,A_i,Y_i)$'s are i.i.d. copies of $(X,A,Y)$. Based on this data, we wish to learn an optimal decision rule to possibly maximize the expected outcome of future subjects using their baseline information. 

Formally speaking, an individualized decision rule (IDR) is a deterministic function $d(\cdot)$ that maps the covariate space $\mathbb{X}$ to the treatment space $[0,1]$. 
The optimal IDR is defined to maximize the expected outcome (value function) $V(d)=\Mean \{Y^*(d(X))\}$ among all IDRs. 
The following assumptions guarantee the optimal IDR is identifiable from the observed data. 

\smallskip

\noindent (A1.) Consistency: $Y= Y^*(A)$, almost surely,\\
\noindent (A2.) No unmeasured confounders: $\{Y^*(a):a\in [0,1]\}\independent A\mid X$,\\ 
\noindent (A3.) Positivity: there exists some constant $c_*>0$ such that $p(a|x)\ge c_*$ for any $x\in \mathbb{X}$ and $a\in [0,1]$.

\smallskip

\change{Assumption (A1) requires the observed outcome to be the same as the potential outcome associated with the observed treatment. Assumption (A2) requires 
the baseline covariates to contain enough confounders given which the 
treatment is conditionally independent of the potential outcomes. 
Assumption (A3) requires the propensity score to be strictly positive for any realization of the baseline covariates. 
Assumptions (A2) and (A3) automatically hold in randomized studies. In observational studies, one can estimate the propensity score from the data to check (A3). Nonetheless, (A2) cannot be verified in general. 
In addition, Assumptions (A1) to (A3) are commonly imposed in the literature \citep[see e.g.,][]{chen2016personalized,zhu2020kernel,schulz2020doubly} to guarantee that the outcome of interest and the optimal IDR are estimable from observed data. 
}
In particular, under (A1)-(A3), 
we have 
$V(d)=\Mean \{Q(X,d(X))\}$
where $Q(x,a)=\Mean (Y|X=x, A=a)$ is the conditional mean of an individual's outcome given their received treatment and baseline covariates. {We refer to this function as the outcome regression function.} 
As a result, the optimal IDR for an individual with covariates $x$ is given by $\argmax_{a\in [0,1]} Q(x,a)$. Let $V^{\tiny{opt}}$ denote the value function under the optimal IDR. We have $V^{\tiny{opt}}=\Mean \{\sup_{a\in [0,1]} Q(X,a)\}$.


\subsection{I2DR}\label{secI2DR}
The focus of this paper is to develop an optimal individualized interval-based decision rule (I2DR). As commented in the introduction, these decision rules are more flexible to implement in practice when compared to single-valued decision rules in personalized dose finding and dynamic pricing.

We define an I2DR as a function $d(\cdot)$ that takes an individual's covariates $x$ as input and outputs an interval $\mathcal{I}\subseteq [0,1]$. Given the recommended interval $\mathcal{I}$, {\color{black}different doctors / agents might assign different treatments to patients / products \change{according to their own preferences. In practice, the decision maker could take the minimum value, the maximum value, the mid-point value, or the value uniformly at random.} 
The actual treatments that subjects receive in the population will have a distribution function $\Pi^*(\cdot;x,\mathcal{I})$. Throughout this paper, we assume $\Pi^*(\cdot;x,\mathcal{I})$ has a bounded density function $\pi^*(\cdot;x,\mathcal{I})$ for any $x$ and $\mathcal{I}$.} 
Apparently, we have $\int_{\mathcal{I}} \pi^*(a;x,\mathcal{I})da=1$, for any interval $\mathcal{I}$ and $x\in \mathbb{X}$. When (A1)-(A3) hold, the associated value function under an I2DR $d(\cdot)$ equals
\begin{eqnarray*}
	V^{\pi^*}(d)=\Mean \left(\int_{d(X)} Q(X,a) \pi^*(a;X,d(X))da\right).
\end{eqnarray*}
Restricting $d(\cdot)$ to be a scalar-valued function, $V^{\pi^*}(d)$ is reduced to $V(d)$. 

Given the dataset, one may estimate $V^{\pi^*}(d)$  nonparametrically for any $d(\cdot)$ and directly search the optimal I2DR based on the estimated value function. However, such a value search method has the following two limitations. First, a nonparametric estimator of $V^{\pi^*}(d)$ requires to specify the preference function $\pi^*$, which might be unknown to us. Second, even though a nonparametric estimator of $V^{\pi^*}(d)$  can be derived, 
{it remains unknown how to efficiently compute the I2DR that maximizes the estimated value (see Section \ref{sec:policysearch} for details).}
To overcome these limitations, we propose a semiparametric model for the outcome regression function and use a model-assisted approach to derive the optimal I2DR. We formally introduce our method in Section \ref{secmethod}.

 \subsection{Working Model Assumptions}\label{secmodels} 

In this section, we introduce two working models for the outcome regression function, corresponding to a piecewise function and a continuous function of the treatment level.  

%


\smallskip

	{\bf Model I (Piecewise Functions).} Suppose
	\begin{eqnarray}\label{main_model}
		Q(x,a)=\sum_{\mathcal{I} \in \mathcal{P}_0} q_{\mathcal{I},0}(x)\mathbb{I}(a \in \mathcal{I})\,\,\,\,\,\,\,\,\forall x\in \mathbb{X},a\in [0,1], 
		 	\end{eqnarray} 
	for some partition $\mathcal{P}_0$ of $[0,1]$ and a collection of continuous functions $(q_{\mathcal{I},0})_{\mathcal{I}\in \mathcal{P}_0}$, where the number of intervals in $\mathcal{P}_0$ is finite. Specifically, a partition $\mathcal{P}$ of $[0,1]$ is defined as a collection of mutually disjoint intervals $\{[\tau_0,\tau_1),[\tau_1,\tau_2),\dots,[\tau_{K-1},\tau_K]\}$ for some $0=\tau_0<\tau_1<\tau_2<\cdots<\tau_{K-1}< \tau_K=1$ and some integer $K\ge 1$. 
	\change{Here, we only require any two consecutive $q$-functions to be different. We do not impose any additional constraints.} As commented in our introduction, we expect the above model assumption holds in real-world examples such as dynamic pricing. 

\smallskip

{\bf Model II (Continuous Functions).} Suppose
$Q(x,a)$ is a continuous function of $a$ and $x$, for any $x\in \mathbb{X}$ and $a\in [0,1]$.

\change{We aim to propose a new method 
that works when either Model I (piecewise function) or Model II (continuous function) holds.}

\section{Methods}\label{secmethod}
In this section, we first present the proposed jump interval-learning and its motivation in Section \ref{sec:JIL}. 
We next introduce two concrete proposals, i.e., linear jump interval-learning and deep jump interval-learning, to detail our methods in Section \ref{sec:linear_dnn}. We then present the dynamic programming algorithm to implement jump interval-learning (see Algorithm \ref{alg1} for an overview) in Section \ref{impl}. Finally, we provide more details on tuning parameter selection in Section \ref{tuning}. 

\subsection{Jump Interval-learning}\label{sec:JIL}
 
We use Model I to present the motivation of our jump interval-learning. In view of \eqref{main_model}, any treatment level within an interval $\mathcal{I}\in \mathcal{P}_0$ will yield the same efficacy to a given individual. The optimal I2DR is then given by 
\begin{eqnarray*}
d^{\tiny{opt}}(x)=\argmax_{\mathcal{I}\in \mathcal{P}_0} q_{\mathcal{I},0}(x),
\end{eqnarray*}
 independent of the preference function $\pi^*$. To see this, notice that
\begin{eqnarray*}
	&&V^{\pi^*}(d^{\tiny{opt}})=\Mean \left(\int_{d^{\tiny{opt}}(X)} \sum_{\mathcal{I} \in \mathcal{P}_0} q_{\mathcal{I},0}(X)\mathbb{I}(a \in \mathcal{I})\pi^*(a;X,d^{\tiny{opt}}(X))da \right)\\
	&=&\Mean  \sum_{\mathcal{I} \in \mathcal{P}_0} q_{\mathcal{I},0}(X)\mathbb{I}(d^{\tiny{opt}}(X) \in \mathcal{I})   \int_{d^{\tiny{opt}}(X)} \pi^*(a;X,d^{\tiny{opt}}(X))da.
	\end{eqnarray*}
	For any I2DR $d(\cdot)$, we have $\int_{d^{\tiny{opt}}(X)} \pi^*(a;X,d^{\tiny{opt}}(X))da = \int_{d(X)} \pi^*(a;X,d(X))da  = 1$ by definition. It follows that
\begin{eqnarray*}
	V^{\pi^*}(d^{\tiny{opt}})&=& \Mean \sum_{\mathcal{I} \in \mathcal{P}_0} q_{\mathcal{I},0}(X)\mathbb{I}(d^{\tiny{opt}}(X) \in \mathcal{I})   \int_{d(X)} \pi^*(a;X,d(X))da\\
 &\ge& \Mean \int_{d(X)} \sum_{\mathcal{I} \in \mathcal{P}_0} q_{\mathcal{I},0}(X)\mathbb{I}(a \in \mathcal{I}) \pi^*(a;X,d(X))da=V^{\pi^*}(d),
\end{eqnarray*}
where the inequality is due to that $Q(X,d^{\tiny{opt}}(X)) = \sum_{\mathcal{I} \in \mathcal{P}_0} q_{\mathcal{I},0}(X)\mathbb{I}(d^{\tiny{opt}}(X) \in \mathcal{I}) \ge \sum_{\mathcal{I} \in \mathcal{P}_0} q_{\mathcal{I},0}(X)\mathbb{I}(a \in \mathcal{I})=Q(X,a)$, almost surely for any $a\in [0,1]$. 
Therefore, to derive the optimal I2DR, it suffices to estimate $q_{\mathcal{I},0}(\cdot)$. For notation simplicity, in the rest of this paper, we denote $V^{\pi^*}(d)$ by $V(d)$ for any decision rule $d$.

From now on, we focus on a subset of intervals in $[0,1]$. By \textit{interval} we always refer to those of the form $[a,b)$ for some $0\le a< b< 1$ or $[a,1]$ for some $0\le a< 1$. For any partition $\mathcal{P}=\{[0,\tau_1),[\tau_1,\tau_2),\dots,[\tau_{K-1},1]\}$, 
we use $J(\mathcal{P})$ to denote the set of change point locations, i.e, $\{\tau_1,\tau_2,\dots,\tau_{K-1}\}$,  and $|\mathcal{P}|$ to denote the number of intervals in $\mathcal{P}$. Our proposed method yields a partition $\widehat{\mathcal{P}}$ and an I2DR $\widehat{d}(\cdot)$ such that $\widehat{d}(x)\in \widehat{\mathcal{P}}$, $\forall x\in \mathbb{X}$. The number of intervals in $\widehat{\mathcal{P}}$ (denoted by $|\widehat{\mathcal{P}}|$) involves a trade-off. If $|\widehat{\mathcal{P}}|$ is too large, then $\widehat{\mathcal{P}}$ will contain many short intervals, making the resulting decision rule hard to implement in practice. Yet, a smaller value of $|\widehat{\mathcal{P}}|$ might result in a smaller value function. Our proposed method adaptively determines $|\widehat{\mathcal{P}}|$ based on jump-penalized regression.

{We next detail our method. Jump interval-learning consists of the following two steps. In the first step, we estimate the outcome regression function using jump penalized least squares regression. Then we derive the corresponding I2DR from the resulting estimator $\widehat{q}_{\mathcal{I}}(\cdot)$. 
To begin with, we cut the entire treatment range into $m$ initial intervals: 
\begin{eqnarray}\label{eqn:initialintervals}
	[0,1/m),[1/m,2/m), \dots, [(m-1)/m,1].
\end{eqnarray}
The integer $m$ is allowed to diverge with the number of observations $n$. For instance, it can be specified by the clinical physician such that the output dose interval for each individual is at least of the length $m^{-1}$. When no prior knowledge is available, we recommend to set $m$ to be proportional to $n$. 
It is worth mentioning that \eqref{eqn:initialintervals} is not the final partition that we recommend. Nor is it equal to $\mathcal{P}_0$ defined in Model I. Given \eqref{eqn:initialintervals}, we are looking for a partition $\widehat{\mathcal{P}}$ such that each interval in $\widehat{\mathcal{P}}$ corresponds to a union of some of the these $m$ intervals. {In other words, we will adaptively combine some of these intervals to form $\widehat{\mathcal{P}}$.}

More specifically, let $\mathcal{B}(m)$ denote the set of partitions $\mathcal{P}$ that satisfy the following requirement: the end-points of each interval $\mathcal{I}\in \mathcal{P}$ lie on the grid $\{j/m:j=0,1,\dots,m\}$.  We associate to each partition $\mathcal{P}\in \mathcal{B}(m)$ a collection of functions $\{q(\cdot;\theta_\mathcal{I})\}_{\mathcal{I}\in \mathcal{P}}\in \prod_{\mathcal{I}\in \mathcal{P}} \mathcal{Q}_{\mathcal{I}}$ for $\mathcal{Q}_{\mathcal{I}}$ as some class of functions, where $\theta_\mathcal{I}$ is the underlying parameter associated to interval $\mathcal{I}$. We propose to estimate $\widehat{\mathcal{P}}$ by solving 
\begin{eqnarray}\label{optimize}
\begin{split}
&~~~~~~~~~~~~~~~(\widehat{\mathcal{P}},\{\widehat{q}_{\mathcal{I}}:\mathcal{I}\in \widehat{\mathcal{P}} \})=\\
 & \argmin_{\substack{\mathcal{P}\in \mathcal{B}(m)\\\{q(\cdot;\theta_\mathcal{I})\in \mathcal{Q}_{\mathcal{I}}: \mathcal{I}\in \mathcal{P} \} }}\Big\{\sum_{\mathcal{I}\in \mathcal{P}} \Big(   {1\over n}\sum_{i=1}^{n} \mathbb{I}(A_i\in \mathcal{I})  \big\{Y_i - q(X_i;\theta_\mathcal{I}) \big\}^2 + \lambda_n |\mathcal{I}| \|\theta_\mathcal{I}\|_2^2\Big)+\gamma_n |\mathcal{P}| \Big\}, 
 \end{split}
\end{eqnarray} 
where $\lambda_n$ and $\gamma_n$ are some nonnegative regularization parameters specified in Section \ref{tuning}, and $\|\theta_\mathcal{I}\|_2^2$ denote the Euclidean norm of the model parameter $\theta_\mathcal{I}$. The purpose of introducing the $\ell_2$-type penalty term $ \lambda_n |\mathcal{I}| \|\theta_\mathcal{I}\|_2^2$ 
is to help to prevent overfitting in large $p$ problems. The purpose of introducing the $\ell_0$-type penalty term $\gamma_n |\mathcal{P}|$ is to control the total number of jumps. When $m=n$, $\lambda_n=0$, $A_i=i/n,\forall 1\le i\le n$, no baseline covariates are collected, the above optimization corresponds to the jump-penalized least square estimator proposed by \cite{boysen2009consistencies}. We refer to this step as jump interval-learning (JIL).

For a fixed $\mathcal{P}$, solving the optimization function in \eqref{optimize} yields
its associated outcome regression functions $\{\widehat{q}_{\mathcal{I}}\}_{\mathcal{I}\in \mathcal{P}}$. This step involves parametric or nonparametric regression and can be solved via existing statistical or machine learning approaches. We provide two concrete study cases below, based on linear regression and deep learning. These estimated outcome regression functions can be viewed as functions of $\mathcal{P}$. As such, $\widehat{\mathcal{P}}$ is adaptively determined by minimizing the penalized least square function in \eqref{optimize}.

To maximize the expected outcome of interest, our proposed I2DR is then given by
\begin{eqnarray}\label{I2DR}
	\widehat{d}(x)=\argmax_{\mathcal{I}\in \widehat{\mathcal{P}}}  \widehat{q}_{\mathcal{I}}(x),\,\,\,\,\,\,\,\,\forall x\in \mathbb{X}.
\end{eqnarray}
When the argmax in \eqref{I2DR} is not unique, $\widehat{d}(\cdot)$ outputs the interval that contains the smallest treatment.

We next evaluate the value function under the proposed I2DR $V(\widehat{d})$ and $V(d^{\tiny{opt}})$. 
For each interval $\mathcal{I}$ in the estimated optimal partition $\widehat{\mathcal{P}}$, we estimate the generalized propensity score function $e(\mathcal{I}|x)\equiv \prob(A\in \mathcal{I}|X=x)$. Let $\widehat{e}(\mathcal{I}|x)$ denote the resulting estimate. Following the estimation strategy in \cite{zhang2012}, we propose the following value estimator under \eqref{I2DR},
\begin{eqnarray}\label{estvalue}
	\widehat{V}=\frac{1}{n}\sum_{i=1}^n  \left[ {\mathbb{I}\{A_i\in \widehat{d}(X_i) \} \over  \widehat{e}( \widehat{d}(X_i) |X_i)}\big\{Y_i -\max_{\mathcal{I}\in \widehat{\mathcal{P}}} \widehat{q}_{\mathcal{I}}(X_i)  \big\} +   \max_{\mathcal{I}\in \widehat{\mathcal{P}}} \widehat{q}_{\mathcal{I}}(X_i) \right].
\end{eqnarray}

Statistical properties of the estimates in \eqref{I2DR} and \eqref{estvalue} are studied in Section \ref{sectheory}.
Although we use the example of piecewise functions to motivate our procedure, the proposed method allows the outcome regression function to be a continuous function of $a$ and $x$ as well. 
\change{Specifically, 
we can approximate any continuous outcome function by a piecewise function with the increasing number of partitions. As such, the proposed jump-interval learning method is applicable to handle Model II as well.} 
See Section \ref{sectheory} for detail. 
 
\subsection{Linear- and Deep-JIL}\label{sec:linear_dnn}

In practice, we consider two concrete proposals to implementing jump interval-learning, by considering a linear function class and a deep neural networks (DNN) class for $\mathcal{Q}_{\mathcal{I}}$ in \eqref{optimize}. \change{In particular, the proposed L-JIL yields a set of linear decision rules that is easy to compute and interpret. See the real data analysis in Section 6 for details. In theory, it achieves a better convergence rate under the correct model specification. We recommend using L-JIL in applications where a simple and interpretable decision rule is preferred. On the contrary, the use of DNN in D-JIL allows us to capture the complicated outcome-covariates relationships that often occur in high-dimensional settings. We recommend using D-JIL in applications with high-dimensional covariates and complicated nonlinear associations.}
 
\change{We also remark that the proposed method is very general and allows a large variety of function approximators. Although we focus on linear models and deep neural nets in this paper, other function approximators such as linear basis expansion, reproducing kernel Hilbert spaces, and random forests are equally applicable. The theoretical properties of the resulting estimated Q-function can be similarly established \citep[see e.g., ][for the convergence rate of these nonparametric estimators]{Burman1989,steinwart2008support,wager2018estimation}.}

\subsubsection{Case 1: Linear-JIL}
	We use a linear regression model for $\mathcal{Q}_{\mathcal{I}}$. Specifically, we set  $q(x,\theta_{\mathcal{I}})$ to $\bar{x}^\top \theta_{\mathcal{I}}$ for any interval $\mathcal{I}$ and $x\in \mathbb{X}$, 
	where $\bar{x}$ is a shorthand for the vector $(1,x^{\top})^{\top}$. 
Adopting the linearity assumption, we have $\widehat{q}_{\mathcal{I}}(x)=\bar{x}^\top \widehat{\theta}_{\mathcal{I}}$ for some $\widehat{\theta}_{\mathcal{I}}$. It follows from \eqref{I2DR} that the proposed I2DR corresponds to a linear decision rule, i.e., $\widehat{d}(x)=\arg\max_{\mathcal{I}\in \widehat{\mathcal{P}}} \bar{x}^\top \widehat{\theta}_{\mathcal{I}}$. 
As such, the linearity assumption ensures our I2DR is interpretable to the domain experts. 

We next discuss how to compute $\widehat{\mathcal{P}}$ and $\{\widehat{\theta}_{\mathcal{I}}:\mathcal{I}\in \widehat{\mathcal{P}}\}$. The objective function in \eqref{optimize} is reduced to
\begin{eqnarray}\label{optimize_ridge}
&&~~~~~~~(\widehat{\mathcal{P}},\{\widehat{\theta}_{\mathcal{I}}:\mathcal{I}\in \widehat{\mathcal{P}} \})=\\\nonumber
		&&=\argmin_{(\mathcal{P}\in \mathcal{B}(m),\{\theta_{\mathcal{I}}: \mathcal{I}\in \mathcal{P} \} )} \left\{\sum_{\mathcal{I}\in \mathcal{P}} \left(\frac{1}{n}\sum_{i=1}^n \mathbb{I}(A_i\in \mathcal{I}) (Y_i-\overline{X}_i^{\top} \theta_{\mathcal{I}})^2+\lambda_n |\mathcal{I}| \|\theta_{\mathcal{I}}\|_2^2\right)+\gamma_n |\mathcal{P}| \right\}, 
\end{eqnarray}
where $\overline{X}_i=(1,X_i^{\top})^{\top}$. 
We refer to this step as linear jump interval-learning (L-JIL).
The ridge penalty $\lambda_n |\mathcal{I}| \|\theta_{\mathcal{I}}\|_2^2$ in \eqref{optimize_ridge} guarantees that for any interval $\mathcal{I}\in \widehat{\mathcal{P}}$, the parameter $\widehat{\theta}_{\mathcal{I}}$ is well defined even when $\sum_{i=1}^n \mathbb{I}(A_i\in \mathcal{I})<p+1$ such that the matrix $\sum_i \mathbb{I}(A_i\in \mathcal{I})\overline{X}_i\overline{X}_i^\top$ is not invertible. It also prevents over-fitting and yields more accurate estimate in high-dimensional settings. 

\subsubsection{Case 2: Deep-JIL}
We next consider using deep neural networks (DNNs) to approximate the outcome regression function, so as to capture the complex dependence between the outcome and covariates. 
Specifically, the network consists of $p$ input units (coloured in blue in Figure \ref{fig:MLP}), corresponding to the covariates $X$. The hidden units (coloured in green) are grouped in a sequence of $L$ layers. Each unit in the hidden layer is determined as a nonlinear transformation of a linear combination of the nodes from the previous layer. The total number of parameters in the network is denoted by $W$. See Figure \ref{fig:MLP} for an illustration. 
The parameters in DNNs can be solved using a stochastic gradient descent algorithm. 
In our implementation, we apply the Multi-layer Perceptron (MLP) regressor \cite{SCIKit2011} for parameter estimation. 
We refer to the resulting optimization as deep jump interval-learning (D-JIL).

\begin{figure} 
\centering
	\includegraphics[width=0.45\textwidth]{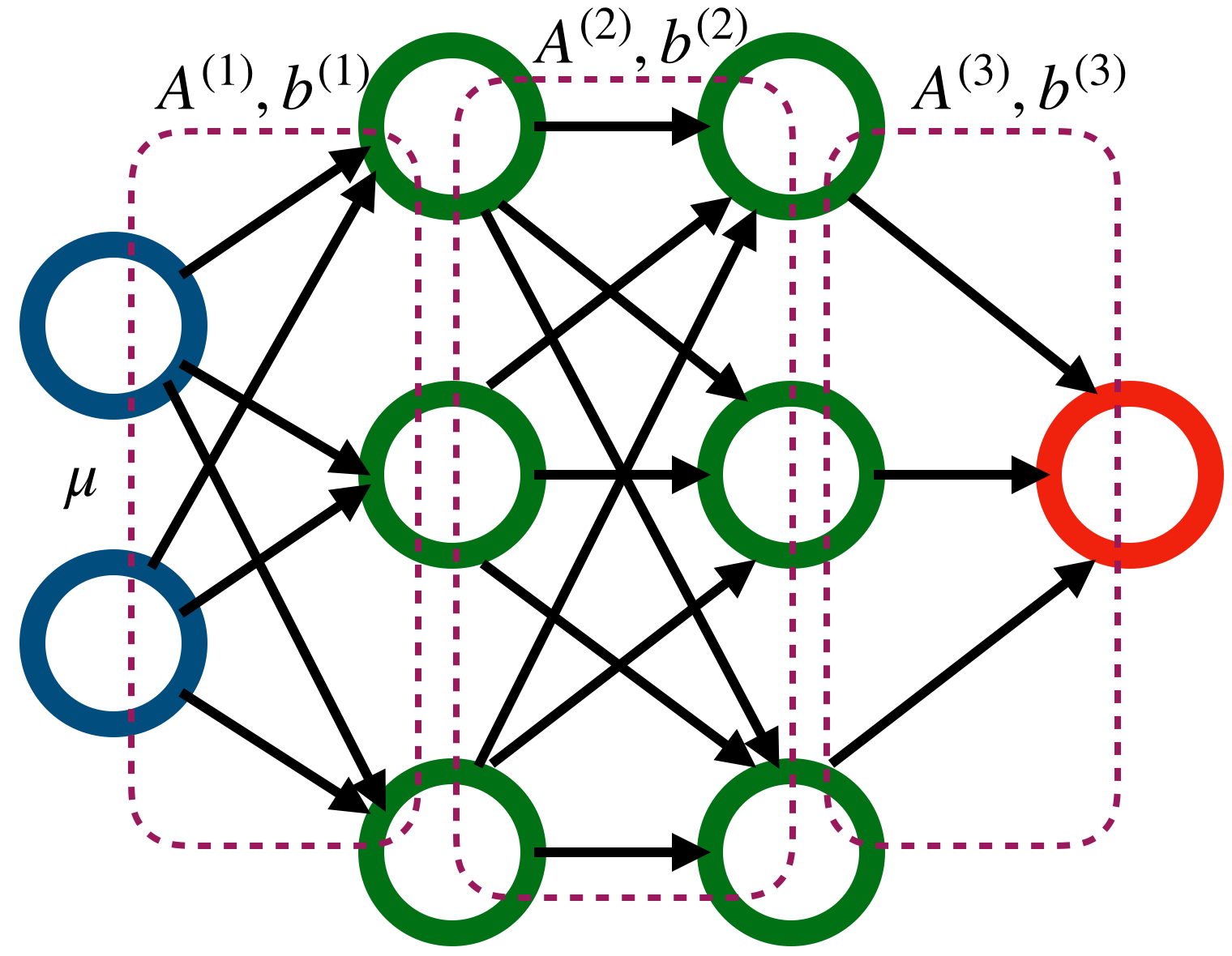} 
	\caption{Illustration of DNN with $L=2$ and $W=25$; here $\mu\in \mathbb{R}^p$ is the input,  the output is given by $A^{(3)}\sigma(A^{(2)}\sigma(A^{(1)}\mu+b^{(1)})+b^{(2)})+b^{(3)}$ where $A^{(l)}$, $b^{(l)}$ denote the corresponding parameters to produce the linear transformation for the $(l-1)$th layer and that $\sigma$ denotes the componentwise rectified linear unit (ReLU) function. In this example, $W=\sum_{j=1}^3 (\|A^{(3)}\|_0+\|b^{(3)}\|_0)=25$ where $\|\bullet\|_0$ denotes the number of nonzero elements in the vector or matrix.}
	\label{fig:MLP}
\end{figure}

 Finally, we remark that alternative to our approach, one may directly apply DNN that takes the covariate-treatment pair $(X,A)$ as the input to learn the outcome regression function. However, the resulting estimator for the outcome regression function is not guaranteed to be a piecewise function of the treatment. As such, it cannot yield an I2DR. 

\subsection{Implementation}\label{impl}
In this section, we present the computational details for jump interval-learning. We employ the dynamic programming algorithm \citep[see e.g.,][]{friedrich2008complexity} to find the optimal partition $\widehat{\mathcal{P}}$ 
that minimizes the objective function \eqref{optimize}. Meanwhile, other algorithms for multi-scale change point detection are equally applicable \citep[see e.g.,][]{scott1974,Harchaoui2010,Piotr2014}.
Specifically, we adopt the PELT method proposed by \cite{killick2012optimal} that includes additional pruning steps within the dynamic programming framework to achieve a linear computational cost. 
Given $\widehat{\mathcal{P}}$, the set of functions $\{\widehat{q}_{\mathcal{I}}:\mathcal{I}\in \widehat{\mathcal{P}} \}$ can be computed via either linear regression or deep neural network. 

To detail our procedure, for any interval $\mathcal{I}\in [0,1]$, we define the cost function
\begin{eqnarray*}
	\hbox{cost}({\mathcal{I}})=   \min_{q(\cdot;\theta_\mathcal{I})\in \mathcal{Q}_{\mathcal{I}}}  \left[   {1\over n}\sum_{i=1}^{n} \mathbb{I}(A_i\in \mathcal{I})  \big\{Y_i - q(X_i;\theta_\mathcal{I}) \big\}^2 + \lambda_n ||\theta_\mathcal{I}||_2^2\right],  
\end{eqnarray*}
where $\mathcal{Q}_{\mathcal{I}}$ is a class of linear functions or deep neural networks, corresponding to L-JIL and D-JIL, respectively.

For any integer $1\le r<m$, denote by $\mathcal{B}(m, r)$ the set consisting of all possible partitions $\mathcal{P}_r$ of $[0,r/m)$ such that the end-points of each interval $\mathcal{I}\in \mathcal{P}_r$ lie on the grid $\{j/m:j=0,1,\dots,r\}$. Set $\mathcal{B}(m,m)=\mathcal{B}(m)$, we define the Bellman function
\begin{eqnarray*}
	B(r)=\inf_{\mathcal{P}_r\in \mathcal{B}(m,r)}\left(\sum_{\mathcal{I}\in \mathcal{P}_r} \hbox{cost}(\mathcal{I})+\gamma_n (|\mathcal{P}_r|-1) \right).
\end{eqnarray*}
Let $B(0)=-\gamma_n$, the dynamic programming algorithm relies on the following recursion formula,
\begin{eqnarray}\label{bellmanrelation}
	B(r)=\min_{j\in \mathcal{R}_{r}} \left\{B(j)+\gamma_n+\hbox{cost}([j/m,r/m))\right\},\,\,\,\,\,\,\,\,\forall r\ge 1.
\end{eqnarray}
 where $\mathcal{R}_{r}$ is the candidate change points list updated by 
\begin{eqnarray}\label{Rvstar}
\{j\in \mathcal{R}_{r-1} \cup\{r-1\}: B(j)+\hbox{cost}([j/m,(r-1)/m))\leq B(r-1)\},
\end{eqnarray}
during each iteration with $\mathcal{R}_0=\{0\}$. The constraint listed in \eqref{Rvstar} iteratively updates the set of candidate change points and removes values that can never be the minima of the objective function. It speeds up the computation, leading to a cost that is linear in the number of observations \citep{killick2012optimal}.

\begin{algorithm}[!t]\label{alg1}
	\caption{Jump interval-learning} \label{alg1}
	\begin{tabbing}
		\enspace \textbf{Global:} data $\{(X_i,A_i,Y_i):i=1,\dots,n\}$; sample size $n$; covariates dimension $p$; \\\qquad \qquad ~ number of initial intervals $m$; penalty terms $\lambda_n$,  $\gamma_n$.\\
		\enspace \textbf{Local:} integers $l, r \in  \mathbb{N}$; cost dictionary $\mathcal{C}$; a vector of integers $\tau\in \mathbb{N}^m$; \\\qquad \quad ~~ Bellman function $B\in  \mathbb{R}^m$; a set of candidate point lists $\mathcal{R}$.\\
		\enspace \textbf{Output:} 
		$\widehat{\mathcal{P}}$ and $\{\widehat{q}_{\mathcal{I}}:\mathcal{I}\in \widehat{\mathcal{P}} \}$.\\
		\enspace I. Initialization. Set $B(0)\gets -\gamma_n$; $\widehat{\mathcal{P}} \gets Null$; $\tau \gets Null$; $\mathcal{R}(0) \gets \{0\}$; \\
		
		\enspace II. Apply the PELT method. For $r=1,\dots,m$:\\ 
		\qquad 1. Compute $B(r)=\min_{j\in \mathcal{R}{(r)}} \{B(j)+\mathcal{C}([j/m,r/m))+ \gamma_n\}$ based on Algorithm \ref{alg2};\\
		\qquad 2. $j^*\gets \argmin_{j\in \mathcal{R}{(r)}} \{B(j)+\mathcal{C}([j/m,r/m))+ \gamma_n\}$;\\
		\qquad 3. $\tau(r)\gets \{j^*,\tau(j^*)\}$;\\ 
		\qquad 4. $\mathcal{R}(r) \gets \{j\in \mathcal{R}(r-1) \cup\{r-1\}: B(j)+\mathcal{C}([j/m,(r-1)/m))\leq B(r-1)\} $;\\
		\enspace III. Get Partitions. $\tau^*\gets \tau(m)$; $r\gets m$; $l\gets \tau^*[r] $; While $r>0$:\\
		\qquad 1. Let $\mathcal{I}=[l/m,r/m)$ if $r<m$ else $\mathcal{I}=[l/m,1]$; \\
		\qquad 2. $\widehat{\mathcal{P}}\gets \widehat{\mathcal{P}}\cup \mathcal{I} $;\\
		\qquad 3. $\widehat{q}_{\mathcal{I} }(\cdot) \gets \arg\min_{q} \sum_i \mathbb{I}(A_i\in \mathcal{I}) \{Y_i-q(X_i)\}^2 $;\\
		\qquad 4.  $r\gets l$; $l\gets \tau^*[r] $;\\		
		 	
		\enspace \Return {$\widehat{\mathcal{P}}$ and $\{\widehat{q}_{\mathcal{I}}:\mathcal{I}\in \widehat{\mathcal{P}} \}$}.
	\end{tabbing}
\end{algorithm}

\begin{algorithm}[!t]
	\caption{Calculation of the cost function} \label{alg2}
	\enspace \textbf{Global:} data $\{(X_i,A_i,Y_i):i=1,\dots,n\}$; cost dictionary $\mathcal{C}$; an interval $\mathcal{I}$; penalty term $\lambda_n$. \\
	\enspace \textbf{Output:} $\mathcal{C}$.\\
	\qquad (I). If $\mathcal{C}(\mathcal{I}) \gets NULL$:\\
	\qquad \quad (1). Apply linear / MLP regression: $$\widehat{q}_{\mathcal{I} }(\cdot) \gets  \arg\min_{q} \sum_i \mathbb{I}(A_i\in \mathcal{I}) \{Y_i-q(X_i;\theta_{\mathcal{I}})\}^2+n\lambda_n |\mathcal{I}| \|\theta_{\mathcal{I}}\|_2^2 ;$$\\
	\qquad \quad (2). Set the cost $\mathcal{C}(\mathcal{I})$ to the objective value;\\ 
	\qquad (II). Return $\mathcal{C}$.
\end{algorithm}

We briefly summarize our algorithm below. For a given integer $r$, we search the optimal change point location $j$ that minimizes the above Bellman function $B(r)$ in \eqref{bellmanrelation}. This requires to apply the linear / MLP regression to learn $\widehat{q}_{[j/m,r/m)}$ and $\hbox{cost} ([j/m,r/m))$ for each $j\in \mathcal{R}_{r}$. Let $j^*$ be the corresponding minimizer. We then define the change points list $\tau(r)=\{j^*,\tau(j^*)\}$. This procedure is iterated to compute $B(r)$ and $\tau(r)$ for $r=1,\dots,m$.
The optimal partition $\widehat{\mathcal{P}}$ is determined by the values stored in $\tau(\cdot)$. A pseudocode containing more details is given in Algorithm 1.

\subsubsection{Analysis of Computational Complexity}\label{add:compu_complex}
\change{We analyze the computational complexity of the proposed methods in this section. The main computation lies in 
the dynamic programming algorithm to find the change points as well as the estimation of the outcome regression function in L- and D-JIL. 

First, recall that we use the PELT method 
to implement the dynamic programming. It requires at least $\mathcal{O}(m)$ computing steps and at most $\mathcal{O}(m^2)$ steps \citep{friedrich2008complexity}. According to Theorem 3.2 in \citet{killick2012optimal}, the expected computational cost is $\mathcal{O}(m)$.

Second, for each step in PELT, 
we need to train the DNN or linear regression model to calculate the cost function. Here, the complexity of training the linear regression is well known and equals $\mathcal{O}(n p^2 +p^3)$ with sample size $n$ and feature dimension $p$. To the contrary, the complexity of training a DNN depends on the model architecture. 
Suppose we use 
a fully connected MLP 
with $w$ width and $d$ depth, 
and set the total number of epochs for training to $e$. Then 
the time complexity is given by $\mathcal{O}\{n e (d-1) w^2 \}$\footnote{See https://ai.stackexchange.com/questions/5728/what-is-the-time-complexity-for-training-a-neural-network-using-back-propagation/5730.}.

To summarize, the expected computational complexities of the proposed linear- and deep-JIL are given by $\mathcal{O}\{m(n p^2 +p^3)\}$ and $\mathcal{O}\{m n e (d-1) w^2\}$, respectively. 
}

\subsection{Tuning Parameters}\label{tuning}
Our proposal requires to specify the tuning parameters $m$, $\lambda_n$ and $\gamma_n$. 
We first discuss the choice of $m$. In practice, we recommend to set $m=n/c$ with some constant $c>0$ such that $m$ and $n$ are of the same order. \change{The choice of $c$ represents a trade-off between the estimation bias and the computational cost. A small value of $c$ would improve the estimation efficiency whereas a larger value of $c$ saves the computation time. We recommend using the smallest possible $c$ whenever the computation is affordable.} In our numerical studies, we tried several different values of $c$ and found the resulting estimated I2DRs have approximately the same value function as long as $c$ is not too large. Thus, the proposed I2DR is not overly sensitive to the choice of this constant. Detailed empirical results can be found in Sections \ref{secchoicem} and \ref{realdata}.

We next discuss the choices of $\lambda_n$ and $\gamma_n$. Selection of these tuning parameters relies on the concrete proposal to approximate the outcome regression function. We elaborate below. 
%

\subsubsection{Tuning in L-JIL}

For L-JIL, we choose $\gamma_n$ and $\lambda_n$ simultaneously via cross-validation. The theoretical requirements of $\gamma_n$ and $\lambda_n$ for L-JIL are imposed in the statement of Theorem \hyperlink{thm1}{1}.  We further develop an algorithm that substantially reduces the computation complexity 
resulting from the use of cross-validation. 

To be more specific, 
let $\Lambda_n=\{\lambda_n^{(1)},\cdots,\lambda_n^{(H)} \}$ and $\Gamma_n=\{\gamma_n^{(1)},\cdots,\gamma_n^{(J)}\}$ be the set of candidate tuning parameters. 
%
%
For a given integer $K_0$, we randomly split the data into $K_0$ equal sized subgroups. Let $\mathbb{G}_k$ denote indices of the subsamples in the $k$th subgroup, for $k=1,\cdots,K_0$. Let $\mathbb{G}_{-k}$ denote the complement of $\mathbb{G}_k$. 
For any $\lambda_n\in \Lambda_n$, $\gamma_n\in \Gamma_n$, $k\in \{1,\cdots,K_0\}$, let $(\widehat{\mathcal{P}}_{\lambda_n,\gamma_n,k},\{\widehat{\theta}_{\mathcal{I},\gamma_n,\lambda_n,k}:\mathcal{I}\in\widehat{\mathcal{P}}_{\lambda_n,\gamma_n,k}\})$ denote the optimizer \eqref{optimize_ridge}, computed based on the data in $\mathbb{G}_{-k}$. We aim to choose $\gamma_n$ and $\lambda_n$ that minimizes
\begin{eqnarray}\label{eqn:solvecv}
	\frac{1}{n}\sum_{k=1}^{K_0} \sum_{i\in \mathbb{G}_{k}} \sum_{\mathcal{I} \in \widehat{\mathcal{P}}_{\lambda_n,\gamma_n,k}}\mathbb{I}(A_i\in \mathcal{I}) \{Y_i-\overline{X}_i^{\top}\widehat{\theta}_{\mathcal{I},\gamma_n,\lambda_n,k}\}^2. 
\end{eqnarray}

To solve \eqref{eqn:solvecv}, we remark that there is no need to apply Algorithm 1 $|\Lambda_n|\times |\Gamma_n|$ times to compute  
the minimizer of \eqref{optimize_ridge} over the set of candidate tuning parameters. We develop an algorithm to facilitate the computation. The key observation is that, for any interval $\mathcal{I}\subseteq [0,1]$ and $k\in \{1,\cdots,K_0\}$, the set of estimators  $\{\widehat{\theta}_{\mathcal{I},\gamma_n,\lambda_n,k}:\gamma_n \in \Gamma_n, \lambda_n\in \Lambda_n\}$ can be obtained simultaneously over the set of candidate tuning parameters. This forms the basis of our algorithm. More details are provided in Section \ref{app_tune_LJIL} of the supplementary article.

\subsubsection{Tuning in D-JIL}

As for D-JIL, we find that the MLP regressor is not overly sensitive to the choice of $\lambda_n$, so we set $\lambda_n=0$. The parameter $\gamma_n$ is chosen based on cross-validation. 
The theoretical requirement of $\gamma_n$ for D-JIL is imposed in the statement of Theorem \hyperlink{thm2}{2}. To implement the cross-validation, we randomly split the data into $K_0$ equal sized subgroups, denoted by $\{(X_i,A_i,Y_i)\}_{i\in \mathbb{G}_1}$, $\{(X_i,A_i,Y_i)\}_{i\in \mathbb{G}_2}$, $\cdots$, $\{(X_i,A_i,Y_i)\}_{i\in \mathbb{G}_{K_0}}$, accordingly. 
For each $\gamma_n$ and $k=1,\dots,K_0$, we compute the estimators $\widehat{\mathcal{P}}_{\gamma_n,k}$ and $\widehat{q}_{\mathcal{I},\gamma_n,k}(\cdot)$ based on the sub-dataset in $\mathbb{G}_{-k}$. Then we choose $\gamma_n$ that minimizes
\begin{eqnarray*}
	\frac{1}{n}\sum_{k=1}^{K_0} \sum_{i\in \mathbb{G}_{k}} \sum_{\mathcal{I} \in \widehat{\mathcal{P}}_{\gamma_n,k}} \mathbb{I}(A_i\in \mathcal{I})\{Y_i- \widehat{q}_{\mathcal{I},\gamma_n,k}(X_i)\}^2.
\end{eqnarray*} 
\change{We also remark that implementing deep neural networks involves some other tuning parameters, such as the learning rate, the numbers of hidden nodes and hidden layers. In our implementation, we set them to the default values of the MLP regressor implementation \citep{SCIKit2011}}. 

\section{Theory}\label{sectheory}

We establish the statistical properties of our proposed method in this section. As we have commented, we allow \change{the  outcome regression function 
to be either a piecewise or continuous function of the treatment}. We first study the statistical properties of L-JIL and D-JIL \change{when Model I holds}, respectively. We next outline a procedure to construct a confidence interval for the value under the proposed I2DR and prove its validity. 
Finally, we investigate the properties of our proposed method \change{when Model II holds}. These theoretical results imply that our method will work when the outcome regression function is either piecewise or continuous function.

\subsection{Properties \change{When Model I Holds}}

\subsubsection{Results for L-JIL}\label{secproperty}
To establish the theoretical properties of the I2DR obtained by L-JIL, we first assume \eqref{main_model} holds with $q_{\mathcal{I},0}(x)=\bar{x}^\top \theta_{\mathcal{I},0}$ for any $x\in \mathbb{X}$ and $\mathcal{I}\in \mathcal{P}_0$. In other words, the outcome regression function $Q(x,a)$ is linear in $x$ and piecewise constant in $a$.   
Without loss of generality, assume $\theta_{0,\mathcal{I}_1}\neq \theta_{0,\mathcal{I}_2}$ for any two adjacent intervals $\mathcal{I}_1,\mathcal{I}_2\in \mathcal{P}_0$. This guarantees that the representation in \eqref{main_model} is unique. 
We write $a_n\asymp b_n$ for two sequences $\{a_n\},\{b_n\}$ if there exists some universal constant $c\ge 1$ such that $c^{-1}b_n\le a_n\le c b_n$. Define $\theta_0(\cdot)=\sum_{\mathcal{I} \in \mathcal{P}_0}\theta_{\mathcal{I},0}\mathbb{I}(\cdot \in \mathcal{I})$. Giving $(\widehat{\mathcal{P}},\{\widehat{\theta}_{\mathcal{I}}:\mathcal{I}\in \widehat{\mathcal{P}} \})$, our estimator for the function $\theta_0(\cdot)$ is defined by
\begin{eqnarray}\label{hatbeta}
	\widehat{\theta}(\cdot)=\sum_{\mathcal{I}\in \widehat{\mathcal{P}}} \widehat{\theta}_{\mathcal{I}} \mathbb{I}(\cdot \in \mathcal{I} ).
\end{eqnarray}
This yields a piecewise constant approximation of $\theta_0(\cdot)$. 
%
%
%
%
We first study the theoretical properties of $\widehat{\theta}(\cdot)$. Toward that end, we need to impose the following condition on the probability tails of $X$ and $Y$. 

\smallskip

\noindent (A4) 
Suppose there exists some constant $\omega>0$ such that $\sup_{a,j} \|X^{(j)}\|_{\psi_2|A=a}\le \omega$ and $\sup_{a} \|Y\|_{\psi_2|A=a}\le \omega$ almost surely, where $X^{(j)}$ denotes the $j$th element of $X$, \change{and that for any random variable $Z$, $\|Z\|_{\psi_2|A=a}$ denotes the conditional Orlicz norm given that $A=a$, i.e.,
\begin{eqnarray*}
	\|Z\|_{\psi_2|A=a}\stackrel{\Delta}{=}\inf_{C>0}\left[\Mean \left\{\left.\exp\left(\frac{|Z|^2}{C^2}\right)\right|A=a\right\}\le 2\right].
\end{eqnarray*}
We remark that Condition (A4) is automatically satisfied when the covariates and the outcomes are bounded.} 
 
 \smallskip
 
\noindent
{\bf \hypertarget{thm1}{Theorem 1}} {\it Assume (A1)-(A4) hold and \eqref{main_model} holds with $q_{\mathcal{I},0}(x)=\bar{x}^\top \theta_{\mathcal{I},0}$. 
	{Assume 
	$A$ has a bounded probability density function on $[0,1]$.} 
	Assume $m\asymp n$,  $\lambda_n=O(n^{-1}\log n)$, $\{\gamma_n\}_{n \in \mathbb{N}}$ satisfies $\gamma_n \to 0$ and $\gamma_n n/\log{n} \to \infty$. Then, there exists some constant $\bar{c}>0$ such that the following events hold with probability at least $1-O(n^{-2})$:
	
	(i) $|\widehat{\mathcal{P}}|=|\mathcal{P}_0|$.
	
	(ii) $\max_{\tau\in J(\mathcal{P}_0)} \min_{\hat{\tau}\in J(\widehat{\mathcal{P}})} |\hat{\tau}-\tau|\le \bar{c}n^{-1}\log n$.
	
	(iii) $\int_{0}^1 \|\widehat{\theta}(a)-\theta_0(a)\|_2^2da\le \bar{c} n^{-1} \log n$.} \hfill$\square$ 

In Theorem \hyperlink{thm1}{1}, results in (i) show the model selection consistency of our jump penalized estimator. Results in (ii) imply that the estimated change point locations converge at a rate of $O_p(n^{-1}\log n)$. 
In (iii), we derive an upper error bound for the integrated $\ell_2$ loss of $\widehat{\theta}(\cdot)$. As discussed in the introduction, derivation of Theorem \hyperlink{thm1}{1} is nontrivial. 
A number of technical lemmas (see Lemma \hyperlink{lemma1}{1}-\hyperlink{lemma5}{4} in Section \ref{secproofthm1}) are established  to prove Theorem \hyperlink{thm1}{1}. 
These results can be easily extended to study general covariate-based change point models.

\smallskip

We next establish the convergence rate of $V^{\tiny{opt}}-V(\widehat{d})$, where $V^{\tiny{opt}} = V(d^{\tiny{opt}})$. The quantity $V^{\tiny{opt}}-V(\widehat{d})$ represents the difference between the optimal value and the value under the proposed I2DR. The smaller the difference, the better the I2DR. 
Notice that $V^{\tiny{opt}}\ge V(d)$ for any I2DR $d(\cdot)$. It suffices to provide an upper bound for $V^{\tiny{opt}}-V(\widehat{d})$. We impose the following condition. 

\smallskip

\noindent (A5.) Assume for any $\mathcal{I}_1,\mathcal{I}_2\in \mathcal{P}_0$, there exist some constants $\gamma,\delta_0>0$ such that $$\hbox{Pr}(0<|q_{\mathcal{I}_1,0}(X)-q_{\mathcal{I}_2,0}(X)|\le t)=O(t^{\gamma}),$$ where the big-$O$ term is uniform in $0<t\le \delta_0$. 

\smallskip

Condition (A5) is commonly assumed in the literature to derive sharp convergence rate for the value function under the estimated optimal IDR \citep{qian2011performance,luedtke2016statistical,shi2020breaking}. It is very similar to the margin condition \citep{Alex2004,Alex2007} used in the classification literature. This condition is automatically satisfied with $\gamma=1$ when $q_{\mathcal{I},0}(X)$ has a bounded probability density function for any $\mathcal{I}\in \mathcal{P}_0$. 

 \smallskip
 
\noindent
{\bf \hypertarget{thm2}{Theorem 2}} {\it Assume the conditions in Theorem \hyperlink{thm1}{1} are satisfied. Further assume (A5) holds. 
	Then, we have 
	\begin{eqnarray}\label{valueconverge0}
		V^{\tiny{opt}}-V(\widehat{d})\le \bar{c} (n^{-1}\log n)^{(1+\gamma)/2}+\bar{c}n^{-1}\log n, 
	\end{eqnarray}
	for some constant $\bar{c}>0$, with probability at least $1-O(n^{-2})$.} \hfill$\square$ 
	
When (A5) holds with $\gamma=1$, Theorem \hyperlink{thm2}{2} suggests that $V(\widehat{d})$ converges to the optimal value at a rate of $O_p(n^{-1})$ up to some logarithmic factor. Notice that the events defined in Theorem \hyperlink{thm1}{1} and \hyperlink{thm2}{2} occur with probability at least $1-O(n^{-2})$. Since $\sum_{n\ge 1} n^{-2}<+\infty$, an application of Borel-Cantelli lemma implies that these events will occur for sufficiently large  $n$ almost surely.

\subsubsection{Results for D-JIL}\label{secstepdnn} 
We study the theoretical properties of the proposed I2DR 
based on D-JIL \change{when Model I is correct}. 
Similar to the linear case, we assume $q_{\mathcal{I}_1,0}\neq q_{\mathcal{I}_2,0}$ for any two adjacent intervals $\mathcal{I}_1,\mathcal{I}_2\in \mathcal{P}_0$. 
For any $\mathcal{I}$, we set the regression class $\mathcal{Q}_{\mathcal{I}}$ to a \change{general class of feedforward architecture with $L_{\mathcal{I}}$ hidden layers,  $W_{\mathcal{I}}$ many number of parameters, and ReLU activation function \citep{farrell2021deep}}. 

To derive the theoretical properties of D-JIL, we assume the outcome regression function is a smooth function of the baseline covariates (see assumption (A6) below). Meanwhile, D-JIL is valid when $Q(x,a)$ is a nonsmooth function of $x$ as well \citep[see e.g.,][]{imaizumi2019deep}. Specifically, define the class of $\beta$-smooth functions (also known as H{\"o}lder smooth functions with exponent $\beta$) as
\begin{eqnarray*}
	\Phi(\beta,c)=\left\{h:\sup_{\|\alpha\|_1\le \floor{\beta}} \sup_{x\in \mathbb{X}} |D^{\alpha} h(x)|\le c, \sup_{\|\alpha\|_1=\floor{\beta}} \sup_{\substack{x,y\in \mathbb{X}\\ x\neq y}} \frac{|D^{\alpha} h(x)-D^{\alpha} h(y)|}{\|x-y\|_2^{\beta-\floor{\beta}}}\le c \right\},
\end{eqnarray*}
for some constant $c>0$, where $\floor{\beta}$ denotes the largest integer that is smaller than $\beta$ and $D^{\alpha}$ denotes the differential operator 
$D^{\alpha}$ denote the differential operator:
\begin{eqnarray*}
	D^{\alpha}h(x)=\frac{\partial^{\|\alpha\|_1} h(x)}{\partial x_1^{\alpha_1}\cdots\partial x_d^{\alpha_d}}.
\end{eqnarray*}
We introduce the following conditions.

\smallskip


\noindent (A6.) Suppose $Q(\bullet,a) \in \Phi(\beta,c)$, and $ p(a|\bullet) \in \Phi(\beta,c)$ for any $a$. 

\smallskip

\noindent (A7.) Functions $\{\widehat{q}_{\mathcal{I}} \}_{\mathcal{I}\in \widehat{\mathcal{P}} }$ are uniformly bounded. 
 
\smallskip


Assumption (A7) ensures that the optimizer would not diverge in the $\ell_{\infty}$ sense. Similar assumptions are commonly imposed in the literature to derive the convergence rates of DNN estimators \citep[see e.g.,][]{farrell2021deep}. Combining (A7) with (A6) allows us to derive the uniform rate of convergence for the class of DNN estimators $\{\widehat{q}_{\mathcal{I}}\}_{\mathcal{I}\in \widehat{\mathcal{P}}}$.  
The following theorem summarizes the theoretical properties of the proposed method via deep neural networks.  

 \smallskip
 
%
%

	
\noindent
{\bf \hypertarget{thm5}{Theorem 3}} {\it Assume (A1)-(A3), (A6), (A7) and Model I hold. 
	{Assume $X$ and $Y$ are bounded variables, and 
	$A$ has a bounded probability density function on $[0,1]$.} 
	Assume $m\asymp n$, $\{\gamma_n\}_{n \in \mathbb{N}}$ satisfies $\gamma_n \to 0$ and $\gamma_n \gg n^{-2\beta/(2\beta+p)}\log^8 n$. Then, there exist some constant $\bar{c}>0$ and DNN classes $\{\mathcal{Q}_{\mathcal{I}}:\mathcal{I}\}$ with $L_{\mathcal{I}}\asymp \log (n|\mathcal{I}|)$ and $W_{\mathcal{I}}\asymp (n|\mathcal{I}|)^{p/(2\beta+p)}\log (n|\mathcal{I}|)$ such that the resulting D-JIL estimator computed by \eqref{optimize} satisfies
	
	(i) $|\widehat{\mathcal{P}}|=|\mathcal{P}_0|$; 
	
	(ii) $\max_{\tau\in J(\mathcal{P}_0)} \min_{\hat{\tau}\in J(\widehat{\mathcal{P}} )} |\hat{\tau}-\tau|\le \bar{c} n^{-2\beta/(2\beta+p)}\log^8 n$;
	
	(iii) $\Mean |Q(X,A)-\sum_{\mathcal{I}\in \widehat{\mathcal{P}}}\mathbb{I}(A\in \mathcal{I})\widehat{q}_{\mathcal{I}}(X)|^2da \le  \bar{c} n^{-2\beta/(2\beta+p)}\log^8 n$,

	\noindent with probability at least $1-O(n^{-2})$. } \hfill$\square$
	
Theorem \hyperlink{thm5}{3} establishes the properties of our method under settings where the $Q(x,a)$ is piecewise function in the treatment. 
Results in (i) imply that D-JIL correctly identifies the number of change points. Results in (ii) imply that any change point in $\mathcal{P}_0$ can be consistently identified at a convergence rate of $O_p( n^{-2\beta/(2\beta+p)})$ up to some logarithmic factors. Notice that we use the piecewise function $\sum_{\mathcal{I}\in \widehat{\mathcal{P}}}\mathbb{I}(a\in \mathcal{I})\widehat{q}_{\mathcal{I}}(x)$ to approximate the outcome regression function. In (iii), {we show our estimator for function $Q(X,A)$ converges at a rate of $O_p( n^{-2\beta/(2\beta+p)})$ up to some logarithmic factors. The theoretical choices of $L_{\mathcal{I}}$ and $W_{\mathcal{I}}$ in Theorem \hyperlink{thm5}{3} are consistent with the literature of  DNN
estimators \citep{imaizumi2019deep,farrell2021deep}. These DNN architectures ensure the convergence rate of our estimator for function $Q(X,A)$}, which achieves the minimax-optimal nonparametric rate of convergence under (A6) \citep[see e.g.,][]{stone1982optimal}.

\smallskip

We next establish the convergence rate of $V^{\tiny{opt}}-V(\widehat{d})$ \change{when Model I holds} in the following theorem.
%
%
%
%
 

 \smallskip
 
\noindent
{\bf \hypertarget{thm7}{Theorem 4}} {\it	Assume the conditions in Theorem \hyperlink{thm5}{3} are satisfied. Further assume (A5) holds. Then,
	we have 
	\begin{eqnarray}\label{valueconverge1}
	V(\widehat{d})\ge V^{\tiny{opt}} -O(1) (n ^{-{2\beta\over 2\beta+p}}\log^8 n+n^{-{2\beta(1+\gamma)\over(2\beta+p)(2+\gamma)}} \log^{8+8\gamma\over 2+\gamma} n),
	\end{eqnarray}
	 with probability at least $1-O(n^{-2})$.} \hfill$\square$
	 
Theorem \hyperlink{thm7}{4} suggests that $V(\widehat{d})$ converges to the optimal value at a rate of $O_p\{n^{-{2\beta(1+\gamma)\over(2\beta+p)(2+\gamma)}}\}$ up to some logarithmic factors. This rate is slower than the rate ($O_p(n^{-1})$ up to some logarithmic factor) we obtained in Theorem \hyperlink{thm2}{2} where we posit a parametric (linear) model.
Suppose the condition ${4\beta(1+\gamma)> (2\beta+p)(2+\gamma)}$ holds, it follows that $V(\widehat{d})=V^{\tiny{opt}}+o_p(n^{-1/2})$. This observation forms the basis of our inference procedure in Section \ref{secevaluate}. 
Here, the extra margin parameter $\gamma$ in our results is introduced by (A5) to bound the bias due to the estimated decision rule $\widehat{d}$. If the margin parameter $\gamma$ goes to infinity, we only require the smooth parameter $\beta>p/2$ to obtain $V(\widehat{d})=V^{\tiny{opt}}+o_p(n^{-1/2})$. This condition ($\beta>p/2$) is commonly assumed in the literature on evaluating average treatment effects \citep[see e.g.,][]{chernozhukov2017double,farrell2021deep}.

\subsubsection{Evaluation of the Value Function}\label{secevaluate}
Suppose Model I holds. 
When L-JIL is used, it follows from Theorem \hyperlink{thm2}{2} that $V(\widehat{d})=V^{\tiny{opt}}+o_p(n^{-1/2})$. When D-JIL is used, if the smoothness parameter $\beta$ (see (A6)) and the margin parameter $\gamma$ (see (A5)) satisfy $4\beta(1+\gamma)> (2\beta+p)(2+\gamma)$, it follows from Theorem \hyperlink{thm7}{4} that $V(\widehat{d})=V^{\tiny{opt}}+o_p(n^{-1/2})$. 
In the following, we derive the asymptotic normality of $\sqrt{n}(\widehat{V}-V^{\tiny{opt}})$. By Slutsky's theorem, this implies that $\sqrt{n}\{\widehat{V}-V(\widehat{d})\}$ is asymptotically normal as well.

\smallskip

\noindent (A8.) 
$[\Mean \{\widehat{e}({\mathcal{I}}|X) - e({\mathcal{I}}|X) \}^2]^{1/2}= o(n^{-1/4})$ and that $\widehat{e}({\mathcal{I}};\bullet)$ belongs to the class of VC-type functions with VC-index upper bounded by $O(n^{1/2})$ \citep[see e.g.][for a detailed definition of the VC-type class]{chernozhukov2017double}, for any ${\mathcal{I}\in  \mathfrak{I}(m) }$.

\smallskip

The first part in Assumption (A8) requires the generalized propensity score function to converge at certain rates. Similar assumptions are commonly imposed in the causal inference literature to derive the asymptotic distribution of the estimated average treatment effect \citep[see e.g.,][]{chernozhukov2017double}. The second part of (A8) essentially controls the model complexity of the estimator $\widehat{e}$. The more complicated $\widehat{e}$ is, the larger the VC index. Under (A6), we can show (A8) holds when DNN is used to model the generalized propensity score. 

 \smallskip
 
\noindent
{\bf \hypertarget{thm3}{Theorem 5}} {\it Assume (A8) holds and suppose functions $\{\widehat{e}_{\mathcal{I}} \}_{\mathcal{I}\in \widehat{\mathcal{P}} }$ are uniformly bounded away from zero. Further assume that for any $\mathcal{I}_1,\mathcal{I}_2\in \mathcal{P}_0$ with $\mathcal{I}_1\neq \mathcal{I}_2$, we have $\hbox{Pr}( q_{\mathcal{I}_1,0}(X) =  q_{\mathcal{I}_2,0} (X))=0$.

(i) Suppose conditions in Theorem \hyperlink{thm2}{2} are satisfied. Then, under L-JIL, we have
		$$\sqrt{n}(\widehat{V}-V^{\tiny{opt}})\stackrel{d}{\to} N(0,\sigma_L^2),$$ 
	for some $\sigma_L^2>0$.

(ii) Suppose conditions in Theorem \hyperlink{thm7}{4} are satisfied with $4\beta(1+\gamma)> (2\beta+p)(2+\gamma)$. 
	Then, under D-JIL, we have
		$$\sqrt{n}(\widehat{V}-V^{\tiny{opt}})\stackrel{d}{\to} N(0,\sigma_D^2),$$ 
	for some $\sigma_D^2>0$.} \hfill$\square$

 \smallskip
 
We now introduce the estimator for the asymptotic variance $\sigma_L^2$ or $\sigma_D^2$, and derive a Wald-type $1-\alpha$ CI for $V^{\tiny{opt}}$. Since $V(\widehat{d})=V^{\tiny{opt}}+o_p(n^{-1/2})$, the proposed CI also covers $V(\widehat{d})$ with probability tending to $1-\alpha$. We estimate $\sigma_L^2$ or $\sigma_D^2$ by
\begin{eqnarray*}
	\widehat{\sigma}^2=\frac{1}{n-1}\sum_{i=1}^n  \left[ {\mathbb{I}\{A_i\in \widehat{d}(X_i) \} \over  \widehat{e}( \widehat{d}(X_i) |X_i)}\big\{Y_i -\max_{\mathcal{I}\in \widehat{\mathcal{P}}} \widehat{q}_{\mathcal{I}}(X_i)  \big\} +  \max_{\mathcal{I}\in \widehat{\mathcal{P}}} \widehat{q}_{\mathcal{I}}(X_i)  - \widehat{V} \right]^2,
\end{eqnarray*}
where $\{\widehat{q}_{\mathcal{I}}(\cdot)\}$ corresponds to the value estimations under L-JIL or D-JIL.

The corresponding $1-\alpha$ CI is given by $\widehat{V}\pm z_{\alpha/2} \widehat{\sigma}$, where $z_{\alpha/2}$ denotes the upper $\alpha/2$-th quantile of a standard normal distribution. Similar to Theorem \hyperlink{thm3}{5}, we can show that $\widehat{\sigma}$ is consistent. This shows the validity of our inference procedure. 
}

 \subsection{Properties \change{When Model II Holds}}\label{secproperty3}

\subsubsection{Properties of L-JIL under Varying Coefficient Model}\label{secproperty2}
We first consider the case when the outcome regression function can be represented by a varying coefficient model and investigate the theoretical properties of the proposed L-JIL. Specifically, suppose the true outcome regression function takes the following form
\begin{eqnarray}\label{md1}
Q(x,a)=\bar{x}^{\top} \theta_0(a),\,\,\,\,\,\,\,\,\forall x\in \mathbb{X},a\in [0,1], 
\end{eqnarray}
where $\bar{x}=(1,x^{\top})^{\top}$ and $\theta_0(\cdot)$ is some continuous $(p+1)$-dimensional function. That is, we assume the conditional mean of the outcome is a linear function of individuals' covariates for any treatment $a\in [0,1]$. Yet, the model is flexible in that 
{$\theta_0(\cdot)$ is allowed to be 
an arbitrary continuous function of $a$ with certain smoothness constraints. 
	Models of this type belong to the class of varying coefficient models popularly applied in many scientific areas \citep[see e.g.,][for an overview]{Fanandzhang2008}}.

\smallskip

Here, we 
consider the following class of H{\"o}lder continuous functions for $\theta_0(\cdot)$. Suppose there exist some constants $L>0$, $0<\alpha_0\le 1$ such that $\theta_0(\cdot)$ satisfies 
\begin{eqnarray}\label{eqn:model2}
\sup_{a_1,a_2\in [0,1]} \|\theta_0(a_1)-\theta_0(a_2)\|_2\le L|a_1-a_2|^{\alpha_0}.
\end{eqnarray}


{We first sketch a few lines to see why our method works under \eqref{eqn:model2}. }For a given integer $k>0$, we define $\theta^*_k(\cdot)$ as
\begin{eqnarray*}
	\theta^*_k(a)=\sum_{j=0}^{k-1} \theta_0\left(\frac{j+1/2}{k+1}\right) \mathbb{I}(j \le (k+1)a<j+1)+\theta_0\left( \frac{k+1/2}{k+1}\right) \mathbb{I}((k+1)a\ge k ).
\end{eqnarray*}
Apparently, $\theta^*_k(\cdot)$ has at most $k$ change points. In addition, with some calculations, we can show that $\sup_{a\in[0,1]} \|\theta^*_k(a)-\theta_0(a)\|_2\le 2^{-\alpha_0}(k+1)^{-\alpha_0}L$. Letting $k\to \infty$, it is immediate to see that $\theta_0(\cdot)$ can be uniformly approximately by a step function as the number of change points increases. 

\smallskip

{In Theorems \hyperlink{thm1}{1} and \hyperlink{thm2}{2}, we have shown the proposed I2DR is consistent under the piecewise linear function assumption. Based on the above discussion}, we expect that jump interval-learning also works when the model \eqref{md1} holds. We formally establish the corresponding theoretical results in the following theorem. 

\smallskip

\noindent
{\bf \hypertarget{thm4}{Theorem 6}} {\it Assume (A1)-(A4) and \eqref{eqn:model2} hold. {Assume 
	$A$ has a bounded probability on $[0,1]$.}   Assume $m\asymp n$, $\lambda_n=O(n^{-1}\log n)$, $\gamma_n$ satisfies $\gamma_n \to 0$ and $\gamma_n \gg n^{-1}\log n$. Under the model \eqref{md1}, there exists some constant $\bar{c}>0$ such that the following holds with probability at least $1-O(n^{-2})$:
	\begin{eqnarray*}
		\int_{0}^1 \|\widehat{\theta}(a)-\theta_0(a)\|_2^2da\le \bar{c} \gamma_n^{2\alpha_0/(1+2\alpha_0)}.
	\end{eqnarray*}
	In addition, assume $\gamma_n\asymp (n^{-1}\log n)^{(1+2\alpha_0)/(1+4\alpha_0)}$. Then 
	there exists some constant $\bar{c}^*>0$ such that the following occurs with probability at least $1-O(n^{-2})$ that
	\begin{eqnarray}\label{valueconverge}
	V^{\tiny{opt}}-V(\widehat{d})\le \bar{c}^* (n^{-1}\log n)^{\alpha_0/(1+4\alpha_0)}.
	\end{eqnarray}
	} \hfill$\square$

It is worth mentioning that with proper choice of $\gamma_n$, the integrated $\ell_2$ loss of $\widehat{\theta}(\cdot)$ converges at a rate of $O_p(n^{-2\alpha_0/(1+2\alpha_0)})$ up to some logarithmic factor. The rate is slower compared to the results in Theorem \hyperlink{thm1}{1}, since $\theta_0(\cdot)$ is only ``approximately" piecewise constant. 
When $\theta_0(\cdot)$ is Lipschitz continuous, it follows from \eqref{valueconverge} that the value under our proposed I2DR will converge to the optimal value function at a rate of $O_p(n^{-1/5}\log^{1/5} n)$. 



\subsubsection{Properties of D-JIL under the Continuous outcome regression function} 

 We next consider the general case when the outcome regression function is specified by model II and study the theoretical properties of the proposed D-JIL. The following theorem proves the consistency of the proposed estimator.
 
\smallskip

\noindent
{\bf \hypertarget{thm6}{Theorem 7}}  {\it Suppose $Q$ is a continuous function of $a$ and $x$. Assume (A1)-(A3) and (A6)-(A7) hold. Assume $X$ and $Y$ are bounded variables, and 
	$A$ has a bounded probability density function on $[0,1]$. Assume $m\asymp n$ and $\{\gamma_n\}_{n \in \mathbb{N}}$ satisfies $\gamma_n \to 0$ and $\gamma_n \gg n^{-2\beta/(2\beta+p)}\log^8 n$. Then, there exist some DNN classes $\{\mathcal{Q}_{\mathcal{I}}:\mathcal{I}\}$ with $L_{\mathcal{I}}\asymp \log (n|\mathcal{I}|)$ and $W_{\mathcal{I}}\asymp (n|\mathcal{I}|)^{p/(2\beta+p)}\log (n|\mathcal{I}|)$  such that the resulting D-JIL estimator computed by \eqref{optimize} satisfies	
 	
 	(i) \change{$\max_{\mathcal{I}\in \widehat{\mathcal{P}} }\sup_{a\in \mathcal{I}}E |\widehat{q}_{\mathcal{I}} (X)-Q(X,a)|^2=O_p(\gamma_n^{\frac{2\alpha_0}{1+2\alpha_0}})+O_p\big( (n\gamma_n)^{-\frac{2\beta}{2\beta+p}}\log^8 n \big)$} where the expectation is taken with respect to the marginal distribution of $X$; 
 	
 	(ii) \change{Suppose $\gamma_n\sim n^{-1/\big(1+\frac{\beta(2\alpha_0+1)}{\alpha_0(2\beta+p)}\big)}$. Then $V^{\tiny{opt}}-V(\widehat{d})=O_p\big(n^{-\alpha_0 \beta/(4\alpha_0\beta+\alpha_0 p+\beta)} \log^4 n\big)$. }} \hfill$\square$
%

 \smallskip
 
 Theorem \hyperlink{thm6}{7} establishes the properties of our method under settings where $Q$ is continuous in $a$. 
 Results in (i) imply that $\widehat{q}_{\mathcal{I}}(x)$ can be used to uniformly approximate $Q(x,a)$ for any $a\in \mathcal{I}$. The consistency of the value in (ii) thus follows. 
 
\change{Finally, we remark that the optimal I2DR is well-defined under Model I. When Model II holds, however, it remains unclear whether the optimal I2DR is uniquely defined or not. Nonetheless, as shown in Theorems 6 and 7, the value under the proposed I2DR converges to that under the optimal IDR. This implies that even when the optimal I2DR is not uniquely defined, our proposal is able to identify one of them asymptotically. 
\subsubsection{Evaluation of the Value Function}
As discussed in Section \ref{secevaluate}, the validity of the proposed CI for the optimal value requires the outcome regression function to satisfy the piecewise model assumption. Under Model II, however, the value under the proposed I2DR might not be $n^{-1/2}$-consistent to the optimal value. As such, in theory, the proposed CI would fail to cover the optimal value. Nonetheless, when the preference function $\pi^*$ is assigned according to the generalized propensity score, i.e.,
\begin{eqnarray*}
	\pi^*(a;x,\mathcal{I})=\frac{p(a|x)}{\int_{\mathcal{I}} p(a|x)da},
\end{eqnarray*}
and other regularity conditions hold, our CI is able to cover the value under the proposed I2DR. We omit the theoretical results to save space. 
}
\section{Simulations}\label{simul}


\subsection{
Confidence interval for the value}\label{secsimustep}
In this section, we focus on scenarios where the outcome regression function takes the form of Model I and examine 
the coverage probability of the proposed CI in Section \ref{secevaluate}.
Simulated data are generated from the following model: 
\begin{eqnarray*}
	Y|X,A \sim N(Q(X,A), 1),\,\,\,\,A|X \sim \hbox{Unif}[0,1]\,\,\,\,\hbox{and}\,\,\,\,X^{(1)},X^{(2)},\dots,X^{(p)}\stackrel{iid}{\sim} \hbox{Unif}[-1,1],
\end{eqnarray*}
where $\hbox{Unif}[a,b]$ denotes the uniform distribution on the interval $[a,b]$. 
We consider the following two scenarios with different choices of $Q(X,A)$.  

\smallskip

\noindent \textbf{Scenario 1}:
\begin{eqnarray*}
	Q(x,a)=\left\{\begin{array}{ll}
		1+x^{(1)} , & a<0.35,\\
		x^{(1)} -x^{(2)} , & 0.35\le a<0.65,\\
		1-x^{(2)} , & a \ge 0.65.
	\end{array}
	\right.
\end{eqnarray*}
Under Scenario 1, the outcome regression function is piecewise constant as a function of $a$, and is linear as a function of $x$. Here, we have $J(\mathcal{P}_0)=\{0.35,0.65\}$ and $|\mathcal{P}_0|=3$. With some calculations, one can show that the optimal value $V^{\tiny{opt}}$ equals $1.34$. 

\smallskip

\noindent \textbf{Scenario 2}:
\begin{eqnarray*}
	Q(x,a)=\left\{\begin{array}{ll}
		1 +(x^{(1)})^3 , & a<0.35,\\
		x^{(1)}- \log(1.5 + x^{(2)}), & 0.35\le a<0.65,\\
		1- \sin(0.5\pi x^{(2)}), & a\ge 0.65.
	\end{array}
	\right.
\end{eqnarray*}
Under Scenario 2, the outcome regression function is piecewise constant as a function of $a$, but is nonlinear as a function of $x$. The change points are $J(\mathcal{P}_0)=\{0.35,0.65\}$ with $|\mathcal{P}_0|=3$. The optimal value equals $1.35$, based on Monte Carlo approximations.  

\begin{table}
\centering
\caption{
\change{The estimated optimal value $\widehat{V}$ with its standard error, the empirical coverage probability of its associated confidence interval, and the averaged number of estimated partitions computed by the proposed L-JIL and D-JIL.}}\label{table:1}
 \scalebox{0.9}{
 \begin{tabular}{ccccc|ccc} 
		\hline
		\hline
		
		 &&\multicolumn{3}{c|}{Scenario 1, $p=4$} &\multicolumn{3}{c}{Scenario 2, $p=4$} \\
		\cline{3-8}
		&& $n=200$ & $n=400$&$n=800$& $n=200$ & $n=400$&$n=800$\\
		\hline
		\hline
		Method &Optimal value $V^{\tiny{opt}}$ &&1.34&&&1.35&\\
		\hline
		\hline
		L-JIL&Estimated optimal value $\widehat{V}$ &1.436& 1.383 & 1.340&1.400& 1.351 &1.421\\
		\cline{2-8}
		 &Mean of standard error $\widehat{\sigma}$ &0.129& 0.091 & 0.066&0.120& 0.085 &0.065\\
		\cline{2-8}
				&Coverage probabilities(\%)& 89.80&  93.20&  95.60& 92.40 &94.60& 95.00\\
		\cline{2-8}
		&Number of partitions $|\widehat{\mathcal{P}}|$& 2.97&  3.01&  3.00& 2.19 &2.81& 3.01\\
		\cline{2-8}
		&Integrated $\ell_2$ loss of $\widehat{\theta}(\cdot)$ &0.371 &0.175 &0.111& 0.786 &0.389& 0.269\\
		\hline
		\hline
		D-JIL&Estimated optimal value $\widehat{V}$ &1.297& 1.338 & 1.345&1.333& 1.331 &1.349\\
		\cline{2-8}
		 &Mean of standard error $\widehat{\sigma}$ &0.160& 0.108 & 0.060&0.166&0.102 &0.060\\
		\cline{2-8}
		&Coverage probabilities(\%)& 90.60&  93.60&  96.00& 95.60 &93.80& 95.00\\
		\cline{2-8}
		&Number of partitions $|\widehat{\mathcal{P}}|$& 2.98&  3.25&  3.18& 2.95 &3.10& 3.08\\
		\hline
	\end{tabular}}
\end{table}

\change{For each scenario, we set $p=4$ and consider three difference choices of the sample size, corresponding to $n=200,400,800$. 
We apply the proposed L-JIL and D-JIL to both scenarios. 
The detailed implementation is discussed in Section \ref{impl}. We set $m=n/5$, $\lambda_n=0$, $\gamma_n=4n^{-1}\log (n)$, 
and construct the CI for $V^{\tiny{opt}}$ based on the procedure described in Section \ref{secevaluate}. 
Reported in Table 1 are the estimated value function $\widehat{V}$ with its standard error $\widehat{\sigma}$, the empirical coverage probabilities of the proposed confidence interval for $V^{\tiny{opt}}$, and the number of estimated partitions $|\widehat{\mathcal{P}}|$, 
aggregated over $500$ simulations. In addition, we include the integrated $\ell_2$ loss of the estimated varying coefficient $\widehat{\theta}(\cdot)$ via L-JIL.


Based on the results, it is clear that the estimated value function approaches the optimal value as the sample size increases for both methods. For instance, when $n=800$, L-JIL obtained an estimated value of $1.340$ under Scenario 1 on average. D-JIL yields an average value of $1.349$ under Scenario 2. These values are very close to the truths $1.34$ and $1.35$, respectively. The performance of our proposed L-JIL and D-JIL are comparable under Scenario 1. 
In addition, 
as the sample size increases, the 
coverage probability of the Wald-type CI approaches to the nominal level. This verifies our theoretical findings in Theorem \hyperlink{thm3}{5}. It is worth noting that the CI computed via L-JIL achieves the nominal coverage under Scenario 2 where the outcome regression function is nonlinear in $x$. We suspect this is due to that the optimal I2DR is close to a linear decision rule despite the nonlinearity of the outcome regression function.  

Moreover, the averaged estimated number of partitions $|\widehat{\mathcal{P}}|$ is approximately 3 for all settings. This demonstrates the consistency of the estimated number of partitions in Theorems \hyperlink{thm1}{1} and \hyperlink{thm5}{3}. In addition, the integrated $\ell_2$ loss of the estimated varying coefficient computed via L-JIL converges to $0$, as the sample size increases. For example, when $n=800$, $\int_0^1 \|\widehat{\theta}(a)-\theta_0(a)\|_2^2da$ equals  $0.111$ for Scenario 1 and  $0.269$ for Scenario 2. These values are fairly small by noting that $\int_0^1 \|\theta_0(a)\|_2^2da=2$. Notice that $\int_0^1 \|\widehat{\theta}(a)-\theta_0(a)\|_2^2da$ decays at a rate that is approximately proportional to $n^{-1}$. This verifies our theoretical findings in Theorem \hyperlink{thm1}{1}.}


\subsection{Value function under the proposed I2DR}\label{secsimugeneral}
In this section, we consider more general settings and compare the proposed procedure with the existing state-of-the-art methods that outputs single-valued decision rule. Similar to Section \ref{secsimustep}, we generate the data from the following model: 
\begin{eqnarray*}
	Y|X,A \sim N(Q(X,A), 1),\,\,\,\,A|X \sim \hbox{Unif}[0,1]\,\,\,\,\hbox{and}\,\,\,\,X^{(1)},X^{(2)},\dots,X^{(p)}\stackrel{iid}{\sim} \hbox{Unif}[-1,1].
\end{eqnarray*}
In addition to Scenarios 1 and 2, we consider several other choices of the outcome regression function, allowing the working model assumption in Model I or Model II to be violated in some scenarios. \change{Specifically, similar to Scenarios 1 to 2, the outcome regression function in Scenario 3 is a piecewise constant function of the treatment.  
As commented earlier, these scenarios are motivated from various applications such as dynamic pricing where the expected demand of a product has jump discontinuities as a function of the charged price. In Scenarios 4 and 5, however, the outcome regression function is continuous in the treatment. In particular, Scenario 4 is known as the varying coefficient model that has been widely applied in many scientific domains. Scenario 5 has been considered by \citet{chen2016personalized} for the personalized dose finding.

\begin{table}
	\centering
	\caption{\textcolor{black}{Simulation scenarios.}}\label{capsimscenario}
	\scalebox{0.85}{
	\begin{tabular}{ccccccc} 
		\hline\hline
		scenarios & piecewise in $a$? & linear in $x$? & $J(\mathcal{P}_0)$ & $|\mathcal{P}_0|$ & optimal rule & optimal value \\ \hline\hline
		1 & $\checkmark$ & $\checkmark$ & $\{0.35, 0.65\}$ & 3 & $\argmax_{I\in \mathcal{P}_0} q_{\mathcal{I}}(x)$ & 1.34 \\ \hline
		2 & $\checkmark$ & $\times$ & $\{0.35, 0.65\}$ & 3 & $\argmax_{I\in \mathcal{P}_0} q_{\mathcal{I}}(x)$ & 1.35 \\ \hline
		3 & $\checkmark$ & $\times$ & $\{0.25, 0.5, 0.75\}$ & 4 & $\argmax_{I\in \mathcal{P}_0} q_{\mathcal{I}}(x)$ & 0.76 \\ \hline
		4 & $\times$ & $\checkmark$ & N.A. & N.A. & $0.5\mathbb{I}(\bar{x}^\top \theta<0)$ & 1.28 \\ \hline
		5 & $\times$ & $\times$ & N.A. & N.A. & $0.5+0.25(x^{(1)}+x^{(2)})$  & 8 \\\hline\hline
	\end{tabular}
	}
\end{table}
}



\smallskip

\noindent \textbf{Scenario 3}:
\begin{eqnarray*}
	Q(x,a)=\left\{\begin{array}{ll}
		\sqrt{x^{(1)}/2 +0.5} , & a<0.25,\\
		\sin(2\pi x^{(2)}), & 0.25\le a<0.5,\\
		0.5 - (x^{(1)}+x^{(2)}-0.75)^2, & 0.5\le a<0.75,\\
		0.5, & a\ge 0.75.
	\end{array}
	\right.
\end{eqnarray*}
 
 \smallskip
 
\noindent \textbf{Scenario 4}:  
	$$Q(x,a)=\bar{x}^{\top} \{{ 2|a-0.5|}\theta^*\},$$
where $\theta^*=(1,2,-2,0_{p-2}^\top)^{\top}$. By setting $\theta_0(a)=2|a-0.5|\theta^*$, it is immediate to see that $Q(x,a)=\bar{x}^{\top} \theta_0(a)$ and satisfies the condition in \eqref{md1}. 

\smallskip

\noindent \textbf{Scenario 5}:
	$$Q(x,a) = 8 + 4x^{(1)} - 2x^{(2)} - 2x^{(3)} - 10( 1 + 0.5x^{(1)} + 0.5x^{(2)} - 2a)^2.$$

We apply the proposed L-JIL and D-JIL to \change{Scenarios 1-5} to estimate the optimal I2DR, 
with $p=20$ and  $n\in \{50,100,200,400,800\}$. The tuning parameters in JILs are specified according to Section \ref{tuning}. Here, we set $m=n/c$ with $c=10$ to save computational cost. In Section \ref{secchoicem}, we report results with $c\in \{6,8\}$ and find the values under the estimated I2DRs are very similar to those with $c=10$. }
\begin{table}[!t]
	\centering
	\caption{\change{The value function under the proposed I2DR and IDRs estimated based on outcome weighted learning (L-O-L and K-O-L) and Q-learning with the linear regression (Q-Linear) for Scenarios 1-5.}}\label{table:2}
	\scalebox{0.9}{
		\begin{tabular}{ccccccc}
			\hline
			\hline
			&$n$ & 50&100&200&400&800\\
			\hline
			\hline
			Scenario $1$& L-JIL &0.783(0.016)&0.832(0.016)&1.080(0.014)&1.259(0.002)&1.297(0.001)\\
			\cline{2-7}
			\cline{2-7}
			 $V$ = 1.34   &  D-JIL &0.914(0.012)&0.967(0.008)&1.050(0.005)&1.071(0.005)&1.138(0.001)\\
			\cline{2-7}
			 $p=20$ & L-O-L&0.558(0.004)&0.574(0.004)&0.600(0.005)&0.597(0.005)&0.583(0.005)\\
			\cline{2-7}
			& K-O-L&0.335(0.008)&0.415(0.006)&0.441(0.006)&0.457(0.005)& 0.489(0.004)\\
			\cline{2-7}
			&Q-Linear&1.026(0.041)&1.055(0.038)&1.080(0.038)&1.048(0.029)&0.829(0.028)\\
			\hline
			\hline
			Scenario $2$& L-JIL &0.741(0.021)&0.854(0.020)&1.180(0.007)&1.266(0.001)&1.299(0.001)\\
			\cline{2-7}
			\cline{2-7}
			 $V$ = 1.35   &  D-JIL &0.900(0.012)&0.978(0.008)&1.074(0.004)&1.102(0.003)&1.141(0.001)\\
			\cline{2-7}
			 $p$ = 20 & L-O-L&0.450(0.009)&0.448(0.006)&0.447(0.005)&0.429(0.004)&0.410(0.003)\\
			\cline{2-7}
			& K-O-L&0.115(0.019)&0.213(0.010)&0.229(0.007)&0.241(0.004)&  0.276(0.002)\\
						\cline{2-7}
			&  Q-Linear&1.048(0.039)&1.071(0.037)&1.080(0.036)&1.042(0.027)&0.772(0.034)\\
			\hline
			\hline
			Scenario $3$& L-JIL  &0.227(0.020)&0.268(0.013)&0.372(0.008)&0.432(0.003)&0.511(0.002)\\
			\cline{2-7}
			 $V$ = 0.76  &  D-JIL &0.453(0.019)&0.469(0.009)&0.511(0.005)&0.526(0.004)&0.545(0.002)\\
			\cline{2-7}
			 $p$ = 20 & L-O-L&0.002(0.010)&-0.009(0.008)&-0.060(0.006)&-0.090(0.005)&-0.107(0.004)\\
			\cline{2-7}
			& K-O-L&-0.268(0.026)&-0.233(0.015)&-0.260(0.009)&-0.251(0.006)&  -0.233(0.003)\\
			\cline{2-7}
			& Q-Linear&0.601(0.039)&0.604(0.032)&0.597(0.022)&0.575(0.015)&  0.315(0.032)\\
			\hline
			\hline
			Scenario 4&L-JIL &0.553(0.013)&0.564(0.011)&0.630(0.011)&0.806(0.006)&0.882(0.002)\\
			\cline{2-7}
			 $V$ = 1.28 &  D-JIL &0.612(0.014)&0.651(0.008)&0.684(0.004)&0.653(0.006)&0.801(0.001)\\
			\cline{2-7}
			$p$ = 20  & L-O-L&0.525(0.016)&0.458(0.010)&0.375(0.004)&0.300(0.002)&0.237(0.001)\\
			\cline{2-7}
			& K-O-L&0.236(0.007)&0.260(0.004)&0.252(0.003)&0.244(0.001)&  0.246(0.001)\\
						\cline{2-7}
			 & Q-Linear&0.995(0.025)&0.995(0.026)&0.999(0.021)&0.998(0.025)&  0.832(0.044)\\
			\hline
			\hline
			Scenario 5& L-JIL & 5.82(0.05)&6.41(0.02)&6.80(0.01)&7.02(0.01)&7.16(0.01)\\
			\cline{2-7}
			$V$ = 8.00   &  D-JIL  & 5.57(0.06)&5.79(0.03)&5.97(0.02)&6.10(0.01)&6.26(0.01)\\
			\cline{2-7}
			$p$ = 20 & L-O-L &5.92(0.07)&6.75(0.03)&7.32(0.02)&7.66(0.01)& 7.81(0.01)\\
			\cline{2-7}
			& K-O-L&6.70(0.02)&7.05(0.02)&7.38(0.01)&7.58(0.01)& 7.56(0.01)\\
			\cline{2-7}
			 & Q-Linear&-0.53(1.27)&1.35(1.05)&3.80(0.57)&6.57(0.21)& 6.57(0.21)\\
			\hline
	\end{tabular}}
\end{table}

To evaluate the proposed I2DRs, we compare its value function $V(\widehat{d})$ with the values under estimated optimal IDRs obtained by \textcolor{black}{the linear outcome weighted learning (L-O-L) and the nonlinear outcome weighted learning based on the Gaussian kernel function (K-O-L) proposed by \cite{chen2016personalized}, and the Q-learning method based on the linear regression (Q-Linear). To implement L-O-L and K-O-L, we fix the parameter $\phi_n=0.1$, and select other tuning parameters by five-fold cross-validation, as in \cite{chen2016personalized}. Finally, to implement Q-Linear, we first fit the outcome on $\{X,X\times X,A,A^2, XA, X\times XA, XA^2, X\times XA^2\}$ via the linear regression where $X\times X$ means the quadratic and cross terms among $X$. Denote the resulting estimator as $\widehat{Q}^L(x,a)$, then the optimal dose for a patient with covariates $X=x$ is given by $\argmax_a \widehat{Q}^L(x,a)$. }
All the value functions are evaluated via Monte Carlo simulations. 
The average value function as well as its standard deviation over 200 replicates are summarized in Table \ref{table:2}.

It can be seen from Table 2 that both L-JIL and D-JIL are very efficient when Model I (Scenarios 1-3) holds,
 and perform reasonably well when Model II (Scenario 4 and 5) holds or the sample size is small. For instance, the proposed L-JIL achieves a value of $1.297$ in Scenario 1 
and {\color{black}$1.299$ in Scenario 2, when $n=800$. \textcolor{black}{These values are} very close to the optimal values, given by $1.34$ and $1.35$.  
\textcolor{black}{In addition, due to the largely increased feature dimension, extra noises compromise the performance of D-JIL in both Scenarios 1 and 2, by comparing the results in Table 1 with that in Table 2. Yet, in Scenario 3, the nonlinear setting is hard to be modeled by the linear pattern, and thus} the proposed D-JIL performs consistently better than L-JIL, due to the capacity of deep neural networks in approximating complicated non-linear relationships. 
Moreover, the value of the proposed I2DR increases with the sample size in most cases. This supports our theoretical findings in Section \ref{sectheory}.

\textcolor{black}{In comparison, the value function under the estimated IDR using L-O-L and K-O-L is no more than half of the optimal value for each setting in Scenarios 1 to 4. \textcolor{black}{This is owing to the `V-structured' non-linear complex optimal decision rule in Scenario 4, which violates the assumption in \cite{chen2016personalized} that the decision rules should be smooth over the entire space of the treatment. While our method still works well for such a varying coefficient model. This supports our theoretical findings in Theorem \hyperlink{thm4}{6}.}
In Scenario 5, L-O-L and K-O-L have better performance, as the true optimal decision rule is linear and the outcome regression function is very sensitive to the change of the treatment level $a$ (by noticing that the coefficient of the quadratic term in Scenario 5 is 10). Our methods perform worse in this scenario. However, the value difference is not large. 
In addition, under Scenarios 1-3, both L-JIL and D-JIL achieve larger value functions than Q-Linear when $n = 800 $, since the linear model misspecifies the true conditional mean function under Model I. 
Under Scenarios 4-5, Q-Linear performs reasonably and comparably well as our proposed methods, because Scenario 4 is linear in $X$ and the conditional mean outcome function under Scenario 5 is correctly specified by Q-Linear. Yet, due to the largely increased feature dimension $p$, the input dimension in Q-Linear is $3p(p-1)/2+3p+2$, which compromises the performance of Q-Linear under Scenario 5 when $n$ is small. 
Among  competing methods, the Q-Linear method has better performance than outcome weighted learning methods by \cite{chen2016personalized} in Scenarios 1-4. Yet, all results of values based on Q-Linear have considerably larger variances than other methods.  }
\change{ More importantly, our methods are able to derive the I2DR, which is more interpretable and easier to implement in practice.

In addition, we notice that D-JIL performs comparably to L-JIL in Scenario 1. This is because the DNN model with ReLU activation function contains the linear model as a special case. Yet, the asymptotic rate of convergence of D-JIL is slower than that of L-JIL due to its complexity. When $n\ge 200$, D-JIL performs worse than L-JIL, as expected. In Scenario 2, although the outcome regression function is nonlinear in $x$, the resulting I2DR can be well-approximated by a linear decision rule. As such, L-JIL and D-JIL achieve similar performance.

Finally, we report the computation time of D-JIL in Table \ref{table:computetime}. It can be seen that the computation time increases approximately linearly with the sample size. The main computation lies in adaptively selecting $\gamma_n$ via cross-validation. We remark that parallel computing can be employed to further reduce the computation time. 

\begin{table}[tb]
	\centering
	\caption{\textcolor{black}{The computation time (in minutes) of the proposed D-JIL.}}\label{table:computetime}
\scalebox{0.9}{
	\begin{tabular}{cccccc}
		\hline
		\hline
		 & 50&100&200&400&800\\
		\hline
		\hline
		Scenario $1$  &0.90(0.04)&1.95(0.03)&4.96(0.05)&14.04(0.12)&35.48(0.21)\\   \hline
		\hline
		Scenario $2$ &0.78(0.02)&1.56(0.02)&4.07(0.04)&13.53(0.07)&35.46(0.12)\\ 
		\hline
		\hline
		Scenario $3$ &0.67(0.01)&1.02(0.02)&2.50(0.04)&7.34(0.04)&23.20(0.06)\\ 
		\hline
		\hline
		Scenario $4$ &0.70(0.01)&1.09(0.02)&3.72(0.04)&10.25(0.06)&15.80(0.19)\\
		\hline
		\hline
		Scenario $5$ &1.01(0.01)&1.51(0.01)&2.30(0.02)&5.37(0.02)&16.32(0.05)\\ 
		\hline
\end{tabular}}
\end{table}
}
\change{
\subsection{Choice of $m$}\label{secchoicem}
Recall that we set $m=n/c$ for some constant $c>0$. In Section \ref{secsimugeneral}, we report our simulation results with $c=10$. In this section, we set $c\in \{6,8\}$ and report the corresponding results in Tables \ref{table:4}-\ref{table:supp4} to investigate the sensitivity of the proposed methods to the choice of $c$. We also include results with $c=10$ for completeness. It can be seen from Tables \ref{table:4} and \ref{table:5} that the value functions are very similar across different choices of $c$. In addition, 
it can be seen from Table \ref{table:supp4} that the averaged number of estimated intervals for the proposed I2DR is very close to the ground truth under Scenarios 1-3 where the underlying models are piecewise constant. Under Scenarios 4-5 however, the number of estimated intervals grows with the sample size, as expected. In all cases, the averaged number of estimated intervals is not overly sensitive to the choice of $m$. Finally, we report the computation time of the proposed D-JIL in Table \ref{table:compute}. It can be seen that the computation time increases with $m$ and $n$, as expected. 

\begin{table}[tb]
	\centering
	\caption{The value function of the proposed I2DR under L-JIL for Scenarios 1-5 with different choice of $m=n/c$.}\label{table:4}
	\scalebox{0.9}{
		\begin{tabular}{ccccccc}
			\hline
			\hline
			&$n$ & 50&100&200&400&800\\
			\hline
			\hline
			Scenario $1$& $c=6$ &0.813(0.019)&0.858(0.017)&1.027(0.014)&1.249(0.003)&1.289(0.001)\\
			\cline{2-7}
			$V$ = 1.34 & $c=8$ &0.836(0.022)&0.870(0.018)&1.024(0.014)&1.238(0.002)&1.295(0.001)\\
			\cline{2-7}
			$p$ = 20 & $c=10$ &0.783(0.016)&0.832(0.016)&1.080(0.014)&1.259(0.002)&1.297(0.001)\\
			\hline
			\hline
			Scenario $2$& $c=6$ &0.804(0.025)&0.891(0.021)&1.132(0.008)&1.257(0.002)&1.290(0.001)\\
			\cline{2-7}
			$V$ = 1.35  & $c=8$ &0.857(0.029)&0.935(0.021)&1.123(0.009)&1.241(0.002)&1.299(0.001)\\
			\cline{2-7}
			$p$ = 20 & $c=10$ &0.741(0.021)&0.854(0.020)&1.180(0.007)&1.266(0.001)&1.299(0.001)\\
			\hline
			\hline
			Scenario 3& $c=6$ &0.280(0.023)&0.310(0.014)&0.339(0.008)&0.422(0.003)&0.504(0.002) \\
			\cline{2-7}
			$V$ = 0.76  & $c=8$ &0.229(0.019)&0.325(0.014)&0.326(0.008)&0.417(0.003)&0.512(0.002)\\
			\cline{2-7}
			$p$ = 20 &  $c=10$ &0.227(0.020)&0.268(0.013)&0.372(0.008)&0.432(0.003)&0.511(0.002)\\
			\hline
			\hline
			Scenario 4& $c=6$ &0.565(0.015)&0.561(0.012)&0.639(0.011)&0.818(0.006)&0.884(0.002)\\
			\cline{2-7}
			$V$ = 1.28 & $c=8$ &0.563(0.015)&0.564(0.012)&0.627(0.011)&0.810(0.006)&0.882(0.002)\\
			\cline{2-7}
			$p$ = 20 & $c=10$ &0.553(0.013)&0.564(0.011)&0.630(0.011)&0.806(0.006)&0.882(0.002)\\
			\hline
			\hline
			Scenario 5& $c=6$ &5.81(0.05)&6.38(0.02)&6.78(0.01)&6.99(0.01)&7.09(0.01)\\
			\cline{2-7}
			$V$ = 8.00  & $c=8$ &5.82(0.05)&6.40(0.02)&6.78(0.01)&7.02(0.01)&7.12(0.01)\\
			\cline{2-7}
			$p$ = 20 & $c=10$ & 5.82(0.05)&6.41(0.02)&6.80(0.01)&7.02(0.01)&7.16(0.01)\\
			\hline
	\end{tabular}}
\end{table}

\begin{table}[tb]
	\centering
	\caption{The value function of the proposed I2DR under D-JIL for Scenarios 1-5 with different choice of $m=n/c$.}\label{table:5}
	\scalebox{0.9}{
		\begin{tabular}{ccccccc}
			\hline
			\hline
			&$n$ & 50&100&200&400&800\\
			\hline
			\hline
			Scenario $1$& $c=6$ &0.941(0.012)&0.972(0.008)&1.028(0.004)&1.065(0.004)&1.127(0.001)\\
			\cline{2-7}
			$V$ = 1.34 & $c=8$ &0.973(0.016)&0.990(0.008)&1.030(0.004)&1.053(0.005)&1.136(0.001)\\
			\cline{2-7}
			$p$ = 20 & $c=10$ &0.914(0.012)&0.967(0.008)&1.050(0.005)&1.071(0.005)&1.138(0.001)\\
			\hline
			\hline
			Scenario $2$& $c=6$ &0.943(0.013)&0.980(0.008)&1.037(0.004)&1.087(0.003)&1.129(0.001)\\
			\cline{2-7}
			$V$ = 1.35  & $c=8$ &1.002(0.015)&1.012(0.008)&1.039(0.004)&1.076(0.003)&1.137(0.001)\\
			\cline{2-7}
			$p$ = 20 & $c=10$ &0.900(0.012)&0.978(0.008)&1.074(0.004)&1.102(0.003)&1.141(0.001)\\
			\hline
			\hline
			Scenario 3& $c=6$ &0.475(0.018)&0.480(0.009)&0.481(0.006)&0.493(0.004)&0.521(0.002)\\
			\cline{2-7}
			$V$ = 0.76  & $c=8$ &0.416(0.019)&0.497(0.009)&0.493(0.006)&0.506(0.003)&0.532(0.002)\\
			\cline{2-7}
			$p$ = 20 &  $c=10$ &0.453(0.019)&0.469(0.009)&0.511(0.005)&0.526(0.004)&0.545(0.002)\\
			\hline
			\hline
			Scenario 4& $c=6$ &0.624(0.014)&0.655(0.008)&0.686(0.004)&0.687(0.005)&0.801(0.001)\\
			\cline{2-7}
			$V$ = 1.28 & $c=8$ &0.622(0.014)&0.651(0.008)&0.684(0.004)&0.676(0.005)&0.801(0.001)\\
			\cline{2-7}
			$p$ = 20 & $c=10$ &0.612(0.014)&0.651(0.008)&0.684(0.004)&0.653(0.006)&0.801(0.001)\\
			\hline
			\hline
			Scenario 5& $c=6$ &5.49(0.06)&5.69(0.03)&5.82(0.02)&5.97(0.01)&6.12(0.01)\\
			\cline{2-7}
			$V$ = 8.00  & $c=8$ &5.58(0.05)&5.77(0.03)&5.91(0.02)&6.04(0.01)&6.20(0.01)\\
			\cline{2-7}
			$p$ = 20 & $c=10$ & 5.57(0.06)&5.79(0.03)&5.97(0.02)&6.10(0.01)&6.26(0.01)\\
			\hline
	\end{tabular}}
\end{table}

\begin{table}[tb]
	\centering
	\caption{\textcolor{black}{The averaged number of estimated intervals computed by L-JIL with different choices of $m=n/c$.}}\label{table:supp4}
	\scalebox{0.9}{
		\begin{tabular}{ccccccc}
			\hline
			\hline
			&$n$ & 50&100&200&400&800\\
			\hline
			\hline
			Scenario $1$& $c=6$ &2.04(0.17)&2.16(0.15)&2.60(0.09)&3.00(0.01)&3.00(0.00)\\
			\cline{2-7}
			$|\mathcal{P}_0|=3$ & $c=8$ &1.98(0.15)&2.15(0.14)&2.47(0.07)&3.01(0.01)&3.00(0.00)\\
			\cline{2-7}
			$p$ = 20 & $c=10$ &1.78(0.14)&1.95(0.13)&2.76(0.10)&3.00(0.00)&3.00(0.00)\\
			\hline
			\hline
			Scenario $2$& $c=6$ &2.38(0.18)&2.76(0.16)&3.17(0.09)&3.00(0.00)&3.00(0.00)\\
			\cline{2-7}
			$|\mathcal{P}_0|=3$  & $c=8$ &2.38(0.15)&3.03(0.16)&3.02(0.04)&3.00(0.00)&3.00(0.00)\\

			\cline{2-7}
			$p$ = 20 & $c=10$ &2.00(0.15)&2.64(0.14)&3.12(0.07)&3.00(0.00)&3.00(0.00)\\
			\hline
			\hline
			Scenario 3& $c=6$ &2.63(0.20)&3.59(0.21)&3.65(0.17)&3.34(0.03)&3.76(0.02)\\			\cline{2-7}
			$|\mathcal{P}_0|=4$  & $c=8$&2.26(0.19)&3.40(0.19)&3.41(0.14)&3.34(0.03)&3.78(0.02)\\
			\cline{2-7}
			$p$ = 20 &  $c=10$ &2.24(0.18)&3.01(0.18)&3.48(0.13)&3.32(0.03)&3.75(0.02)\\
			\hline
			\hline
			Scenario 4& $c=6$ &1.68(0.15)&1.59(0.13)&1.86(0.07)&2.86(0.04)&3.12(0.01)\\ 			\cline{2-7}
			/ & $c=8$ &1.61(0.14)&1.55(0.11)&1.79(0.07)&2.80(0.04)&3.15(0.02)\\ 
			\cline{2-7}
			$p$ = 20 & $c=10$ &1.58(0.13)&1.62(0.13)&1.82(0.07)&2.79(0.04)&3.13(0.02)\\ 
			\hline
			\hline
			Scenario 5& $c=6$ &4.78(0.17)&6.56(0.12)&9.56(0.08)&15.08(0.08)&25.71(0.07)\\
			\cline{2-7}
			/   & $c=8$ &4.29(0.13)&6.28(0.13)&9.30(0.09)&14.36(0.08)&23.91(0.08)\\
			\cline{2-7}
			$p$ = 20 & $c=10$ &3.91(0.11)&5.96(0.11)&8.67(0.08)&13.62(0.08)&22.00(0.08)\\ 
			\hline
	\end{tabular}}
\end{table}

\begin{table}[tb]
	\centering
	\caption{\textcolor{black}{The computation time (in minutes) of D-JIL with different choices of $m=n/c$.}}\label{table:compute}
\scalebox{0.9}{
	\begin{tabular}{ccccccc}
		\hline
		\hline
		&$n$ & 50&100&200&400&800\\
		\hline
		\hline
		Scenario $1$& $c=6$ &0.69(0.03)&2.36(0.05)&10.38(0.10)&33.00(0.22)&87.56(0.36)\\
		\cline{2-7}
		& $c=8$ &1.21(0.03)&2.23(0.05)&6.33(0.09)&19.96(0.16)&53.75(0.29)\\
		\cline{2-7}
		$p$ = 20 & $c=10$ &0.90(0.04)&1.95(0.03)&4.96(0.05)&14.04(0.12)&35.48(0.21)\\   \hline
		\hline
		Scenario $2$& $c=6$ &0.71(0.04)&2.35(0.06)&10.35(0.11)&32.08(0.21)&86.53(0.26)\\
		\cline{2-7}
		& $c=8$ &1.02(0.02)&2.27(0.03)&5.79(0.08)&18.54(0.11)&54.36(0.24)\\
		\cline{2-7}
		$p$ = 20 &$c=10$ &0.78(0.02)&1.56(0.02)&4.07(0.04)&13.53(0.07)&35.46(0.12)\\ 
		\hline
		\hline
		Scenario $3$& $c=6$ &0.71(0.02)&2.33(0.03)&6.12(0.05)&17.23(0.10)&52.96(0.24)\\
		\cline{2-7}
		& $c=8$ &0.79(0.01)&1.59(0.02)&4.28(0.03)&9.61(0.06)&32.52(0.17)\\
		\cline{2-7}
		$p$ = 20 &$c=10$ &0.67(0.01)&1.02(0.02)&2.50(0.04)&7.34(0.04)&23.20(0.06)\\ 
		\hline
		\hline
		Scenario $4$& $c=6$ &1.15(0.02)&2.67(0.05)&7.45(0.08)&21.38(0.19)&52.07(0.38)\\
		\cline{2-7}
		& $c=8$ &0.88(0.02)&1.99(0.02)&4.92(0.04)&13.38(0.06)&32.77(0.19)\\
		\cline{2-7}
		$p$ = 20 &$c=10$ &0.70(0.01)&1.09(0.02)&3.72(0.04)&10.25(0.06)&15.80(0.19)\\ 
		\hline
		\hline
		Scenario $5$& $c=6$ &1.24(0.02)&2.49(0.03)&4.47(0.04)&8.91(0.05)&28.52(0.09)\\
		\cline{2-7}
		& $c=8$ &0.86(0.01)&1.37(0.01)&3.53(0.02)&7.39(0.03)&20.80(0.09)\\
		\cline{2-7}
		$p$ = 20 &$c=10$ &1.01(0.01)&1.51(0.01)&2.30(0.02)&5.37(0.02)&16.32(0.05)\\ 
		\hline
\end{tabular}}
\end{table}
}

\section{Real Data Analysis}\label{realdata}
In this section, we illustrate the empirical performance of our proposed method on a real data from the International Warfarin Pharmacogenetics Consortium \citep{international2009estimation}. \change{Warfarin is a medication that is commonly used for preventing blood clots such as thrombosis and thromboembolism. Its effect is evaluated by the international normalized ratio (INR), which is a measurement of the time it takes for the blood to clot, with an ideal number of $2.5$.  High doses of Warfarin are more beneficial than its lower doses, but may lead to a high risk of bleeding as well. Proper dosing of Warfarin is thus of significant importance. Yet, this problem is particularly challenging due to the complex interactions between Warfarin and many commonly used medications \citep{holbrook2005systematic}. Nonetheless, existing methods are not able to recommend individualized interval-based dose rule for Warfarin.}
  


 \change{To develop the optimal I2DR for Warfarin dosing, we use the dataset provided by the International Warfarin Pharmacogenetics \cite{international2009estimation} for analysis. We choose $6$ baseline covariates, including age, height, weight, gender, the VKORC1.AG genotype, and the VKORC1.AA genotype. This yields a total of $3848$ with complete records of baseline information. Here, the VKORC1 genotype has been shown to play a particularly large role in response to Warfarin \citep{wadelius2005common}. The outcome is defined as the negative absolute distance between the INR after the treatment and the ideal number of $2.5$, i.e, $Y=-|\hbox{INR}-2.5|$. Thus, a larger outcome represents a better performance of preventing blood clots, with the optimal value of 0.} We use the min-max normalization to convert the range of the dose level $A$ into $[0,1]$. 
%

To implement L-JIL and D-JIL, we set $c=5$, i.e., $m=n/5$, and select $\gamma_n$ and $\lambda_n$ via cross validation, as in Section \ref{secsimugeneral}. 
To further evaluate the empirical performance of the proposed I2DRs, we compare their values with the value under the IDR estimated by K-O-L. Specifically, we randomly select $70\%$ of the data to compute the proposed I2DR and the IDR obtained by K-O-L, and evaluate their value functions using the remaining dataset. We then iterate this procedure $50$ times to calculate the average value function. For each iteration, 
the value function is estimated based on the nonparametric estimator proposed by \cite{zhu2020kernel}. 

Specifically, let $\mathbb{G}_{\tiny{test}}$ denote observations in the testing dataset. 
For the IDR $\widetilde{d}$ computed by K-O-L, we consider the following nonparametric estimator for its value function,
\begin{eqnarray*}
	\widetilde{V}(\widetilde{d})=\int_x \frac{\sum_{i\in \mathbb{G}_{\tiny{test}}} Y_iK(h_x^{-1}(x-X_i)) K(h_a^{-1}(\widetilde{d}(x)-A_i))}{\sum_{i\in \mathbb{G}_{\tiny{test}}} K(h_x^{-1}(x-X_i)) K(h_a^{-1}(\widetilde{d}(x)-A_i))}\bigg\{\sum_{i\in \mathbb{G}_{\tiny{test}}} \frac{K(h_x^{-1}(x-X_i))}{|\mathbb{G}_{\tiny{test}}|h_x^p}\bigg\} dx,
\end{eqnarray*}
where $K(\cdot)$ denotes the Gaussian kernel function, and $h_x$ and $h_a$ are some bandwidth parameters. 
The tuning parameters $h_x$ and $h_a$ are chosen according to the numerical results in Section 5 of \cite{zhu2020kernel}.

\change{Notice that the value function under the proposed I2DR $\widehat{d}(\cdot)$ depends on the preference function $\pi^*$. To evaluate $\widehat{d}$, we consider multiple preference functions, including the maximum value, the minimum value, the mid-point value, and the value uniformly at random. In particular, when $\pi^*$ is set to the uniform density function, we compute $\widetilde{V}^{\pi^*}(\widehat{d})$ as
\begin{eqnarray*}
	\int_x \int_{\widehat{d}(x)} \frac{1}{|\widehat{d}(x)|} \frac{\sum_{i\in \mathbb{G}_{\tiny{test}}} Y_iK(h_x^{-1}(x-X_i)) K(h_a^{-1}(d(x)-A_i))}{\sum_{i\in \mathbb{G}_{\tiny{test}}} K(h_x^{-1}(x-X_i)) K(h_a^{-1}(d(x)-A_i))}\bigg\{\sum_{i\in \mathbb{G}_{\tiny{test}}} \frac{K(h_x^{-1}(x-X_i))}{|\mathbb{G}_{\tiny{test}}|h_x^p}\bigg\}da dx,
\end{eqnarray*}
Reported in Table \ref{table:diff_pi} are the averaged values under the proposed I2DRs computed via L-JIL and D-JIL, with 
the aforementioned four choices of $\pi^*$, 
aggregated over 10 replications. It can be seen that our method is not sensitive to the choice of $\pi^*$.}

\begin{table}[!t]
\centering
\caption{\color{black} The averaged value and their standard deviation  under four different choices of $\pi^*(\cdot;x,\mathcal{I})$ in the real data application.}\label{table:diff_pi} 
\scalebox{0.9}{
	\begin{tabular}{c|cccc}
		\hline
		\hline
		Choice of $\pi^*(\cdot;x,\mathcal{I})$&Minimum Dose&Maximum Dose& Mid-point Dose& Uniformly Sample\\
		\hline
		Under  L-JIL &-0.329(0.008)& -0.329(0.007)&-0.328(0.008)& -0.328(0.008)\\
		\hline
		Under  D-JIL  &-0.333(0.012)& -0.332(0.011)&-0.333(0.012)&-0.333(0.012)\\
		\hline
	\end{tabular}}
\end{table}

Over 50 iterations, the average value functions of our proposed I2DRs computed by L-JIL and D-JIL are $-0.332$ and $-0.331$, larger than the value $-0.344$ of the IDR obtained by K-O-L. \change{These values mean that the INR of patients following the recommended I2DR is around 0.33 far from the ideal amount of 2.5. In contrast, using K-O-L would lead to a greater departure from the ideal amount. We also set $c$ to 6, 8, or 10 when employing L-JIL and report the average value functions and their standard deviations in Table \ref{table:realc}.}  \change{ It can be seen that the performance is similar across difference choices of $c$. In addition, among the 50 iterations, $|\widehat{\mathcal{P}}|$ computed by L-JIL equals 3 for 40 iterations. Let $\widehat{\theta}_1$, $\widehat{\theta}_2$ and $\widehat{\theta}_3$ denote the corresponding regression coefficients associated with these three subintervals. We report the means and standard deviations of the estimated regression coefficients across these 40 iterations in Table \ref{table:3_add}. It can be seen that except for the intercept term associated with the first subinterval, the standard deviations of other parameters are fairly small. In the rest 10 iterations, $|\widehat{\mathcal{P}}|$ is either 2 or 4. As such, the change points and parameter estimates are relatively stable across 50 iterations.} 

\begin{table}[t]
	\centering
	\caption{{\color{black} The averaged value with their standard deviation under L-JIL with different choice of $c$ in real data analysis.}}\label{table:realc} 
	\scalebox{1}{
		\begin{tabular}{c|ccc}
			\hline
			\hline
			Choice of $c$&$c=6$&$c=8$& $c=10$\\
			\hline
			Estimated Value &-0.329(0.009)& -0.329(0.009)&-0.328(0.008)\\ 
			\hline
	\end{tabular}}
\end{table}



\begin{table}[!t]
	\centering
	\caption{\textcolor{black}{The means and standard deviations of regression coefficients associated with the three subintervals computed by L-JIL over 40 iterations.}}\label{table:3_add}
	\scalebox{0.85}{
		\begin{tabular}{c|ccccccc}
			\hline
			\hline
			&Intercept&Age& Weight& Height& Gender& VKORC1.AG& VKORC1.AA\\
			\hline
			$\widehat{\theta}_1$ &-1.289(0.841)& 0.005(0.036)&  0.003(0.005)& 0.005(0.005)& -0.123(0.105)& -0.493(0.140)& -0.533(0.141)\\
			\hline
			$\widehat{\theta}_2$ & -1.900(0.163)&0.021(0.006)&  0.004(0.001)& 0.008(0.001)&-0.203(0.021)&-0.024(0.027)& -0.125(0.026)\\ 
			\hline
			$\widehat{\theta}_3$ &  -0.450(0.063)& 0.014(0.002)& 0.001(0.001)&0.001(0.001)&-0.026(0.009)&-0.007(0.006)&-0.116(0.012)\\
			\hline 
	\end{tabular}}
\end{table}


\begin{table}[!t]
\centering
\caption{\color{black} The regression coefficients associated with the three subintervals computed by applying L-JIL to the entire data without sample-splitting.}\label{table:3}
\scalebox{0.9}{
	\begin{tabular}{c|ccccccc}
		\hline
		\hline
		&Intercept&Age& Weight& Height& Gender& VKORC1.AG& VKORC1.AA\\
		\hline
		$\widehat{\theta}_1$ &-1.229& 0.013&  0.003& 0.004& -0.130& -0.466& -0.541\\
		\hline
		$\widehat{\theta}_2$ & -2.008&0.021&  0.004& 0.008&-0.212&-0.030& -0.127\\	
		\hline
		$\widehat{\theta}_3$ &  -0.518& 0.015& 0.001&0.001&-0.032&-0.001&-0.118\\
		\hline
	\end{tabular}}
\end{table}

 \begin{figure}[tb]
	\centering
	\includegraphics[width=5in]{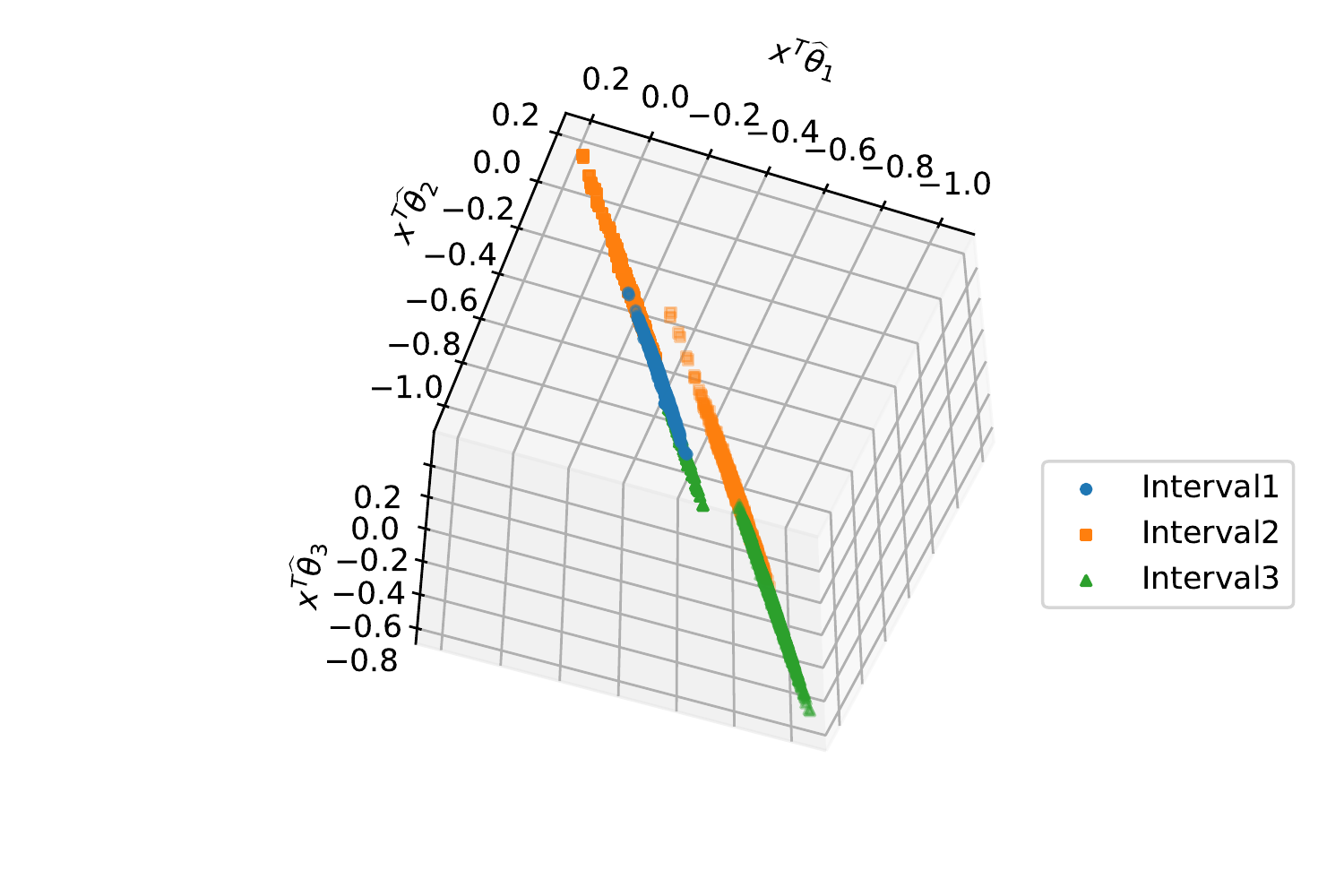}
	\caption{3D plot of the proposed I2DR computed by L-JIL.}\label{fig:p3d_bs}
\end{figure}
} 

 \change{Finally, we apply L-JIL to the entire data without sample-splitting to compute an I2DR and illustrate its interpretability. 
It turns out that L-JIL partitions $[0,1]$ into three subintervals: $[0, 0.02)$, $[0.02, 0.17)$, and $[0.17,1]$. 
We report these regression coefficients in Table \ref{table:3}. According to Table \ref{table:3}, the proposed I2DR based on L-JIL gives us a clear interpretation about the effect of baseline information on the dose assignment rule. For instance, patients whose genotype VKORC1 is AG or AA are more likely to receive low doses of Warfarin to prevent bleeding; older patients with larger weight shall be treated with higher dose levels. Future experiments are warranted to confirm these scientific findings. In Figure \ref{fig:p3d_bs}, we give a virtual representation of the proposed I2DR under L-JIL. Specifically, for each of the 3848 patients in the whole dataset, we plot a 3-dimensional vector $(\bar{x}^{\top} \widehat{\theta}_1, \bar{x}^{\top} \widehat{\theta}_2, \bar{x}^{\top} \widehat{\theta}_3)^{\top}$ based on his/her covariates $x$. Patients that are recommended to receive dose level in $[0, 0.02)$, $[0.02, 0.17)$, and $[0.17,1]$ are colored in blue, orange and green, respectively. That is, we classify patients into three groups according to the recommended dose interval. It can be seen from Figure \ref{fig:p3d_bs} that these three subgroups are well separated and have comparable sample sizes. In Figure \ref{fig:LJIL_real}, we further plot the histograms of the recommended treatments (uniformly randomly sampled from the computed I2DR) and the received treatments.}

\begin{figure}[tb]
	\centering
	\includegraphics[width=5in]{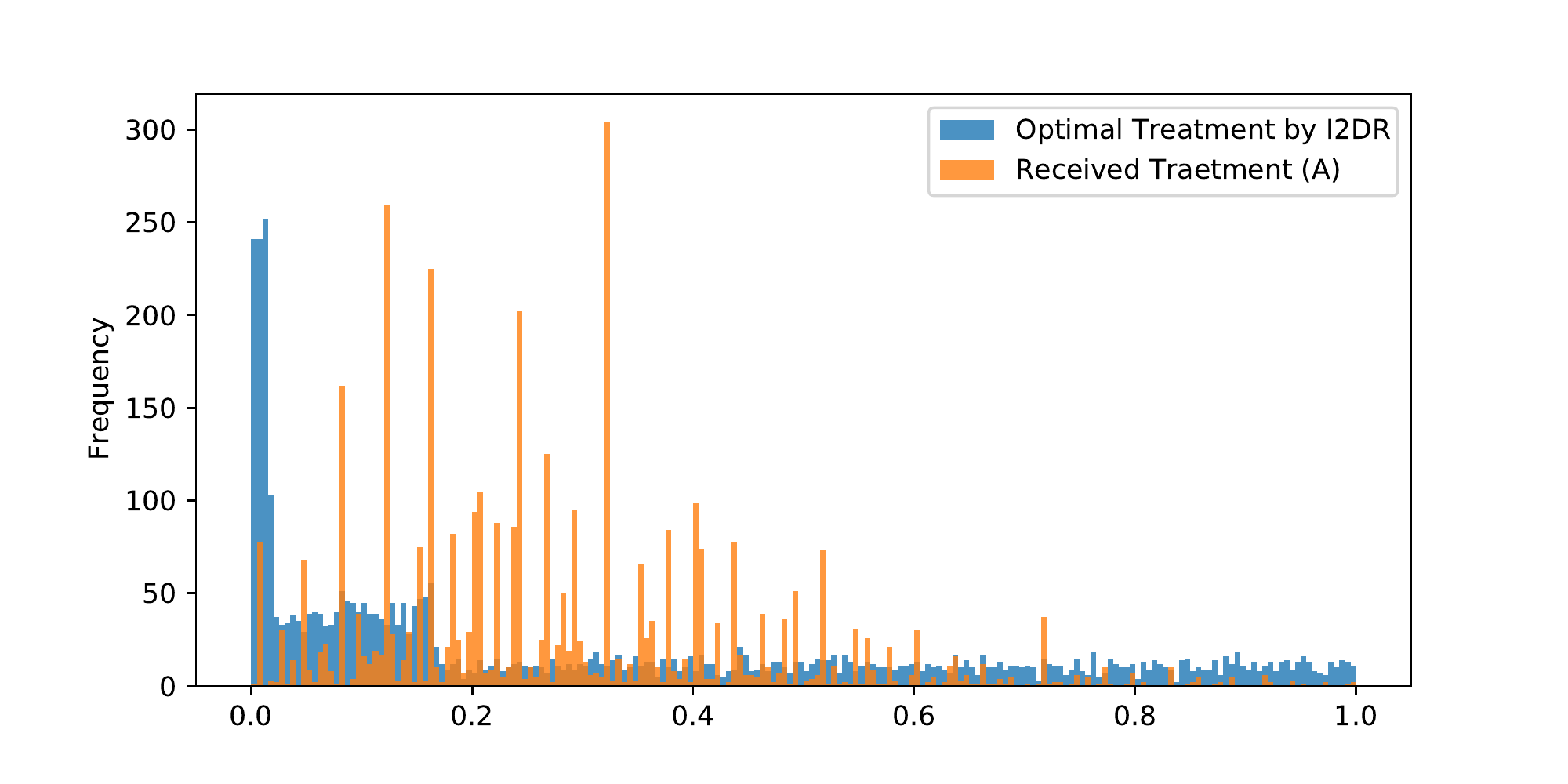}
	\caption{\color{black} Histograms of the recommended treatments (uniformly randomly sampled from the computed I2DR) and the received treatments.}\label{fig:LJIL_real}
\end{figure}

%
%

{\color{black}\section{Discussions}\label{secdis}

\subsection{Diverging Number of Change Points}	
\textcolor{black}{When Model I is true}, 
we assume $|\mathcal{P}_0|$ is fixed {to simplify the results in Theorems \hyperlink{thm1}{1}, \hyperlink{thm2}{2}, \hyperlink{thm5}{3}, and \hyperlink{thm7}{4}}. 
Our theoretical results can be generalized to the situation where $|\mathcal{P}_0|$ diverges with $n$ as well. Take L-JIL as an example. 
Similar to Theorem \hyperlink{thm1}{1}, we can show  that the $\ell_2$ integrated loss satisfies $\int_{0}^1 \|\widehat{\theta}(a)-\theta_0(a)\|_2^2da=O_p(|\mathcal{P}_0|n^{-1}\log n)$. 
Compared to the results in Theorem \hyperlink{thm1}{1}, the convergence rate here is slower by a factor $|\mathcal{P}_0|$. In addition, $|\mathcal{P}_0|=o(n/\log n)$ is required to guarantee the consistency of $\widehat{\theta}$.

We next present more technical details. In the proof of Theorem \hyperlink{thm4}{6} (see Section \ref{secproofthm4} for details), we consider a more general framework and establish the $\ell_2$ integrated loss of $\widehat{\theta}(\cdot)$ by assuming $\theta_0$ satisfies $\limsup_{k\to \infty} k^{\alpha_0} \hbox{AE}_k(\theta_0)<\infty$ where
\begin{eqnarray*}
	\hbox{AE}_k(\theta_0)=\inf_{\substack{\mathcal{P}: |\mathcal{P}|\le k+1 \\ (\theta_{\mathcal{I}})_{\mathcal{I}\in \mathcal{P}}\in \prod_{\mathcal{I}\in \mathcal{P}} \mathbb{R}^{p+1} } } \left\{ \sup_{a\in [0,1]}\left\|\theta_0(a)-\sum_{\mathcal{I}\in \mathcal{P}} \theta_{\mathcal{I}} \mathbb{I}(a\in \mathcal{I}) \right\|_2  \right\}.
\end{eqnarray*}
By definition, $\hbox{AE}_k(\theta_0)$ describes how well $\theta_0(\cdot)$ can be approximated by a step function.

When $\theta_0(\cdot)$ is a step function with number of jumps equal to $|\mathcal{P}_0|$, we have $\hbox{AE}_k(\theta_0)=0$ for any $k \ge |\mathcal{P}_0|$. As a result, $\theta_0$ satisfies the condition $\limsup_{k\to \infty} k^{\alpha_0} \hbox{AE}_k(\theta_0)<\infty$ for any $\alpha_0>0$. As a result, the assertion \eqref{proofthm3finaleq} in the proof of Theorem \hyperlink{thm4}{6} also holds for $\theta_0(\cdot)$ and we have with probability at least $1-O(n^{-2})$ that
\begin{eqnarray*}
	\int_0^1 \|\widehat{\theta}(a)-\theta_0(a)\|_2^2da\le O(1)(|\mathcal{P}_0|^{-\alpha_0}+\gamma_n |\mathcal{P}_0|),
\end{eqnarray*}
where $O(1)$ denotes some positive constant. As $|\mathcal{P}_0|\to \infty$ and $\alpha_0$ can be made arbitrarily large, we have with probability at least $1-O(n^{-2})$ that
\begin{eqnarray*}
	\int_0^1 \|\widehat{\theta}(a)-\theta_0(a)\|_2^2da\le O(1)(\gamma_n |\mathcal{P}_0|),
\end{eqnarray*}
where $O(1)$ denotes some positive constant.

In Theorem \hyperlink{thm4}{6}, we require $\gamma_n\gg n^{-1}\log n$. However, this condition can be relaxed to $\gamma_n\ge \mathbb{M}_0 n^{-1}\log n$ for some sufficiently large constant $\mathbb{M}_0>0$. Under the latter condition, we have
\begin{eqnarray*}
	\int_0^1 \|\widehat{\theta}(a)-\theta_0(a)\|_2^2da=O_p(|\mathcal{P}_0|n^{-1}\log n).
\end{eqnarray*}
This yields the convergence rate of the $\ell_2$ integrated loss of $\widehat{\theta}(\cdot)$. 
\subsection{Potential alternative approaches}
In this paper, {we focus on modeling the outcome regression function to derive I2DR.} 
Below, we outline two other potential approaches and discuss their weaknesses.
\subsubsection{A-learning Type Methods}\label{secalearning}
Let's assume $q_{\mathcal{I}}(\cdot)$ satisfies \eqref{main_model} and the partition $\mathcal{P}_0$ is known to us. In order to eliminate the baseline function $u_0(\cdot)$, we can apply Robinson's transformation \citep[see for example,][and the references therein]{Robinson1988,Zhao2017,Chernozhukov2018} and compute $\widetilde{q}_{\mathcal{I}}$ by minimizing
\begin{eqnarray*}
	\argmin_{ \{q_\mathcal{I} \in \mathcal{Q}_{\mathcal{I}}: \mathcal{I}\in \mathcal{P}_0 \}   } \frac{1}{n}\sum_{i=1}^n [Y_i-\widehat{\mu}(X_i)- \sum_{\mathcal{I} \in \mathcal{P}_0}\{\mathbb{I}(A_i\in \mathcal{I})-\widehat{e}(\mathcal{I}|X_i) \}   q_{\mathcal{I}}(X) ]^2,
\end{eqnarray*}
where $\widehat{\mu}(x)$ correspond tos some nonparametric estimators for $\Mean (Y|X=x)$. 
Both $\widehat{\mu}$ and $\widehat{e}$ can be obtained by some generic machine learning methods with good prediction performance. 

When $\mathcal{P}_0$ is unknown, one might consider estimating $\mathcal{P}_0$ and $ \{q_{\mathcal{I}}:\mathcal{I}\in \mathcal{P}_0\}$ jointly by
\begin{eqnarray*}
	\argmin_{ \substack{ \mathcal{P}\in \mathcal{B}(m), \\ \{q_\mathcal{I} \in \mathcal{Q}_{\mathcal{I}}: \mathcal{I}\in \mathcal{P} \}  } } \frac{1}{n}\sum_{i=1}^n \left[ Y_i-\widehat{\mu}(X_i)- \sum_{\mathcal{I} \in \mathcal{P}}\{\mathbb{I}(A_i\in \mathcal{I})-\widehat{e}(\mathcal{I}|X_i) \} q_{\mathcal{I}}(X_i) \right]^2+ \gamma_n |\mathcal{P}|,
\end{eqnarray*}
for some tuning parameter $\gamma_n$. However, different from the objective function in \eqref{optimize}, for a given partition $\mathcal{P}$, all the functions $\{q_{\mathcal{I}}:\mathcal{I}\in \mathcal{P}\}$ need to be jointed estimated. As a result, standard 
change point detection algorithms such as dynamic programming or binary segmentation  \citep{scott1974} cannot be applied. Exhaustive search among all possible partitions is computationally infeasible. It remains unknown how to efficiently solve the above optimization problem. We leave it for future research.

\subsubsection{Policy Search}\label{sec:policysearch}
As commented in Section \ref{secI2DR}, to apply value search, we need to specify a preference function $\pi^*$. To better illustrate the idea, let us suppose $\pi^*(\cdot;x,\mathcal{I})=p(a|x)/\int_{x'\in \mathcal{I}} p(a|x')dx'$. That is, the preference function is the same as the one we observe in our data. Then, for a given I2DR $d$, we can consider the following inverse propensity score weighted estimator for $V^{\pi^*}(d)$, 
\begin{eqnarray*}
	\widehat{V}^{\pi^*}(d)=\frac{1}{n}\sum_{i=1}^n  \frac{\mathbb{I}(A_i\in d(X_i))}{\widehat{e}( d(X_i)|X_i)}Y_i.
\end{eqnarray*}

For a given partition $\mathcal{P}$, let $\mathcal{D}_{\mathcal{P}}$ denote the space of I2DRs that we consider. Then $\widehat{d}$ can be computed by maximizing
\begin{eqnarray*}
	&&\argmax_{ \mathcal{P}\in \mathcal{B}(m)} \argmax_{d\in \mathcal{D}_{\mathcal{P}}} \frac{1}{n}\sum_{i=1}^n  \frac{\mathbb{I}(A_i\in d(X_i))}{\widehat{e}( d(X_i)|X_i)}Y_i\\&=&\argmax_{ \mathcal{P}\in \mathcal{B}(m)} \argmax_{d\in \mathcal{D}_{\mathcal{P}}} \frac{1}{n}\sum_{i=1}^n \sum_{\mathcal{I}\in \mathcal{P}} \frac{\mathbb{I}(A_i\in \mathcal{I})}{\widehat{e}( d(X_i)|X_i)}Y_i\mathbb{I}(d(X_i)=\mathcal{I}).
\end{eqnarray*}
Suppose we consider the class of linear decision rules, i.e., 
\begin{eqnarray*}
	\mathcal{D}_{\mathcal{P}}=\{d:d(x)=\argmax_{\mathcal{I}\in \mathcal{P}} \theta_{\mathcal{I}}^\top \bar{x}\}.
\end{eqnarray*}
It suffices to maximize
\begin{eqnarray*}
	\argmax_{ \substack{ \mathcal{P}\in \mathcal{B}(m) \\ \{q_\mathcal{I} \in \mathcal{Q}_{\mathcal{I}}: \mathcal{I}\in \mathcal{P} \}} } \frac{1}{n}\sum_{i=1}^n \sum_{\mathcal{I}\in \mathcal{P}} \frac{\mathbb{I}(A_i\in \mathcal{I})}{\widehat{e}( d(X_i)|X_i)}Y_i\mathbb{I}\{d(X_i)=\argmax_{\mathcal{I}\in \mathcal{P}} q_{\mathcal{I}} ({X}_i)\}.
\end{eqnarray*}
Similar to Section \ref{secalearning}, for a given partition $\mathcal{P}$, all the functions $\{q_{\mathcal{I}}:\mathcal{I}\in \mathcal{P}\}$ need to be jointed estimated. As a result, dynamic programming cannot be applied. It remains unknown how to efficiently solve the above optimization problem. We leave it for future research.

\change{
\subsection{Other approaches}
Recently, \cite{meng2020near} developed set-valued decision rules that contain equally beneficial treatments, borrowing ideas from multicategory classification with reject and refine options. Their method is developed under a discrete treatment setting with finitely many treatment options. It remains unclear whether it can be extended to our continuous treatment setting or not. We leave it for future research. 

Next, in addition to the jump penalized regression formulation, one can alternatively consider the following constrained optimization function that directly restricts the number of estimated intervals to smaller than or equal to some integer $M$, 
\begin{eqnarray*}
	(\widehat{\mathcal{P}},\{\widehat{\theta}_{\mathcal{I}}:\mathcal{I}\in \widehat{\mathcal{P}} \})=
	\argmin_{(|\mathcal{P}|\le M,\{\theta_{\mathcal{I}}: \mathcal{I}\in \mathcal{P} \} )} \sum_{\mathcal{I}\in \mathcal{P}} \left(\frac{1}{n}\sum_{i=1}^n \mathbb{I}(A_i\in \mathcal{I}) (Y_i-q(X_i; \theta_{\mathcal{I}})^2+\lambda_n |\mathcal{I}| \|\theta_{\mathcal{I}}\|_2^2\right).
\end{eqnarray*}
It would be interesting to further investigate the theoretical and numerical properties of this method. However, this is beyond the scope of the current paper and we leave it for future research.

Finally, our estimated partition is independent of the patient's baseline information. It might be practically more useful to consider patient-specific partitions and allow $\widehat{\mathcal{P}}$ to be a function of the baseline information. We leave it for future research.}

\subsection{Other penalty functions in L-JIL}

In L-JIL, We use a ridge penalty in \eqref{optimize_ridge} to prevent overfitting in large $p$ problems. When the true regression coefficient $\theta_0(\cdot)$ is sufficiently sparse, one can consider replacing the ridge penalty with the LASSO \cite{Tibs96} to improve the estimation accuracy. However, optimizing the resulting objective function requires to compute the LASSO estimator $m(m-1)/2$ times. This is far more computationally expensive than the proposed method. 
It remains unknown whether the computation can be simplified. Finally, 
We leave it for future research.

\bibliographystyle{agsm}
\bibliography{mycite} 

\newpage
\appendix
This appendix is organized as follows. In Section \ref{app_tune_LJIL}, we discuss more details on tuning parameters in L-JIL.
Technical proofs are given in Section \ref{sectechproof}. 

\section{More on Tuning in L-JIL}\label{app_tune_LJIL}

For L-JIL, we choose $\gamma_n$ and $\lambda_n$ simultaneously via cross-validation. As we will show below, the use of cross-validation will not increase the computation complexity substantially in L-JIL. 

To elaborate, let us revisit the proposed jump interval-learning in Algorithm 1. The most time consuming part lies in computing the ridge-type estimator
\begin{eqnarray}\label{linearequation}
	~~~~~~\widehat{\theta}_{\mathcal{I}}(\lambda_n)=\left(\sum_{i\in \mathbb{G}_{-k}} \overline{X}_i \overline{X}_i^\top \mathbb{I}(A_i\in \mathcal{I}) + n\lambda_n |\mathcal{I}| \mathbb{E}_{p+1} \right)^{-1} \left(\sum_{i\in \mathbb{G}_{-k}} \overline{X}_i Y_i\mathbb{I}(A_i\in \mathcal{I}) \right),
\end{eqnarray}
where $\mathbb{E}_{p+1}$ is the identity matrix with dimension $p+1$, and the cost function
\begin{eqnarray*}
	\hbox{cost}(\mathcal{I},\lambda_n)=\frac{1}{n} \sum_{i\in \mathbb{G}_{k}} \mathbb{I}(A_i\in \mathcal{I}) \left\{Y_i-\overline{X}_i^\top \widehat{\theta}_{\mathcal{I}}(\lambda_n)\right\}^2 ,
\end{eqnarray*}
for any $\mathcal{I}\in \{[l/m,r/m): 1\le l<r<m  \}\cup \{[l/m,1]: 1\le l<m \}$.

To compute $\{\widehat{\theta}_{\mathcal{I},\gamma_n,\lambda_n,k}:\gamma_n \in \Gamma_n, \lambda_n\in \Lambda_n\}$, we need to calculate $\{\widehat{\theta}_{\mathcal{I}}(\lambda_n):\lambda_n\in \Lambda_n\}$ and $\{\hbox{cost}(\mathcal{I},\lambda_n): \lambda_n\in \Lambda_n\}$ for any $\mathcal{I}$. We first factorize the matrix $\sum_{i\in \mathbb{G}_{-k}} \overline{X}_i \overline{X}_i^\top \mathbb{I}(A_i\in \mathcal{I})$ as
\begin{eqnarray*}
	\sum_{i\in \mathbb{G}_{-k}} \overline{X}_i \overline{X}_i^\top \mathbb{I}(A_i\in \mathcal{I})=U \mathcal{T} U^\top,
\end{eqnarray*}
according to the eigendecomposition, where $U$ is some $(p+1)\times (p+1)$ orthogonal matrix and $\mathcal{T}=\diag(\tau_0,\tau_1,\cdots,\tau_p)$ is some diagonal matrix. Let $\phi=U^\top \{\sum_{i\in \mathbb{G}_{-k}} \overline{X}_i Y_i\mathbb{I}(A_i\in \mathcal{I})\}$. Then the set of estimators $\{\widehat{\theta}_{\mathcal{I}}(\lambda_n):\lambda\in \Lambda_n\}$ can be calculated by
\begin{eqnarray*}
	\widehat{\theta}_{\mathcal{I}}(\lambda_n)=U \diag\left\{ (\tau_0+n\lambda_n |\mathcal{I}|)^{-1},(\tau_1+n\lambda_n |\mathcal{I}|)^{-1},\cdots,(\tau_p+n\lambda_n |\mathcal{I}|)^{-1} \right\}\phi,
\end{eqnarray*}
simultaneously for all $\lambda_n$.

Compared to separately inverting the matrix $\sum_{i\in \mathbb{G}_{-k}} \overline{X}_i \overline{X}_i^\top \mathbb{I}(A_i\in \mathcal{I}) + n\lambda_n |\mathcal{I}| \mathbb{E}_{p+1}$ in  \eqref{linearequation} for each $\lambda_n$ to compute $\{\widehat{\theta}_{\mathcal{I}}(\lambda_n):\lambda_n\in \Lambda_n\}$, the proposed method saves a lot of time especially for large $p$. Similarly, based on the eigendecomposition, we have
\begin{eqnarray}\label{cost} 
~~~	~~~~~~ n\hbox{cost}&&(\mathcal{I},\lambda_n)=\sum_{i\in \mathbb{G}_{-k}} Y_i^2\mathbb{I}(A_i\in \mathcal{I})\\\nonumber
	&&-2\phi^\top \diag\left\{ (\tau_0+n\lambda_n |\mathcal{I}|)^{-1},(\tau_1+n\lambda_n |\mathcal{I}|)^{-1},\cdots,(\tau_p+n\lambda_n |\mathcal{I}|)^{-1} \right\}\phi\\\nonumber
	&&+\phi^\top \diag\left\{ \tau_0(\tau_0+n\lambda_n |\mathcal{I}|)^{-2},\tau_1(\tau_1+n\lambda_n |\mathcal{I}|)^{-2},\cdots,\tau_p(\tau_p+n\lambda_n |\mathcal{I}|)^{-2} \right\}\phi,
\end{eqnarray}
for all $\lambda_n\in \Lambda_n$. This facilitates the computation of $\{\hbox{cost}(\mathcal{I},\lambda_n): \lambda_n\in \Lambda_n\}$.

After obtaining these cost functions, we can recursively compute the Bellman function $B(r,\lambda_n,\gamma_n)$ by
\begin{eqnarray*}
	B(r,\lambda_n,\gamma_n)=\min_{j\in \mathcal{R}_{r}} \left\{B(j,\lambda_n,\gamma_n)+\gamma_n+\hbox{cost}([j/m,r/m),\lambda_n)\right\}, 
\end{eqnarray*} 
for all $r\ge 1$, $\lambda_n\in \Lambda_n$ and $\gamma_n\in \Gamma_n$. Given the Bellman function, the set of estimators $\{\widehat{\theta}_{\mathcal{I},\gamma_n,\lambda_n,k}:\gamma_n \in \Gamma_n, \lambda_n\in \Lambda_n\}$ thus can be computed efficiently.

\section{Technical proofs}\label{sectechproof}
In the proofs, we use $c,C>0$ to denote some universal constants whose values are allowed to change from place to place. 
For any vector $\phi\in \mathbb{R}^{q}$, we use $\phi^{(j)}$ to denote the $j$-th element of $\phi$, for any $j\in \{1,\dots,q\}$. For any two positive sequences $\{a_n\}$, $\{b_n\}$, $a_n\propto b_n$ means that $a_n\le cb_n$ for some universal constant $c>0$

\subsection{Proof of Theorem 1}\label{secproofthm1}
We provide the proof for Theorem \hyperlink{thm1}{1} in this section. 
%
%
%
%
%
%
%
We present an outline of the proof first. {Let $\delta_{\min}=\min_{\mathcal{I}\in \mathcal{P}_0} |\mathcal{I}|/3>0$.} We divide the proof into four parts. In Part 1, we show that the following event occurs with probability at least $1-O(n^{-2})$,
\begin{eqnarray}\label{locationconv}
\max_{\tau\in J(\mathcal{P}_0)} \min_{\hat{\tau}\in J(\widehat{\mathcal{P}})} |\hat{\tau}-\tau|< \delta_{\min}.
\end{eqnarray}
By the definition of $\delta_{\min}$, this implies that
\begin{eqnarray}\label{underfit}
\hbox{Pr}(|\widehat{\mathcal{P}}|\ge |\mathcal{P}_0|)\ge 1-O(n^{-2}). 
\end{eqnarray}
In Part 2, we show that 
\begin{eqnarray}\label{locationlocalconv}
\max_{\tau\in J(\mathcal{P}_0)} \min_{\hat{\tau}\in J(\widehat{\mathcal{P}})} |\hat{\tau}-\tau|=O(n^{-1}\log n),
\end{eqnarray}
with probability at least $1-O(n^{-2})$. This proves (ii) in Theorem \hyperlink{thm1}{1}. In Part 3, we prove
\begin{eqnarray}\label{overfit}
\hbox{Pr}(|\widehat{\mathcal{P}}|\le |\mathcal{P}_0|)\ge 1-O(n^{-2}). 
\end{eqnarray}
This together with \eqref{underfit} proves (i) in Theorem \hyperlink{thm1}{1}. In the last part, we show (iii) holds.

In the following, we first introduce some notations and auxiliary lemmas. Then, we present the proofs for Part 1, 2, 3 and 4.

\textit{Notations and technical lemmas:} For any interval $\mathcal{I}\subseteq [0,1]$, define
\begin{eqnarray*}
	&&\widehat{\theta}_{\mathcal{I}}=\left( \frac{1}{n}\sum_{i=1}^n \mathbb{I}(A_i\in \mathcal{I})\overline{X}_i \overline{X}_i^\top+\lambda_n |\mathcal{I}| \mathbb{E}_{p+1} \right)^{-1}\left( \frac{1}{n}\sum_{i=1}^n \mathbb{I}(A_i\in \mathcal{I})\overline{X}_i Y_i \right),\\
	&&\theta_{0,\mathcal{I}}=\left(\Mean \mathbb{I}(A\in \mathcal{I}) \overline{X} \overline{X}^\top\right)^{-1} \{\Mean \mathbb{I}(A\in \mathcal{I}) \overline{X} Y\},
\end{eqnarray*}
where $\overline{X}=(1,X^\top)^\top$. It is immediate to see that the definition of $\widehat{\theta}_{\mathcal{I}}$ here is consistent with the one defined in \eqref{optimize} for any $\mathcal{I}\in \widehat{\mathcal{P}}$. In addition, under the model assumption in \eqref{md1}, the definition of $\theta_{0,\mathcal{I}}$ here is consistent with the one defined in step function model $\theta_0(a)=\sum_{\mathcal{I}\in \mathcal{P}_0} \theta_{0,\mathcal{I}}\mathbb{I}(a\in \mathcal{I})$ for any $\mathcal{I}\in \mathcal{P}_0$. 

Let $\mathfrak{I}(m)$ denote the set of intervals
\begin{eqnarray*}
	\mathfrak{I}(m)&=&\{ [i_1/m,i_2/m):\hbox{for~some~integers}~i_1\hbox{~and~}i_2~\hbox{that~satisfy}~0\le i_1<i_2<m\}\\
	&\cup& \{ [i_3/m,1]:\hbox{for~some~integers}~i_3~\hbox{that~satisfy}~0\le i_3<m\}.
\end{eqnarray*}
Let $\{\tau_{0,k}\}_{k=1}^{K-1}$ with $0<\tau_{0,1}<\tau_{0,2}<\cdots<\tau_{0,K-1}<1$ be the locations of the true change points of $\theta_0(\cdot)$. Set $\tau_{0,0}=0$, $\tau_{0,K}=1$. We introducing the following lemmas.

\smallskip

\noindent
{\bf \hypertarget{lemma1}{Lemma 1}} {\it	Assume conditions in Theorem \hyperlink{thm1}{1} are satisfied. Then there exist some constants $\bar{c}_0>0$, $c_0\ge 1$ such that the following events occur with probability at least $1-O(n^{-2})$: for any interval $\mathcal{I}\in \mathfrak{I}(m)$ that satisfies $|\mathcal{I}|\ge \bar{c}_0 n^{-1}\log n$, we have
	\begin{align}\label{event1}
	&\|\widehat{\theta}_{\mathcal{I}}-\theta_{0,\mathcal{I}}\|_2\le \frac{c_0\sqrt{\log n}}{\sqrt{|\mathcal{I}|n}},\\ \label{event2}
	&\left\|\frac{1}{n}\sum_{i=1}^{n}\mathbb{I}(A_i\in \mathcal{I})(Y_i-\overline{X}_i^\top\theta_{0,\mathcal{I}})\overline{X}_i\right\|_2\le \frac{c_0\sqrt{|\mathcal{I}|\log n}}{\sqrt{n}},\\ \label{event3}
	&\left|\frac{1}{n}\sum_{i=1}^n \mathbb{I}(A_i\in\mathcal{I})\{Y_i-\overline{X}^\top\theta_0(A_i)\}\overline{X}_i^\top \{\theta_0(A_i)-\theta_{0,\mathcal{I}}\} \right|\le \frac{c_0\sqrt{|\mathcal{I}|\log n}}{\sqrt{n}},
\\
	\label{event4}
	 &\frac{1}{n}\sum_{i=1}^n \mathbb{I}(A_i\in \mathcal{I})[\overline{X}_i^\top \{\theta_0(A_i)-\theta_{0,\mathcal{I}}\}]^2\ge \frac{1}{c_0} \int_{\mathcal{I}} \|\theta_0(a)-\theta_{0,\mathcal{I}}\|_2^2da- \frac{c_0\sqrt{|\mathcal{I}|\log n}}{\sqrt{n}},\\ \label{event5}
	&\frac{1}{n}\sum_{i=1}^n \mathbb{I}(A_i\in \mathcal{I})(|Y_i|^2+\|\overline{X}_i\|_2^2) \le c_0\left( \frac{\sqrt{|\mathcal{I}|\log n}}{\sqrt{n}}+|\mathcal{I}| \right).  
	\end{align}   
	In addition, for any $\mathcal{I}\in [0,1]$, we have
	\begin{eqnarray}\label{boundbeta0I}
	\|\theta_{0,\mathcal{I}}\|_2\le c_0.
	\end{eqnarray}} \hfill$\square$

\smallskip
\noindent
{\bf \hypertarget{lemma2}{Lemma 2}}  {\it	Assume conditions in Theorem 1 are satisfied. Then there exist some constants $\bar{c}_1>0,c_1\ge 1$ such that the following events occur with probability at least $1-O(n^{-2})$: for any interval $\mathcal{I}\in \mathfrak{I}(m)$ that satisfies $\int_{\mathcal{I}} \|\theta_0(a)-\theta_{0,\mathcal{I}}\|_2^2da\ge \bar{c}_1 n^{-1}\log n$, 
	\begin{eqnarray}\nonumber
	&&\left|\sum_{i=1}^n \mathbb{I}(A_i\in\mathcal{I})\{Y_i-\overline{X}^\top_i\theta_0(A_i)\}\overline{X}_i^\top \{\theta_0(A_i)-\theta_{0,\mathcal{I}}\} \right|\\\label{event6}
	&\le& c_1 \sqrt{n\int_{\mathcal{I}} \|\theta_0(a)-\theta_{0,\mathcal{I}}\|_2^2da \log n},\\ \nonumber
	&&\sum_{i=1}^n \mathbb{I}(A_i\in \mathcal{I})[\overline{X}_i^\top \{\theta_0(A_i)-\theta_{0,\mathcal{I}}\}]^2\\ \label{event7}&\ge& 
	\frac{n}{c_1} \int_{\mathcal{I}} \|\theta_0(a)-\theta_{0,\mathcal{I}}\|_2^2da- c_1\sqrt{n\int_{\mathcal{I}} \|\theta_0(a)-\theta_{0,\mathcal{I}}\|_2^2da \log n}.
	\end{eqnarray}} \hfill$\square$

\smallskip

\noindent
{\bf \hypertarget{lemma3}{Lemma 3}} {\it		Assume conditions in Theorem 1 are satisfied. Then for sufficiently large $n$ and any interval $\mathcal{I}\subseteq [0,1]$ of the form $[i_1,i_2)$ or $[i_1,i_2]$ with $i_2=1$ that satisfies $\int_{\mathcal{I}} \|\theta_0(a)-\theta_{0,\mathcal{I}}\|_2^2da=c_n$ for some sequence $\{c_n\}_n$ such that $c_n\ge 0, \forall n$ and $c_n \to 0$ as $n\to \infty$, we have either $\tau_{0,k-1}\le i_1\le i_2\le \tau_{0,k}$ for some integer $k$ such that $1\le k\le K$ or 
	\begin{eqnarray*}
		\tau_{0,k-2}\le i_1<\tau_{0,k-1} <i_2\le \tau_{0,k}\,\,\,\,\hbox{and}\,\,\,\,\min_{j\in \{1,2\}} |i_j-\tau_{0,k-1}|\le c_2 c_n,
	\end{eqnarray*}
	for some integer $k$ such that $2\le k\le K$ and some constant $c_2>0$, or 
	\begin{eqnarray*}
		\tau_{0,k-3}\le i_1<\tau_{0,k-2}<\tau_{0,k-1} <i_2\le \tau_{0,k}\,\,\,\,\hbox{and}\,\,\,\,\max_{j\in\{1,2\}} |i_j-\tau_{0,k-3+j}|\le \bar{c}_2 c_n,
	\end{eqnarray*}
	for some integer $k$ such that $3\le k\le K$ and some constant $c_2>0$. 
	
	In addition, the following events occur with probability at least $1-O(n^{-2})$: for any interval $\mathcal{I}\in \mathfrak{I}(m)$ that satisfies $\int_{\mathcal{I}} \|\theta_0(a)-\theta_{0,\mathcal{I}}\|_2^2da\le \bar{c}_1 n^{-1}\log n$, we have
	\begin{eqnarray}\label{event8}
	\left|\sum_{i=1}^n \mathbb{I}(A_i\in\mathcal{I})\{Y_i-\overline{X}^\top_i\theta_0(A_i)\}\overline{X}_i^\top \{\theta_0(A_i)-\theta_{0,\mathcal{I}}\} \right|\le \bar{c}_2 \log n,
	\end{eqnarray}
	for some constant $\bar{c}_2>0$.} \hfill$\square$

\smallskip

\noindent
{\bf \hypertarget{lemma5}{Lemma 4}}  {\it		Under the conditions in Theorem \hyperlink{thm1}{1}, the following events occur with probability at least $1-O(n^{-2})$: there exists some constant $\bar{c}_3>0$ such that $\min_{\mathcal{I}\in \widehat{\mathcal{P}}}|\mathcal{I}|\ge \bar{c}_3 \gamma_n$. } \hfill$\square$

\smallskip

\noindent \textit{Part 1:} Assume $|\mathcal{P}_0|>1$. Otherwise, \eqref{locationconv} trivially hold. 
Consider the partition $\mathcal{P}=\{[0,1]\}$ which consists of a single interval and a zero vector $\bm{0}_{p+1}$. By definition, we have
\begin{eqnarray*}
	&&\sum_{\mathcal{I}\in \widehat{\mathcal{P}}} \left( \sum_{i=1}^n \mathbb{I}(A_i\in \mathcal{I})(Y_i-\overline{X}_i^\top \widehat{\theta}_{\mathcal{I}})^2+n\lambda_n |\mathcal{I}|\|\widehat{\theta}_{\mathcal{I}}\|_2^2 \right)+n\gamma_n|\widehat{\mathcal{P}}|\\
	&\le& \sum_{i=1}^n (Y_i-\overline{X}_i^\top \bm{0}_{p+1})^2+n\lambda_n \|\bm{0}_{p+1}\|_2^2 +n\gamma_n=\sum_{i=1}^n Y_i^2+ n\gamma_n.
\end{eqnarray*}
In view of \eqref{event5}, we obtain with probability at least $1-O(n^{-2})$,
\begin{eqnarray*}
	\sum_{\mathcal{I}\in \widehat{\mathcal{P}}} \left( \sum_{i=1}^n \mathbb{I}(A_i\in \mathcal{I})(Y_i-\overline{X}_i^\top \widehat{\theta}_{\mathcal{I}})^2+n\lambda_n |\mathcal{I}|\|\widehat{\theta}_{\mathcal{I}}\|_2^2 \right)+n\gamma_n(|\widehat{\mathcal{P}}|-1)\le c_0n\left( \frac{\sqrt{\log n}}{\sqrt{n}}+1 \right).
\end{eqnarray*}
This implies that under the event defined in \eqref{event5}, we have for sufficiently large $n$,
\begin{eqnarray*}
	\gamma_n (|\widehat{\mathcal{P}}|-1)\le c_0 \left( \frac{\sqrt{\log n}}{\sqrt{n}}+1 \right),
\end{eqnarray*}
and hence
\begin{eqnarray}\label{upperPhat}
|\widehat{\mathcal{P}}|\le 2c_0\gamma_n^{-1},
\end{eqnarray}
for sufficiently large $n$.

Under the event defined in Lemma \hyperlink{lemma5}{4}, we have $\min_{\mathcal{I}\in \widehat{\mathcal{P}}} |\mathcal{I}|\ge \bar{c}_0 n^{-1}\log n$ for sufficiently large $n$, since $\gamma_n\gg n^{-1}\log n$. Thus, with probability at least $1-O(n^{-2})$, the events defined in \eqref{event1}-\eqref{event5} hold for any interval $\mathcal{I}\in \widehat{\mathcal{P}}$.

Notice that
\begin{eqnarray}\label{someinequality1}
&&~~~~~~~~~~~~~\sum_{\mathcal{I}\in \widehat{\mathcal{P}}} \left( \sum_{i=1}^n \mathbb{I}(A_i\in \mathcal{I})(Y_i-\overline{X}_i^\top \widehat{\theta}_{\mathcal{I}})^2+n\lambda_n |\mathcal{I}|\|\widehat{\theta}_{\mathcal{I}}\|_2^2 \right)+n\gamma_n|\widehat{\mathcal{P}}|\\ \nonumber
&\ge&\sum_{\substack{\mathcal{I}\in \widehat{\mathcal{P}} }} \sum_{i=1}^n \mathbb{I}(A_i\in \mathcal{I})(Y_i-\overline{X}_i^\top \widehat{\theta}_{\mathcal{I}})^2+n\gamma_n|\widehat{\mathcal{P}}|\ge n\gamma_n|\widehat{\mathcal{P}}|+ \underbrace{ \sum_{\substack{\mathcal{I}\in \widehat{\mathcal{P}}} }\sum_{i=1}^n  \mathbb{I}(A_i\in \mathcal{I})(Y_i-\overline{X}_i^\top \theta_{0,\mathcal{I}})^2}_{\eta_1}\\ \nonumber
&+&\underbrace{\sum_{\substack{\mathcal{I}\in \widehat{\mathcal{P}} }}\sum_{i=1}^n \mathbb{I}(A_i\in \mathcal{I}) \{\overline{X}_i^\top (\widehat{\theta}_{\mathcal{I}}-\theta_{0,\mathcal{I}})\}^2}_{\eta_2}
-\underbrace{2\sum_{\substack{\mathcal{I}\in \widehat{\mathcal{P}} }}\left|\sum_{i=1}^n \mathbb{I}(A_i\in \mathcal{I})(Y_i-\overline{X}_i^\top \theta_{0,\mathcal{I}}) \overline{X}_i^\top (\widehat{\theta}_{\mathcal{I}}-\theta_{0,\mathcal{I}}) \right|}_{\eta_3}.
\end{eqnarray}
By \eqref{event1} and \eqref{event2}, we obtain that
\begin{eqnarray}\label{someinequality2}
\eta_3&\le& 2\sum_{\substack{\mathcal{I}\in \widehat{\mathcal{P}} }}\left\|\sum_{i=1}^n \mathbb{I}(A_i\in \mathcal{I})(Y_i-\overline{X}_i^\top \theta_{0,\mathcal{I}}) \overline{X}_i\right\|_2 \|\widehat{\theta}_{\mathcal{I}}-\theta_{0,\mathcal{I}}\|_2\\\nonumber
&\le&\sum_{\substack{\mathcal{I}\in \widehat{\mathcal{P}} }} 2c_0^2 \log n\le 2c_0^2\log n|\widehat{\mathcal{P}}|,
\end{eqnarray}
with probability at least $1-O(n^{-2})$. 
Since $\gamma_n\gg n^{-1}\log n$, $\eta_2\ge 0$, for sufficiently large $n$, we have with probability at least $1-O(n^{-2})$, 
\begin{eqnarray}\label{lowerboundeq1}
\sum_{\mathcal{I}\in \widehat{\mathcal{P}}} \left( \sum_{i=1}^n \mathbb{I}(A_i\in \mathcal{I})(Y_i-\overline{X}_i^\top \widehat{\theta}_{\mathcal{I}})^2+n\lambda_n |\mathcal{I}|\|\widehat{\theta}_{\mathcal{I}}\|_2^2 \right)+n\gamma_n|\widehat{\mathcal{P}}|\ge \eta_1.
\end{eqnarray}

Notice that
\begin{eqnarray*}
	&&\eta_1=\sum_{\substack{\mathcal{I}\in \widehat{\mathcal{P}}}}\sum_{i=1}^n \mathbb{I}(A_i\in \mathcal{I}) \{Y_i- \overline{X}_i^\top \theta_0(A_i)+\overline{X}_i^\top \theta_0(A_i)-\overline{X}_i^\top \theta_{0,\mathcal{I}}\}^2\\
	&=&\underbrace{\sum_{\substack{\mathcal{I}\in \widehat{\mathcal{P}}}}\sum_{i=1}^n \mathbb{I}(A_i\in \mathcal{I}) \{Y_i- \overline{X}_i^\top \theta_0(A_i)\}^2}_{\eta_4}+\sum_{\substack{\mathcal{I}\in \widehat{\mathcal{P}}}}\sum_{i=1}^n \mathbb{I}(A_i\in \mathcal{I}) \{\overline{X}_i^\top \theta_0(A_i)- \overline{X}_i^\top \theta_{0,\mathcal{I}}\}^2\\
	&+&2\sum_{\substack{\mathcal{I}\in \widehat{\mathcal{P}}}}\sum_{i=1}^n \mathbb{I}(A_i\in \mathcal{I}) \{Y_i- \overline{X}_i^\top \theta_0(A_i)\} \{\overline{X}_i^\top \theta_0(A_i)-\overline{X}_i^\top \theta_{0,\mathcal{I}}\}.
\end{eqnarray*}
Under the events defined in \eqref{event3} and \eqref{event4}, it follows that
\begin{eqnarray}\nonumber
\eta_1&\ge& \eta_4+n\sum_{\substack{\mathcal{I}\in \widehat{\mathcal{P}}}} \frac{1}{c_0} \int_{\mathcal{I}} \|\theta_0(a)-\theta_{0,\mathcal{I}}\|_2^2da-2c_0\sum_{\substack{\mathcal{I}\in \widehat{\mathcal{P}}}}\sqrt{|\mathcal{I}|n\log n}\\ \nonumber
&\ge & \eta_4+n\sum_{\substack{\mathcal{I}\in \widehat{\mathcal{P}}}} \frac{1}{c_0} \int_{\mathcal{I}} \|\theta_0(a)-\theta_{0,\mathcal{I}}\|_2^2da-2c_0 \sqrt{|\widehat{\mathcal{P}}|n\log n},
\end{eqnarray}
where the last inequality is due to Cauchy-Schwarz inequality. By \eqref{upperPhat} and the condition that $\gamma_n\gg n^{-1}\log n$, we obtain
\begin{eqnarray}\label{lowerboundeq1.5}
\eta_1\ge \eta_4+n\sum_{\substack{\mathcal{I}\in \widehat{\mathcal{P}}}} \frac{1}{c_0} \int_{\mathcal{I}} \|\theta_0(a)-\theta_{0,\mathcal{I}}\|_2^2da+o(n),
\end{eqnarray}
with probability at least $1-O(n^{-2})$. 
Notice that
\begin{eqnarray*}
	\eta_4=\sum_{i=1}^n \{Y_i-\overline{X}_i^\top \theta_0(A_i) \}^2
	=\sum_{\mathcal{I}\in \mathcal{P}_0} \sum_{i=1}^n \mathbb{I}(A_i\in \mathcal{I}) (Y_i-\overline{X}_i^\top \theta_{0,\mathcal{I}} )^2.
\end{eqnarray*}
Combining \eqref{lowerboundeq1} with \eqref{lowerboundeq1.5}, 
we've shown that
\begin{eqnarray*}
	&&\sum_{\mathcal{I}\in \widehat{\mathcal{P}}} \left( \sum_{i=1}^n \mathbb{I}(A_i\in \mathcal{I})(Y_i-\overline{X}_i^\top \widehat{\theta}_{\mathcal{I}})^2+n\lambda_n |\mathcal{I}|\|\widehat{\theta}_{\mathcal{I}}\|_2^2 \right)+n\gamma_n|\widehat{\mathcal{P}}|\\
	&\ge&\sum_{\mathcal{I}\in \mathcal{P}_0} \sum_{i=1}^n \mathbb{I}(A_i\in \mathcal{I}) (Y_i-\overline{X}_i^\top \theta_{0,\mathcal{I}} )^2+n\sum_{\substack{\mathcal{I}\in \widehat{\mathcal{P}}}} \frac{1}{c_0} \int_{\mathcal{I}} \|\theta_0(a)-\theta_{0,\mathcal{I}}\|_2^2da+o(n),
\end{eqnarray*}
with probability at least $1-O(n^{-2})$. By \eqref{boundbeta0I} and the condition that $\lambda_n=O(n^{-1}\log n)$, $\gamma_n=o(1)$, this further implies
\begin{eqnarray}\nonumber
&&\sum_{\mathcal{I}\in \widehat{\mathcal{P}}} \left( \sum_{i=1}^n \mathbb{I}(A_i\in \mathcal{I})(Y_i-\overline{X}_i^\top \widehat{\theta}_{\mathcal{I}})^2+n\lambda_n |\mathcal{I}|\|\widehat{\theta}_{\mathcal{I}}\|_2^2 \right)+n\gamma_n|\widehat{\mathcal{P}}|\\ \nonumber
&\ge&\sum_{\mathcal{I}\in \mathcal{P}_0} \left(\sum_{i=1}^n \mathbb{I}(A_i\in \mathcal{I}) (Y_i-\overline{X}_i^\top \theta_{0,\mathcal{I}} )^2+n\lambda_n|\mathcal{I}| \|\theta_{0,\mathcal{I}}\|_2^2\right)+n\gamma_n |\mathcal{P}_0|\\ \label{lowerboundeq1.7}
&+&n\sum_{\substack{\mathcal{I}\in \widehat{\mathcal{P}}}} \frac{1}{c_0} \int_{\mathcal{I}} \|\theta_0(a)-\theta_{0,\mathcal{I}}\|_2^2da+o(n).
\end{eqnarray}

For any integer $k$ such that $1\le k\le K-1$, let $\tau_{0,k}^*$ be the change point location that satisfies $\tau_{0,k}^*=i/m$ for some integer $i$ and that $|\tau_{0,k}-\tau_{0,k}^*|<m^{-1}$. Denoted by $\mathcal{P}^*$ the oracle partition formed by the change point locations $\{\tau_{0,k}^*\}_{k=1}^{K-1}$. Set $\tau_{0,0}^*=0$, $\tau_{0,K}^*=1$ and $\theta_{[\tau_{0,k-1}^*,\tau_{0,k}^*)}^*=\theta_{0,[\tau_{0,k-1},\tau_{0,k})}$ for $1\le k\le K-1$ and $\theta_{[\tau_{0,K-1}^*,1]}^*=\theta_{0,[\tau_{0,K-1},1]}$. Let $\Delta_k=[\tau_{0,k-1}^*,\tau_{0,k}^*) \cap [\tau_{0,k-1},\tau_{0,k})^c$ for $1\le k\le K-1$ and $\Delta_{K}=[\tau_{0,K-1}^*,1] \cap [\tau_{0,K-1},1]^c$. The length of each interval $\Delta_k$ is at most $m^{-1}$. Since $m\asymp n$, we have $m^{-1}\ll \bar{c}_0 n^{-1} \log n$. For any $k$ and sufficiently large $n$, we can find an interval $\mathcal{I}\in \mathfrak{I}(m)$ with length between $\bar{c}_0n^{-1}\log n$ and $2\bar{c}_0n^{-1}\log n$ that covers $\Delta_k$. It follows that
\begin{eqnarray}\label{lowerboundeq2}
&&~~~\left\{ \sum_{\mathcal{I}\in \mathcal{P}^*}\left( \sum_{i=1}^n \mathbb{I}(A_i\in \mathcal{I})(Y_i-\overline{X}_i^\top \theta_{\mathcal{I}}^*)^2+n\lambda_n |\mathcal{I}|\|\theta_{\mathcal{I}}^*\|_2^2 \right)+n\gamma_n |\mathcal{P}^*| \right\}\\ \nonumber
&-&\left\{ \sum_{\mathcal{I}\in \mathcal{P}_0}\left( \sum_{i=1}^n \mathbb{I}(A_i\in \mathcal{I})(Y_i-\overline{X}_i^\top \theta_{0,\mathcal{I}})^2+n\lambda_n |\mathcal{I}|\|\theta_{0,\mathcal{I}}\|_2^2 \right)+n\gamma_n |\mathcal{P}_0| \right\}\\ \nonumber
&\le& n\lambda_n \sup_{\mathcal{I}\subseteq [0,1]} \|\theta_{0,\mathcal{I}}\|_2^2+\sum_{k=1}^K \sum_{i=1}^n \mathbb{I}(A_i\in \Delta_k) \left(Y_i^2+\sup_{\mathcal{I}\subseteq [0,1]}\|\theta_{0,\mathcal{I}}\|_2^2 \|\overline{X}_i\|_2^2\right)\\ \nonumber
&\le& n\lambda_n \sup_{\mathcal{I}\subseteq [0,1]} \|\theta_{0,\mathcal{I}}\|_2^2+K \sup_{\substack{\mathcal{I}\in \mathfrak{I}(m)\\ 1\le \bar{c}_0^{-1} |\mathcal{I}|n  \log^{-1}n \le 2 } } \sum_{i=1}^n \mathbb{I}(A_i\in \mathcal{I})\left(Y_i^2+\sup_{\mathcal{I}\subseteq [0,1]}\|\theta_{0,\mathcal{I}}\|_2^2 \|\overline{X}_i\|_2^2\right).
\end{eqnarray} 
Since $\lambda_n=O(n^{-1}\log n)$, combining \eqref{lowerboundeq2} together with \eqref{event5} and \eqref{boundbeta0I}, we obtain with probability at least $1-O(n^{-2})$,
\begin{eqnarray}\nonumber
&&\left\{ \sum_{\mathcal{I}\in \mathcal{P}^*}\left( \sum_{i=1}^n \mathbb{I}(A_i\in \mathcal{I})(Y_i-\overline{X}_i^\top \theta_{\mathcal{I}}^*)^2+n\lambda_n |\mathcal{I}|\|\theta_{\mathcal{I}}^*\|_2^2 \right)+n\gamma_n |\mathcal{P}^*| \right\}\\\nonumber&-&\left\{ \sum_{\mathcal{I}\in \mathcal{P}_0}\left( \sum_{i=1}^n \mathbb{I}(A_i\in \mathcal{I})(Y_i-\overline{X}_i^\top \theta_{0,\mathcal{I}})^2+n\lambda_n |\mathcal{I}|\|\theta_{0,\mathcal{I}}\|_2^2 \right)+n\gamma_n |\mathcal{P}_0| \right\}
\\\label{lowerboundeq2.5}&\le& c_0^2 n \lambda_n + K(c_0^2+1) c_0 (\sqrt{2\bar{c}_0}+2\bar{c}_0) \log n=O(\log n)=o(n).
\end{eqnarray}
By definition, we have
\begin{eqnarray*}
	&&\sum_{\mathcal{I}\in \widehat{\mathcal{P}}} \left( \sum_{i=1}^n \mathbb{I}(A_i\in \mathcal{I})(Y_i-\overline{X}_i^\top \widehat{\theta}_{\mathcal{I}})^2+n\lambda_n |\mathcal{I}|\|\widehat{\theta}_{\mathcal{I}}\|_2^2 \right)+n\gamma_n|\widehat{\mathcal{P}}|\\
	&\le& \sum_{\mathcal{I}\in \mathcal{P}^*}\left( \sum_{i=1}^n \mathbb{I}(A_i\in \mathcal{I})(Y_i-\overline{X}_i^\top \theta_{\mathcal{I}}^*)^2+n\lambda_n |\mathcal{I}|\|\theta_{\mathcal{I}}^*\|_2^2 \right)+n\gamma_n |\mathcal{P}^*|.
\end{eqnarray*}
In view of \eqref{lowerboundeq1.7} and \eqref{lowerboundeq2.5}, we obtain that
\begin{eqnarray}\label{lowerboundeq3}
\sum_{\substack{\mathcal{I}\in \widehat{\mathcal{P}}}} \int_{\mathcal{I}} \|\theta_0(a)-\theta_{0,\mathcal{I}}\|_2^2da=o(1),
\end{eqnarray}
with probability at least $1-O(n^{-2})$. We now show \eqref{locationconv} holds under the event defined in \eqref{lowerboundeq3}. Otherwise, there exists some $\tau_0\in J(\mathcal{P}_0)$ such that $|\hat{\tau}-\tau_0|\ge \delta_{\min}$, for all $\hat{\tau}\in J(\widehat{\mathcal{P}})$. 
Under the event defined in \eqref{lowerboundeq3}, we obtain that
\begin{eqnarray}\label{lowerboundeq3.5}
\int_{\tau_0-\delta_{\min}}^{\tau_0+\delta_{\min}} \|\theta_0(a)-\theta_{0,[\tau_0-\delta_{\min}, \tau_0+\delta_{\min})}\|_2^2da=o(1).
\end{eqnarray}
On the other hand, since $\theta_0(a)$ is a constant function on $[\tau_0-\delta_{\min},\tau_0)$ or $[\tau_0, \tau_0+\delta_{\min})$, we have
\begin{eqnarray*}
	&&\int_{\tau_0-\delta_{\min}}^{\tau_0+\delta_{\min}} \|\theta_0(a)-\theta_{0,[\tau_0-\delta_{\min}, \tau_0+\delta_{\min})}\|_2^2da\\
	&\ge& \min_{\theta\in \mathbb{R}^{p+1}}\left(\delta_{\min} \|\theta_{0,[\tau_0-\delta_{\min},\tau_0)}-\theta\|_2^2+\delta_{\min} \|\theta_{0,[\tau_0,\tau_0+\delta_{\min})}-\theta\|_2^2\right)\\
	&\ge&\frac{\delta_{\min}}{2} \|\theta_{0,[\tau_0-\delta_{\min},\tau_0)}-\theta_{0,[\tau_0,\tau_0+\delta_{\min})}\|_2^2\ge \frac{\delta_{\min} \kappa_0^2}{2},
\end{eqnarray*}
where
\begin{eqnarray*}
	\kappa_0\equiv \min_{\substack{\mathcal{I}_1,\mathcal{I}_2\in \mathcal{P}_0\\ \mathcal{I}_1~\hbox{and}~\mathcal{I}_2~\hbox{are}~\hbox{adjacent} }} \|\theta_{0,\mathcal{I}_1}-\theta_{0,\mathcal{I}_2}\|_2>0.
\end{eqnarray*}
This apparently violates \eqref{lowerboundeq3.5}. \eqref{locationconv} thus holds with probability at least $1-O(n^{-2})$. 

\smallskip

\noindent \textit{Part 2:} By \eqref{someinequality1} and \eqref{someinequality2}, we have with probability at least $1-O(n^{-2})$ that
\begin{eqnarray*}
	\sum_{\mathcal{I}\in \widehat{\mathcal{P}}} \left( \sum_{i=1}^n \mathbb{I}(A_i\in \mathcal{I})(Y_i-\overline{X}_i^\top \widehat{\theta}_{\mathcal{I}})^2+n\lambda_n |\mathcal{I}|\|\widehat{\theta}_{\mathcal{I}}\|_2^2 \right)+n\gamma_n|\widehat{\mathcal{P}}|
	\ge \eta_1+n\gamma_n |\widehat{\mathcal{P}}|-2c_0^2 |\widehat{\mathcal{P}}|\log n.
\end{eqnarray*}
Notice that
\begin{eqnarray*}
	\eta_1&=&\eta_4+2\sum_{\substack{\mathcal{I}\in \widehat{\mathcal{P}}}}\sum_{i=1}^n \mathbb{I}(A_i\in \mathcal{I}) \{Y_i- \overline{X}_i^\top \theta_0(A_i)\} \{\overline{X}_i^\top \theta_0(A_i)-\overline{X}_i^\top \theta_{0,\mathcal{I}}\}\\
	&+&\sum_{\substack{\mathcal{I}\in \widehat{\mathcal{P}}}}\sum_{i=1}^n \mathbb{I}(A_i\in \mathcal{I}) \{\overline{X}_i^\top \theta_0(A_i)- \overline{X}_i^\top \theta_{0,\mathcal{I}}\}^2.
\end{eqnarray*}

Denoted by $\mathfrak{T}(m)$ the set of intervals $\mathcal{I}\in \mathfrak{I}(m)$ with $\int_{\mathcal{I}} \|\theta_0(a)-\theta_{0,\mathcal{I}}\|_2^2da\ge \bar{c}_1 n^{-1}\log n$. Under the events defined in Lemma \hyperlink{lemma2}{2} and \hyperlink{lemma3}{3}, we have
\begin{eqnarray*}
	\eta_1&\ge&\eta_4+2\sum_{\substack{\mathcal{I}\in \widehat{\mathcal{P}}}}\sum_{i=1}^n \mathbb{I}(A_i\in \mathcal{I}) \{Y_i- \overline{X}_i^\top \theta_0(A_i)\} \{\overline{X}_i^\top \theta_0(A_i)-\overline{X}_i^\top \theta_{0,\mathcal{I}}\}\\
	&+&\sum_{\substack{\mathcal{I}\in \widehat{\mathcal{P}}, \mathcal{I}\in \mathfrak{T}(m) }}\sum_{i=1}^n \mathbb{I}(A_i\in \mathcal{I}) \{\overline{X}_i^\top \theta_0(A_i)- \overline{X}_i^\top \theta_{0,\mathcal{I}}\}^2\ge \eta_4-2\bar{c}_2 |\widehat{\mathcal{P}}| \log n\\
	&+&\sum_{\substack{\mathcal{I}\in \widehat{\mathcal{P}}, \mathcal{I}\in \mathfrak{T}(m) }}\left(\frac{n}{c_1} \int_{\mathcal{I}} \|\theta_0(a)-\theta_{0,\mathcal{I}}\|_2^2da- 3c_1\sqrt{n\int_{\mathcal{I}} \|\theta_0(a)-\theta_{0,\mathcal{I}}\|_2^2da \log n}\right).
\end{eqnarray*}
To summarize, we've shown that with probability at least $1-O(n^{-2})$,
\begin{eqnarray}\label{someresults0}
&&\sum_{\mathcal{I}\in \widehat{\mathcal{P}}} \left( \sum_{i=1}^n \mathbb{I}(A_i\in \mathcal{I})(Y_i-\overline{X}_i^\top \widehat{\theta}_{\mathcal{I}})^2+n\lambda_n |\mathcal{I}|\|\widehat{\theta}_{\mathcal{I}}\|_2^2 \right)+n\gamma_n|\widehat{\mathcal{P}}|\\ \nonumber
&\ge& \sum_{\substack{\mathcal{I}\in \widehat{\mathcal{P}}, \mathcal{I}\in \mathfrak{T}(m) }}\left(\frac{n}{c_1} \int_{\mathcal{I}} \|\theta_0(a)-\theta_{0,\mathcal{I}}\|_2^2da- 3c_1\sqrt{n\int_{\mathcal{I}} \|\theta_0(a)-\theta_{0,\mathcal{I}}\|_2^2da \log n}\right)\\ \nonumber
&+& \eta_4+n\gamma_n |\widehat{\mathcal{P}}|-2(c_0^2+\bar{c}_2) |\widehat{\mathcal{P}}|\log n.
\end{eqnarray}
It follows from \eqref{lowerboundeq2} and \eqref{lowerboundeq2.5} that
\begin{eqnarray*}
	&&\eta_4+n\lambda_n \sup_{\mathcal{I}\in \mathcal{P}_0}\|\theta_{0,\mathcal{I}}\|_2^2 +n\gamma_n |\mathcal{P}_0|\\
	&\ge& \left\{ \sum_{\mathcal{I}\in \mathcal{P}^*}\left( \sum_{i=1}^n \mathbb{I}(A_i\in \mathcal{I})(Y_i-\overline{X}_i^\top \theta_{\mathcal{I}}^*)^2+n\lambda_n |\mathcal{I}|\|\theta_{\mathcal{I}}^*\|_2^2 \right)+n\gamma_n |\mathcal{P}^*| \right\}-c_0^*\log n,
\end{eqnarray*}
for some constants $c_0^*>0$, with probability at least $1-O(n^{-2})$. By \eqref{boundbeta0I} and the condition that $\lambda_n=O(n^{-1}\log n)$, there exists some constant $c_1^*>c_0^*$ such that
\begin{eqnarray*}
	&&\eta_4 +n\gamma_n |\mathcal{P}_0|\\
	&\ge& \left\{ \sum_{\mathcal{I}\in \mathcal{P}^*}\left( \sum_{i=1}^n \mathbb{I}(A_i\in \mathcal{I})(Y_i-\overline{X}_i^\top \theta_{\mathcal{I}}^*)^2+n\lambda_n |\mathcal{I}|\|\theta_{\mathcal{I}}^*\|_2^2 \right)+n\gamma_n |\mathcal{P}^*| \right\}-c_1^*\log n,
\end{eqnarray*}
with probability at least $1-O(n^{-2})$.
In view of \eqref{someresults0}, we've shown that with probability at least $1-O(n^{-2})$,
\begin{eqnarray}\label{someresults3}
&&\sum_{\mathcal{I}\in \widehat{\mathcal{P}}} \left( \sum_{i=1}^n \mathbb{I}(A_i\in \mathcal{I})(Y_i-\overline{X}_i^\top \widehat{\theta}_{\mathcal{I}})^2+n\lambda_n |\mathcal{I}|\|\widehat{\theta}_{\mathcal{I}}\|_2^2 \right)+n\gamma_n|\widehat{\mathcal{P}}|\\ \nonumber
&\ge& \sum_{\substack{\mathcal{I}\in \widehat{\mathcal{P}}, \mathcal{I}\in \mathfrak{T}(m) }}\left(\frac{n}{c_1} \int_{\mathcal{I}} \|\theta_0(a)-\theta_{0,\mathcal{I}}\|_2^2da- 3c_1\sqrt{n\int_{\mathcal{I}} \|\theta_0(a)-\theta_{0,\mathcal{I}}\|_2^2da \log n}\right) \\ \nonumber
&+& \left\{ \sum_{\mathcal{I}\in \mathcal{P}^*}\left( \sum_{i=1}^n \mathbb{I}(A_i\in \mathcal{I})(Y_i-\overline{X}_i^\top \theta_{\mathcal{I}}^*)^2+n\lambda_n |\mathcal{I}|\|\theta_{\mathcal{I}}^*\|_2^2 \right)+n\gamma_n |\mathcal{P}^*| \right\}\\ \nonumber
&+&n\gamma_n |\widehat{\mathcal{P}}|-(2c_0^2+2\bar{c}_2) |\widehat{\mathcal{P}}|\log n-c_1^* \log n-n\gamma_n |\mathcal{P}_0|.
\end{eqnarray}
By definition, 
\begin{eqnarray*}
	&&\sum_{\mathcal{I}\in \widehat{\mathcal{P}}} \left( \sum_{i=1}^n \mathbb{I}(A_i\in \mathcal{I})(Y_i-\overline{X}_i^\top \widehat{\theta}_{\mathcal{I}})^2+n\lambda_n |\mathcal{I}|\|\widehat{\theta}_{\mathcal{I}}\|_2^2 \right)+n\gamma_n|\widehat{\mathcal{P}}|\\
	&\le& \sum_{\mathcal{I}\in \mathcal{P}^*}\left( \sum_{i=1}^n \mathbb{I}(A_i\in \mathcal{I})(Y_i-\overline{X}_i^\top \theta_{\mathcal{I}}^*)^2+n\lambda_n |\mathcal{I}|\|\theta_{\mathcal{I}}^*\|_2^2 \right)+n\gamma_n |\mathcal{P}^*|.
\end{eqnarray*}
Thus, we have with probability at least $1-O(n^{-2})$,
\begin{eqnarray*}
	\sum_{\substack{\mathcal{I}\in \widehat{\mathcal{P}}, \mathcal{I}\in \mathfrak{T}(m) }}\left(\frac{n}{c_1} \int_{\mathcal{I}} \|\theta_0(a)-\theta_{0,\mathcal{I}}\|_2^2da- 3c_1\sqrt{n\int_{\mathcal{I}} \|\theta_0(a)-\theta_{0,\mathcal{I}}\|_2^2da \log n}\right)\\
	\le (2c_0^2+2\bar{c}_2) |\widehat{\mathcal{P}}|\log n+c_1^* \log n+n\gamma_n |\mathcal{P}_0|-n\gamma_n |\widehat{\mathcal{P}}|,
\end{eqnarray*}
and hence,
\begin{eqnarray}\label{someresults1}
\sum_{\substack{\mathcal{I}\in \widehat{\mathcal{P}}, \mathcal{I}\in \mathfrak{T}(m) }} \frac{n}{c_1} \left(\sqrt{\int_{\mathcal{I}} \|\theta_0(a)-\theta_{0,\mathcal{I}}\|_2^2da}-\frac{3c_1}{2}n^{-1/2} \sqrt{\log n}\right)^2\\ \nonumber
\le (2c_0^2+2\bar{c}_2+9c_1/4) |\widehat{\mathcal{P}}|\log n+c_1^* \log n+n\gamma_n |\mathcal{P}_0|-n\gamma_n |\widehat{\mathcal{P}}|.
\end{eqnarray}
Under the event defined in \eqref{underfit}, we have either $|\widehat{\mathcal{P}}|\ge 2|\mathcal{P}_0|$, or $|\mathcal{P}_0|\le |\widehat{\mathcal{P}}|\le 2|\mathcal{P}_0|$. When $|\widehat{\mathcal{P}}|\ge 2|\mathcal{P}_0|$, it follows from the condition $n\gamma_n\gg \log n$ that for sufficiently large $n$, $\gamma_n/4\ge 2c_0^2+2\bar{c}_2+9c_1/4$, $n|\mathcal{P}_0|\gamma_n\ge 2 c_1^*\log n$ and hence
\begin{eqnarray*}
	&&(2c_0^2+2\bar{c}_2+9c_1/4) |\widehat{\mathcal{P}}|\log n+c_1^* \log n+n\gamma_n |\mathcal{P}_0|-n\gamma_n |\widehat{\mathcal{P}}|\\
	&\le& (2c_0^2+2\bar{c}_2+9c_1/4) |\widehat{\mathcal{P}}|\log n+c_1^* \log n-n\gamma_n|\widehat{\mathcal{P}}|/4-n\gamma_n |\mathcal{P}_0|/2\\
	&\le& (2c_0^2+2\bar{c}_2+9c_1/4) |\widehat{\mathcal{P}}|\log n-n\gamma_n|\widehat{\mathcal{P}}|/4\le 0,
\end{eqnarray*}
When $|\mathcal{P}_0|\le |\widehat{\mathcal{P}}|\le 2|\mathcal{P}_0|$, we have
\begin{eqnarray*}
	&&(2c_0^2+2\bar{c}_2+9c_1/4) |\widehat{\mathcal{P}}|\log n+c_1^* \log n+n\gamma_n |\mathcal{P}_0|-n\gamma_n |\widehat{\mathcal{P}}|\\
	&\le& 2(2c_0^2+2\bar{c}_2+9c_1/4) |\mathcal{P}_0|\log n+c_1^* \log n.
\end{eqnarray*}
In view of \eqref{someresults1}, we have with probability at least $1-O(n^{-2})$,
\begin{eqnarray*}
	\sum_{\substack{\mathcal{I}\in \widehat{\mathcal{P}}, \mathcal{I}\in \mathfrak{T}(m) }} \frac{n}{c_1} \left(\sqrt{\int_{\mathcal{I}} \|\theta_0(a)-\theta_{0,\mathcal{I}}\|_2^2da}-\frac{3c_1}{2}n^{-1/2} \sqrt{\log n}\right)^2\le c\log n,
\end{eqnarray*}
for some constant $c>0$. Thus, with probability at least $1-O(n^{-2})$, we have 
\begin{eqnarray*}
	\int_{\mathcal{I}} \|\theta_0(a)-\theta_{0,\mathcal{I}}\|_2^2da=O(n^{-1}\log n),
\end{eqnarray*}
for any $\mathcal{I}\in \widehat{\mathcal{P}}\cap  \mathfrak{T}(m)$. By the definition of $\mathfrak{T}(m)$, we obtain that with probability at least $1-O(n^{-2})$, 
\begin{eqnarray}\label{someresults2}
\int_{\mathcal{I}} \|\theta_0(a)-\theta_{0,\mathcal{I}}\|_2^2da=O(n^{-1}\log n), \,\,\,\,\forall \mathcal{I}\in \widehat{\mathcal{P}}. 
\end{eqnarray}

Consider a given change point $\tau \in \mathcal{P}_0$, there exists an interval $\mathcal{I}\in \widehat{\mathcal{P}}$ of the form $[i_1,i_2)$ or $[i_1,i_2]$ with $i_2=1$ such that $i_1\le \tau<i_2$. Under the event defined in \eqref{someresults2}, it follows from Lemma \hyperlink{lemma3}{3} such that $\min (|i_1-\tau|,|i_2-\tau|)=O(n^{-1}\log n)$. This proves \eqref{locationlocalconv}.

\noindent \textit{Part 3:} Using similar arguments in proving \eqref{event8}, we can show that the following events occur with probability at least $1-O(n^{-2})$: for any interval $\mathcal{I}\in \widehat{\mathcal{P}}$, we have
\begin{eqnarray*}
	\left|\sum_{i=1}^n \mathbb{I}(A_i\in\mathcal{I})\{Y_i-\overline{X}^\top_i\theta_0(A_i)\}\overline{X}_i^\top \{\theta_0(A_i)-\theta_{0,\mathcal{I}}\} \right|\le C \log n,
\end{eqnarray*}
for some constant $C>0$.

By \eqref{someresults2}, using similar arguments in proving \eqref{someresults0} and \eqref{someresults3}, we can show the following event occurs with probability at least $1-O(n^{-2})$,
\begin{eqnarray*}
	&&\sum_{\mathcal{I}\in \widehat{\mathcal{P}}} \left( \sum_{i=1}^n \mathbb{I}(A_i\in \mathcal{I})(Y_i-\overline{X}_i^\top \widehat{\theta}_{\mathcal{I}})^2+n\lambda_n |\mathcal{I}|\|\widehat{\theta}_{\mathcal{I}}\|_2^2 \right)+n\gamma_n|\widehat{\mathcal{P}}|\\
	&\ge & \left\{ \sum_{\mathcal{I}\in \mathcal{P}^*}\left( \sum_{i=1}^n \mathbb{I}(A_i\in \mathcal{I})(Y_i-\overline{X}_i^\top \theta_{\mathcal{I}}^*)^2+n\lambda_n |\mathcal{I}|\|\theta_{\mathcal{I}}^*\|_2^2 \right)+n\gamma_n |\mathcal{P}^*| \right\}\\ \nonumber
	&+&n\gamma_n |\widehat{\mathcal{P}}|-C |\widehat{\mathcal{P}}|\log n-C \log n-n\gamma_n |\mathcal{P}_0|,
\end{eqnarray*}
for some constant $C>0$. By definition,
\begin{eqnarray*}
	&&\sum_{\mathcal{I}\in \widehat{\mathcal{P}}} \left( \sum_{i=1}^n \mathbb{I}(A_i\in \mathcal{I})(Y_i-\overline{X}_i^\top \widehat{\theta}_{\mathcal{I}})^2+n\lambda_n |\mathcal{I}|\|\widehat{\theta}_{\mathcal{I}}\|_2^2 \right)+n\gamma_n|\widehat{\mathcal{P}}|\\
	&\le& \sum_{\mathcal{I}\in \mathcal{P}^*}\left( \sum_{i=1}^n \mathbb{I}(A_i\in \mathcal{I})(Y_i-\overline{X}_i^\top \theta_{\mathcal{I}}^*)^2+n\lambda_n |\mathcal{I}|\|\theta_{\mathcal{I}}^*\|_2^2 \right)+n\gamma_n |\mathcal{P}^*|.
\end{eqnarray*}
Thus, we have with probability at least $1-O(n^{-2})$,
\begin{eqnarray*}
	n\gamma_n |\widehat{\mathcal{P}}|-C |\widehat{\mathcal{P}}|\log n-C \log n-n\gamma_n |\mathcal{P}_0|\le 0.
\end{eqnarray*}
Since $\gamma_n\gg n^{-1}\log n$, the above event occurs only when $|\widehat{\mathcal{P}}|\le |\mathcal{P}_0|$. To see this, notice that if $|\widehat{\mathcal{P}}|>|\mathcal{P}_0|$, we have
\begin{eqnarray*}
	n\gamma_n-C \log n-\frac{C \log n}{|\widehat{\mathcal{P}}|}-n\gamma_n \frac{|\mathcal{P}_0|}{|\widehat{\mathcal{P}}|}
	\ge n\gamma_n-C \log n-\frac{C \log n}{|\mathcal{P}_0|+1}-n\gamma_n \frac{|\mathcal{P}_0|}{|\mathcal{P}_0|+1}\\
	= \frac{n\gamma_n}{|\mathcal{P}_0|+1}-C \log n-\frac{C \log n}{|\mathcal{P}_0|+1}\gg 0,
\end{eqnarray*}
since $\gamma_n\gg n^{-1}\log n$. This proves \eqref{overfit}.

\noindent \textit{Part 4: }In the first three parts, we've shown that
\begin{eqnarray}\label{someresults4}
|\widehat{\mathcal{P}}|=|\mathcal{P}_0|\,\,\,\,\hbox{and}\,\,\,\,\max_{\tau\in J(\mathcal{P}_0)} \min_{\hat{\tau}\in J(\widehat{\mathcal{P}})} |\hat{\tau}-\tau|=O(n^{-1}\log n),
\end{eqnarray}
with probability tending to $1$. For sufficiently large $n$, the event defined in \eqref{someresults4} implies that $|\mathcal{I}|\ge \bar{c}_0 n^{-1}\log n$ for any $\mathcal{I}\in \widehat{\mathcal{P}}$. Thus, it follows from Lemma \hyperlink{lemma1}{1} that the following occurs with probability at least $1-O(n^{-2})$: for any $\mathcal{I}\in \widehat{\mathcal{P}}$, we have
\begin{eqnarray}\label{someresults5}
|\mathcal{I}|\|\widehat{\theta}_{\mathcal{I}}-\theta_{0,\mathcal{I}}\|_2^2\le c_0^2 n^{-1}\log n.
\end{eqnarray}

Under the events defined in \eqref{someresults2}, \eqref{someresults4} and \eqref{someresults5}, we have
\begin{eqnarray*}
	\int_0^1 \|\widehat{\theta}(a)-\theta_0(a)\|_2^2 da=\sum_{\mathcal{I} \in \widehat{\mathcal{P}}} \int_{\mathcal{I}} \|\widehat{\theta}_{\mathcal{I}}-\theta_0(a)\|_2^2da=\sum_{\mathcal{I} \in \widehat{\mathcal{P}}} \int_{\mathcal{I}} \|\widehat{\theta}_{\mathcal{I}}-\theta_{0,\mathcal{I}}+\theta_{0,\mathcal{I}}-\theta_0(a)\|_2^2da\\
	=\sum_{\mathcal{I} \in \widehat{\mathcal{P}}} \int_{\mathcal{I}} \|\widehat{\theta}_{\mathcal{I}}-\theta_{0,\mathcal{I}}\|_2^2 da+\sum_{\mathcal{I} \in \widehat{\mathcal{P}}} \int_{\mathcal{I}} \|\theta_{0,\mathcal{I}}-\theta_0(a)\|_2^2 da+2\sum_{\mathcal{I} \in \widehat{\mathcal{P}}} \int_{\mathcal{I}} (\widehat{\theta}_{0,\mathcal{I}}-\theta_{0,\mathcal{I}})^\top \{\theta_{0,\mathcal{I}}-\theta_0(a)\}da\\
	\le2\sum_{\mathcal{I} \in \widehat{\mathcal{P}}} \int_{\mathcal{I}} \|\widehat{\theta}_{\mathcal{I}}-\theta_{0,\mathcal{I}}\|_2^2 da+2\sum_{\mathcal{I} \in \widehat{\mathcal{P}}} \int_{\mathcal{I}} \|\theta_{0,\mathcal{I}}-\theta_0(a)\|_2^2 da=O(|\widehat{\mathcal{P}}| n^{-1}\log n)=O(|\mathcal{P}_0|n^{-1}\log n),
\end{eqnarray*}
where the first inequality is due to Cauchy-Schwarz inequality. This proves (iii). The proof is hence completed.

\subsection{Proof of Lemma 1}\index{Proof of Lemma 1}


\noindent \textit{Proof of \eqref{event1}: } By definition, we have
\begin{eqnarray*}
	&&\|\widehat{\theta}_{\mathcal{I}}-\theta_{0,\mathcal{I}}\|_2\\
	&\le& \left\|\left( \frac{1}{n}\sum_{i=1}^n \mathbb{I}(A_i\in \mathcal{I})\overline{X}_i \overline{X}_i^\top+\lambda_n |\mathcal{I}| \mathbb{E}_{p+1} \right)^{-1}\left( \frac{1}{n}\sum_{i=1}^n \mathbb{I}(A_i\in \mathcal{I})\overline{X}_i Y_i-\Mean \mathbb{I}(A\in \mathcal{I}) \overline{X} Y \right)\right\|_2\\
	&+&\left\|\left\{\left(\frac{1}{n}\sum_{i=1}^n \mathbb{I}(A_i\in \mathcal{I})\overline{X}_i \overline{X}_i^\top+\lambda_n |\mathcal{I}| \mathbb{E}_{p+1}\right)^{-1}-\left(\Mean \mathbb{I}(A\in \mathcal{I}) \overline{X} \overline{X}^\top \right)^{-1}\right\}\{\Mean \mathbb{I}(A\in \mathcal{I}) \overline{X} Y\}\right\|_2\\
	&\le& \underbrace{\left\|\left( \frac{1}{n}\sum_{i=1}^n \mathbb{I}(A_i\in \mathcal{I})\overline{X}_i \overline{X}_i^\top+\lambda_n |\mathcal{I}| \mathbb{E}_{p+1} \right)^{-1}\right\|_2}_{\eta_1(\mathcal{I})}\underbrace{\left\|\left( \frac{1}{n}\sum_{i=1}^n \mathbb{I}(A_i\in \mathcal{I})\overline{X}_i Y_i-\Mean \mathbb{I}(A\in \mathcal{I}) \overline{X} Y \right)\right\|_2}_{\eta_2(\mathcal{I})}\\
	&+&\underbrace{\left\|\left(\frac{1}{n}\sum_{i=1}^n \mathbb{I}(A_i\in \mathcal{I})\overline{X}_i \overline{X}_i^\top+\lambda_n |\mathcal{I}| \mathbb{E}_{p+1}\right)^{-1}-\left(\Mean \mathbb{I}(A\in \mathcal{I}) \overline{X} \overline{X}^\top \right)^{-1}\right\|_2}_{\eta_3(\mathcal{I})}\underbrace{\|\Mean \mathbb{I}(A\in \mathcal{I}) \overline{X} Y\|_2}_{\eta_4(\mathcal{I})}.
\end{eqnarray*}
It follows from Cauchy-Schwarz inequality that
\begin{eqnarray}\label{betaeta}
\|\widehat{\theta}_{\mathcal{I}}-\theta_{0,\mathcal{I}}\|_2^2\le 2\eta_1^2(I) \eta_2^2(I)+2\eta_3^2(I) \eta_4^2(I).
\end{eqnarray}
In the following, we provides upper bounds for
\begin{eqnarray*}
	\max_{\substack{\mathcal{I}\in \mathfrak{I}(m)\\ |\mathcal{I}|\ge \bar{c}_0 n^{-1}\log n }} \eta_j(\mathcal{I}),
\end{eqnarray*}
for $j=1,2,3,4$, where the constant $\bar{c}_0$ will be specified later. The uniform convergence rates of $\|\widehat{\theta}_{\mathcal{I}}-\theta_{0,\mathcal{I}}\|_2$ can thus be derived. 

Without loss of generality, assume the constant $\omega$ in Condition (A4) is greater than or equal to $\log^{-1/2} 2$. Then, we have $\exp(1/\omega^2)\le \exp(\log 2)=2$ and hence $\|1\|_{\psi_2|A}\le \omega$. 
Therefore, we have $\max_{j\in \{1,\dots,p+1\}}\|\overline{X}^{(j)}\|_{\psi_2|A}\le \omega$, almost surely. By the definition of the conditional Orlicz norm, this implies that
\begin{eqnarray*}
	\Mean \left\{\left.1+\sum_{q=1}^{+\infty} \frac{|\overline{X}^{(j)}|^{2q}}{\omega^{2q} q!} \right|A\right\}\le 2,\,\,\,\,\,\,\,\,\forall j\in \{1,\dots,p+1\},
\end{eqnarray*}
almost surely, and hence
\begin{eqnarray}\label{meanXcA0}
\Mean (|\overline{X}^{(j)}|^{2q}|A)\le q! \omega^{2q}, \,\,\,\,\,\,\,\,\forall j\in \{1,\dots,p+1\},q=1,2,\dots
\end{eqnarray}
By Cauchy-Schwarz inequality, we obtain that
\begin{eqnarray}\label{meanXcA}
\Mean (|\overline{X}^{(j_1)}\overline{X}^{(j_2)}|^{q}|A)\le \sqrt{\Mean (|\overline{X}^{(j_1)}|^{2q}|A) \Mean (|\overline{X}^{(j_2)}|^{2q}|A)}\le q! \omega^{2q}, 
\end{eqnarray}
for any $j_1,j_2\in \{1,\dots,p+1\}$ and any integer $q\ge 1$, almost surely.

Since $A$ has a bounded probability density function $p_A(\cdot)$ in $[0,1]$, there exists some constant $C_0>0$ such that
\begin{eqnarray}\label{boundedA}
\sup_{a\in[0,1]}p_A(a)\le C_0\,\,\,\,\hbox{and}\,\,\,\,\hbox{Pr}(A\in \mathcal{I})\le C_0 |\mathcal{I}|,
\end{eqnarray}
for any interval $\mathcal{I}\in [0,1]$. This together with \eqref{meanXcA} yields that for any integer $q\ge 1$, $j_1,j_2\in \{1,\dots,p+1\}$ and any interval $\mathcal{I}\in [0,1]$, we have
\begin{eqnarray*}
	\Mean |\mathbb{I}(A\in \mathcal{I}) \overline{X}^{(j_1)}\overline{X}^{(j_2)}|^q=\Mean \{\mathbb{I}(A\in \mathcal{I}) \Mean (|\overline{X}^{(j_1)}\overline{X}^{(j_2)}|^q|A)\}\le q!\omega^{2q} \Mean \mathbb{I}(A\in \mathcal{I})\le C_0 q!\omega^{2q} |\mathcal{I}|.
\end{eqnarray*}
It follows that
\begin{eqnarray*}
	&&\Mean |\mathbb{I}(A\in \mathcal{I})\overline{X}^{(j_1)}\overline{X}^{(j_2)}-\Mean \mathbb{I}(A\in \mathcal{I})\overline{X}^{(j_1)}\overline{X}^{(j_2)}|^q\\
	&\le& \Mean |\mathbb{I}(A_1\in \mathcal{I})\overline{X}_1^{(j_1)}\overline{X}^{(j_2)}_1-\mathbb{I}(A_2\in \mathcal{I})\overline{X}_2^{(j_1)}\overline{X}^{(j_2)}_2|^q\\
	&\le& 2^{q-1} \Mean |\mathbb{I}(A_1\in \mathcal{I})\overline{X}^{(j_1)}_1\overline{X}^{(j_2)}_1|^{q}+2^{q-1} \Mean |\mathbb{I}(A_2\in \mathcal{I})\overline{X}^{(j_1)}_2\overline{X}^{(j_2)}_2|^{q}\\&=&2^q \Mean |\mathbb{I}(A\in \mathcal{I})\overline{X}^{(j_1)}\overline{X}^{(j_2)}|^q\le C_0q! (2\omega^2)^q |\mathcal{I}|,
\end{eqnarray*}
where the second inequality follows from Jensen's inequality and the third inequality is due to that $|a+b|^q\le 2^{q-1} |a|^q+2^{q-1} |b|^{q-1}$, for any $a,b\in \mathbb{R}$ and $q\ge 1$.

By the Bernstein's inequality \citep[see Lemma 2.2.11,][]{van1996}, we obtain that
\begin{eqnarray} \label{bernsteininequality1}
&&~~~~~~\hbox{Pr}\left( \left| \sum_{i=1}^n \mathbb{I}(A_i\in \mathcal{I}) \overline{X}_i^{(j_1)} \overline{X}_i^{(j_2)}-n\Mean \mathbb{I}(A\in \mathcal{I}) \overline{X}^{(j_1)} \overline{X}^{(j_2)} \right|\ge t\omega^2 \sqrt{|\mathcal{I}|n\log n} \right)\\\nonumber
&\le & 2\exp\left(-\frac{1}{2} \frac{\omega^4 t^2 |\mathcal{I}| n\log n}{8nC_0\omega^4|\mathcal{I}|+2\omega^4 t\sqrt{|\mathcal{I}|n\log n}} \right)\le 2\exp\left(-\frac{t^2 \log n}{16C_0 + 4t(n|\mathcal{I}|)^{-1/2}\sqrt{\log n}} \right),
\end{eqnarray}
for any $t>0$, any integers $j_1,j_2\in \{1,\dots,p+1\}$ and any interval $\mathcal{I}\in [0,1]$. Set $t=20\sqrt{C_0}$, for any interval $\mathcal{I}$ with $|\mathcal{I}|\ge C_0^{-1}n^{-1}\log n$, we have
\begin{eqnarray*}
	t^2 \log n\ge 4 \{16 C_0+4t(n|\mathcal{I}|)^{-1/2}\sqrt{\log n}\}.
\end{eqnarray*}
It follows from \eqref{bernsteininequality1} that
\begin{eqnarray*}
	\hbox{Pr}\left( \left| \sum_{i=1}^n \mathbb{I}(A_i\in \mathcal{I}) \overline{X}_i^{(j_1)} \overline{X}_i^{(j_2)}-n\Mean \mathbb{I}(A\in \mathcal{I}) \overline{X}^{(j_1)} \overline{X}^{(j_2)} \right|\ge 20\omega^2 \sqrt{C_0|\mathcal{I}|n\log n} \right)\le 2n^{-4},
\end{eqnarray*}
for any integers $j_1,j_2\in \{1,\dots,p+1\}$ and any interval $\mathcal{I}$ that satisfies $|\mathcal{I}|\ge C_0^{-1}n^{-1}\log n$. Notice that the number of elements in $\mathfrak{I}(m)$ is bounded by $(m+1)^2$. Since $m\asymp n$, it follows from Bonferroni's inequality that
\begin{eqnarray}\label{firstevent}
\hbox{Pr}(\mathcal{A}_1)\ge 1-2(m+1)^2 (p+1)^2 n^{-4}=1-O(n^{-2}),
\end{eqnarray}
where the event $\mathcal{A}_1$ is defined as
\begin{eqnarray*}
	\bigcap_{\substack{j_1,j_2\in \{1,\dots,p+1\}\\ \mathcal{I}\in \mathfrak{I}(m)\\ |\mathcal{I}|\ge C_0^{-1}n^{-1}\log n }}\left\{ \left| \sum_{i=1}^n \mathbb{I}(A_i\in \mathcal{I}) \overline{X}_i^{(j_1)} \overline{X}_i^{(j_2)}-n\Mean \mathbb{I}(A\in \mathcal{I}) \overline{X}^{(j_1)} \overline{X}^{(j_2)} \right|\le 20 \omega^2 \sqrt{C_0|\mathcal{I}|n\log n} \right\}.
\end{eqnarray*}
For any symmetric matrix $\bm{A}$, we have $\|\bm{A}\|_{2}\le \sqrt{\|\bm{A}\|_{\infty} \|\bm{A}\|_{1} }=\|\bm{A}\|_{\infty}$. Thus, under the event defined in $\mathcal{A}_1$, we have
\begin{eqnarray*}
	\left\| \sum_{i=1}^n \mathbb{I}(A_i\in \mathcal{I}) \overline{X}_i \overline{X}_i^\top-n\Mean \mathbb{I}(A\in \mathcal{I}) \overline{X}\overline{X}^\top \right\|_2\le \left\| \sum_{i=1}^n \mathbb{I}(A_i\in \mathcal{I}) \overline{X}_i \overline{X}_i^\top-n\Mean \mathbb{I}(A\in \mathcal{I}) \overline{X}\overline{X}^\top \right\|_{\infty}\\
	\le 20\omega^2 (p+1) \sqrt{C_0|\mathcal{I}|n\log n},
\end{eqnarray*}
for any $\mathcal{I}\in \mathfrak{I}(m)$ with $|\mathcal{I}|\ge C_0^{-1} n^{-1}\log n$. Since $\lambda_n=O(n^{-1}\log n)$, we obtain
\begin{eqnarray*}
	\left\| \sum_{i=1}^n \mathbb{I}(A_i\in \mathcal{I}) \overline{X}_i \overline{X}_i^\top-n\Mean \mathbb{I}(A\in \mathcal{I}) \overline{X}\overline{X}^\top +n\lambda_n \mathbb{E}|\mathcal{I}| \right\|_2\\
	\le n\lambda_n|\mathcal{I}|+\left\| \sum_{i=1}^n \mathbb{I}(A_i\in \mathcal{I}) \overline{X}_i \overline{X}_i^\top-n\Mean \mathbb{I}(A\in \mathcal{I}) \overline{X}\overline{X}^\top\right\|_2\le c\sqrt{|\mathcal{I}|n\log n},
\end{eqnarray*}
for some constant $c>0$. To summarize, under the event defined in $\mathcal{A}_1$, we've shown that 
\begin{eqnarray}\label{diff1}
\left\| \sum_{i=1}^n \mathbb{I}(A_i\in \mathcal{I}) \overline{X}_i \overline{X}_i^\top-n\Mean \mathbb{I}(A\in \mathcal{I}) \overline{X}\overline{X}^\top +n\lambda_n \mathbb{E}|\mathcal{I}| \right\|_2\le c\sqrt{|\mathcal{I}|n\log n},
\end{eqnarray}
for any interval $\mathcal{I}\in \mathfrak{I}(m)$ with $|\mathcal{I}|\ge C_0^{-1} n^{-1}\log n$. 

Let $\Sigma=\Mean \overline{X} \overline{X}^\top$. If $\Sigma$ is singular, there exists some nonzero vector $a\in \mathbb{R}^{p}$ and some $b\in \mathbb{R}$ such that $a^\top X=b$, almost surely. As a result, the covariance matrix of $X$ is degenerate. Thus, we've reached a contraction. Therefore, $\Sigma$ is nonsingular. There exists some constant $\bar{c}_*>0$ such that
\begin{eqnarray}\label{minieigensigma}
\lambda_{\min}(\Sigma)\ge \bar{c}_*.
\end{eqnarray}
By (A3), we have
\begin{eqnarray*}
	\hbox{Pr}(A\in \mathcal{I}|X)\ge c_* |\mathcal{I}|,
\end{eqnarray*}
for any interval $\mathcal{I}\in [0,1]$. This together with \eqref{minieigensigma} implies that
\begin{eqnarray}\label{prooflemma1eq1}
\lambda_{\min}\left(\Mean \mathbb{I}(A\in \mathcal{I}) \overline{X} \overline{X}^\top\right)&=&\lambda_{\min}\left(\Mean \hbox{Pr}(A\in \mathcal{I}|X) \overline{X} \overline{X}^\top\right)\\\nonumber
&\ge& c_{*} \lambda_{\min}(\Mean \overline{X} \overline{X}^\top) |\mathcal{I}|\ge c_* \bar{c}_* |\mathcal{I}|. 
\end{eqnarray}
For any interval $\mathcal{I}$ with $|\mathcal{I}|\ge 4c^2 (c_* \bar{c}_*)^{-2} n^{-1}\log n$, we have
\begin{eqnarray*}
	c_* \bar{c}_* |\mathcal{I}|-c\sqrt{|\mathcal{I}|n^{-1}\log n}\ge \frac{c_* \bar{c}_* |\mathcal{I}|}{2}.
\end{eqnarray*}
In view of \eqref{diff1} and \eqref{prooflemma1eq1}, we obtain that
\begin{eqnarray}\label{prooflemmaeq2}
&&\lambda_{\min}\left( \frac{1}{n}\sum_{i=1}^n \mathbb{I}(A_i\in \mathcal{I}) \overline{X}_i \overline{X}_i^\top+\lambda_n \mathbb{E}|\mathcal{I}|\right)\ge \lambda_{\min}\left(\Mean \mathbb{I}(A\in \mathcal{I}) \overline{X} \overline{X}^\top\right)\\ \nonumber
&-&\frac{1}{n} \left\| \sum_{i=1}^n \mathbb{I}(A_i\in \mathcal{I}) \overline{X}_i \overline{X}_i^\top-n\Mean \mathbb{I}(A\in \mathcal{I}) \overline{X}\overline{X}^\top +n\lambda_n \mathbb{E}|\mathcal{I}| \right\|_2\ge \frac{c_* \bar{c}_* |\mathcal{I}|}{2}.
\end{eqnarray}
Set $\bar{c}_0=\max(4c^2 (c_*\bar{c}_*)^{-1}, C_0^{-1})$, it is immediate to see that
\begin{eqnarray}\label{eta1}
\max_{\substack{\mathcal{I}\in \mathfrak{I}(m)\\ |\mathcal{I}|\ge \bar{c}_0 n^{-1}\log n }} \eta_1(\mathcal{I})\le \frac{2}{c_*\bar{c}_* |\mathcal{I}|},
\end{eqnarray}
under the event defined in $\mathcal{A}_1$. 

For any $\mathcal{I}\in [0,1]$, we have
\begin{eqnarray*}
	&&\left\| \left(\frac{1}{n}\sum_{i=1}^n \mathbb{I}(A_i\in \mathcal{I}) \overline{X}_i \overline{X}_i^\top+\lambda_n \mathbb{E}|\mathcal{I}|\right)^{-1}-\left(\Mean \mathbb{I}(A\in \mathcal{I}) \overline{X} \overline{X}^\top\right)^{-1}\right\|_2\\
	&\le & \left\|\left(\frac{1}{n}\sum_{i=1}^n \mathbb{I}(A_i\in \mathcal{I}) \overline{X}_i \overline{X}_i^\top+\lambda_n \mathbb{E}|\mathcal{I}|\right)^{-1}\right\|_2 \left\| \left(\Mean \mathbb{I}(A\in \mathcal{I}) \overline{X} \overline{X}^\top\right)^{-1}\right\|_2\\
	&\times & \left\| \sum_{i=1}^n \mathbb{I}(A_i\in \mathcal{I}) \overline{X}_i \overline{X}_i^\top-n\Mean \mathbb{I}(A\in \mathcal{I}) \overline{X}\overline{X}^\top +n\lambda_n \mathbb{E}|\mathcal{I}| \right\|_2
\end{eqnarray*}
This together with \eqref{diff1}, \eqref{prooflemma1eq1} and \eqref{prooflemmaeq2} yields
\begin{eqnarray}\label{eta3}
\max_{\substack{\mathcal{I}\in \mathfrak{I}(m)\\ |\mathcal{I}|\ge \bar{c}_0 n^{-1}\log n }} \eta_3(\mathcal{I})\le \frac{2c\sqrt{n^{-1}\log n}}{c_*^2\bar{c}_*^2 |\mathcal{I}|^{3/2}},
\end{eqnarray}
under the event defined in $\mathcal{A}_1$.

Similar to \eqref{meanXcA}, we can show that for any integer $q\ge 1$ and $j\in \{1,\dots,p+1\}$,
\begin{eqnarray}\label{meanYcA}
\Mean (|\overline{X}^{(j)}Y|^q|A)\le q! \omega^{2q}, 
\end{eqnarray}
almost surely. Specifically, set $q=1$, we obtain $\Mean (|\overline{X}^{(j)}Y| |A)\le \omega^2$.
By \eqref{boundedA}, we have that
\begin{eqnarray*}
	\|\Mean \mathbb{I}(A\in \mathcal{I}) \overline{X} Y\|_2\le \left(\sum_{j=1}^{p+1} |\Mean \mathbb{I}(A\in \mathcal{I}) \overline{X}^{(j)} Y|^2\right)^{1/2}\le \left(\sum_{j=1}^{p+1} |\Mean \{\mathbb{I}(A\in \mathcal{I}) \Mean (|\overline{X}^{(j)} Y| |A)\} |^2\right)^{1/2}\\
	\le \left(\sum_{j=1}^{p+1} | \omega^{2}\Mean (A\in \mathcal{I})|^2\right)^{1/2}\le C_0 \sqrt{p+1} \omega^{2}|\mathcal{I}|.
\end{eqnarray*}
for any $\mathcal{I}\in [0,1]$. This implies that
\begin{eqnarray}\label{eta4}
\max_{\substack{\mathcal{I}\in \mathfrak{I}(m)\\ |\mathcal{I}|\ge \bar{c}_0 n^{-1}\log n }} \eta_4(\mathcal{I})\le C_0 \sqrt{p+1}\omega^2|\mathcal{I}|.
\end{eqnarray}
Moreover, in view of \eqref{bernsteininequality1} and \eqref{firstevent}, we can similarly show that 
\begin{eqnarray}\label{secondevent}
\hbox{Pr}(\mathcal{A}_2)\ge 1-2(m+1)^2(p+1)n^{-4}=1-O(n^{-2}),
\end{eqnarray}
where the event $\mathcal{A}_2$ is defined as
\begin{eqnarray*}
	\bigcap_{\substack{j\in \{1,\dots,p+1\}\\ \mathcal{I}\in \mathfrak{I}(m)\\ |\mathcal{I}|\ge \bar{c}_0 n^{-1}\log n }}\left\{ \left| \sum_{i=1}^n \mathbb{I}(A_i\in \mathcal{I}) \overline{X}_i^{(j)} Y_i-n\Mean \mathbb{I}(A\in \mathcal{I}) \overline{X}^{(j)} Y \right|\le 20\omega^2 \sqrt{C_0|\mathcal{I}|n\log n} \right\}.
\end{eqnarray*}
Under the event defined in $\mathcal{A}_2$, we have
\begin{eqnarray}\label{eta2}
\max_{\substack{\mathcal{I}\in \mathfrak{I}(m)\\ |\mathcal{I}|\ge \bar{c}_0 n^{-1}\log n }} \eta_2(\mathcal{I})\le 20\omega^2 \sqrt{(p+1)C_0|\mathcal{I}|n^{-1}\log n}.
\end{eqnarray}
Combining \eqref{eta1} together with \eqref{eta3}, \eqref{eta4}, \eqref{eta2} yields
\begin{eqnarray*}
	\max_{\substack{\mathcal{I}\in \mathfrak{I}(m)\\ |\mathcal{I}|\ge \bar{c}_0 n^{-1}\log n }} |\mathcal{I}|\|\widehat{\theta}_{\mathcal{I}}-\theta_{0,\mathcal{I}}\|_2^2=O\left(\frac{\log n}{n}\right),
\end{eqnarray*}
under the events defined in $\mathcal{A}_1$ and $\mathcal{A}_2$. The proof is thus completed based on \eqref{firstevent} and \eqref{secondevent}. 

\smallskip

\noindent \textit{Proofs of \eqref{event2}, \eqref{event3} and \eqref{boundbeta0I}: }
We first prove \eqref{boundbeta0I}. 
By the definition of $\theta_{0,\mathcal{I}}$, we have
\begin{eqnarray*}
	\|\theta_{0,\mathcal{I}}\|_2\le \left\|\left(\Mean \overline{X} \overline{X}^\top \mathbb{I}(A\in \mathcal{I})\right)^{-1}\right\|_2 \left\|\Mean \overline{X} \overline{Y} \mathbb{I}(A\in \mathcal{I}) \right\|_2.
\end{eqnarray*}
It follows from \eqref{boundedA}, \eqref{prooflemma1eq1} and \eqref{meanYcA} that
\begin{eqnarray*}
	\|\theta_{0,\mathcal{I}}\|_2\le (c_* \bar{c}_* |\mathcal{I}|)^{-1}\sqrt{\sum_{j=1}^{p+1} \left[\Mean \left\{ \left( \Mean |\overline{X}^{(j)}Y| |A \right) \mathbb{I}(A\in \mathcal{I})\right\}\right]^2}\le \sqrt{p+1} (c_* \bar{c}_*)^{-1} C_0 \omega^2, 
\end{eqnarray*}
for any $\mathcal{I}\in [0,1]$. Assertion \eqref{boundbeta0I} thus follows. 

Consider \eqref{event2}. 
Since $p$ is fixed, it suffices to show for any $j\in \{1,\dots,p+1\}$, the following event occurs with probability at least $1-O(n^{-2})$:
\begin{eqnarray}\label{inequality3}
\left|\frac{1}{n}\sum_{i=1}^{n}\mathbb{I}(A_i\in \mathcal{I})(Y_i-\overline{X}_i^\top\theta_{0,\mathcal{I}})\overline{X}_i^{(j)}\right|_2=O\left(\frac{\sqrt{|\mathcal{I}|\log n}}{\sqrt{n}}\right).
\end{eqnarray}
By \eqref{boundbeta0I}, \eqref{inequality3} can be proven in a similar manner as \eqref{firstevent} and \eqref{secondevent}. \eqref{event3} can be similarly proven.

\smallskip

\noindent \textit{Proof of \eqref{event4}: }
Similar to \eqref{event2}, we can show that the following event occurs with probability at least $1-O(n^{-2})$: for any $\mathcal{I}\in \mathfrak{I}(m)$ such that $|\mathcal{I}|\ge \bar{c}_0 n^{-1}\log n$,
\begin{eqnarray*}
	\left|\frac{1}{n}\sum_{i=1}^n \mathbb{I}(A_i\in \mathcal{I})[\overline{X}_i^\top \{\theta_0(A_i)-\theta_{0,\mathcal{I}}\}]^2-\Mean \mathbb{I}(A\in \mathcal{I})[\overline{X}^\top \{\theta_0(A)-\theta_{0,\mathcal{I}}\}]^2 \right|=O\left(\frac{\sqrt{|\mathcal{I}|\log n}}{\sqrt{n}}\right).
\end{eqnarray*}
Notice that
\begin{eqnarray}\nonumber
&&\Mean \mathbb{I}(A\in \mathcal{I})[\overline{X}^\top \{\theta_0(A)-\theta_{0,\mathcal{I}}\}]^2=\Mean \int_{\mathcal{I}} [\overline{X}^\top \{\theta_0(a)-\theta_{0,\mathcal{I}}\}]^2\pi(a|X)da \\ \nonumber
&\ge&c_* \Mean \int_{\mathcal{I}} [\overline{X}^\top \{\theta_0(a)-\theta_{0,\mathcal{I}}\}]^2da=c_* \int_{\mathcal{I}} \{\theta_0(a)-\theta_{0,\mathcal{I}}\}^\top \Sigma \{\theta_0(a)-\theta_{0,\mathcal{I}}\}da \\ \label{someinequality0}
&\ge& c_* \lambda_{\min}(\Sigma) \int_{\mathcal{I}} \|\theta_0(a)-\theta_{0,\mathcal{I}}\|_2^2 da\ge c_* \bar{c}_* \int_{\mathcal{I}} \|\theta_0(a)-\theta_{0,\mathcal{I}}\|_2^2 da,
\end{eqnarray}
where the first inequality is due to Condition (A3) and the last inequality is due to \eqref{minieigensigma}.

It follows that
\begin{eqnarray*}
	\frac{1}{n}\sum_{i=1}^n \mathbb{I}(A_i\in \mathcal{I})[\overline{X}_i^\top \{\theta_0(A_i)-\theta_{0,\mathcal{I}}\}]^2\ge c_* \bar{c}_* \int_{\mathcal{I}} \|\theta_0(a)-\theta_{0,\mathcal{I}}\|_2^2 da-O\left(\frac{\sqrt{|\mathcal{I}|\log n}}{\sqrt{n}}\right),
\end{eqnarray*}
for any $\mathcal{I}\in \mathfrak{I}(m)$ such that $|\mathcal{I}|\ge \bar{c}_0 n^{-1}\log n$, with probability at least $1-O(n^{-2})$. This completes the proof.

\smallskip

\noindent \textit{Proof of \eqref{event5}: }
Similar to \eqref{event2}, we can show that the following event occurs with probability at least $1-O(n^{-2})$: for any $\mathcal{I}\in \mathfrak{I}(m)$ such that $|\mathcal{I}|\ge \bar{c}_0 n^{-1}\log n$,
\begin{eqnarray}\label{event5eq1}
~~~~~~~~~\left|\frac{1}{n}\sum_{i=1}^n \mathbb{I}(A_i\in \mathcal{I})(|Y_i|^2+\|\overline{X}_i\|_2^2)-\Mean \mathbb{I}(A\in \mathcal{I})(Y^2+\|\overline{X}\|_2^2) \right|=O\left( \frac{\sqrt{|\mathcal{I}|\log n}}{\sqrt{n}} \right).
\end{eqnarray}
By \eqref{boundedA} and \eqref{meanXcA0}, we have
\begin{eqnarray*}
	\Mean \mathbb{I}(A\in \mathcal{I}) \|\overline{X}\|_2^2\le \sum_{j=1}^{p+1} \Mean \mathbb{I}(A\in \mathcal{I}) |\overline{X}^{(j)}|^2\le (p+1)C_0 \omega^2|\mathcal{I}|.
\end{eqnarray*}
Similarly, we can show
\begin{eqnarray*}
	\Mean \mathbb{I}(A\in \mathcal{I}) Y^2\le C_0\omega^2|\mathcal{I}|,
\end{eqnarray*}
and thus
\begin{eqnarray*}
	\Mean \mathbb{I}(A\in \mathcal{I})(Y^2+\|\overline{X}\|_2^2)\le (p+2)C_0\omega^2|\mathcal{I}|.
\end{eqnarray*}
This together with \eqref{event5eq1} yields \eqref{event5}. 

\subsection{Proof of Lemma 2}\index{Proof of Lemma 2}
We first prove \eqref{event6}. By \eqref{boundbeta0I}, we have $\sup_{a\in [0,1]}\|\theta_0(a)\|_2\le c_0$ and hence 
\begin{eqnarray}\label{boundbeta0I2}
\sup_{a\in [0,1]}\|\theta_0(a)-\theta_{0,\mathcal{I}}\|_2\le 2c_0.
\end{eqnarray}
Similar to \eqref{meanXcA0}, we can show that for any integer $q\ge 1$,
\begin{eqnarray}\label{meanYcA0}
\Mean (|Y|^{2q}|A)\le q! \omega^{2q}.
\end{eqnarray}
For any $\mathcal{I}\subseteq \mathfrak{I}(m)$ and integer $q\ge 2$, it follows from \eqref{meanYcA}, \eqref{boundbeta0I2} and \eqref{meanYcA0} that
\begin{eqnarray}\label{prooflemma2eq1}
&&~~~~~~~~\Mean \left([Y \overline{X}^\top \{\theta_0(A)-\theta_{0,\mathcal{I}}\}]^q |A\right)\le \|\theta_0(A)-\theta_{0,\mathcal{I}}\|_2^q \Mean ( |Y|^q\| \overline{X} \|_2^q|A)\\ \nonumber
&\le& \frac{1}{2}\|\theta_0(A)-\theta_{0,\mathcal{I}}\|_2^q\Mean \left(\left. |Y|^{2q}+ \left|\sum_{j=1}^{p+1} (\overline{X}^{(j)})^2\right|^q \right| A\right)\le \frac{1}{2}\|\theta_0(A)-\theta_{0,\mathcal{I}}\|_2^q q!\omega^{2q}\\\nonumber
&+&\frac{1}{2}\|\theta_0(A)-\theta_{0,\mathcal{I}}\|_2^q (p+1)^{q-1} \sum_{j=1}^{p+1} \Mean (|\overline{X}^{(j)}|^{2q}|A)\le \frac{q!\omega^{2q}}{2}\{1+ (p+1)^q\} \|\theta_0(A)-\theta_{0,\mathcal{I}}\|_2^q\\\nonumber
&\le& q! \omega^{2q} (p+1)^q \|\theta_0(A)-\theta_{0,\mathcal{I}}\|_2^q\le q! \omega^{2q} (p+1)^q (2c_0)^{q-2} \|\theta_0(A)-\theta_{0,\mathcal{I}}\|_2^2.
\end{eqnarray}
Similarly, we can show
\begin{eqnarray*}
	\Mean \left([ \{\overline{X}^\top \theta_0(A)\} \overline{X}^\top \{\theta_0(A)-\theta_{0,\mathcal{I}}\}]^q |A\right)\le q!\omega^{2q} (p+1)^{q} 2^{q-2}c_0^{2q-2} \|\theta_0(A)-\theta_{0,\mathcal{I}}\|_2^2.
\end{eqnarray*}
This together with \eqref{prooflemma2eq1} yields that for any integer $q\ge 2$, $\mathcal{I}\subseteq [0,1]$, we have
\begin{eqnarray}\label{prooflemma2eq2}
\Mean \left([ \{Y-\overline{X}^\top\theta_0(A)\} \overline{X}^\top \{\theta_0(A)-\theta_{0,\mathcal{I}}\}]^q |A\right)\le q! c^q \|\theta_0(A)-\theta_{0,\mathcal{I}}\|_2^2,
\end{eqnarray}
for some constant $c>0$. Combining \eqref{boundedA} together with \eqref{prooflemma2eq2}, we obtain that for any integer $q\ge 2$, $\mathcal{I}\subseteq [0,1]$, 
\begin{eqnarray*}
	\Mean [\mathbb{I}(A\in \mathcal{I}) \{Y-\overline{X}^\top\theta_0(A)\} \overline{X}^\top \{\theta_0(A)-\theta_{0,\mathcal{I}}\}]^q\le C_0 q! c^q \int_{\mathcal{I}} \|\theta_0(a)-\theta_{0,\mathcal{I}}\|_2^2 p_A(a)da\\
	\le C_0 q! c^q \int_{\mathcal{I}} \|\theta_0(a)-\theta_{0,\mathcal{I}}\|_2^2 da.
\end{eqnarray*}
Applying the Bernstein's inequality (using similar arguments in \eqref{bernsteininequality1} and \eqref{firstevent}), we can show that with probability at least $1-O(n^{-2})$, we have for any interval $\mathcal{I}$ that satisfies $\int_{\mathcal{I}} \|\theta_0(a)-\theta_{0,\mathcal{I}}\|_2^2 da\ge (C_0)^{-1}n^{-1}\log n$ and $\mathcal{I}\in \mathfrak{I}(m)$,
\begin{eqnarray*}
	\left|\sum_{i=1}^n \mathbb{I}(A_i\in \mathcal{I})\{Y_i-\overline{X}_i^\top \theta_0(A_i)\} \overline{X}_i^\top \{\theta_0(A_i)-\theta_{0,\mathcal{I}}\}\right|\le O(1) \sqrt{n\log n} \left( \int_{\mathcal{I}} \|\theta_0(a)-\theta_{0,\mathcal{I}}\|_2^2 da \right)^{1/2},
\end{eqnarray*}
where $O(1)$ denotes some positive constant. This proves \eqref{event6}. 

Similarly, we can show that with probability at least $1-O(n^{-2})$, there exists some constant $C>0$ such that for any interval $\mathcal{I}$ that satisfies $\int_{\mathcal{I}} \|\theta_0(a)-\theta_{0,\mathcal{I}}\|_2^2 da\ge (C_0)^{-1}n^{-1}\log n$ and $\mathcal{I}\in \mathfrak{I}(m)$, we have
\begin{eqnarray}\nonumber
\left|\sum_{i=1}^n \mathbb{I}(A_i\in \mathcal{I}) [\overline{X}_i^\top \{\theta_0(A_i)-\theta_{0,\mathcal{I}}\}]^2-n\Mean \mathbb{I}(A\in \mathcal{I}) [\overline{X}^\top \{\theta_0(A)-\theta_{0,\mathcal{I}}\}]^2 \right|\\ \nonumber
\le O(1) \sqrt{n\log n}\left( \int_{\mathcal{I}} \|\theta_0(a)-\theta_{0,\mathcal{I}} \|_2^2da \right)^{1/2},
\end{eqnarray}
for some postive constant $O(1)$. This together with \eqref{someinequality0} yields \eqref{event7}. 

\subsection{Proof of Lemma 3}\index{Proof of Lemma 3}
Consider the following three categories of intervals. 

\noindent \textit{Category 1:} Suppose $i_1$ and $i_2$ satisfy $\tau_{0,k-1}\le i_1\le i_2\le \tau_{0,k}$ for some integer $k$ such that $1\le k\le K$. Then apparently, we have $\theta_{0,\mathcal{I}}=\theta_0(a)$, $\forall a\in \mathcal{I}$, and hence $\int_{\mathcal{I}} \|\theta_0(a)-\theta_{0,\mathcal{I}}\|_2^2da=0$. The assertion $\int_{\mathcal{I}} \|\theta_0(a)-\theta_{0,\mathcal{I}}\|_2^2da\le c_n$ is thus automatically satisfied. 

\noindent \textit{Category 2:} Suppose there exists some integer $k$ such that $2\le k\le K$ and $i_1,i_2$ satisfy $\tau_{0,k-2}\le i_1<\tau_{0,k-1} <i_2\le \tau_{0,k}$. Assume we have
\begin{eqnarray*}
	\min_{j\in\{1,2\}} |i_j-\tau_{0,k-1}|\ge \frac{3}{\kappa_0^2} c_n.
\end{eqnarray*}
where
\begin{eqnarray*}
	\kappa_0\equiv \min_{\substack{\mathcal{I}_1,\mathcal{I}_2\in \mathcal{P}_0\\ \mathcal{I}_1~\hbox{and}~\mathcal{I}_2~\hbox{are}~\hbox{adjacent} }} \|\theta_{0,\mathcal{I}_1}-\theta_{0,\mathcal{I}_2}\|_2>0.
\end{eqnarray*}
Since $c_n\to 0$, for sufficiently large $n$, we have $\tau_{0,k}> \tau_{0,k-1}+3\kappa_0^{-2} c_n$ and $\tau_{0,k-2}+3\kappa_0^{-2} c_n<\tau_{0,k-1}$. Then, we have
\begin{eqnarray*}
	&&\int_{\mathcal{I}} \|\theta_0(a)-\theta_{0,\mathcal{I}}\|_2^2da\ge \min_{\theta\in \mathbb{R}^{p+1}} \int_{\tau_{0,k-1}-3\kappa_0^{-2}c_n}^{\tau_{0,k-1}+3\kappa_0^{-2}c_n} \|\theta-\theta_0(a)\|_2^2da\\
	&\ge & \frac{6}{\kappa_0^{-2}} c_n \min_{\theta\in \mathbb{R}^{p+1}} \left( \|\theta-\theta_{0,[\tau_{0,k-2}, \tau_{0,k-1})}\|_2^2, \|\theta-\theta_{0,[\tau_{0,k-1},\tau_{0,k} )} \|_2^2 \right)\ge \frac{6}{\kappa_0^{-2}} c_n \frac{\kappa_0^{-2}}{4}>c_n.
\end{eqnarray*}
This violates the assertion that $\int_{\mathcal{I}} \|\theta_0(a)-\theta_{0,\mathcal{I}}\|_2^2da\le c_n$. We've thus reached a contradiction. As a result, we have
\begin{eqnarray*}
	\min_{j\in\{1,2\}} |i_j-\tau_{0,k-1}|\le \frac{3}{\kappa_0^2} c_n.
\end{eqnarray*}

\noindent \textit{Category 3:} Suppose there exists some integer $k$ such that $3\le k\le K$ and $i_1,i_2$ satisfy $\tau_{0,k-3}\le i_1<\tau_{0,k-2}<\tau_{0,k-1} <i_2\le \tau_{0,k}$. Assume we have 
\begin{eqnarray*}
	|i_1-\tau_{0,k-2}| \ge \frac{3}{\kappa_0^2} c_n.
\end{eqnarray*}
Then for sufficiently large $n$, we have
\begin{eqnarray*}
	&&\int_{\mathcal{I}} \|\theta_0(a)-\theta_{0,\mathcal{I}}\|_2^2da\ge \min_{\theta\in \mathbb{R}^{p+1}} \int_{\tau_{0,k-2}-3\kappa_0^{-2}c_n}^{\tau_{0,k-2}+3\kappa_0^{-2}c_n} \|\theta-\theta_0(a)\|_2^2da\\
	&\ge&\frac{6}{\kappa_0^{-2}} c_n \min_{\theta\in \mathbb{R}^{p+1}} \left( \|\theta-\theta_{0,[\tau_{0,k-2}, \tau_{0,k-1})}\|_2^2, \|\theta-\theta_{0,[\tau_{0,k-1},\tau_{0,k} )} \|_2^2 \right)\ge \frac{6}{\kappa_0^{-2}} c_n \frac{\kappa_0^{-2}}{4}>c_n.
\end{eqnarray*}
This violates the assertion that $\int_{\mathcal{I}} \|\theta_0(a)-\theta_{0,\mathcal{I}}\|_2^2da\le c_n$. We've thus reached a contradiction. As a result, we have $|i_1-\tau_{0,k-2}|\le 3\kappa_0^{-2} c_n$. Similarly, we can show $|i_2-\tau_{0,k-1}|\le 3\kappa_0^{-2} c_n$. Therefore, we obtain
\begin{eqnarray*}
	\max_{j\in\{1,2\}} |i_j-\tau_{0,k-3+j}|\le \frac{3}{\kappa_0^2} c_n.
\end{eqnarray*}

If $\mathcal{I}$ belongs to none of these categories, then there exists some integer $k$ such that $2\le k\le K$ and $i_1,i_2$ satisfy $i_1\le \tau_{0,k-2}$ and $i_2\ge \tau_{0,k}$. Using similar arguments, we can show that
\begin{eqnarray*}
	\int_{\mathcal{I}} \|\theta_0(a)-\theta_{0,\mathcal{I}}\|_2^2da\ge \int_{\tau_{0,k-2}}^{\tau_{0,k}} \|\theta_0(a)-\theta_{0,\mathcal{I}}\|_2^2da\ge \frac{\kappa_0^2}{4} \min_{\mathcal{I} \in \mathcal{P}_0 } |\mathcal{I}|. 
\end{eqnarray*}
For sufficiently large $n$, this violates the assertion that $\int_{\mathcal{I}} \|\theta_0(a)-\theta_{0,\mathcal{I}}\|_2^2da\le c_n$. We've thus reached a contradiction. Therefore, we shall have $\tau_{0,k-2}\le i_1< i_2\le \tau_{0,k}$. This completes the first part of the proof. 

We now show \eqref{event8}. Take $c_n=\bar{c}_1 n^{-1} \log n$ and consider any interval $\mathcal{I}\in \mathfrak{I}(m)$ that satisfies $\int_{\mathcal{I}} \|\theta_0(a)-\theta_{0,\mathcal{I}}\|_2^2da\le \bar{c}_1 n^{-1}\log n$. 

If $\mathcal{I}$ belongs to Category 1, then $\theta_0(a)=\theta_{0,\mathcal{I}}$ for any $a\in \mathcal{I}$. As a result, we have 
\begin{eqnarray*}
	\sum_{i=1}^n \mathbb{I}(A_i\in\mathcal{I})\{Y_i-\overline{X}^\top_i\theta_0(A_i)\}\overline{X}_i^\top \{\theta_0(A_i)-\theta_{0,\mathcal{I}}\}=0.
\end{eqnarray*}

If $\mathcal{I}$ belongs to Category 2, then there exists some integer $k$ such that $2\le k\le K$ and $i_1,i_2$ satisfy $\tau_{0,k-2}\le i_1<\tau_{0,k-1} <i_2\le \tau_{0,k}$. Thus, we have
\begin{eqnarray*}
	&&\sum_{i=1}^n \mathbb{I}(A_i\in\mathcal{I})\{Y_i-\overline{X}^\top_i\theta_0(A_i)\}\overline{X}_i^\top \{\theta_0(A_i)-\theta_{0,\mathcal{I}}\}\\
	&=&\underbrace{\sum_{i=1}^n \mathbb{I}(A_i\in [i_1,\tau_{0,k-1}))\{Y_i-\overline{X}^\top_i\theta_0(A_i)\}\overline{X}_i^\top \{\theta_0(A_i)-\theta_{0,\mathcal{I}}\}}_{\zeta_1}\\ 
	&+&\underbrace{\sum_{i=1}^n \mathbb{I}(A_i\in [\tau_{0,k-1},i_2))\{Y_i-\overline{X}^\top_i\theta_0(A_i)\}\overline{X}_i^\top \{\theta_0(A_i)-\theta_{0,\mathcal{I}}\}}_{\zeta_2}.
\end{eqnarray*}
Notice that we've shown
\begin{eqnarray*}
	\min_{j\in\{1,2\}} |i_j-\tau_{0,k-1}|\le \frac{3\bar{c}_1}{\kappa_0^2} n^{-1}\log n.
\end{eqnarray*}
Without loss of generality, suppose $|i_1-\tau_{0,k-1}|\le 3\bar{c}_1 \kappa_0^{-2} n^{-1}\log n$. Using similar arguments in \eqref{lowerboundeq2} and \eqref{lowerboundeq2.5}, we can show that $\zeta_1=O(\log n)$, with probability at least $1-O(n^{-2})$. 

As for $\zeta_2$, consider intervals of the form $[\tau_{0,j}, (m+1)^{-1} i)$ for $j=0,1,\dots,K-1$, $i=1,\dots,m+1$. Denoted by $\mathfrak{J}(m)$ the set consisting of all such intervals. Similar to Lemma \hyperlink{lemma1}{1}, we can show that the following event occurs with probability at least $1-O(n^{-2})$: 
\begin{eqnarray}\label{someevent}
\left\|\sum_{i=1}^n \mathbb{I}(A_i\in \mathcal{I})\{Y_i-\overline{X}^\top_i\theta_0(A_i)\}\overline{X}_i\right\|_2=O(\sqrt{n|\mathcal{I}|\log n}),
\end{eqnarray}
for any $\mathcal{I}\in \mathfrak{J}(m)$ with $|\mathcal{I}|\ge c n^{-1}\log n$ for some constant $c>0$. Suppose $i_2-\tau_{0,k-1}\ge c n^{-1}\log n$. Under the event defined in \eqref{someevent}, it follows that 
\begin{eqnarray}\label{someevent1}
\left\|\sum_{i=1}^n \mathbb{I}(A_i\in [\tau_{0,k-1},i_2))\{Y_i-\overline{X}^\top_i\theta_0(A_i)\}\overline{X}_i\right\|_2=O(\sqrt{n|\mathcal{I}|\log n}),
\end{eqnarray}
Since $\int_{\mathcal{I}} \|\theta_0(a)-\theta_{0,\mathcal{I}}\|_2^2da \le \bar{c}_1 n^{-1}\log n$, we have $\int_{\tau_{0,k-1}}^{i_2} \|\theta_0(a)-\theta_{0,\mathcal{I}}\|_2^2da \le \bar{c}_1 n^{-1}\log n$, and hence $(i_2-\tau_{0,k-1}) \|\theta_0(a)-\theta_{0,\mathcal{I}}\|_2^2\le \bar{c}_1 n^{-1}\log n$, for any $a\in [\tau_{0,k-1},i_2)$. This together with \eqref{someevent1} yields that
\begin{eqnarray*}
	&&\left| \sum_{i=1}^n \mathbb{I}(A_i\in [\tau_{0,k-1},i_2))\{Y_i-\overline{X}^\top_i\theta_0(A_i)\}\overline{X}_i\{\theta_0(A_i)-\theta_{0,\mathcal{I}} \} \right|\\
	&\le& \left\|\sum_{i=1}^n \mathbb{I}(A_i\in [\tau_{0,k-1},i_2))\{Y_i-\overline{X}^\top_i\theta_0(A_i)\}\overline{X}_i\right\|_2 \|\theta_{0}(\tau_{0,k-1})-\theta_{0,\mathcal{I}}\|_2=O(\log n),
\end{eqnarray*}
and hence $\zeta_2=O(\log n)$. When $i_2-\tau_{0,k-1}\le c n^{-1}\log n$, using similar arguments in \eqref{lowerboundeq2} and \eqref{lowerboundeq2.5}, we can show that $\zeta_2=O(\log n)$, with probability at least $1-O(n^{-2})$. Thus, we've shown that with probability at least $1-O(n^{-2})$, for any interval $\mathcal{I}$ that belongs to the Category 2 with $\int_{\mathcal{I}} \|\theta_0(a)-\theta_{0,\mathcal{I}}\|_2^2\le \bar{c}_1 n^{-1}\log n$, we have
\begin{eqnarray*}
	\left| \sum_{i=1}^n \mathbb{I}(A_i\in \mathcal{I})\{Y_i-\overline{X}^\top_i\theta_0(A_i)\}\overline{X}_i\{\theta_0(A_i)-\theta_{0,\mathcal{I}} \} \right|=O(\log n).
\end{eqnarray*}	

Similarly, one can show that with probability at least $1-O(n^{-2})$, for any interval $\mathcal{I}$ that belongs to the Category 3 with $\int_{\mathcal{I}} \|\theta_0(a)-\theta_{0,\mathcal{I}}\|_2^2\le \bar{c}_1 n^{-1}\log n$, we have
\begin{eqnarray*}
	\left| \sum_{i=1}^n \mathbb{I}(A_i\in \mathcal{I})\{Y_i-\overline{X}^\top_i\theta_0(A_i)\}\overline{X}_i\{\theta_0(A_i)-\theta_{0,\mathcal{I}} \} \right|=O(\log n).
\end{eqnarray*}	
The proof is thus completed. 

\subsection{Proof of Lemma 4}\index{Proof of Lemma 4}
Consider a given interval $\mathcal{I}\in \widehat{\mathcal{P}}$. Suppose $|\mathcal{I}|<\bar{c}_3 \gamma_n$. The value of the constant $\bar{c}_3$ will be determined later. Then, for sufficiently large $n$, we can find some interval $\mathcal{I}' \in \mathfrak{I}(m)\cap \widehat{\mathcal{P}}$ that is adjacent to $\mathcal{I}$. Thus, we have $\mathcal{I}\cup \mathcal{I}'\in \mathfrak{I}(m)$, and hence
\begin{eqnarray}\label{event00}
&&~~\frac{1}{n}\sum_{i=1}^n \mathbb{I}(A_i\in \mathcal{I})(Y_i-\overline{X}_i^\top \widehat{\theta}_{\mathcal{I}})^2+\lambda_n |\mathcal{I}| \| \widehat{\theta}_{\mathcal{I}}\|_2^2+\frac{1}{n}\sum_{i=1}^n \mathbb{I}(A_i\in \mathcal{I}') (Y_i-\overline{X}_i^\top \widehat{\theta}_{\mathcal{I}'})^2\\ \nonumber
&+&\lambda_n |\mathcal{I}'| \| \widehat{\theta}_{\mathcal{I}'}\|_2^2\le \frac{1}{n}\sum_{i=1}^n \mathbb{I}(A_i\in \mathcal{I}\cup \mathcal{I}') (Y_i-\overline{X}_i^\top \widehat{\theta}_{\mathcal{I}\cup \mathcal{I}'})^2+\lambda_n |\mathcal{I}\cup \mathcal{I}'| \|\widehat{\theta}_{\mathcal{I}\cup \mathcal{I}'}\|_2^2-\gamma_n. 
\end{eqnarray}
Notice that the left-hand-side of the above expression is nonnegative. It follows that
\begin{eqnarray*}
	\gamma_n\le \frac{1}{n}\sum_{i=1}^n \mathbb{I}(A_i\in \mathcal{I}\cup \mathcal{I}') (Y_i-\overline{X}_i^\top \widehat{\theta}_{\mathcal{I}\cup \mathcal{I}'})^2+\lambda_n |\mathcal{I}\cup \mathcal{I}'| \|\widehat{\theta}_{\mathcal{I}\cup \mathcal{I}'}\|_2^2.
\end{eqnarray*}
By definition, we have
\begin{eqnarray}\label{minimizer}
\widehat{\theta}_{\mathcal{I}\cup \mathcal{I}'}=\argmin_{\theta\in \mathbb{R}^{p+1}} \left(\frac{1}{n}\sum_{i=1}^n \mathbb{I}(A_i\in \mathcal{I}\cup \mathcal{I}') (Y_i-\overline{X}_i^\top \theta)^2+\lambda_n |\mathcal{I}\cup \mathcal{I}'|\|\theta\|_2^2\right).
\end{eqnarray}
Therefore, we obtain that
\begin{eqnarray}\label{result1}
\gamma_n&&\le \sum_{i=1}^n \frac{\mathbb{I}(A_i\in \mathcal{I}\cup \mathcal{I}') (Y_i-\overline{X}_i^\top \bm{0}_{p+1})^2}{n}	+\lambda_n |\mathcal{I}\cup \mathcal{I}'| \|\bm{0}_{p+1} \|_2^2\\\nonumber
&&=\sum_{i=1}^n \frac{\mathbb{I}(A_i\in \mathcal{I}\cup \mathcal{I}') Y_i^2}{n}.
\end{eqnarray}
Suppose
\begin{eqnarray}\label{assertion1}
|\mathcal{I}\cup \mathcal{I}'|\le \frac{\gamma_n}{8c_0},
\end{eqnarray}
where the constant $c_0$ is defined in Lemma 1. 

Since $\gamma_n\gg n^{-1}$ and $m\asymp n$, we can find some interval $\mathcal{I}^*\in \mathfrak{I}(m)$ that covers $\mathcal{I}\cup \mathcal{I}'$ and satisfies $(8c_0)^{-1} \gamma_n\le |\mathcal{I}^*|\le (4c_0)^{-1} \gamma_n$. Under the event defined in \eqref{event5}, it follows from the condition $\gamma_n\gg n^{-1}\log n$ that
\begin{eqnarray*}
	\frac{1}{n}\sum_{i=1}^n \mathbb{I}(A_i\in \mathcal{I}\cup \mathcal{I}') Y_i^2\le \frac{1}{n}\sum_{i=1}^n \mathbb{I}(A_i\in \mathcal{I}^*) Y_i^2\le c_0\left( \frac{\sqrt{|(4c_0)^{-1}\gamma_n|\log n}}{\sqrt{n}}+(4c_0)^{-1}\gamma_n \right)\\
	\le 2c_0 (4c_0)^{-1}\gamma_n=\frac{\gamma_n}{2},
\end{eqnarray*}
for sufficiently large $n$. This apparently violates the results in \eqref{result1}. Thus, Assertion \eqref{assertion1} doesn't hold. Therefore, we obtain that
\begin{eqnarray}\label{assertion2}
|\mathcal{I}\cup \mathcal{I}'|\ge \frac{\gamma_n}{8c_0},
\end{eqnarray}
with probability at least $1-O(n^{-2})$. 

Suppose the constant $\bar{c}_3$ satisfies $\bar{c}_3\le (16c_0)^{-1}$. Under the event defined in \eqref{assertion2}, we have $|\mathcal{I}'|\ge \gamma_n (16c_0)^{-1}$. By \eqref{event1}, we have with probability at least $1-O(n^{-2})$ that $\|\widehat{\theta}_{\mathcal{I}'}-\theta_{0,\mathcal{I}'}\|_2\le c_0\sqrt{n^{-1}\log n}|\mathcal{I}'|^{-1/2}\le 4c_0^{3/2} \sqrt{n^{-1}\log n}\gamma_n^{-1}\ll 1$. By \eqref{boundbeta0I}, we have with probability at least $1-O(n^{-2})$ that 
\begin{eqnarray}\label{boundwidehatbeta0I}
\|\widehat{\theta}_{\mathcal{I}'}\|_2\le 2c_0,
\end{eqnarray}
for sufficiently large $n$. 

In addition, it follows from \eqref{minimizer} that
\begin{eqnarray*}
	\frac{1}{n}\sum_{i=1}^n \mathbb{I}(A_i\in \mathcal{I}\cup \mathcal{I}') (Y_i-\overline{X}_i^\top \widehat{\theta}_{\mathcal{I}\cup \mathcal{I}'})^2+\lambda_n |\mathcal{I}\cup \mathcal{I}'|\|\widehat{\theta}_{\mathcal{I}\cup \mathcal{I}'}\|_2^2\\
	\le  \frac{1}{n}\sum_{i=1}^n \mathbb{I}(A_i\in \mathcal{I}\cup \mathcal{I}') (Y_i-\overline{X}_i^\top \widehat{\theta}_{\mathcal{I}'})^2+\lambda_n |\mathcal{I}\cup \mathcal{I}'|\|\widehat{\theta}_{ \mathcal{I}'}\|_2^2. 
\end{eqnarray*}
By \eqref{event00}, this further implies that
\begin{eqnarray*}
	\frac{1}{n}\sum_{i=1}^n \mathbb{I}(A_i\in \mathcal{I})(Y_i-\overline{X}_i^\top \widehat{\theta}_{\mathcal{I}})^2+\lambda_n |\mathcal{I}| \| \widehat{\theta}_{\mathcal{I}}\|_2^2\le \frac{1}{n}\sum_{i=1}^n \mathbb{I}(A_i\in \mathcal{I})(Y_i-\overline{X}_i^\top \widehat{\theta}_{\mathcal{I}'})^2+\lambda_n |\mathcal{I}| \| \widehat{\theta}_{\mathcal{I}'}\|_2^2-\gamma_n,
\end{eqnarray*}
and hence
\begin{eqnarray*}
	\gamma_n\le \frac{1}{n}\sum_{i=1}^n \mathbb{I}(A_i\in \mathcal{I})(Y_i-\overline{X}_i^\top \widehat{\theta}_{\mathcal{I}'})^2+\lambda_n |\mathcal{I}| \| \widehat{\theta}_{\mathcal{I}'}\|_2^2.
\end{eqnarray*}
By \eqref{boundwidehatbeta0I} and the conditions that $\lambda_n=O(n^{-1}\log n)$, $\gamma_n\gg n^{-1}\log n$, we have for sufficiently large $n$,
\begin{eqnarray*}
	\frac{\gamma_n}{2}\le \frac{1}{n}\sum_{i=1}^n \mathbb{I}(A_i\in \mathcal{I})(Y_i-\overline{X}_i^\top \widehat{\theta}_{\mathcal{I}'})^2.
\end{eqnarray*}
It thus follows from Cauchy-Schwarz inequality and \eqref{boundwidehatbeta0I} that
\begin{eqnarray*}
	\frac{\gamma_n}{2}\le \frac{2}{n}\sum_{i=1}^n \mathbb{I}(A_i\in \mathcal{I})(Y_i^2+\|\overline{X}_i^\top\|_2^2 \|\widehat{\theta}_{\mathcal{I}'}\|_2^2)\le \frac{2(1+4c_0^2)}{n}\sum_{i=1}^n \mathbb{I}(A_i\in \mathcal{I})(Y_i^2+\|\overline{X}_i\|_2^2).
\end{eqnarray*}
Using similar arguments in showing \eqref{assertion2}, we obtain that
\begin{eqnarray*}
	|\mathcal{I}|\ge \frac{\gamma_n}{32(1+4c_0^2)c_0}.
\end{eqnarray*}
with probability at least $1-O(n^{-2})$. Set $\bar{c}_3=32^{-1} (1+4c_0^2)^{-1} c_0^{-1}$, this violates the assumption that $|\mathcal{I}|<\bar{c}_3 \gamma_n$. Thus, with probability at least $1-O(n^{-2})$, we obtain that $|\mathcal{I}|\ge \bar{c}_3\gamma_n$, for any $\mathcal{I}\in \widehat{\mathcal{P}}$. The proof is hence completed.

\subsection{Proof of Theorem 2}
Let $\{\widehat{\tau}_1,\widehat{\tau}_2,\dots,\widehat{\tau}_{\widehat{K}-1}\}$ be the set of change points in $J(\widehat{\mathcal{P}})$.  Under the events defined in Theorem \hyperlink{thm1}{1}, we have $\widehat{K}=K$, and
\begin{eqnarray}\label{localdist}
\max_{k\in \{1,\dots,K-1\}} |\widehat{\tau}_k-\tau_{0,k}|\le c n^{-1}\log n,
\end{eqnarray}
for some constant $c>0$. Set $\widehat{\tau}_0=0$ and $\widehat{\tau}_{K}=1$. 

Under the event defined in \eqref{localdist}, we have for sufficiently large $n$ that 
\begin{eqnarray}\label{anotherevent}
\widehat{\tau}_{k}-\widehat{\tau}_{k-1}\ge \delta_{\min},\,\,\,\, \,\,\,\,\forall k\in \{1,\dots,K\}.
\end{eqnarray}
Since $\pi^*$ satisfies $\sup_{\mathcal{I}\subseteq [0,1]} \sup_{\substack{ a\in \mathcal{I},x\in \mathbb{X} }} |\mathcal{I}|\pi^*(a;x,\mathcal{I})\asymp 1$, there exists some constant $\bar{c}_4>0$ such that $\pi^*(a;x,\widehat{d}(x))\le \bar{c}_4 |\widehat{d}(x)|^{-1}$ for all $a$ and $x$. This together with \eqref{anotherevent} yields that
\begin{eqnarray}\label{boundpistar}
\pi^*(a;x,\widehat{d}(x))\le \bar{c}_4 \delta_{\min}^{-1},\,\,\,\,\,\,\,\,\forall a\in [0,1], x\in \mathbb{X}.
\end{eqnarray}

The rest of our proof is divided into three parts. In the first part, we show that there exists some constant $C>0$ such that
\begin{eqnarray}\label{anotherevent2}
\|\widehat{\theta}_{[\widehat{\tau}_{k-1}, \widehat{\tau}_k)}-\widehat{\theta}_{[\tau_{0,k-1},\tau_{0,k})}\|_2\le \frac{C \log n}{n},\,\,\,\,\,\,\,\,\forall k\in \{1,\dots,K\},
\end{eqnarray}
with probability at least $1-O(n^{-2})$. Using similar arguments in Lemma \hyperlink{lemma1}{1}, we can show that there exists some constant $c_3>0$ such that the following events occur with probability at least $1-O(n^{-2})$:
\begin{eqnarray*}
	\|\widehat{\theta}_{[\tau_{0,k-1}, \tau_{0,k})}-\theta_{0,[\tau_{0,k-1}, \tau_{0,k})}\|_2\le \frac{c_3\sqrt{\log n}}{\sqrt{n\delta_{\min}}},\,\,\,\,\,\,\,\forall k\in \{1,\dots,K\}. 
\end{eqnarray*}
This together with \eqref{anotherevent2} implies that
\begin{eqnarray}\label{anotherevent3}
\|\widehat{\theta}_{[\widehat{\tau}_{k-1},\widehat{\tau}_{k} )}-\theta_{0,[\tau_{0,k-1}, \tau_{0,k})}\|_2\le \frac{2c_3\sqrt{\log n}}{\sqrt{n\delta_{\min}}},\,\,\,\,\,\,\,\forall k\in \{1,\dots,K\},
\end{eqnarray}
for sufficiently large $n$, with probability at least $1-O(n^{-2})$. 

In the second part, we define an integer-valued function $\widehat{\mathbb{K}}(x)$ as follows. We set $\widehat{\mathbb{K}}(x)=k$ if $\widehat{d}(x)=[\widehat{\tau}_{k-1}, \widehat{\tau}_k)$ for some integer $k$ such that $1\le k\le K-1$, and set $\widehat{\mathbb{K}}(x)=K$ if $\widehat{d}(x)=[\widehat{\tau}_{K-1}, 1]$. By the definition of $\widehat{\theta}_{\mathcal{I}}$ and $\theta_{0,\mathcal{I}}$, we have almost surely $\widehat{\theta}_{[\widehat{\tau}_{K-1},1)}=\widehat{\theta}_{[\widehat{\tau}_{K-1},1]}$ and $\theta_{0,[\tau_{0,K-1},1)}=\theta_{0,[\tau_{0,K-1},1]}$.
It is immediate to see that 
\begin{eqnarray}\label{widehatK}
\widehat{\mathbb{K}}(x)=\sargmax_{k\in\{1,\dots,K\}} \bar{x}^\top \widehat{\theta}_{[\widehat{\tau}_{k-1}, \widehat{\tau}_k)},
\end{eqnarray}
where $\sargmax$ denotes the smallest maximizer when the argmax is not unique. In Part 2, we focus on proving
\begin{eqnarray}\label{middlestep1}
V^{\pi^*}(\widehat{d})\ge \Mean \left(\overline{X}^\top \theta_{0,[\tau_{0,\widehat{\mathbb{K}}(X)-1}, \tau_{0,\widehat{\mathbb{K}}(X)})}\right)-O(1)n^{-1}\log n,
\end{eqnarray}
with probability at least $1-O(n^{-2})$, where $O(1)$ denotes some positive constant. 

In the last part, we provide an opper bound for
\begin{eqnarray*}
	V^{\tiny{opt}}-\Mean \left(\overline{X}^\top \theta_{0,[\tau_{0,\widehat{\mathbb{K}}(X)-1}, \tau_{0,\widehat{\mathbb{K}}(X)})}\right).
\end{eqnarray*}
This together with \eqref{middlestep1} yields the desired results.

\smallskip

\noindent \textit{Proof of Part 1:} Let $\widehat{\Delta}_k=[\widehat{\tau}_{k-1}, \widehat{\tau}_k) \cup [\tau_{0,k-1},\tau_{0,k})^c+[\widehat{\tau}_{k-1}, \widehat{\tau}_k)^c \cup [\tau_{0,k-1},\tau_{0,k})$. With some calculations, we can show that
\begin{eqnarray*}
	\|\widehat{\theta}_{[\widehat{\tau}_{k-1}, \widehat{\tau}_k)}-\widehat{\theta}_{[\tau_{0,k-1},\tau_{0,k})}\|_2\le \zeta_1(k) \zeta_2(k)+\zeta_3(k) \zeta_4(k),
\end{eqnarray*}
where
\begin{eqnarray*}
	\zeta_1(k)&=& \left\| \left(\frac{1}{n}\sum_{i=1}^n \mathbb{I}(\tau_{0,k-1}\le A_i<\tau_{0,k}) \overline{X}_i \overline{X}_i^\top+\lambda_n (\tau_{0,k}-\tau_{0,k-1}) \mathbb{E}_{p+1} \right)^{-1}\right\|_2, \\
	\zeta_2(k)&=& \left\| \frac{1}{n}\sum_{i=1}^n \mathbb{I}(A_i\in \widehat{\Delta}_k) \overline{X}_i Y_i \right\|_2,\,\,\,\,\zeta_3(k)= \left\| \frac{1}{n}\sum_{i=1}^n \mathbb{I}(\tau_{0,k-1}\le A_i<\tau_{0,k}) \overline{X}_i Y_i \right\|_2,\\
	\zeta_4(k)&=&\left\| \left(\frac{1}{n}\sum_{i=1}^n \mathbb{I}(\tau_{0,k-1}\le A_i<\tau_{0,k}) \overline{X}_i \overline{X}_i^\top+\lambda_n (\tau_{0,k}-\tau_{0,k-1}) \mathbb{E}_{p+1} \right)^{-1}\right.\\
	&-&\left. \left(\frac{1}{n}\sum_{i=1}^n \mathbb{I}(\widehat{\tau}_{k-1}\le A_i<\widehat{\tau}_{k}) \overline{X}_i \overline{X}_i^\top+\lambda_n (\widehat{\tau}_{k}-\widehat{\tau}_{k-1}) \mathbb{E}_{p+1} \right)^{-1}\right\|_2.
\end{eqnarray*}

Similar to \eqref{eta1}, we can show with probability at least $1-O(n^{-2})$ that
\begin{eqnarray}\label{zeta1k}
\max_{k\in\{1,\dots,K\}}\zeta_1(k)=O(1)\,\,\,\,\hbox{and}\max_{k\in\{1,\dots,K\}}\zeta_5(k)=O(1),
\end{eqnarray}
where
\begin{eqnarray*}
	\zeta_5(k)=\left\|\left(\frac{1}{n}\sum_{i=1}^n \mathbb{I}(\widehat{\tau}_{k-1}\le A_i<\widehat{\tau}_{k}) \overline{X}_i \overline{X}_i^\top+\lambda_n (\widehat{\tau}_{k}-\widehat{\tau}_{k-1}) \mathbb{E}_{p+1} \right)^{-1}\right\|_2.
\end{eqnarray*}
Under the event defined in \eqref{localdist}, the Lebesgue measure of $\widehat{\Delta}_k$ is uniformly bounded by $2c n^{-1}\log n$, for any $k\in\{1,\dots,K\}$. Using similar arguments in \eqref{lowerboundeq2} and \eqref{lowerboundeq2.5}, we can show with probability at least $1-O(n^{-2})$ that
\begin{eqnarray}\label{zeta2k}
\max_{k\in\{1,\dots,K\}}\zeta_2(k)=O(n^{-1}\log n).
\end{eqnarray}
Similar to \eqref{eta4}, we can show with probability at least $1-O(n^{-2})$ that
\begin{eqnarray}\label{zeta3k}
\max_{k\in\{1,\dots,K\}}\zeta_3(k)=O(1).
\end{eqnarray}
Notice that $\zeta_4(k)$ can be upper bounded by
\begin{eqnarray*}
	\zeta_4(k)&\le& \zeta_1(k)\zeta_5(k)\left\| \frac{1}{n}\sum_{i=1}^n \mathbb{I}(\tau_{0,k-1}\le A_i<\tau_{0,k}) \overline{X}_i \overline{X}_i^\top+\lambda_n (\tau_{0,k}-\tau_{0,k-1}) \mathbb{E}_{p+1} \right.\\
	&-&\left. \frac{1}{n}\sum_{i=1}^n \mathbb{I}(\widehat{\tau}_{k-1}\le A_i<\widehat{\tau}_{k}) \overline{X}_i \overline{X}_i^\top-\lambda_n (\widehat{\tau}_{k}-\widehat{\tau}_{k-1}) \mathbb{E}_{p+1}\right\|_2\\
	&\le& \zeta_1(k)\zeta_5(k) \left\| \frac{1}{n}\sum_{i=1}^n \mathbb{I}(A_i\in \widehat{\Delta}_k) \overline{X}_i \overline{X}_i^\top+\lambda_n (\tau_{0,k}-\tau_{0,k-1}-\widehat{\tau}_k+\widehat{\tau}_{k-1}) \mathbb{E}_{p+1}\right\|_2.
\end{eqnarray*}
Under the condition $\lambda_n=O(n^{-1}\log n)$, using similar arguments in \eqref{lowerboundeq2} and \eqref{lowerboundeq2.5}, we can show that with probability at least $1-O(n^{-2})$, the absolute value of each element in the matrix
\begin{eqnarray*}
	\frac{1}{n}\sum_{i=1}^n \mathbb{I}(A_i\in \widehat{\Delta}_k) \overline{X}_i \overline{X}_i^\top+\lambda_n (\tau_{0,k}-\tau_{0,k-1}-\widehat{\tau}_k+\widehat{\tau}_{k-1}) \mathbb{E}_{p+1}
\end{eqnarray*}
is upper bounded by $O(n^{-1}\log n)$, uniformly for any $k\in \{1,\dots,K\}$. It follows that
\begin{eqnarray*}
	\left\| \frac{1}{n}\sum_{i=1}^n \mathbb{I}(A_i\in \widehat{\Delta}_k) \overline{X}_i \overline{X}_i^\top+\lambda_n (\tau_{0,k}-\tau_{0,k-1}-\widehat{\tau}_k+\widehat{\tau}_{k-1}) \mathbb{E}_{p+1}\right\|_2=O(n^{-1}\log n).
\end{eqnarray*}
In view of \eqref{zeta1k}, we obtain that
\begin{eqnarray}\label{zeta4k}
\max_{k\in \{1,\dots,K\}}\zeta_4(k)=O(n^{-1}\log n),
\end{eqnarray}
with probability at least $1-O(n^{-2})$. Combining \eqref{zeta1k}-\eqref{zeta4k} yields \eqref{anotherevent2}. 

\smallskip

\noindent \textit{Proof of Part 2:} It follows from Condition (A4) and the definition of the conditional Orlicz norm that
\begin{eqnarray*}
	\Mean \left\{\exp\left(\frac{|X^{(j)}|^2}{\omega^2}\right)\right\}=\Mean \left[\Mean \left\{\left.\exp\left(\frac{|X^{(j)}|^2}{\omega^2}\right)\right|A\right\}\right]\le 2,
\end{eqnarray*}
for any $j\in\{1,\dots,p\}$. Without loss of generality, suppose $\omega\ge \log^{-1/2} 2$. Then, we have
\begin{eqnarray*}
	\Mean \left\{\exp\left(\frac{|\overline{X}^{(j)}|^2}{\omega^2}\right)\right\}=\Mean \left[\Mean \left\{\left.\exp\left(\frac{|\overline{X}^{(j)}|^2}{\omega^2}\right)\right|A\right\}\right]\le 2,
\end{eqnarray*}
for any $j\in\{1,\dots,p+1\}$. As a result, it follows from Bonferroni's inequality and Markov's inequality that
\begin{eqnarray*}
	&&\hbox{Pr}\left( \|\overline{X}\|_2>\omega\sqrt{2(p+1)\log n}\right)\le \sum_{j=1}^{p+1} \hbox{Pr}(|\overline{X}^{(j)}|>\omega\sqrt{2\log n})\\ 
	&\le& \sum_{j=1}^{p+1} \Mean \left\{\exp\left(\frac{|\overline{X}^{(j)}|^2}{\omega^2}\right)\right\} / \exp\left( \frac{2\omega^2\log n}{\omega^2} \right)\le \frac{2(p+1)}{n^2}. 
\end{eqnarray*}
Thus, we obtain
\begin{eqnarray}\label{boundX}
\hbox{Pr}(\mathcal{A}^*)\ge 1-\frac{2(p+1)}{n^2},
\end{eqnarray}
where
\begin{eqnarray*}
	\mathcal{A}^*=\{ \|\overline{X}\|_2\le \omega\sqrt{2(p+1)\log n} \}. 
\end{eqnarray*}

Consider the event
\begin{eqnarray*}
	\mathcal{A}_0=\bigcup_{\mathcal{I}_1,\mathcal{I}_2\in \mathcal{P}_0} \left\{0<\left|\overline{X}^\top (\theta_{0,\mathcal{I}_1}-\theta_{0,\mathcal{I}_2})\right|\le \frac{4\sqrt{2(p+1)}c_3 \omega \log n}{\sqrt{n\delta_{\min}}}\right\}.
\end{eqnarray*}
By Condition (A5) and Bonferroni's inequality, we have
\begin{eqnarray}\label{eventA0}
\hbox{Pr}(\mathcal{A}_0)\le \sum_{\substack{\mathcal{I}_1,\mathcal{I}_2\in \mathcal{P}_0\\ \mathcal{I}_1\neq \mathcal{I}_2 }} \hbox{Pr}\left( 0<\left|\overline{X}^\top (\theta_{0,\mathcal{I}_1}-\theta_{0,\mathcal{I}_2})\right|\le \frac{4\sqrt{2(p+1)}c_3 \omega \log n}{\sqrt{n\delta_{\min}}} \right)\\ \nonumber
\le K^2 \left(\frac{4\sqrt{2(p+1)}c_3 \omega \log n}{\sqrt{n\delta_{\min}}}\right)^{\gamma}.
\end{eqnarray}


By the definition of $V^{\pi^*}(\cdot)$, we have
\begin{eqnarray*}
	V^{\pi^*}(\widehat{d})=\Mean \left(\int_{\widehat{d}(X)} \overline{X}^\top \theta_0(a)  \pi^*(a;X,\widehat{d}(X))da\right).
\end{eqnarray*}
Notice that the expectation in the above expression is taken with respect to $X$. Define an interval-valued function $\widehat{d}_0(x)=[\tau_{0,\widehat{\mathbb{K}}(x)-1}, \tau_{0,\widehat{\mathbb{K}}(x)})$ and set $\widehat{\Delta}(x)=\widehat{d}(x)\cap \{\widehat{d}_0(x)\}^c$. It follows that
\begin{eqnarray*}
	V^{\pi^*}(\widehat{d})= \Mean \left(\int_{\widehat{d}_0(X)\cap \widehat{d}(X)} \overline{X}^\top \theta_0(a)  \pi^*(a;X,\widehat{d}(X))da\right)+\underbrace{\Mean \left(\int_{\widehat{\Delta}(X)} \overline{X}^\top \theta_0(a)  \pi^*(a;X,\widehat{d}(X))da\right)}_{\chi_1}\\
	=\Mean \left(\int_{\widehat{d}_0(X)} \overline{X}^\top \theta_0(a)  \pi^*(a;X,\widehat{d}(X))da\right)+\chi_1.
\end{eqnarray*}
Here, the second equality is due to that $\pi^*(a;X,\widehat{d}(X))=0$, for any $a\in \{\widehat{d}(X)\}^c$. By \eqref{boundbeta0I} and \eqref{boundpistar}, we have
\begin{eqnarray*}
	|\chi_1| \le c_0 \bar{c}_4 \delta_{\min}^{-1} \Mean \left(\int_{\widehat{\Delta}(X)} \|\overline{X}\|_2 da\right)=c_0 \bar{c}_4 \delta_{\min}^{-1} \Mean \|\overline{X}\|_2 \lambda(\widehat{\Delta}(X)),
\end{eqnarray*}
where $\lambda(\widehat{\Delta}(X))$ denotes the Lebesgue measure of $\widehat{\Delta}(X)$. Under the event defined in \eqref{localdist}, we have $\lambda(\widehat{\Delta}(X))\le 2cn^{-1}\log n$, for any realization of $X$. It follows that
\begin{eqnarray}\label{chi10}
|\chi_1|\le 2c c_0 \bar{c}_4 \delta_{\min}^{-1} (n^{-1}\log n) \Mean \|\overline{X}\|_2.
\end{eqnarray} 
By \eqref{meanXcA0}, we have
\begin{eqnarray}\label{meanXsecondorder}
\Mean \|\overline{X}\|_2^2=\sum_{j=1}^{p+1} \Mean |\overline{X}^{(j)}|^2=\sum_{j=1}^{p+1} \Mean(\Mean |\overline{X}^{(j)}|^2 |A)\le \omega^2 (p+1).
\end{eqnarray}
By Cauchy-Schwarz inequality, this further implies that
\begin{eqnarray*}
	\Mean \|\overline{X}\|_2\le \sqrt{\Mean \|\overline{X}\|_2^2}\le \omega\sqrt{p+1}. 
\end{eqnarray*}
This together with \eqref{chi10} yields
\begin{eqnarray}\label{chi1}
|\chi_1|\le 2c c_0 \bar{c}_4 \omega\sqrt{p+1} \delta_{\min}^{-1} n^{-1}\log n,
\end{eqnarray}
with probability at least $1-O(n^{-2})$. 

Notice that $\theta_0(\cdot)$ is a constant on $\widehat{d}_0(x)$ for any $x$. It follows that
\begin{eqnarray*}
	\Mean \left(\int_{\widehat{d}_0(X)} \overline{X}^\top \theta_0(a)  \pi^*(a;X,\widehat{d}(X))da\right)=\Mean \left(\overline{X}^\top \theta_{0,[\tau_{0,\widehat{\mathbb{K}}(x)-1}, \tau_{0,\widehat{\mathbb{K}}(x)})}\right) \int_{\widehat{d}_0(X)} \pi^*(a;X,\widehat{d}(X))da\\
	=\Mean \left(\overline{X}^\top \theta_{0,[\tau_{0,\widehat{\mathbb{K}}(x)-1}, \tau_{0,\widehat{\mathbb{K}}(x)})}\right) \int_{\widehat{d}_0(X) \cap \widehat{d}(X) } \pi^*(a;X,\widehat{d}(X))da\\
	=\Mean \left(\overline{X}^\top \theta_{0,[\tau_{0,\widehat{\mathbb{K}}(x)-1}, \tau_{0,\widehat{\mathbb{K}}(x)})}\right) \int_{\widehat{d}(X) } \pi^*(a;X,\widehat{d}(X))da-\chi_2=\Mean \left(\overline{X}^\top \theta_{0,[\tau_{0,\widehat{\mathbb{K}}(x)-1}, \tau_{0,\widehat{\mathbb{K}}(x)})}\right)-\chi_2,
\end{eqnarray*}
where
\begin{eqnarray*}
	\chi_2=\Mean \left(\overline{X}^\top \theta_{0,[\tau_{0,\widehat{\mathbb{K}}(x)-1}, \tau_{0,\widehat{\mathbb{K}}(x)})}\right) \int_{\widehat{\Delta}(X) } \pi^*(a;X,\widehat{d}(X))da. 
\end{eqnarray*}
Similar to \eqref{chi1}, we can show that
\begin{eqnarray*}
	|\chi_2|=O(n^{-1}\log n),
\end{eqnarray*}
with probability at least $1-O(n^{-2})$. This together with \eqref{chi1} yields \eqref{middlestep1}. 

\smallskip

\noindent \textit{Proof of Part 3}: Similar to the definition of $\widehat{\mathbb{K}}$, we define 
\begin{eqnarray}\label{K0}
\mathbb{K}_0(x)=\sargmax_{k\in\{1,\dots,K\}} \bar{x}^\top \theta_{0,[\tau_{0,k-1}, \tau_{0,k})}. 
\end{eqnarray}
Let
\begin{eqnarray*}
	\mathbb{K}^*(x)=\left\{k_0: k_0=\argmax_{k\in\{1,\dots,K\}} \bar{x}^\top \theta_{0,[\tau_{0,k-1}, \tau_{0,k})}\right\},
\end{eqnarray*}
denote the set that consists of all the maximizers. Apparently, $\mathbb{K}_0(x)\in \mathbb{K}^*(x)$, $\forall x\in \mathbb{X}$. 

We now claim that
\begin{eqnarray}\label{correctK}
\widehat{\mathbb{K}}(X)\in \mathbb{K}^*(X),
\end{eqnarray}
under the events defined in $\mathcal{A}_0^c\cap \mathcal{A}^*$ and \eqref{anotherevent3}. Otherwise, suppose there exists some $k_0\in \{1,\dots,K\}$ such that
\begin{eqnarray}\label{inequality1}
&&\overline{X}^\top \widehat{\theta}_{[\widehat{\tau}_{k_0-1}, \widehat{\tau}_{k_0})}\ge \max_{k\neq k_0} \overline{X}^\top \widehat{\theta}_{[\widehat{\tau}_{k-1}, \widehat{\tau}_{k})},\\ \label{inequality2}
&&\max_{k\neq k_0}\overline{X}^\top \theta_{0,[\tau_{0,k-1}, \tau_{0,k})}> \overline{X}^\top \theta_{0,[\tau_{0,k_0-1}, \tau_{0,k_0})}.
\end{eqnarray}
Under $\mathcal{A}_0^c$, it follows from \eqref{inequality2} that
\begin{eqnarray}\label{inequality4}
\max_{k\neq k_0}\overline{X}^\top \theta_{0,[\tau_{0,k-1}, \tau_{0,k})}> \overline{X}^\top \theta_{0,[\tau_{0,k_0-1}, \tau_{0,k_0})}+\frac{4\sqrt{2(p+1)}c_3 \omega \log n}{\sqrt{n\delta_{\min}}}.
\end{eqnarray}
Under the events defined in $\mathcal{A}^*$ and \eqref{anotherevent3}, we have
\begin{eqnarray*}
	\max_{k\in\{1,\dots,K\}} |\overline{X}^\top (\widehat{\theta}_{[\widehat{\tau}_{k-1},\widehat{\tau}_{k} )}-\theta_{0,[\tau_{0,k-1}, \tau_{0,k})}) | \le \|\overline{X}\|_2 \max_{k\in\{1,\dots,K\}} \|\widehat{\theta}_{[\widehat{\tau}_{k-1},\widehat{\tau}_{k} )}-\theta_{0,[\tau_{0,k-1}, \tau_{0,k})}\|_2\\
	\le \frac{2\sqrt{2(p+1)}c_3 \omega \log n}{\sqrt{n\delta_{\min}}}.
\end{eqnarray*}
This together with \eqref{inequality4} yields that
\begin{eqnarray*}
	\max_{k\neq k_0}\overline{X}^\top \widehat{\theta}_{[\widehat{\tau}_{k-1}, \widehat{\tau}_{k})}> \overline{X}^\top \widehat{\theta}_{[\widehat{\tau}_{k_0-1}, \widehat{\tau}_{k_0})}.
\end{eqnarray*}
In view of \eqref{inequality1}, we have reached an contradiction. Therefore, \eqref{correctK} holds under the events defined in $\mathcal{A}_0^c\cap \mathcal{A}^*$ and \eqref{anotherevent3}. When \eqref{correctK} holds, it follows from the definition of $\mathbb{K}^*(\cdot)$ that $\overline{X}^\top \theta_{0,[\widehat{\mathbb{K}}(X)-1, \widehat{\mathbb{K}}(X))}=\overline{X}^\top \theta_{0,[\mathbb{K}_0(X)-1, \mathbb{K}_0(X))}$. Therefore, under the event defined in \eqref{anotherevent3}, we have
\begin{eqnarray}\label{inequality5}
&&~~~~~~~\Mean \left(\overline{X}^\top \theta_{0,[\tau_{0,\widehat{\mathbb{K}}(X)-1}, \tau_{0,\widehat{\mathbb{K}}(X)})}\right)=\Mean \left(\overline{X}^\top \theta_{0,[\tau_{0,\widehat{\mathbb{K}}(X)-1}, \tau_{0,\widehat{\mathbb{K}}(X)})}\right)\mathbb{I}(\mathcal{A}_0^c \cap \mathcal{A}^*)\\ \nonumber
&+&\underbrace{\Mean \left(\overline{X}^\top \theta_{0,[\tau_{0,\widehat{\mathbb{K}}(X)-1}, \tau_{0,\widehat{\mathbb{K}}(X)})}\right)\mathbb{I}(\mathcal{A}_0 \cup \mathcal{A}^{*c})}_{\chi_3}=\Mean \left(\overline{X}^\top \theta_{0,[\tau_{0,\mathbb{K}_0(X)-1}, \tau_{0,\mathbb{K}_0(X)})}\right)\mathbb{I}(\mathcal{A}_0^c \cap \mathcal{A}^*)\\ \nonumber
&+&\chi_3=\Mean \left(\overline{X}^\top \theta_{0,[\tau_{0,\mathbb{K}_0(X)-1}, \tau_{0,\mathbb{K}_0(X)})}\right)+\chi_3-\underbrace{\Mean \left(\overline{X}^\top \theta_{0,[\tau_{0,\mathbb{K}_0(X)-1}, \tau_{0,\mathbb{K}_0(X)})}\right)\mathbb{I}(\mathcal{A}_0 \cup \mathcal{A}^{*c})}_{\chi_4}.
\end{eqnarray}
Notice that
\begin{eqnarray*}
	\chi_3-\chi_4=\Mean \overline{X}^\top \left(\theta_{0,[\tau_{0,\widehat{\mathbb{K}}(X)-1}, \tau_{0,\widehat{\mathbb{K}}(X)})}-\theta_{0,[\tau_{0,\mathbb{K}_0(X)-1}, \tau_{0,\mathbb{K}_0(X)})}\right)\mathbb{I}(\mathcal{A}_0 \cup \mathcal{A}^{*c}).
\end{eqnarray*}
Using similar arguments in showing \eqref{correctK}, we can show that under the event defined in \eqref{anotherevent3}, 
\begin{eqnarray*}
	\overline{X}^\top \left(\theta_{0,[\tau_{0,\widehat{\mathbb{K}}(X)-1}, \tau_{0,\widehat{\mathbb{K}}(X)})}-\theta_{0,[\tau_{0,\mathbb{K}_0(X)-1}, \tau_{0,\mathbb{K}_0(X)})}\right)\neq 0,
\end{eqnarray*}
only when 
\begin{eqnarray*}
	0<\left|\overline{X}^\top \left(\theta_{0,[\tau_{0,\widehat{\mathbb{K}}(X)-1}, \tau_{0,\widehat{\mathbb{K}}(X)})}-\theta_{0,[\tau_{0,\mathbb{K}_0(X)-1}, \tau_{0,\mathbb{K}_0(X)})}\right)\right|\le \frac{4\sqrt{2(p+1)}c_3 \omega \log n}{\sqrt{n\delta_{\min}}}.
\end{eqnarray*}
Therefore, under the event defined in \eqref{anotherevent3}, we have
\begin{eqnarray*}
	|\chi_3-\chi_4|\le \frac{4\sqrt{2(p+1)}c_3 \omega \log n}{\sqrt{n\delta_{\min}}} \hbox{Pr}(\mathcal{A}_0\cup \mathcal{A}^{*c}).
\end{eqnarray*}
It follows from \eqref{boundX} and \eqref{eventA0}, we have
\begin{eqnarray*}
	|\chi_3-\chi_4|\le \frac{4\sqrt{2(p+1)}c_3 \omega \log n}{\sqrt{n\delta_{\min}}} \left\{\frac{2(p+1)}{n^2}+K^2 \left(\frac{4\sqrt{2(p+1)}c_3 \omega \log n}{\sqrt{n\delta_{\min}}}\right)^{\gamma} \right\}. 
\end{eqnarray*}
For sufficiently large $n$, this together with \eqref{middlestep1} and \eqref{inequality5} implies that we have with probability at least $1-O(n^{-2})$, 
\begin{eqnarray*}
	V^{\pi^*}(\widehat{d})\ge \Mean \left(\overline{X}^\top \theta_{0,[\tau_{0,\mathbb{K}_0(X)-1}, \tau_{0,\mathbb{K}_0(X)})}\right)-O(1) (n^{-1}\log n+n^{-(1+\gamma)/2} \log^{1+\gamma} n),
\end{eqnarray*}
for some positive constant $O(1)$. The proof is hence completed by noting that
\begin{eqnarray*}
	V^{\tiny{opt}}=\Mean \left(\overline{X}^\top \theta_{0,[\tau_{0,\mathbb{K}_0(X)-1}, \tau_{0,\mathbb{K}_0(X)})}\right).
\end{eqnarray*}

\subsection{Proof of Theorem 3}

%
%
%
We first introduce some technical lemmas. We remark that the key ingredient of the proof lies in Lemma \hyperlink{lemma1_thm5}{5}, which establishes a uniform upper bound on the mean squared error of $\widehat{q}_{\mathcal{I}}$. 
Proofs of these lemmas can be found in Sections E.1 - E.3 of \cite{cai2021deep}\footnote{See \url{https://openreview.net/attachment?id=rvKD3iqtBdk&name=supplementary_material}} and we omit them for brevity. 
The rest of the proof can be similarly proven as Theorem \hyperlink{thm1}{1}. Specifically, we first show the consistency of the estimated change point locations. We then derive the rate of convergence of the estimated change point locations and the estimated outcome regression function. 

\smallskip
\noindent
{\bf \hypertarget{lemma1_thm5}{Lemma 5}}  {\it 	Assume conditions in Theorem \hyperlink{thm5}{3} are satisfied. Then there exists some constant $\bar{C}>0$ such that  the following holds with probability at least $1-O(n^{-2})$: For any $ \mathcal{I}\in \mathfrak{I}(m)$ and $|\mathcal{I}|\ge c\gamma_n $,
	\begin{eqnarray}\label{eqn:event1}
	\Mean  |q_{\mathcal{I},0}(X)-\widehat{q}_{\mathcal{I}} (X)|^2 \le \bar{C} (n|\mathcal{I}|)^{-2\beta/(2\beta+p)}\log^8 n,
	\end{eqnarray}
	where $q_{\mathcal{I},0}=\Mean (Y|A\in \mathcal{I},X)$ for any interval $\mathcal{I}$.} \hfill$\square$

\smallskip

\noindent 
{\bf \hypertarget{lemma3_thm5}{Lemma 6}} {\it Assume conditions in Theorem \hyperlink{thm5}{3} are satisfied. Then there exists some constant $\bar{C}>0$ such that the followings hold with probability at least $1-O(n^{-2})$: For any $\mathcal{I}\in  \mathfrak{I}(m)$ and $|\mathcal{I}|\ge c\gamma_n$,
\begin{eqnarray*}
\sum_{\substack{\mathcal{I}\in \widehat{\mathcal{P}} }}\left|\sum_{i=1}^n \mathbb{I}(A_i\in \mathcal{I})\{Y_i-q_{\mathcal{I},0}(X_i)\}  \{\widehat{q}_{\mathcal{I}}(X_i)-q_{\mathcal{I},0}(X_i)\} \right|\le \bar{C} (n|\mathcal{I}|)^{p/(2\beta+p)} \log^8 n,
\end{eqnarray*}
for any $\mathcal{I}\in \mathfrak{I}(m)$ such that $|\mathcal{I}|\ge c\gamma_n$ for any positive constant $c>0$. } \hfill$\square$

\smallskip

\noindent
{\bf \hypertarget{lemma2_thm5}{Lemma 7}} {\it	Under the conditions in Theorem \hyperlink{thm5}{3}, the following events occur with probability at least $1-O(n^{-2})$: there exists some constant $C>0$ such that $\min_{\mathcal{I}\in \widehat{\mathcal{P}}}|\mathcal{I}|\ge C \gamma_n$. } \hfill$\square$

\smallskip

We next show the consistency of the estimated change-point locations. Using similar arguments in proving \eqref{upperPhat}, we can show that
\begin{eqnarray}\label{upperPhat_thm5}
|\widehat{\mathcal{P}}|\le  {C}_0\gamma_n^{-1},
\end{eqnarray}
for sufficiently large $n$ and some constant $  {C}_0>0$.

Notice that
\begin{eqnarray*}
\begin{split}
\sum_{\mathcal{I}\in \widehat{\mathcal{P}}}  \sum_{i=1}^n \mathbb{I}(A_i\in \mathcal{I})\{Y_i-\widehat{q}_{\mathcal{I}}(X_i)\}^2 \ge \underbrace{ \sum_{\substack{\mathcal{I}\in \widehat{\mathcal{P}}} }\sum_{i=1}^n \mathbb{I}(A_i\in \mathcal{I})\{Y_i-q_{\mathcal{I},0}(X_i)\}^2}_{\eta_1^*}\\
+\sum_{\substack{\mathcal{I}\in \widehat{\mathcal{P}} }}\sum_{i=1}^n \mathbb{I}(A_i\in \mathcal{I}) \{\widehat{q}_{\mathcal{I}}(X_i)-q_{\mathcal{I},0}(X_i)\}^2\\
-2\sum_{\substack{\mathcal{I}\in \widehat{\mathcal{P}} }}\left|\sum_{i=1}^n \mathbb{I}(A_i\in \mathcal{I})\{Y_i-q_{\mathcal{I},0}(X_i)\}  \{\widehat{q}_{\mathcal{I}}(X_i)-q_{\mathcal{I},0}(X_i)\} \right|.
\end{split}
\end{eqnarray*}
The second line is non-negative. Under Lemmas \hyperlink{lemma3_thm5}{6} and \hyperlink{lemma2_thm5}{7}, the third line is lower bounded by $-C_1\sum_{\mathcal{I}\in \widehat{\mathcal{P}}} (n|\mathcal{I}| )^{p/(2\beta+p)}\log^8 n$ for some constant $C_1>0$. By H{\"o}lder's inequality, it can be further lower bounded by $-C_1 |\widehat{\mathcal{P}}|^{2\beta/(2\beta+p)} n^{p/(2\beta+p)}\log^8 n$. 
By \eqref{upperPhat_thm5} and the given condition on $\gamma_n$, the third line is $o(n)$. It follows that
\begin{eqnarray}\label{eqn:some00}
	\sum_{\mathcal{I}\in \widehat{\mathcal{P}}}  \sum_{i=1}^n \mathbb{I}(A_i\in \mathcal{I})\{Y_i-\widehat{q}_{\mathcal{I}}(X_i)\}^2 \ge \eta_1^*+o(n),
\end{eqnarray}
with probability at least $1-O(n^{-2})$. 

Similar to \eqref{event3} and \eqref{event4}, we can show that the following events occur with probability at least $1-O(n^{-2})$, 
\begin{eqnarray*}
	\left|\frac{1}{n}\sum_{i=1}^n \mathbb{I}(A_i\in\mathcal{I})\{Y_i-Q(X_i,A_i)\}\{Q(X_i,A_i)-q_{\mathcal{I},0}(X_i)\} \right|\\
	\le c_0\left[n^{-1/2}\sqrt{ \Mean \mathbb{I}(A\in \mathcal{I})\{Q(X,A)-q_{\mathcal{I},0}(X)\}^2\log n}+n^{-1}\log n\right],
	\\
	\left|\frac{1}{n}\sum_{i=1}^n \mathbb{I}(A_i\in \mathcal{I})\{Q(X_i,A_i)-q_{\mathcal{I},0}(X_i)\}^2- \Mean \mathbb{I}(A\in \mathcal{I})|Q(X,A)-q_{\mathcal{I},0}(X)|^2\right|\\\le c_0\left[n^{-1/2}\sqrt{ \Mean \mathbb{I}(A\in \mathcal{I})\{Q(X,A)-q_{\mathcal{I},0}(X)\}^2\log n}+n^{-1}\log n\right],
\end{eqnarray*}
for some constant $c_0>0$ and any $\mathcal{I}$. 
The two upper bounds are $o(1)$. Similar to \eqref{lowerboundeq1.5}, we can show that
\begin{eqnarray*}
	\eta_1^*=\sum_{i=1}^n |Y_i-Q(X_i,A_i)|^2+n\sum_{\mathcal{I}\in \widehat{\mathcal{P}}}\Mean \mathbb{I}(A\in \mathcal{I})|Q(X,A)-q_{\mathcal{I},0}(X)|^2+o(n),
\end{eqnarray*}
with probability at least $1-O(n^{-2})$. It follows from \eqref{eqn:some00} that
\begin{eqnarray}\label{eqn:some000}
\begin{split}
	\sum_{\mathcal{I}\in \widehat{\mathcal{P}}}  \sum_{i=1}^n \mathbb{I}(A_i\in \mathcal{I})\{Y_i-\widehat{q}_{\mathcal{I}}(X_i)\}^2 \ge \underbrace{\sum_{i=1}^n |Y_i-Q(X_i,A_i)|^2}_{\eta_2^*}\\+n\sum_{\mathcal{I}\in \widehat{\mathcal{P}}}\Mean \mathbb{I}(A\in \mathcal{I})|Q(X,A)-q_{\mathcal{I},0}(X)|^2+o(n),
\end{split}	
\end{eqnarray}
with probability at least $1-O(n^{-2})$. 

Let us consider $\eta_2^*$. We observe that
\begin{eqnarray*}
	\eta_2^*=\sum_{\mathcal{I}\in \mathcal{P}_0}\sum_{i=1}^n \mathbb{I}(A_i\in \mathcal{I})|Y_i-q_{\mathcal{I},0}(X_i)|^2.
\end{eqnarray*}
By the uniform approximation property of DNN, there exists some $q_{\mathcal{I}}^*\in \mathcal{Q}_{\mathcal{I}}$ such that
\begin{eqnarray*}
	\sum_{i=1}^n |q_{\mathcal{I},0}(X_i)-q_{\mathcal{I}}^*(X_i)|^2\propto n(n|\mathcal{I}|)^{-2\beta/(2\beta+p)}.
\end{eqnarray*}
See Part 1 of the proof of Lemma \hyperlink{lemma1_thm5}{5} for details. Similar to \eqref{event3} and \eqref{event4}, we can show that the following events occur with probability at least $1-O(n^{-2})$, 
\begin{eqnarray*}
	\left|\frac{1}{n}\sum_{i=1}^n \mathbb{I}(A_i\in\mathcal{I})\{Y_i-q_{\mathcal{I},0}(X_i)\}\{q_{\mathcal{I},0}(X_i)-q_{\mathcal{I}}^*(X_i)\} \right|\le \frac{c_0\sqrt{|\mathcal{I}|\log n}}{\sqrt{n}}(n|\mathcal{I}|)^{-\beta/(2\beta+p)},
\end{eqnarray*}
for some constant $c_0>0$ and any $\mathcal{I}\in \mathcal{P}_0$. It follows that
\begin{eqnarray*}
	\eta_2^*-\sum_{\mathcal{I}\in \mathcal{P}_0}\sum_{i=1}^n \mathbb{I}(A_i\in \mathcal{I})|Y_i-q_{\mathcal{I}}^*(X_i)|^2\ge -\sum_{\mathcal{I}\in \mathcal{P}_0}\sum_{i=1}^n \mathbb{I}(A_i\in \mathcal{I})|q_{\mathcal{I},0}(X_i)-q_{\mathcal{I}}^*(X_i)|^2\\
	-2\sum_{\mathcal{I}\in \mathcal{P}_0}\left|\sum_{i=1}^n \mathbb{I}(A_i\in\mathcal{I})\{Y_i-q_{\mathcal{I},0}(X_i)\}\{q_{\mathcal{I},0}(X_i)-q_{\mathcal{I}}^*(X_i)\} \right|\ge -\bar{c} n^{p/(2\beta+p)},
\end{eqnarray*}
for some constant $\bar{c}>0$. This together with \eqref{eqn:some000} yields that
\begin{eqnarray*}
	\sum_{\mathcal{I}\in \widehat{\mathcal{P}}}  \sum_{i=1}^n \mathbb{I}(A_i\in \mathcal{I})\{Y_i-\widehat{q}_{\mathcal{I}}(X_i)\}^2 \ge \sum_{\mathcal{I}\in \mathcal{P}_0}\sum_{i=1}^n \mathbb{I}(A_i\in \mathcal{I})|Y_i-q_{\mathcal{I}}^*(X_i)|^2\\+n\sum_{\mathcal{I}\in \widehat{\mathcal{P}}}\Mean \mathbb{I}(A\in \mathcal{I})|Q(X,A)-q_{\mathcal{I},0}(X)|^2+o(n)+O(n^{p/(2\beta+p)}),
\end{eqnarray*}
with probability at least $1-O(n^{-2})$. 

Next, using similar arguments in proving \eqref{lowerboundeq3}, we can show that there exist a partition $\mathcal{P}^*\in \mathcal{B}(m)$ and a set of functions $\{q_{\mathcal{I}}^{**}:\mathcal{I}\in \mathcal{P}^*\}$ with $|\mathcal{P}^*|=|\mathcal{P}_0|$ such that
\begin{eqnarray*}
\sum_{\mathcal{I}\in \mathcal{P}_0}\sum_{i=1}^n \mathbb{I}(A_i\in \mathcal{I})|Y_i-q_{\mathcal{I}}^{**}(X_i)|^2\ge \sum_{\mathcal{I}\in \mathcal{P}^*}\sum_{i=1}^n \mathbb{I}(A_i\in \mathcal{I})|Y_i-q_{\mathcal{I}}^{**}(X_i)|^2+O(1).
\end{eqnarray*}
It follows that
\begin{eqnarray}\label{eqn:some0000}
\begin{split}
	\sum_{\mathcal{I}\in \widehat{\mathcal{P}}}  \sum_{i=1}^n \mathbb{I}(A_i\in \mathcal{I})\{Y_i-\widehat{q}_{\mathcal{I}}(X_i)\}^2 \ge \sum_{\mathcal{I}\in \mathcal{P}^*}\sum_{i=1}^n \mathbb{I}(A_i\in \mathcal{I})|Y_i-q_{\mathcal{I}}^{**}(X_i)|^2\\+n\sum_{\mathcal{I}\in \widehat{\mathcal{P}}}\Mean \mathbb{I}(A\in \mathcal{I})|Q(X,A)-q_{\mathcal{I},0}(X)|^2+o(n)+O(n^{p/(2\beta+p)}),
\end{split}	
\end{eqnarray}
with probability at least $1-O(n^{-2})$.
Since 
\begin{eqnarray}\label{eqn:some00000}
\begin{split}
	\sum_{\mathcal{I}\in \widehat{\mathcal{P}}}  \sum_{i=1}^n \mathbb{I}(A_i\in \mathcal{I})\{Y_i-\widehat{q}_{\mathcal{I}}(X_i)\}^2+n\gamma_n |\widehat{\mathcal{P}}|\\\le \sum_{\mathcal{I}\in \mathcal{P}^*}\sum_{i=1}^n \mathbb{I}(A_i\in \mathcal{I})|Y_i-q_{\mathcal{I}}^{**}(X_i)|^2+n\gamma_n |\mathcal{P}_0|,
\end{split}	
\end{eqnarray}
and that $\gamma_n\to0$, we obtain that
\begin{eqnarray*}
	\sum_{\mathcal{I}\in \widehat{\mathcal{P}}}\Mean \mathbb{I}(A\in \mathcal{I})|Q(X,A)-q_{\mathcal{I},0}(X)|^2=o(1).
\end{eqnarray*}
Under the condition that $q_{\mathcal{I}_1,0}\neq q_{\mathcal{I}_2,0}$ for any adjacent $\mathcal{I}_1,\mathcal{I}_2\in \mathcal{P}_0$, we have $\Mean |q_{\mathcal{I}_1,0}(X)- q_{\mathcal{I}_2,0}(X)|^2>0$. Using similar arguments in the Part 1 of the proof of Theorem \hyperlink{thm1}{1}, we obtain that $\max_{\tau\in J(\mathcal{P}_0)} \min_{\hat{\tau}\in J(\widehat{\mathcal{P}})} |\hat{\tau}-\tau|\le \delta$ for any constant $\delta>0$. This further implies that $|\widehat{\mathcal{P}}|\ge |\mathcal{P}_0|$.

We next derive the rate of convergence of the estimated change point locations and the estimated outcome regression function. Similar to \eqref{eqn:some0000}, with a more refined analysis (see e.g., Step 2 of the proof of Theorem \hyperlink{thm1}{1}), we obtain that 
\begin{eqnarray*}
\begin{split}
\sum_{\mathcal{I}\in \widehat{\mathcal{P}}}  \sum_{i=1}^n \mathbb{I}(A_i\in \mathcal{I})\{Y_i-\widehat{q}_{\mathcal{I}}(X_i)\}^2 \ge \sum_{\mathcal{I}\in \mathcal{P}^*}\sum_{i=1}^n \mathbb{I}(A_i\in \mathcal{I})|Y_i-q_{\mathcal{I}}^{**}(X_i)|^2\\+n\sum_{\mathcal{I}\in \widehat{\mathcal{P}}}\Mean \mathbb{I}(A\in \mathcal{I})|Q(X,A)-q_{\mathcal{I},0}(X)|^2-C_1|\widehat{\mathcal{P}}|^{2\beta/(2\beta+p)} n^{p/(2\beta+p)}\log^8 n+O(n^{p/(2\beta+p)}),
\end{split}	
\end{eqnarray*}
with probability at least $1-O(n^{-2})$. 
This together with \eqref{eqn:some00000} yields that
\begin{eqnarray*}
	n\sum_{\mathcal{I}\in \widehat{\mathcal{P}}}\Mean \mathbb{I}(A\in \mathcal{I})|Q(X,A)-q_{\mathcal{I},0}(X)|^2\le C_1|\widehat{\mathcal{P}}|^{2\beta/(2\beta+p)} n^{p/(2\beta+p)}\log^8 n\\+O(n^{p/(2\beta+p)})+n\gamma_n (|\mathcal{P}_0|-|\widehat{\mathcal{P}}|).
\end{eqnarray*}
Under the given condition on $\gamma_n$, we obtain that $|\mathcal{\widehat{P}}|\le |\mathcal{P}_0|$. Combining this together with $|\mathcal{\widehat{P}}|\ge |\mathcal{P}_0|$, we obtain that $|\mathcal{\widehat{P}}|=|\mathcal{P}_0|$. This proves the results in (i). 

Consequently, we obtain that
\begin{eqnarray*}
	n\sum_{\mathcal{I}\in \widehat{\mathcal{P}}}\Mean \mathbb{I}(A\in \mathcal{I})|Q(X,A)-q_{\mathcal{I},0}(X)|^2= O(n^{p/(2\beta+p)}\log^8 n),
\end{eqnarray*}
As such, we have that
\begin{eqnarray*}
	\sum_{\mathcal{I}\in \widehat{\mathcal{P}}}\Mean \mathbb{I}(A\in \mathcal{I})|Q(X,A)-q_{\mathcal{I},0}(X)|^2= O(n^{-2\beta/(2\beta+p)}\log^8 n),
\end{eqnarray*}
This together with Lemma \hyperlink{lemma1_thm5}{5} proves the result in (iii). Using similar arguments in Part 2 of the proof of Theorem \hyperlink{thm1}{1}, we can show the result in (ii) holds. This completes the proof.

\subsection{Proof of Theorem 4} 

The proof of Theorem \hyperlink{thm7}{4} is similar to that of Theorem \hyperlink{thm2}{2}. We provide the outline as below and omit the duplicated arguments for brevity. 

\smallskip

Under the events defined in Theorem \hyperlink{thm5}{3}, we have $\widehat{K}=K$, and
\begin{eqnarray}\label{localdist_thm5}
\max_{k\in \{1,\dots,K-1\}} |\widehat{\tau}_k-\tau_{0,k}|\le c n ^{-2\beta/(2\beta+p)}\log^8 n,
\end{eqnarray}
for some constant $c>0$. By similar arguments in the proof of Theorem \hyperlink{thm2}{2}, there exists some constant $\bar{C}_4>0$ such that 
\begin{eqnarray}\label{boundpistar_thm5}
\pi^*(a;x,\widehat{d}(x))\le \bar{C}_4 \delta_{\min}^{-1},\,\,\,\,\,\,\,\,\forall a\in [0,1], x\in \mathbb{X}.
\end{eqnarray}

The rest of our proof is divided into two parts. In the first part,  we focus on proving
\begin{eqnarray}\label{middlestep1_thm5}
V^{\pi^*}(\widehat{d})\ge \Mean \left(q_{[\tau_{0,\widehat{\mathbb{K}}(x)-1}, \tau_{0,\widehat{\mathbb{K}}(x)})}(X)\right) -O(1)n ^{-2\beta/(2\beta+p)}\log^8 n,
\end{eqnarray}
with probability at least $1-O(n^{-2})$, where $O(1)$ denotes some positive constant.

In Part 2, we provide an upper bound for
\begin{eqnarray*}
	V^{\tiny{opt}}- \Mean \left(q_{[\tau_{0,\widehat{\mathbb{K}}(x)-1}, \tau_{0,\widehat{\mathbb{K}}(x)})}(X)\right).
\end{eqnarray*}
This together with \eqref{middlestep1_thm5} yields the desired results.

\smallskip

\noindent \textit{Proof of Part 1:} Recall the integer-valued function 
\begin{eqnarray}\label{widehatK_thm5}
\widehat{\mathbb{K}}(x)=\sargmax_{k\in\{1,\dots,K\}}  \widehat{q}_{[\widehat{\tau}_{k-1}, \widehat{\tau}_k)}(x),
\end{eqnarray}
where $\sargmax$ denotes the smallest maximizer when the argmax is not unique. Similarly, we have $\widehat{\mathbb{K}}(x)=k$ if $\widehat{d}(x)=[\widehat{\tau}_{k-1}, \widehat{\tau}_k)$ for some integer $k$ such that $1\le k\le K-1$, and set $\widehat{\mathbb{K}}(x)=K$ if $\widehat{d}(x)=[\widehat{\tau}_{K-1}, 1]$.   
 \smallskip
 
Let $\widehat{\Delta}_k=[\widehat{\tau}_{k-1}, \widehat{\tau}_k) \cup [\tau_{0,k-1},\tau_{0,k})^c+[\widehat{\tau}_{k-1}, \widehat{\tau}_k)^c \cup [\tau_{0,k-1},\tau_{0,k})$. Using similar arguments in the proof of Theorem \hyperlink{thm2}{2}, we have 
\begin{eqnarray*}
	V^{\pi^*}(\widehat{d})= \Mean \left(\int_{\widehat{d}_0(X)\cap \widehat{d}(X)} Q(X,a)  \pi^*(a;X,\widehat{d}(X))da\right)+\underbrace{\Mean \left(\int_{\widehat{\Delta}(X)} Q(X,a) \pi^*(a;X,\widehat{d}(X))da\right)}_{\chi_1^*}\\
	=\Mean \left(\int_{\widehat{d}_0(X)} Q(X,a) \pi^*(a;X,\widehat{d}(X))da\right)+\chi_1^*,
\end{eqnarray*}
where $\widehat{d}_0(x)=[\tau_{0,\widehat{\mathbb{K}}(x)-1}, \tau_{0,\widehat{\mathbb{K}}(x)})$ and $\widehat{\Delta}(x)=\widehat{d}(x)\cap \{\widehat{d}_0(x)\}^c$.

By \eqref{boundpistar_thm5} and the assumption that $Y$ is bounded, we have
\begin{eqnarray*}
	|\chi_1^*| \le c_0  \bar{C}_4 \delta_{\min}^{-1}   \lambda(\widehat{\Delta}(X)),
\end{eqnarray*}
where $\lambda(\widehat{\Delta}(X))$ denotes the Lebesgue measure of $\widehat{\Delta}(X)$. Under the event defined in \eqref{localdist_thm5}, we have $\lambda(\widehat{\Delta}(X))\le 2cn ^{-2\beta/(2\beta+p)}\log^8 n$, for any realization of $X$. It follows that 
\begin{eqnarray}\label{chi1_thm5}
|\chi_1^*|\le  \bar{C}_0 \delta_{\min}^{-1} n ^{-2\beta/(2\beta+p)}\log^8 n,
\end{eqnarray}
for some constant $\bar{C}_0$ with probability at least $1-O(n^{-2})$. 

\smallskip

Using similar arguments in the proof of Theorem \hyperlink{thm2}{2}, we have
\begin{eqnarray*}
\Mean \left(\int_{\widehat{d}_0(X)}Q(X,a) \pi^*(a;X,\widehat{d}(X))da\right) = \Mean \left(q_{[\tau_{0,\widehat{\mathbb{K}}(x)-1}, \tau_{0,\widehat{\mathbb{K}}(x)})}(X)\right)-\chi_2^*,
\end{eqnarray*}
where
\begin{eqnarray*}
	\chi_2^*=\Mean \left(q_{[\tau_{0,\widehat{\mathbb{K}}(x)-1}, \tau_{0,\widehat{\mathbb{K}}(x)}),0}(X)\right) \int_{\widehat{\Delta}(X) } \pi^*(a;X,\widehat{d}(X))da. 
\end{eqnarray*}
Similar to \eqref{chi1_thm5}, we can show that
\begin{eqnarray*}
	|\chi_2^*|=O(n ^{-2\beta/(2\beta+p)}\log^8 n),
\end{eqnarray*}
with probability at least $1-O(n^{-2})$. This together with \eqref{chi1_thm5} yields \eqref{middlestep1_thm5}. 

\smallskip

\noindent \textit{Proof of Part 2:}  
Let $\epsilon_n = \bar{C}_1 (n\delta_{\min})^{-2\beta/\{(2\beta+p)(2+\gamma)\}}\log^{8/(2+\gamma)}  n$ for some constant $\bar{C}_1$. Define an event 
$$\mathcal{A}_\epsilon = \bigcup_k \left \{ |q_{[\tau_{0,k-1},\tau_{0,k})}(X)-\widehat{q}_{[\widehat{\tau}_{k-1}, \widehat{\tau}_k)} (X)| \le \epsilon_n \right\}.
$$
Based on Lemma \hyperlink{lemma1_thm5}{5}, by Markov's inequality, we can show that there exists some constant $\bar{c}>0$ such that
 \begin{eqnarray}\label{anotherevent1_thm5}
 &&\hbox{Pr}\{|q_{[\tau_{0,k-1},\tau_{0,k}),0}(X)-\widehat{q}_{[\widehat{\tau}_{k-1}, \widehat{\tau}_k)} (X)| >\epsilon_n \} \\\nonumber
 \le&& \bar{C}_2 (n\delta_{\min})^{-2\beta(1+\gamma)/\{(2\beta+p)(2+\gamma)\}}\log^{8(1+\gamma)/(2+\gamma)}  n,\forall k\in \{1,\dots,K\},
   \end{eqnarray}
with probability at least $1-O(n^{-2})$ for some constant $\bar{C}_2$. Thus, by Bonferroni's inequality, we have 
\begin{eqnarray}\label{anotherevent2_thm5}
\hbox{Pr}\{\mathcal{A}_\epsilon^c \} \le  \bar{C}_3 (n\delta_{\min})^{-2\beta(1+\gamma)/\{(2\beta+p)(2+\gamma)\}}\log^{8(1+\gamma)/(2+\gamma)}  n
  \end{eqnarray}
 holds with probability at least $1-O(n^{-2})$  for some constant $\bar{C}_3$.

\smallskip

Consider the event
\begin{eqnarray*}
	\mathcal{A}_0=\bigcup_{\mathcal{I}_1,\mathcal{I}_2\in \mathcal{P}_0} \left\{0<\left|q_{\mathcal{I}_1,0}(X)-q_{\mathcal{I}_2,0}(X)\right|\le 2 \epsilon_n \right\}.
\end{eqnarray*}
By Condition (A5) and Bonferroni's inequality, we have
\begin{eqnarray}\label{eventA0_thm5}
\hbox{Pr}(\mathcal{A}_0)\le \sum_{\substack{\mathcal{I}_1,\mathcal{I}_2\in \mathcal{P}_0\\ \mathcal{I}_1\neq \mathcal{I}_2 }} \hbox{Pr}\left( 0<\left|q_{\mathcal{I}_1,0}(X)-q_{\mathcal{I}_2,0}(X)\right|\le  2 \epsilon_n \right) \le K^2 \left( 2 \epsilon_n  \right)^{\gamma}.
\end{eqnarray}

\smallskip

Similar to the definition of $\widehat{\mathbb{K}}$, we define 
\begin{eqnarray}\label{K0_thm5}
\mathbb{K}_0(x)=\sargmax_{k\in\{1,\dots,K\}}  q_{[\tau_{0,k-1}, \tau_{0,k}),0}(x). 
\end{eqnarray}
Let
\begin{eqnarray*}
	\mathbb{K}^*(x)=\left\{k_0: k_0=\argmax_{k\in\{1,\dots,K\}} q_{[\tau_{0,k-1}, \tau_{0,k}),0}(x)\right\},
\end{eqnarray*}
denote the set that consists of all the maximizers. Apparently, $\mathbb{K}_0(x)\in \mathbb{K}^*(x)$, $\forall x\in \mathbb{X}$.


We now claim that 
\begin{eqnarray}\label{correctK_thm5}
\widehat{\mathbb{K}}(X)\in \mathbb{K}^*(X),
\end{eqnarray}
under the events defined in $\mathcal{A}_0^c $ and $\mathcal{A}_\epsilon$. Otherwise, suppose there exists some $k_0\in \{1,\dots,K\}$ such that
\begin{eqnarray}\label{inequality1_thm5}
&&  \widehat{q}_{[\widehat{\tau}_{k_0-1}, \widehat{\tau}_{k_0})}(X)\ge \max_{k\neq k_0}  \widehat{q}_{[\widehat{\tau}_{k-1}, \widehat{\tau}_{k})}(X),\\ \label{inequality2_thm5}
&&\max_{k\neq k_0}  {q}_{[\tau_{0,k-1}, \tau_{0,k}),0}(X)>  {q}_{[\tau_{0,k_0-1}, \tau_{0,k_0}),0}(X).
\end{eqnarray}
Under $\mathcal{A}_0^c$, it follows from \eqref{inequality2_thm5} that
\begin{eqnarray}\label{inequality4_thm5}
\max_{k\neq k_0} {q}_{[\tau_{0,k-1}, \tau_{0,k}),0}(X)>q_{[\tau_{0,k_0-1}, \tau_{0,k_0}),0}+ 2 \epsilon_n.
\end{eqnarray}
Under the event $\mathcal{A}_\epsilon$, we have
\begin{eqnarray*}
	\max_{k\in\{1,\dots,K\}} | \widehat{q}_{[\widehat{\tau}_{k-1},\widehat{\tau}_{k} )}(X)-q_{[\tau_{0,k-1}, \tau_{0,k}),0}(X) |  \le \epsilon_n. 
\end{eqnarray*}
This together with \eqref{inequality4_thm5} yields that
\begin{eqnarray*}
	\max_{k\neq k_0} \widehat{q}_{[\widehat{\tau}_{k-1}, \widehat{\tau}_{k})}(X)>  \widehat{q}_{[\widehat{\tau}_{k_0-1}, \widehat{\tau}_{k_0})}(X).
\end{eqnarray*}
In view of \eqref{inequality1_thm5}, we have reached a contradiction. Therefore, \eqref{correctK_thm5} holds under the events defined in $\mathcal{A}_0^c $ and $\mathcal{A}_\epsilon$. 

By the definition of $\mathbb{K}^*(\cdot)$ that $q_{[\tau_{0,\widehat{\mathbb{K}}(X)-1}, \tau_{0,\widehat{\mathbb{K}}(X)}),0}(X)=q_{[\tau_{0,\mathbb{K}_0(X)-1}, \tau_{0,\mathbb{K}_0(X)}),0}(X)$ when \eqref{correctK_thm5} holds.
Using the similar augments in \eqref{inequality5}, we have
\begin{eqnarray}\label{inequality5_thm5}
 \Mean \left(q_{[\tau_{0,\widehat{\mathbb{K}}(x)-1}, \tau_{0,\widehat{\mathbb{K}}(x)}),0}(X)\right)  =\Mean \left(q_{[\tau_{0,\mathbb{K}_0(X)-1}, \tau_{0,\mathbb{K}_0(X)}),0}(X)\right)   +  \chi_3 + \chi_4,
\end{eqnarray} 
where
\begin{eqnarray*}
	\chi_3=\Mean \left(q_{[\tau_{0,\widehat{\mathbb{K}}(X)-1}, \tau_{0,\widehat{\mathbb{K}}(X)}),0}(X)-q_{[\tau_{0,\mathbb{K}_0(X)-1}, \tau_{0,\mathbb{K}_0(X)}),0}(X)\right)\mathbb{I}(\mathcal{A}_0 ) \mathbb{I}(\mathcal{A}_\epsilon),
\end{eqnarray*}
and
 \begin{eqnarray*}
	\chi_4=\Mean \left(q_{[\tau_{0,\widehat{\mathbb{K}}(X)-1}, \tau_{0,\widehat{\mathbb{K}}(X)}),0}(X)-q_{[\tau_{0,\mathbb{K}_0(X)-1}, \tau_{0,\mathbb{K}_0(X)}),0}(X)\right) \mathbb{I}(\mathcal{A}_\epsilon^c),
\end{eqnarray*}

Therefore, under the event $\mathcal{A}_\epsilon $, it follows from \eqref{eventA0_thm5} that
\begin{eqnarray}\label{chi3_thm5}
	|\chi_3|\le K^2 \left(2\epsilon_n\right)^{\gamma+1}  . 
\end{eqnarray}

Similarly, by Condition (A7) and the outcome is bounded, following Markov's inequality, we have
\begin{eqnarray}\label{chi4_thm5}
	|\chi_4|&&\le \bar{C}_3  \hbox{Pr}\{\mathcal{A}_\epsilon^c \} 
\end{eqnarray}

Based on \eqref{anotherevent2_thm5} and $\epsilon_n = \bar{C}_1 (n\delta_{\min})^{-2\beta/\{(2\beta+p)(2+\gamma)\}}\log^{8/(2+\gamma)}  n$, for sufficiently large $n$, the above \eqref{chi4_thm5} and \eqref{chi3_thm5} together with \eqref{middlestep1_thm5} and \eqref{inequality5_thm5} implies that we have with probability at least $1-O(n^{-2})$, 
\begin{eqnarray*}
V^{\pi^*}(\widehat{d})\ge V^{\tiny{opt}} -O(1) (n ^{-{2\beta\over 2\beta+p}}\log^8 n+n^{-{2\beta(1+\gamma)\over(2\beta+p)(2+\gamma)}} \log^{8+8\gamma\over 2+\gamma} n),
\end{eqnarray*}
for some positive constant $O(1)$. The proof is hence completed.

\subsection{Proof of Theorem 5}\index{Proof of Theorem 4}
We focus on proving Theorem \hyperlink{thm3}{5} (ii) when conditions in Theorem \hyperlink{thm7}{4} are satisfied with $4\beta(1+\gamma)> (2\beta+p)(2+\gamma)$, where D-JIL is applied. Since the piecewise linear case requires weaker conditions (when conditions in Theorem \hyperlink{thm2}{2} are satisfied), one can similarly derive the asymptotic normality of $\widehat{V}$ under L-JIL. 

\smallskip

We present an outline of the proof first, which can be divided into two parts. Define $d_0(x)=\argmax_{\mathcal{I}\in \mathcal{P}_0} q_{\mathcal{I},0}(x)$, $\forall x\in \mathbb{X}$. Under the given conditions, the maximizers $d_0(X_i)$'s are almost surely unique. By the definition of $\widehat{\mathbb{K}}(\cdot)$ in \eqref{widehatK_thm5}, we have  
\begin{eqnarray*}
 \widehat{V}=\frac{1}{n}\sum_{i=1}^n  \left[ {\mathbb{I}\{A_i\in \widehat{d}(X_i) \} \over  \widehat{e}( \widehat{d}(X_i) |X_i)}\big\{Y_i - \widehat{q}_{[\widehat{\tau}_{\widehat{\mathbb{K}}(X_i)-1}, \widehat{\tau}_{\widehat{\mathbb{K}}(X_i)})}(X_i) \big\} +    \widehat{q}_{[\widehat{\tau}_{\widehat{\mathbb{K}}(X_i)-1}, \widehat{\tau}_{\widehat{\mathbb{K}}(X_i)})}(X_i) \right].
\end{eqnarray*}
Given $\mathbb{K}_0(\cdot)$ defined in \eqref{K0_thm5}, the above value estimator can be decomposed by 
\begin{eqnarray*}
 \widehat{V}= \widehat{V}_1 + \underbrace{\frac{1}{n}\sum_{i=1}^n  \left[ \left\{{\mathbb{I}\{A_i\in \widehat{d}(X_i) \} \over  \widehat{e}( \widehat{d}(X_i) |X_i)} -1 \right\}\big\{ q_{[\tau_{0,\mathbb{K}_0(X_i)-1},\tau_{0,\mathbb{K}_0(X_i)}),0}(X_i) -  \widehat{q}_{[\widehat{\tau}_{\widehat{\mathbb{K}}(X_i)-1}, \widehat{\tau}_{\widehat{\mathbb{K}}(X_i)})}(X_i) \big\}   \right]}_{\eta_7},
\end{eqnarray*}
where
\begin{eqnarray*}
	 \widehat{V}_1= \frac{1}{n}\sum_{i=1}^n  \left[ {\mathbb{I}\{A_i\in \widehat{d}(X_i) \} \over  \widehat{e}( \widehat{d}(X_i) |X_i)}\big\{Y_i - q_{[\tau_{0,\mathbb{K}_0(X_i)-1},\tau_{0,\mathbb{K}_0(X_i)}),0}(X_i) \big\} +    q_{[\tau_{0,\mathbb{K}_0(X_i)-1},\tau_{0,\mathbb{K}_0(X_i)}),0}(X_i) \right].
\end{eqnarray*} 

\smallskip

In Part 1, we first establish the following result that
\begin{eqnarray}\label{eta7_res}
\eta_7= o_p(n^{-1/2}).
\end{eqnarray} 
This implies 
\begin{eqnarray}\label{prt1_res}
 \widehat{V} =  \widehat{V}_1 +o_p(n^{-1/2}).
\end{eqnarray}
 
\smallskip
 
In the second step, we further decompose $ \widehat{V}_1$ as
\begin{eqnarray*}
	   \widehat{V}_1&= \widehat{V}_2+ \underbrace{\frac{1}{n}\sum_{i=1}^n  \left[ \left\{ {\mathbb{I}\{A_i\in \widehat{d}(X_i) \}  \over  \widehat{e}( \widehat{d}(X_i) |X_i)} - {\mathbb{I}\{A_i\in d_0(X_i) \} \over  e(d_0(X_i) |X_i)}\right \}\big\{Y_i - q_{[\tau_{0,\mathbb{K}_0(X_i)-1},\tau_{0,\mathbb{K}_0(X_i)}),0}(X_i) \big\}  \right]}_{\eta_8},
\end{eqnarray*}
where
\begin{eqnarray*}
 \widehat{V}_2 = \frac{1}{n}\sum_{i=1}^n  \left[ {\mathbb{I}\{A_i\in d_0(X_i) \} \over  e( d_0(X_i) |X_i)}\big\{Y_i - q_{[\tau_{0,\mathbb{K}_0(X_i)-1},\tau_{0,\mathbb{K}_0(X_i)}),0}(X_i) \big\} +    q_{[\tau_{0,\mathbb{K}_0(X_i)-1},\tau_{0,\mathbb{K}_0(X_i)}),0}(X_i) \right].
\end{eqnarray*}

We focus on proving 
\begin{eqnarray}\label{eta8_res}
\eta_8=  o_p(n^{-1/2}).
\end{eqnarray}
This together with \eqref{prt1_res} leads to
\begin{eqnarray}\label{prt2_res}
 \widehat{V} =  \widehat{V}_2 +o_p(n^{-1/2}).
\end{eqnarray}

\smallskip

Combing the results in the first two steps, it follows from the definition of $d_0(\cdot)$ that
\begin{eqnarray*}
	  \widehat{V}=&&\frac{1}{n}\sum_{i=1}^n \sum_{\mathcal{I}\in  {\mathcal{P}}_0} \mathbb{I}(\mathcal{I}=d_0(X_i))  \left[ {\mathbb{I}\{A_i\in {d}_0(X_i) \} \over   {e}( {d}_0(X_i) |X_i)}\big\{Y_i -  {q}_{\mathcal{I},0}(X_i) \big\} +     {q}_{\mathcal{I},0}(X_i)\right] + o_p(n^{-1/2}),
\end{eqnarray*}
almost surely. Notice that the first term at RHS corresponds to a sum of i.i.d random variables. Hence, based on Lindeberg-Feller central limit theorem, one can show the asymptotic normality result of the value estimator under the proposed I2DR.

\smallskip

\noindent \textit{Proof of Part 1:} We aim to show \eqref{eta7_res}. Toward that end, we define 
\begin{eqnarray*}
	\widehat{V}_3=\frac{1}{n}\sum_{i=1}^n  \left[ {\mathbb{I}\{A_i\in \widehat{d}(X_i) \} \over  \widehat{e}( \widehat{d}(X_i) |X_i)}\big\{Y_i - q_{[\tau_{0,\widehat{\mathbb{K}}(X_i)-1},\tau_{0,\widehat{\mathbb{K}}(X_i)}),0}(X_i) \big\} +    q_{[\tau_{0,\widehat{\mathbb{K}}(X_i)-1},\tau_{0,\widehat{\mathbb{K}}(X_i)}),0}(X_i) \right].
\end{eqnarray*}
The difference $|\eta_7|$ can be upper bounded by $|\widehat{V}_1-\widehat{V}_3|+|\widehat{V}-\widehat{V}_3|$. Consider $|\widehat{V}_1-\widehat{V}_3|$ first. Under the given conditions, the term $\left\{{\mathbb{I}\{A_i\in \widehat{d}(X_i) \} \over  \widehat{e}( \widehat{d}(X_i) |X_i)} -1 \right\}$ is bounded, it suffices to show that
\begin{eqnarray*}
 \frac{1}{n}\sum_{i=1}^n  \left| q_{[\tau_{0,\mathbb{K}_0(X_i)}-1,\tau_{0,\mathbb{K}_0(X_i)}),0}(X_i) -  q_{[\widehat{\tau}_{\widehat{\mathbb{K}}(X_i)-1}, \widehat{\tau}_{\widehat{\mathbb{K}}(X_i)}),0}(X_i)   \right| = o_p(n^{-1/2}),
\end{eqnarray*}
where $\mathbb{K}_0(\cdot)$ and $\widehat{\mathbb{K}}(\cdot)$ are defined in \eqref{K0_thm5} and \eqref{widehatK_thm5}, respectively.
Under the margin-type condition, the above expression can be proven using similar arguments in the proof of Theorem \hyperlink{thm7}{4}. We omit the details to save space. 
 
 It remains to show $|\widehat{V}-\widehat{V}_3|=o_p(n^{1/2})$. Notice that $|\widehat{V}-\widehat{V}_3|$ can be further upper bounded by 
 \begin{eqnarray*}
 	\left|\frac{1}{n}\sum_{i=1}^n  \left\{{\mathbb{I}\{A_i\in \widehat{d}(X_i) \} \over e( \widehat{d}(X_i) |X_i)} -1 \right\}\big\{ q_{[\tau_{0,\widehat{\mathbb{K}}(X_i)-1},\tau_{0,\widehat{\mathbb{K}}(X_i)}),0}(X_i) -  \widehat{q}_{[\widehat{\tau}_{\widehat{\mathbb{K}}(X_i)-1}, \widehat{\tau}_{\widehat{\mathbb{K}}(X_i)})}(X_i) \big\}\right| \\
 	+ \left|\frac{1}{n}\sum_{i=1}^n  \left\{{\mathbb{I}\{A_i\in \widehat{d}(X_i) \} \over e( \widehat{d}(X_i) |X_i)} -{\mathbb{I}\{A_i\in \widehat{d}(X_i) \} \over \widehat{e}( \widehat{d}(X_i) |X_i)} \right\}\big\{ q_{[\tau_{0,\widehat{\mathbb{K}}(X_i)-1},\tau_{0,\widehat{\mathbb{K}}(X_i)}),0}(X_i) -  \widehat{q}_{[\widehat{\tau}_{\widehat{\mathbb{K}}(X_i)-1}, \widehat{\tau}_{\widehat{\mathbb{K}}(X_i)})}(X_i) \big\}\right|.
 \end{eqnarray*}
Consider the first line. Notice that it can be represented by
\begin{eqnarray*}
	\left|\frac{1}{n}\sum_{\mathcal{I}\in \widehat{\mathcal{P}}}\sum_{i=1}^n  \left\{{\mathbb{I}\{A_i\in \mathcal{I} \} \over e( \mathcal{I} |X_i)} -1 \right\}\big\{ q_{\mathcal{I},0}(X_i) -  \widehat{q}_{\mathcal{I}}(X_i) \big\}\mathbb{I}(\mathcal{I}=\widehat{d}(X_i))\right|.
\end{eqnarray*}
Since the number of intervals in $\widehat{\mathcal{P}}$ is finite with probability tending to 1 (see Results (i) in Theorem \hyperlink{thm1}{1}), to show the above expression is $o_p(n^{-1/2})$, it suffices to show
\begin{eqnarray*}
	\sup_{\mathcal{I}\in \mathfrak{I}(m)}\left|\frac{1}{n}\sum_{i=1}^n  \left\{{\mathbb{I}\{A_i\in \mathcal{I} \} \over e( \mathcal{I} |X_i)} -1 \right\}\big\{ q_{\mathcal{I},0}(X_i) -  \widehat{q}_{\mathcal{I}}(X_i) \big\}\mathbb{I}(\mathcal{I}=\widehat{d}(X_i))\right|=o_p(n^{-1/2}).
\end{eqnarray*}
The key observation is that, by Corollary A.1 of \cite{chernozhukov2014gaussian}, the above empirical sum forms a VC-type class. Using similar arguments in bounding the stochastic error in Step 2 of the proof of Lemma \hyperlink{lemma1_thm5}{5}, we can show the above assertion holds. 

To bound the second line, notice that by Cauchy-Schwarz inequality, it is smaller than or equal to the square root of
\begin{eqnarray*}
  \underbrace{	\frac{1}{n}\sum_{i=1}^n \left|{\mathbb{I}\{A_i\in \widehat{d}(X_i) \} \over e( \widehat{d}(X_i) |X_i)} -{\mathbb{I}\{A_i\in \widehat{d}(X_i) \} \over \widehat{e}( \widehat{d}(X_i) |X_i)}\right|^2}_{\eta_{7}^{(1)}} \underbrace{\frac{1}{n}\sum_{i=1}^n |q_{[\tau_{0,\widehat{\mathbb{K}}(X_i)-1},\tau_{0,\widehat{\mathbb{K}}(X_i)}),0}(X_i) -  \widehat{q}_{[\widehat{\tau}_{\widehat{\mathbb{K}}(X_i)-1}, \widehat{\tau}_{\widehat{\mathbb{K}}(X_i)})}(X_i)|^2}_{\eta_{7}^{(2)}}
\end{eqnarray*}
Using similar arguments in establishing the uniform convergence rate of $\widehat{q}_{\mathcal{I}}$, we can show that $\eta_{7}^{(2)}=o_p(n^{-c})$ for some $c>1/2$. To prove the second line is $o_p(n^{-1/2})$, it remains to show $\eta_7^{(1)}=O_p(n^{-1/2}\log n)$. Under the positivity assumption on $e$ and $\widehat{e}$, it suffices to show
\begin{eqnarray}\label{eqn:someterm}
	\frac{1}{n}\sum_{i=1}^n |e(\widehat{d}(X_i) |X_i)-\widehat{e}(\widehat{d}(X_i) |X_i)|^2=O_p(n^{1/2}\log n). 
\end{eqnarray}
The left-hand-side can be further upper bounded by
\begin{eqnarray*}
	\frac{1}{n}\sum_{\mathcal{I}\in \widehat{\mathcal{P}}}\sum_{i=1}^n |e(\mathcal{I} |X_i)-\widehat{e}(\mathcal{I} |X_i)|^2\\
	\le \sum_{\mathcal{I}\in \widehat{\mathcal{P}}} \Mean |e(\mathcal{I} |X)-\widehat{e}(\mathcal{I} |X)|^2+\sum_{\mathcal{I}\in \widehat{\mathcal{P}}}\left[\frac{1}{n}\sum_{i=1}^n |e(\mathcal{I} |X_i)-\widehat{e}(\mathcal{I} |X_i)|^2-\Mean |e(\mathcal{I} |X)-\widehat{e}(\mathcal{I} |X)|^2\right].
\end{eqnarray*}
The first term on the second line is $O_p(n^{-1/2})$ under Condition (A8) and the fact that $|\widehat{\mathcal{P}}|=O(1)$ with probability tending to $1$. To prove \eqref{eqn:someterm}, by the boundedness of $|\widehat{\mathcal{P}}|$, it suffices to show the supremum of the empirical process term
\begin{eqnarray*}
	\sup_{\mathcal{I}\in \mathfrak{I}(m)}\left[\frac{1}{n}\sum_{i=1}^n |e(\mathcal{I} |X_i)-\widehat{e}(\mathcal{I} |X_i)|^2-\Mean |e(\mathcal{I} |X)-\widehat{e}(\mathcal{I} |X)|^2\right]=O_p(n^{-1/2}\log n). 
\end{eqnarray*}
Under Condition (A8), this can be proven in a similar manner as Step 2 of the proof of Lemma \hyperlink{lemma1_thm5}{5}. We omit the details to save space. 

\smallskip 

\noindent \textit{Proof of Part 2:} We next focus on proving \eqref{eta8_res}. We notice that $|\eta_8|$ can be upper bounded by 
\begin{eqnarray*}
\left|\frac{1}{n}\sum_{i=1}^n  \left[ \left\{ {\mathbb{I}\{A_i\in \widehat{d}(X_i) \}  \over  \widehat{e}( \widehat{d}(X_i) |X_i)} - {\mathbb{I}\{A_i\in \widehat{d}(X_i) \} \over  e(\widehat{d}(X_i) |X_i)}\right \}\big\{Y_i - q_{[\tau_{0,\mathbb{K}_0(X_i)-1},\tau_{0,\mathbb{K}_0(X_i)}),0}(X_i) \big\}  \right]\right|\\
+\left|\frac{1}{n}\sum_{i=1}^n  \left[ \left\{ {\mathbb{I}\{A_i\in d_0(X_i) \}  \over  e(d_0(X_i) |X_i)} - {\mathbb{I}\{A_i\in \widehat{d}(X_i) \} \over  e(\widehat{d}(X_i) |X_i)}\right \}\big\{Y_i - q_{[\tau_{0,\mathbb{K}_0(X_i)-1},\tau_{0,\mathbb{K}_0(X_i)}),0}(X_i) \big\}  \right]\right|.
\end{eqnarray*}
The first line can be shown to be $o_p(n^{-1/2})$ using similar arguments in the proof of Part 1. The second line can be shown to be $o_p(n^{1/2})$ by noting that the difference between $d_0$ and $\widehat{d}$ is asymptotically negligible. This completes the proof.

\subsection{Proof of Theorem 6}\label{secproofthm4}
Before proving Theorem \hyperlink{thm4}{6}, it is worth mentioning that results in Lemma \hyperlink{lemma1}{1} and Lemma \hyperlink{lemma5}{4} do not rely on the assumption that $\theta_0(\cdot)$ is piecewise constant. These lemmas hold under the conditions in Theorem \hyperlink{thm4}{6} as well. The proof is divided into two parts. In the first part, we derive the convergence rate of the integrated $\ell_2$ loss for $\widehat{\theta}$. Then, we establish the convergence rate of the value under our I2DR. 

\noindent \textit{Convergence rate of the integrated $\ell_2$ loss: }
We first establish the upper error bound on the integrated $\ell_2$ loss of $\widehat{\theta}(\cdot)$. Here, we consider a more general framework. Specifically, define 
\begin{eqnarray*}
	\hbox{AE}_k(\theta_0)=\inf_{\substack{\mathcal{P}: |\mathcal{P}|\le k+1 \\ (\theta_{\mathcal{I}})_{\mathcal{I}\in \mathcal{P}}\in \prod_{\mathcal{I}\in \mathcal{P}} \mathbb{R}^{p+1} } } \left\{ \sup_{a\in [0,1]}\left\|\theta_0(a)-\sum_{\mathcal{I}\in \mathcal{P}} \theta_{\mathcal{I}} \mathbb{I}(a\in \mathcal{I}) \right\|_2  \right\}.
\end{eqnarray*}
It describes how well $\theta_0(\cdot)$ can be approximated by a step function with at most $k$ change points. Consider the following class of functions
\begin{eqnarray*}
	\mathbb{B}^{\alpha_0}=\left\{\theta_0(\cdot): \limsup_{k\to \infty} k^{\alpha_0} \hbox{AE}_k(\theta_0)<\infty \right\},
\end{eqnarray*} 
for some $\alpha_0>0$. The parameter $\alpha_0$ characterizes the speed of approximation as the number of change points increases. According to the discussion in Section \ref{secproperty2}, the class of H{\"o}lder continuous functions in Model II belongs to $\mathbb{B}^{\alpha_0}$. In the following, we show with probability at least $1-O(n^{-2})$ that $\int_{0}^{1} \|\widehat{\theta}(a)-\theta_0(a)\|_2^2da\le \bar{c} \gamma_n^{2\alpha_0/(1+2\alpha_0)}$ for any $\theta_0(\cdot) \in \mathbb{B}^{\alpha_0}$. 

Since $\theta_0(\cdot)\in \mathbb{B}^{\alpha_0}$, for some sequence $\{k_n\}_n$ that satisfies $k_n\to \infty$ as $n\to \infty$, there exists a piecewise constant function $\theta^*(\cdot)$ such that
\begin{eqnarray*}
	\theta^*(a)=\sum_{\mathcal{I} \in \mathcal{P}^*} \theta_{\mathcal{I}}^* \mathbb{I}(a\in \mathcal{I}),\,\,\,\,\,\,\,\,\forall a\in [0,1],
\end{eqnarray*}
for some partition $\mathcal{P}^*$ of $[0,1]$ with $|\mathcal{P}^*|\le k_n+1$ and some $(\theta_{\mathcal{I}}^*)_{\mathcal{I}\in \mathcal{P}^*}\in \prod_{\mathcal{I}\in \mathcal{P}^*} \mathbb{R}^{p+1}$, and
\begin{eqnarray}\label{infinityconvrate}
\sup_{\mathcal{I}\in \mathcal{P}^*}\sup_{a\in \mathcal{I}} \|\theta_0(a)-\theta_{\mathcal{I}}^*\|_2\le \frac{c_4}{k_n^{\alpha_0}},
\end{eqnarray}
for some constant $c_4>0$. 
Detailed choice of $k_n$ will be given later. Combining \eqref{infinityconvrate} together with \eqref{boundbeta0I}, we obtain that
\begin{eqnarray}\label{boundbetastar}
\sup_{\mathcal{I}\in \mathcal{P}^*}\| \theta_{\mathcal{I}}^*\|_2\le 2c_0,
\end{eqnarray}
for sufficiently large $n$. 

Let $\{\tau_k^*\}_{k=1}^{|\mathcal{P}^*|-1}$ with $0<\tau_1^*<\tau_2^*<\cdots<\tau_{|\mathcal{P}^*|-1}^*<1$ be the locations of the change points in $J(\mathcal{P}^*)$. For $1\le k\le |\mathcal{P}^*|-1$, define $\tau_k^{**}$ such that $0\le \tau_{k}^{**}- \tau_{k}^*<1/m$ and $\tau_k^{**}\in \{1/m,2/m,\dots,1\}$. Let $k_n^*$ be the largest integer that satisfies $k_n^*\le |\mathcal{P}^*|-1$ and $\tau_{k_n^*}^{**}<1$. Apparently, $k_n^*\le k_n$. Set $\tau_0^*=\tau_0^{**}=0$ and $\tau_{k_n^*+1}^*=\tau_{k_n^*+1}^{**}=1$. Define a new partition $\mathcal{P}^{**}\in \mathcal{B}(m)$ and the set of vectors $(\theta_{\mathcal{I}}^{**})_{\mathcal{I}\in \mathcal{P}^{**}}$ as follows,
\begin{eqnarray*}
	&&\mathcal{P}^{**}=\{[\tau_{0}^{**}, \tau_1^{**}), [\tau_1^{**},\tau_2^{**}),\cdots,[\tau_{k_n^*}^{**},\tau_{k_n^*+1}^{**} ] \},\\
	&&\theta_{[\tau_k^{**},\tau_{k+1}^{**})}^{**}=\theta_{[\tau_k^{*}, \tau_{k+1}^{*})}^*,\,\,\,\,\,\,\,\,\forall k\in\{0,1,\dots,k_n^*-1\}\,\,\hbox{and}\,\,\theta_{[\tau_{k_n^*}^{**}, 1]}^{**}=\theta_{[\tau_{k_n^*}^{*}, \tau_{k_n^*+1}^{*})}^*~~(\hbox{or}~~\theta_{[\tau_{k_n^*}^{*}, 1]}^*).
\end{eqnarray*}
Notice that it is possible that $[\tau_{k}^{**},\tau_{k+1}^{**})=\emptyset$ for some $k< k_n^*$. 

Then, it follows from \eqref{boundbetastar} that
\begin{eqnarray}\label{boundbetastar2}
\sup_{\mathcal{I}\in \mathcal{P}^{**}}\| \theta_{\mathcal{I}}^{**}\|_2\le 2c_0,
\end{eqnarray}
Moreover, it follows from \eqref{boundbeta0I}, \eqref{boundbetastar2} and the condition $m\asymp n$ that
\begin{eqnarray}\nonumber
\sum_{\mathcal{I}\in \mathcal{P}^{**}}\int_{\mathcal{I}} \|\theta_0(a)-\theta_{\mathcal{I}}^{**} \|_2^2da&\le& \sum_{\mathcal{I}\in \mathcal{P}^{*}}\int_{\mathcal{I}} \|\theta_0(a)-\theta_{\mathcal{I}}^{*} \|_2^2da+\frac{|\mathcal{P}^{**}|}{m} \sup_{a\in[0,1],\mathcal{I}\in \mathcal{P}^{**}} \|\theta_0(a)-\theta_{\mathcal{I}}^{**} \|_2^2\\\label{star0L2norm}
&\le&  c_4^2 k_n^{-2\alpha_0}+9c_0^2 (k_n+1) m^{-1}\le O(1) (k_n^{-2\alpha_0}+n^{-1} k_n),
\end{eqnarray}
for sufficiently large $n$, where $O(1)$ denotes some positive constant. 

Notice that
\begin{eqnarray*}
	&&\sum_{i=1}^n \sum_{\mathcal{I} \in \widehat{\mathcal{P}}} \mathbb{I}(A_i\in \mathcal{I}) (Y_i-\overline{X}_i^\top \widehat{\theta}_{\mathcal{I}})^2=\sum_{i=1}^n \sum_{\mathcal{I}_1 \in \widehat{\mathcal{P}}} \sum_{\mathcal{I}_2 \in \mathcal{P}^{**}} \mathbb{I}(A_i\in \mathcal{I}_1\cap \mathcal{I}_2) (Y_i-\overline{X}_i^\top \widehat{\theta}_{\mathcal{I}_1})^2\\
	&=& \sum_{i=1}^n \sum_{\mathcal{I}_1 \in \widehat{\mathcal{P}}} \sum_{\mathcal{I}_2 \in \mathcal{P}^{**}} \mathbb{I}(A_i\in \mathcal{I}_1\cap \mathcal{I}_2) (Y_i-\overline{X}_i^\top \theta_{\mathcal{I}_2}^{**} +\overline{X}_i^\top\theta_{\mathcal{I}_2}^{**} -\overline{X}_i^\top \widehat{\theta}_{\mathcal{I}_1})^2\\
	&=&\sum_{i=1}^n \sum_{\mathcal{I}_2 \in \mathcal{P}^{**}} \mathbb{I}(A_i\in \mathcal{I}_2) (Y_i-\overline{X}_i^\top \theta_{\mathcal{I}_2}^{**})^2+\underbrace{\sum_{i=1}^n \sum_{\mathcal{I}_1 \in \widehat{\mathcal{P}}} \sum_{\mathcal{I}_2 \in \mathcal{P}^{**}} \mathbb{I}(A_i\in \mathcal{I}_1\cap \mathcal{I}_2) (\overline{X}_i^\top\theta_{\mathcal{I}_2}^{**} -\overline{X}_i^\top \widehat{\theta}_{\mathcal{I}_1})^2}_{\chi_5}\\
	&+&2\underbrace{\sum_{i=1}^n \sum_{\mathcal{I}_1 \in \widehat{\mathcal{P}}} \sum_{\mathcal{I}_2 \in \mathcal{P}^{**}} \mathbb{I}(A_i\in \mathcal{I}_1\cap \mathcal{I}_2) (Y_i-\overline{X}_i^\top \theta_{\mathcal{I}_2}^{**})\overline{X}_i^\top(\theta_{\mathcal{I}_2}^{**} - \widehat{\theta}_{\mathcal{I}_1})}_{\chi_6}.
\end{eqnarray*}
By definition, we have
\begin{eqnarray*}
	\sum_{i=1}^n \sum_{\mathcal{I} \in \widehat{\mathcal{P}}} \mathbb{I}(A_i\in \mathcal{I}) (Y_i-\overline{X}_i^\top \widehat{\theta}_{\mathcal{I}})^2+n\gamma_n|\widehat{\mathcal{P}}|
	\le \sum_{i=1}^n \sum_{\mathcal{I} \in \mathcal{P}^{**}} \mathbb{I}(A_i\in \mathcal{I}) (Y_i-\overline{X}_i^\top \theta_{\mathcal{I}}^{**})^2+n(k_n+1) \gamma_n.
\end{eqnarray*}
It follows that
\begin{eqnarray}\label{proofimmediatestep1}
\chi_5+2\chi_6+n\gamma_n|\widehat{\mathcal{P}}|\le n(k_n+1) \gamma_n.
\end{eqnarray} 

We now give a lower bound for $\chi_5$. Similar to \eqref{prooflemma1eq1} and \eqref{prooflemmaeq2}, we can show that the following event occurs with probability at least $1-O(n^{-2})$:
\begin{eqnarray}\label{someevent2}
\lambda_{\min}\left( \sum_{i=1}^n \mathbb{I}(A_i\in \mathcal{I}) \overline{X}_i \overline{X}_i^\top \right)\ge c_5n|\mathcal{I}|,
\end{eqnarray}
for some constant $c_5>0$, and any interval $\mathcal{I}\in \mathfrak{I}(m)$ that satisfies $|\mathcal{I}|\ge \bar{c}_0 n^{-1}\log n$ where the constant $\bar{c}_0$ is defined in Lemma \hyperlink{lemma1}{1}. Under the event defined in \eqref{someevent2}, we obtain that
\begin{eqnarray}\nonumber
\chi_5\ge \sum_{\mathcal{I}_1 \in \widehat{\mathcal{P}}} \sum_{\mathcal{I}_2 \in \mathcal{P}^{**}} \sum_{i=1}^n \mathbb{I}(A_i\in \mathcal{I}_1\cap \mathcal{I}_2) \mathbb{I}(|\mathcal{I}_1\cap \mathcal{I}_2|\ge \bar{c}_0 n^{-1}\log n) (\overline{X}_i^\top\theta_{\mathcal{I}_2}^{**} -\overline{X}_i^\top \widehat{\theta}_{\mathcal{I}_1})^2\\\label{chi51}  \ge c_5 n \sum_{\mathcal{I}_1 \in \widehat{\mathcal{P}}} \sum_{\mathcal{I}_2 \in \mathcal{P}^{**}} \mathbb{I}(|\mathcal{I}_1\cap \mathcal{I}_2|\ge \bar{c}_0 n^{-1}\log n) |\mathcal{I}_1\cap \mathcal{I}_2| \|\theta_{\mathcal{I}_2}^{**} -\widehat{\theta}_{\mathcal{I}_1}\|_2^2.
\end{eqnarray}
In addition, under the events defined in \eqref{event1} and Lemma 4, we have
\begin{eqnarray*}
	\sup_{\mathcal{I}\in \widehat{\mathcal{P}}} \|\widehat{\theta}_{\mathcal{I}}-\theta_{0,\mathcal{I}}\|_2\le \sup_{\mathcal{I}\in \widehat{\mathcal{P}}}\frac{c_0\sqrt{\log n}}{\sqrt{|\mathcal{I}|n}} \le  \frac{c_0 \sqrt{\log n}}{\sqrt{\bar{c}_3 n \gamma_n}}=o(1),
\end{eqnarray*}
since $\gamma_n\gg n^{-1}\log n$. In view of \eqref{boundbeta0I}, we obtain that 
\begin{eqnarray}\label{boundbetaIhat}
\sup_{\mathcal{I}\in \widehat{\mathcal{P}}} \|\widehat{\theta}_{\mathcal{I}}\|_2\le 2c_0,
\end{eqnarray}
for sufficiently large $n$. 
This together with \eqref{boundbetastar2} yields that
\begin{eqnarray*}
	&&\sum_{\mathcal{I}_1 \in \widehat{\mathcal{P}}} \sum_{\mathcal{I}_2 \in \mathcal{P}^{**}} \mathbb{I}(|\mathcal{I}_1\cap \mathcal{I}_2|\le \bar{c}_0 n^{-1}\log n) |\mathcal{I}_1\cap \mathcal{I}_2| \|\theta_{\mathcal{I}_2}^{**} -\widehat{\theta}_{\mathcal{I}_1}\|_2^2\\
	&\le& (4c_0^2 \bar{c}_0 n^{-1}\log n) \sum_{\mathcal{I}_1 \in \widehat{\mathcal{P}}} \sum_{\mathcal{I}_2 \in \mathcal{P}^{**}} \mathbb{I}(|\mathcal{I}_1\cap \mathcal{I}_2|\le \bar{c}_0 n^{-1}\log n),
\end{eqnarray*}
with probability at least $1-O(n^{-2})$. Recall that $\mathcal{P}^{**}$ has at most $k_n$ change points. The number of nonempty intervals $\mathcal{I}_1\cap \mathcal{I}_2$ is at most $k_n+1+|\widehat{\mathcal{P}}|$. Thus, we obtain that
\begin{eqnarray*}
	\sum_{\mathcal{I}_1 \in \widehat{\mathcal{P}}} \sum_{\mathcal{I}_2 \in \mathcal{P}^{**}} \mathbb{I}(|\mathcal{I}_1\cap \mathcal{I}_2|\le \bar{c}_0 n^{-1}\log n) |\mathcal{I}_1\cap \mathcal{I}_2| \|\theta_{\mathcal{I}_2}^{**} -\widehat{\theta}_{\mathcal{I}_1}\|_2^2\le (k_n+1+|\widehat{\mathcal{P}}|) (4c_0^2 \bar{c}_0 n^{-1}\log n),
\end{eqnarray*}
with probability at least $1-O(n^{-2})$. This together with \eqref{chi51} yields that
\begin{eqnarray*}
	\chi_5\ge c_5 n \sum_{\mathcal{I}_1 \in \widehat{\mathcal{P}}} \sum_{\mathcal{I}_2 \in \mathcal{P}^{**}} |\mathcal{I}_1\cap \mathcal{I}_2| \|\theta_{\mathcal{I}_2}^{**} -\widehat{\theta}_{\mathcal{I}_1}\|_2^2-c_5(k_n+1+|\widehat{\mathcal{P}}|) (4c_0^2 \bar{c}_0 \log n),
\end{eqnarray*}
with probability at least $1-O(n^{-2})$, or equivalently,
\begin{eqnarray}\label{chi5}
\chi_5\ge c_5 n \int_{0}^1 \|\widehat{\theta}(a)-\theta^{**}(a) \|_2^2da-c_5(k_n+1+|\widehat{\mathcal{P}}|) (4c_0^2 \bar{c}_0 \log n),
\end{eqnarray}
with probability at least $1-O(n^{-2})$, where
\begin{eqnarray*}
	\theta^{**}(a)=\sum_{\mathcal{I}\in \mathcal{P}^{**}} \theta^{**}_{\mathcal{I}} \mathbb{I}(a\in \mathcal{I}).
\end{eqnarray*}

We now provide an upper bound for $|\chi_6|$. Notice that
\begin{eqnarray}\label{chi6}
~~~\chi_6&=&\sum_{i=1}^n \sum_{\mathcal{I}_1 \in \widehat{\mathcal{P}}} \sum_{\mathcal{I}_2 \in \mathcal{P}^{**}} \mathbb{I}(A_i\in \mathcal{I}_1\cap \mathcal{I}_2) (Y_i-\overline{X}_i^\top \theta_{\mathcal{I}_2}^{**})\overline{X}_i^\top(\theta_{\mathcal{I}_2}^{**} - \widehat{\theta}_{\mathcal{I}_1})\\ \nonumber
&=&\underbrace{\sum_{i=1}^n \sum_{\mathcal{I}_1 \in \widehat{\mathcal{P}}} \sum_{\mathcal{I}_2 \in \mathcal{P}^{**}} \mathbb{I}(A_i\in \mathcal{I}_1\cap \mathcal{I}_2) \{Y_i-\overline{X}_i^\top \theta_0(A_i)\}\overline{X}_i^\top(\theta_{\mathcal{I}_2}^{**} - \widehat{\theta}_{\mathcal{I}_1})}_{\chi_7}\\ \nonumber
&+&\underbrace{\sum_{i=1}^n \sum_{\mathcal{I}_1 \in \widehat{\mathcal{P}}} \sum_{\mathcal{I}_2 \in \mathcal{P}^{**}} \mathbb{I}(A_i\in \mathcal{I}_1\cap \mathcal{I}_2) \{\overline{X}_i^\top \theta_0(A_i)-\overline{X}_i^\top \theta_{\mathcal{I}_2}^{**}\}\overline{X}_i^\top(\theta_{\mathcal{I}_2}^{**} - \widehat{\theta}_{\mathcal{I}_1})}_{\chi_8}.
\end{eqnarray}
It suffices to provide upper bounds for $|\chi_7|$ and $|\chi_8|$. 

Under the event defined in \eqref{event2}, we obtain that
\begin{eqnarray*}
	\left|\sum_{i=1}^n \sum_{\mathcal{I}_1 \in \widehat{\mathcal{P}}} \sum_{\mathcal{I}_2 \in \mathcal{P}^{**}} \mathbb{I}(A_i\in \mathcal{I}_1\cap \mathcal{I}_2) \mathbb{I}(|\mathcal{I}_1\cap \mathcal{I}_2|\ge \bar{c}_0 n^{-1}\log n) \{Y_i-\overline{X}_i^\top \theta_0(A_i)\}\overline{X}_i^\top(\theta_{\mathcal{I}_2}^{**} - \widehat{\theta}_{\mathcal{I}_1})\right|\\
	\le \sum_{\mathcal{I}_1 \in \widehat{\mathcal{P}}} \sum_{\mathcal{I}_2 \in \mathcal{P}^{**}} \left\|\sum_{i=1}^n \mathbb{I}(A_i\in \mathcal{I}_1\cap \mathcal{I}_2) \mathbb{I}(|\mathcal{I}_1\cap \mathcal{I}_2|\ge \bar{c}_0 n^{-1}\log n) \{Y_i-\overline{X}_i^\top \theta_0(A_i)\}\overline{X}_i\right\|_2 \|\theta_{\mathcal{I}_2}^{**} - \widehat{\theta}_{\mathcal{I}_1}\|_2\\
	\le \sum_{\mathcal{I}_1 \in \widehat{\mathcal{P}}} \sum_{\mathcal{I}_2 \in \mathcal{P}^{**}} \sqrt{c_0 |\mathcal{I}_1 \cap \mathcal{I}_2| n\log n}\|\theta_{\mathcal{I}_2}^{**} - \widehat{\theta}_{\mathcal{I}_1}\|_2\le \frac{c_5n}{16} \int_{0}^1 \|\widehat{\theta}(a)-\theta^{**}(a) \|_2^2da\\
	+\frac{4c_0\log n}{c_5} \sum_{\mathcal{I}_1 \in \widehat{\mathcal{P}}} \sum_{\mathcal{I}_2 \in \mathcal{P}^{**}} |\mathcal{I}_1 \cap \mathcal{I}_2|\le \frac{c_5n}{16} \int_{0}^1 \|\widehat{\theta}(a)-\theta^{**}(a) \|_2^2da+4c_0 c_5^{-1} \log n,
\end{eqnarray*}
where the third inequality is due to Cauchy-Schwarz inequality. 

In addition, using similar arguments in \eqref{lowerboundeq2} and \eqref{lowerboundeq2.5}, we have with probability at least $1-O(n^{-2})$ that, for any interval $\mathcal{I}\in \mathfrak{I}(m)$ that satisfies $|\mathcal{I}|\le \bar{c}_0 n^{-1}\log n$, 
\begin{eqnarray}\label{someevent3}
\left\|\sum_{i=1}^n  \mathbb{I}(A_i\in \mathcal{I}) \{Y_i-\overline{X}_i^\top \theta_0(A_i)\}\overline{X}_i\right\|_2\le \bar{c}_5 \log n,
\end{eqnarray}
for some constant $\bar{c}_5>0$. Since the number of nonempty intervals $\mathcal{I}_1\cap \mathcal{I}_2$ is at most $k_n+1+|\widehat{\mathcal{P}}|$, we obtain that
\begin{eqnarray*}
	\left|\sum_{i=1}^n \sum_{\mathcal{I}_1 \in \widehat{\mathcal{P}}} \sum_{\mathcal{I}_2 \in \mathcal{P}^{**}} \mathbb{I}(A_i\in \mathcal{I}_1\cap \mathcal{I}_2) \mathbb{I}(|\mathcal{I}_1\cap \mathcal{I}_2|\le \bar{c}_0 n^{-1}\log n) \{Y_i-\overline{X}_i^\top \theta_0(A_i)\}\overline{X}_i^\top(\theta_{\mathcal{I}_2}^{**} - \widehat{\theta}_{\mathcal{I}_1})\right|\\
	\le \sum_{\mathcal{I}_1 \in \widehat{\mathcal{P}}} \sum_{\mathcal{I}_2 \in \mathcal{P}^{**}} \left\|\sum_{i=1}^n \mathbb{I}(A_i\in \mathcal{I}_1\cap \mathcal{I}_2) \mathbb{I}(|\mathcal{I}_1\cap \mathcal{I}_2|\le \bar{c}_0 n^{-1}\log n) \{Y_i-\overline{X}_i^\top \theta_0(A_i)\}\overline{X}_i\right\|_2 \|\theta_{\mathcal{I}_2}^{**} - \widehat{\theta}_{\mathcal{I}_1}\|_2\\
	\le (k_n+1+|\widehat{\mathcal{P}}|)(\bar{c}_5 \log n) \sup_{\mathcal{I}_1\in \widehat{\mathcal{P}}} \sup_{\mathcal{I}_2\in \mathcal{P}^{**}}\|\theta_{\mathcal{I}_2}^{**} - \widehat{\theta}_{\mathcal{I}_1}\|_2\le 4c_0(k_n+1+|\widehat{\mathcal{P}}|)(\bar{c}_5 \log n),
\end{eqnarray*}
with probability at least $1-O(n^{-2})$. It follows that
\begin{eqnarray}\label{chi7}
~~~~~~~~~~~~|\chi_7|\le \frac{c_5n}{16} \int_{0}^1 \|\widehat{\theta}(a)-\theta^{**}(a) \|_2^2da+4c_0 c_5^{-1} \log n+ 4c_0(k_n+1+|\widehat{\mathcal{P}}|)(\bar{c}_5 \log n),
\end{eqnarray}
with probability at least $1-O(n^{-2})$. 

As for $|\chi_8|$, it follows from Cauchy-Schwarz inequality that
\begin{eqnarray}\nonumber
|\chi_8|&\le& \frac{1}{4}\sum_{i=1}^n \sum_{\mathcal{I}_1 \in \widehat{\mathcal{P}}} \sum_{\mathcal{I}_2 \in \mathcal{P}^{**}} \mathbb{I}(A_i\in \mathcal{I}_1\cap \mathcal{I}_2) (\overline{X}_i^\top\theta_{\mathcal{I}_2}^{**} -\overline{X}_i^\top \widehat{\theta}_{\mathcal{I}_1})^2\\ \label{chi8}
&+&\underbrace{\sum_{i=1}^n \sum_{\mathcal{I}_1 \in \widehat{\mathcal{P}}} \sum_{\mathcal{I}_2 \in \mathcal{P}^{**}} \mathbb{I}(A_i\in \mathcal{I}_1\cap \mathcal{I}_2) \{\overline{X}_i^\top \theta_0(A_i)-\overline{X}_i^\top \theta_{\mathcal{I}_2}^{**}\}^2}_{\chi_9}=\frac{\chi_5}{4}+\chi_9.
\end{eqnarray}
Notice that
\begin{eqnarray*}
	\chi_9=\sum_{i=1}^n \sum_{\mathcal{I} \in \mathcal{P}^{**}} \mathbb{I}(A_i\in \mathcal{I}) \{\overline{X}_i^\top \theta_0(A_i)-\overline{X}_i^\top \theta_{\mathcal{I}}^{**}\}^2
\end{eqnarray*}
It follows from \eqref{meanXcA0}, \eqref{boundedA}, \eqref{star0L2norm} and Cauchy-Schwarz inequality that
\begin{eqnarray*}
	\Mean (\chi_9)=n\sum_{\mathcal{I} \in \mathcal{P}^{**}} \Mean \mathbb{I}(A\in \mathcal{I}) \{\overline{X}^\top \theta_0(A)-\overline{X}^\top \theta_{\mathcal{I}}^{**}\}^2\le n\sum_{\mathcal{I} \in \mathcal{P}^{**}} \Mean \|\overline{X}\|_2^2 \mathbb{I}(A\in \mathcal{I}) |\theta_0(A)-\theta_{\mathcal{I}}^{**}|_2^2\\
	\le n\sum_{\mathcal{I} \in \mathcal{P}^{**}} \Mean (\Mean \|\overline{X}\|_2^2|A) \mathbb{I}(A\in \mathcal{I}) |\theta_0(A)-\theta_{\mathcal{I}}^{**}|_2^2\le \omega^2 n \sum_{\mathcal{I} \in \mathcal{P}^{**}} \Mean \mathbb{I}(A\in \mathcal{I}) \|\theta_0(A)-\theta_{\mathcal{I}}^{**}\|_2^2\\
	\le C_0 \omega^2 n \sum_{\mathcal{I} \in \mathcal{P}^{**}} \int_{\mathcal{I}} \|\theta_0(a)-\theta_{\mathcal{I}}^{**}\|_2^2da\le O(1) (n\kappa_n^{-2\alpha_0}+\kappa_n),
\end{eqnarray*}
where $O(1)$ denotes some positive constant. Using similar arguments in \eqref{prooflemma2eq1}, we have for any integer $q\ge 2$ that
\begin{eqnarray*}
	\Mean \left(\sum_{\mathcal{I} \in \mathcal{P}^{**}} \mathbb{I}(A\in \mathcal{I}) \{\overline{X}^\top \theta_0(A)-\overline{X}^\top \theta_{\mathcal{I}}^{**}\}^2\right)^q\le \sum_{\mathcal{I} \in \mathcal{P}^{**}} \Mean \mathbb{I}(A\in \mathcal{I}) \{\overline{X}^\top \theta_0(A)-\overline{X}^\top \theta_{\mathcal{I}}^{**}\}^{2q}\\
	\le q! c^q \sum_{\mathcal{I} \in \mathcal{P}^{**}}\int_{\mathcal{I}} \|\theta_0(a)-\theta_{\mathcal{I}}^{**}\|_2^2da\le q! C^q (n\kappa_n^{-2\alpha_0}+\kappa_n),
\end{eqnarray*}
for some constants $c,C>0$. Using Bernstein's inequality, we have for any $t>0$ that
\begin{eqnarray*}
	\hbox{Pr}(\chi_9\ge \Mean \chi_9+t)\le \exp\left(-\frac{1}{2} \frac{t^2}{tC+2C^2 (nk_n^{-2\alpha_0}+k_n)}\right).
\end{eqnarray*}
We will require the sequence $\{k_n\}_n$ to satisfy $k_n\gg \log n$. Set $t_0=4C \sqrt{(nk_n^{-2\alpha_0}+k_n)\log n}$, we have
\begin{eqnarray*}
	\frac{t_0^2}{t_0 C+2C^2 (nk_n^{-2\alpha_0}+k_n)}=\frac{8 \sqrt{nk_n^{-2\alpha_0}+k_n}\log n}{2\sqrt{\log n}+\sqrt{nk_n^{-2\alpha_0}+k_n}}\ge 2\log n,
\end{eqnarray*}
for sufficiently large $n$. Therefore, we obtain with probability at least $1-O(n^{-2})$ that
\begin{eqnarray*}
	\chi_9\le O(1) (n\kappa_n^{-2\alpha_0}+\kappa_n)+4C \sqrt{(nk_n^{-2\alpha_0}+k_n)\log n}=O(nk_n^{-2\alpha_0}+k_n).
\end{eqnarray*}
This together with \eqref{chi6}, \eqref{chi7} and \eqref{chi8} yields that 
\begin{eqnarray*}
	|\chi_6|\le \frac{c_5n}{16} \int_{0}^1 \|\widehat{\theta}(a)-\theta^{**}(a) \|_2^2da+c_6 \{(k_n+|\widehat{\mathcal{P}}|)\log n+nk_n^{-2\alpha_0}\}+\frac{\chi_5}{4}.
\end{eqnarray*}
with probability at least $1-O(n^{-2})$, for some constant $c_6>0$ and sufficiently large $n$. In view of \eqref{proofimmediatestep1} and \eqref{chi5}, we obtain with probability at least $1-O(n^{-2})$ that $\chi_5\le n\gamma_n (k_n+1-\widehat{\mathcal{P}})+2|\chi_6|$ and hence
\begin{eqnarray}\label{someevent4}
&&\frac{3c_5n}{8} \int_{0}^1 \|\widehat{\theta}(a)-\theta^{**}(a) \|_2^2da\\\nonumber
\le&& 2c_6 \{(k_n+|\widehat{\mathcal{P}}|)\log n+nk_n^{-2\alpha_0}\}+n\gamma_n (k_n+1-\widehat{\mathcal{P}}).
\end{eqnarray}
Suppose $|\widehat{\mathcal{P}}|\ge 2k_n+1$. Under the event defined in \eqref{someevent4}, it follows from the condition $n\gamma_n \gg \log n$ that
\begin{eqnarray*}
	n\gamma_n (k_n+1-|\widehat{\mathcal{P}}|)+2c_6 (k_n+|\widehat{\mathcal{P}}|)\log n\le 3c_6 |\widehat{\mathcal{P}}|\log n -2^{-1} n\gamma_n |\widehat{\mathcal{P}}|\le 0,
\end{eqnarray*}
for sufficiently large $n$, and hence
\begin{eqnarray}\label{someevent5}
\frac{3c_5n}{8} \int_{0}^1 \|\widehat{\theta}(a)-\theta^{**}(a) \|_2^2da\le 2c_6 nk_n^{-2\alpha_0}.
\end{eqnarray}
Otherwise, suppose $|\widehat{\mathcal{P}}|\le 2k_n$. It follows from \eqref{someevent4} that
\begin{eqnarray*}
	\frac{3c_5n}{8} \int_{0}^1 \|\widehat{\theta}(a)-\theta^{**}(a) \|_2^2da\le 6c_6 (k_n\log n+nk_n^{-2\alpha_0})+n\gamma_n k_n,
\end{eqnarray*}
with probability at least $1-O(n^{-2})$. This together with \eqref{someevent5} yields that
\begin{eqnarray}\label{someevent6}
\int_{0}^1 \|\widehat{\theta}(a)-\theta^{**}(a) \|_2^2da\le 16 c_5^{-1} c_6 n^{-1} (k_n\log n+nk_n^{-2\alpha_0})+3\gamma_n k_n,
\end{eqnarray}
with probability at least $1-O(n^{-2})$. 

By Cauchy-Schwarz inequality, we have
\begin{eqnarray*}
	\int_{0}^1 \|\widehat{\theta}(a)-\theta_0(a) \|_2^2da=\int_{0}^1 \|\widehat{\theta}(a)-\theta^{**}(a)+\theta^{**}(a)-\theta_0(a) \|_2^2da\\
	\le 2 \int_{0}^1 \|\widehat{\theta}(a)-\theta^{**}(a)\|_2^2da+2\int_0^1 \|\theta^{**}(a)-\theta_0(a) \|_2^2da.
\end{eqnarray*}
In view of \eqref{star0L2norm} and \eqref{someevent6}, we obtain that
\begin{eqnarray}\label{proofthm3finaleq}
~~~~~~~~	\int_{0}^1 \|\widehat{\theta}(a)-\theta_0(a) \|_2^2da=O(n^{-1}k_n\log n+k_n^{-2\alpha_0}+\gamma_n k_n)=O(k_n^{-2\alpha_0}+\gamma_n k_n),
\end{eqnarray}
with probability at least $1-O(n^{-2})$, where the last equality is due to the condition that $\gamma_n\gg n^{-1}\log n$. Set $k_n=\floor{\gamma_n^{-(1+2\alpha_0)}}$ (the largest integer that is smaller than $\gamma_n^{-(1+2\alpha_0)}$), we obtain that
\begin{eqnarray*}
	\int_{0}^1 \|\widehat{\theta}(a)-\theta_0(a) \|_2^2da=O(\gamma_n^{2\alpha_0/(1+2\alpha_0)}),
\end{eqnarray*}
with probability at least $1-O(n^{-2})$. The proof is hence completed. 

\noindent {\textit {Convergence rate of the value function: }}
To derive the convergence rate of the value function under the proposed I2DR, we introduce the following lemma. 

\smallskip

\noindent
{\bf \hypertarget{lemma6}{Lemma 8}} {\it		Assume conditions in Theorem \hyperlink{thm4}{6} hold. Then for any interval $\mathcal{I}\in \mathfrak{I}(m)$ with $|\mathcal{I}|\ge \bar{c}_0 n^{-1}\log n$ and any interval $\mathcal{I}'\in \widehat{\mathcal{P}}$ with $\mathcal{I}\subseteq \mathcal{I}'$, we have with probability at least $1-O(n^{-2})$ that
	\begin{eqnarray*}
		\|\theta_{0,\mathcal{I}}-\theta_{0,\mathcal{I}'}\|_2\le 3\sqrt{c_5^{-1}\gamma_n |\mathcal{I}|^{-1}},
	\end{eqnarray*}
	where the constant $c_5$ is defined in \eqref{someevent2}.} \hfill$\square$
	
\smallskip

Recall by the definition of the value function that
\begin{eqnarray}\label{valuefundefi}
~~~~~~~~~~~~~~V^{\tiny{opt}}-V^{\pi^*}(\widehat{d})=\Mean \left(\sup_{a\in [0,1]} \overline{X}^\top \theta_0(a) \right)-\Mean \left( \int_{\widehat{d}(X)} \overline{X}^\top \theta_0(a)\pi^*(a;X,\widehat{d}(X))da \right).
\end{eqnarray}
We begin by providing an upper bound for
\begin{eqnarray*}
	\chi_{11}=\Mean \left( \sup_{a\in [0,1]} \overline{X}^\top \theta_0(a) \right)-\Mean \left( \sup_{\mathcal{I}\in \widehat{\mathcal{P}}} \overline{X}^\top \theta_{0,\mathcal{I}} \right). 
\end{eqnarray*}
It follows from \eqref{meanXcA0} and Cauchy-Schwarz inequality that
\begin{eqnarray}\label{Holder4} 
\begin{split}
\chi_{11}&=&\Mean \left( \sup_{\mathcal{I}\in \widehat{\mathcal{P}}} \sup_{a\in \mathcal{I}}\overline{X}^\top \theta_0(a) \right)-\Mean \left( \sup_{\mathcal{I}\in \widehat{\mathcal{P}}} \overline{X}^\top \theta_{0,\mathcal{I}} \right)\le \Mean \|\overline{X}\|_2 \sup_{\mathcal{I}\in \widehat{\mathcal{P}}}  \sup_{a\in \mathcal{I}} \|\theta_0(a)-\theta_{0,\mathcal{I}}\|_2\\
&\le& \sqrt{\Mean \sum_{j=1}^{p+1} |\overline{X}^{(j)}|^2 } \sup_{\mathcal{I}\in \widehat{\mathcal{P}}}  \sup_{a\in \mathcal{I}} \|\theta_0(a)-\theta_{0,\mathcal{I}}\|_2
\le (p+1)^{1/2}\omega \sup_{\mathcal{I}\in \widehat{\mathcal{P}}}  \sup_{a\in \mathcal{I}} \|\theta_0(a)-\theta_{0,\mathcal{I}}\|_2.
\end{split}
\end{eqnarray}
Consider a sequence $\{d_n\}_n$ that satisfies $d_n\ge 0,\forall n$, $d_n\to 0$ as $n\to \infty$ and $d_n\gg n^{-1}\log n$. By the definition of H{\"o}lder continuous functions, we have for any $\mathcal{I}$ with $|\mathcal{I}|\le d_n$ that
\begin{eqnarray*}
	\sup_{a_1,a_2\in \mathcal{I}} \|\theta_0(a_1)-\theta_0(a_2)\|_2\le L\sup_{a_1,a_2\in \mathcal{I}} |a_1-a_2|^{\alpha_0}\le L d_n^{\alpha_0}. 
\end{eqnarray*}
It follows that
\begin{eqnarray}\nonumber
\sup_{a\in \mathcal{I}} \|\theta_0(a)-\theta_{0,\mathcal{I}}\|_2\le \sup_{a\in \mathcal{I}} \left\|\theta_0(a)- \{\Mean \overline{X}\overline{X}^\top \mathbb{I}(A\in \mathcal{I})\}^{-1} \Mean \overline{X} \overline{X}^\top \theta_0(A) \mathbb{I}(A\in \mathcal{I}) \right\|_2\\  \label{Holder1}
\le \sup_{a\in \mathcal{I}} \left\| \{\Mean \overline{X}\overline{X}^\top \mathbb{I}(A\in \mathcal{I})\}^{-1} \Mean \overline{X} \overline{X}^\top\mathbb{I}(A\in \mathcal{I}) \{\theta_0(a)-\theta_0(A)\}  \right\|_2\\  \nonumber
\le \sup_{a\in \mathcal{I}} \left\| \{\Mean \overline{X}\overline{X}^\top \mathbb{I}(A\in \mathcal{I})\}^{-1} \Mean \overline{X} \overline{X}^\top\mathbb{I}(A\in \mathcal{I})\right\|_2 \sup_{a,a^*\in \mathcal{I}} \|\theta_0(a)-\theta_0(a^*)\|_2\le Ld_n^{\alpha_0},
\end{eqnarray}
for any $\mathcal{I}$ that satisfies $|\mathcal{I}|\le d_n$. 

Consider an interval $\mathcal{I}\in \mathfrak{I}(m)$ that satisfies $|\mathcal{I}|>d_n$. For any $a\in \mathcal{I}$, we can find an interval $\mathcal{I}' \subseteq \mathcal{I}$ with $d_n/2\le |\mathcal{I}'|\le d_n$ and $\mathcal{I}'\in \mathfrak{I}(m)$ that covers $a$. Similar to \eqref{Holder1}, we have
\begin{eqnarray}\label{Holder2}
\|\theta_{0,\mathcal{I}'}-\theta_0(a)\|_2\le Ld_n^{\alpha_0}. 
\end{eqnarray}
Since $d_n\gg n^{-1}\log n$, by Lemma \hyperlink{lemma6}{8}, we have with probability at least $1-O(n^{-2})$ that 
\begin{eqnarray*}
	\|\theta_{0,\mathcal{I}}-\theta_{0,\mathcal{I}'}\|_2\le 3\sqrt{2c_5^{-1} \gamma_n d_n^{-1} }.
\end{eqnarray*}
This together with \eqref{Holder2} yields that
\begin{eqnarray*}
	\sup_{a\in \mathcal{I}} \|\theta_0(a)-\theta_{0,\mathcal{I}}\|_2\le Ld_n^{\alpha_0}+3\sqrt{2c_5^{-1} \gamma_n d_n^{-1} },
\end{eqnarray*}
for any $\mathcal{I}\in \mathfrak{I}(m)$ that satisfies $|\mathcal{I}|>d_n$. Combining this together with \eqref{Holder1}, we obtain that
\begin{eqnarray*}
	\sup_{a\in \mathcal{I}} \|\theta_0(a)-\theta_{0,\mathcal{I}}\|_2\le Ld_n^{\alpha_0}+3\sqrt{2c_5^{-1} \gamma_n d_n^{-1} },
\end{eqnarray*}
for any $\mathcal{I}\in \mathfrak{I}(m)$, with probability at least $1-O(n^{-2})$. Set $d_n\asymp \gamma_n^{(1+2\alpha_0)^{-1}}$, we have with probability at least $1-O(n^{-2})$ that
\begin{eqnarray*}
	\sup_{a\in \mathcal{I}} \|\theta_0(a)-\theta_{0,\mathcal{I}}\|_2\le O(1) \gamma_n^{\alpha_0/(1+2\alpha_0)},
\end{eqnarray*}
for any $\mathcal{I}\in \mathfrak{I}(m)$, where $O(1)$ denotes some positive constant. 

Therefore, we obtain with probability at least $1-O(n^{-2})$ that
\begin{eqnarray}\label{chi11}
\chi_{11}\le O(1) \gamma_n^{\alpha_0/(1+2\alpha_0)},
\end{eqnarray}
where $O(1)$ denotes some positive constant. Similarly, we can show with probability at least $1-O(n^{-2})$ that
\begin{eqnarray*}
	\chi_{12}=\Mean \overline{X}^\top \theta_{0,\widehat{d}(X)}-\Mean \left( \int_{\widehat{d}(X)} \overline{X}^\top \theta_0(a)\pi^*(a;X,\widehat{d}(X))da \right)\le O(1) \gamma_n^{\alpha_0/(1+2\alpha_0)},
\end{eqnarray*}
where $O(1)$ denotes some positive constant. This together with \eqref{valuefundefi} and \eqref{chi11} yields that,
\begin{eqnarray}\label{Holder3}
V^{\tiny{opt}}-V^{\pi^*}(\widehat{d})\le \Mean \left(\sup_{\mathcal{I}\in \widehat{\mathcal{P}}} \overline{X}^\top \theta_{0,\mathcal{I}} \right)-\Mean \overline{X}^\top \theta_{0,\widehat{d}(X)}+O(1) \gamma_n^{\alpha_0/(1+2\alpha_0)},
\end{eqnarray}
with probability at least $1-O(n^{-2})$, where $O(1)$ denotes some positive constant. 

Using similar arguments in \eqref{Holder4}, we can show that
\begin{eqnarray*}
	\Mean \left( \sup_{\mathcal{I}\in \widehat{\mathcal{P}}} \overline{X}^\top \theta_{0,\mathcal{I}} \right)-\Mean \left( \sup_{\mathcal{I}\in \widehat{\mathcal{P}}} \overline{X}^\top \widehat{\theta}_{\mathcal{I}} \right)\le (p+1)^{1/2}\omega \sup_{\mathcal{I}\in \widehat{\mathcal{P}}} \|\theta_{0,\mathcal{I}}-\widehat{\theta}_{\mathcal{I}}\|_2,
\end{eqnarray*}
and
\begin{eqnarray*}
	\Mean \overline{X}^\top \theta_{0,\widehat{d}(X)}-\Mean \overline{X}^\top \widehat{\theta}_{\widehat{d}(X)}\le (p+1)^{1/2}\omega \sup_{\mathcal{I}\in \widehat{\mathcal{P}}} \|\theta_{0,\mathcal{I}}-\widehat{\theta}_{\mathcal{I}}\|_2.
\end{eqnarray*}
Since $\sup_{\mathcal{I}\in \widehat{\mathcal{P}}} \overline{X}^\top \widehat{\theta}_{\mathcal{I}}=\overline{X}^\top \widehat{\theta}_{\widehat{d}(X)}$, we have
\begin{eqnarray}\label{Holder5}
V^{\tiny{opt}}-V^{\pi^*}(\widehat{d})\le 2 (p+1)^{1/2}\omega \sup_{\mathcal{I}\in \widehat{\mathcal{P}}} \|\theta_{0,\mathcal{I}}-\widehat{\theta}_{\mathcal{I}}\|_2+O(1) \gamma_n^{\alpha_0/(1+2\alpha_0)},
\end{eqnarray}
under the event defined in \eqref{Holder3}. It follows from Lemma \hyperlink{lemma1}{1} and \hyperlink{lemma5}{4} that
\begin{eqnarray*}
	\sup_{\mathcal{I}\in \widehat{\mathcal{P}}}\|\theta_{0,\mathcal{I}}-\widehat{\theta}_{\mathcal{I}}\|_2\le \frac{\sqrt{\log n}}{\sqrt{n\gamma_n}},
\end{eqnarray*}
with probability at least $1-O(n^{-2})$. This together with \eqref{Holder5} yields that
\begin{eqnarray*}
	V^{\tiny{opt}}-V^{\pi^*}(\widehat{d})\le O(1) \left( \gamma_n^{\alpha_0/(1+2\alpha_0)}+\frac{\sqrt{\log n}}{\sqrt{n\gamma_n}} \right),
\end{eqnarray*}
with probability at least $1-O(n^{-2})$, where $O(1)$ denotes some positive constant. Set $\gamma_n\asymp (n^{-1/2} \log^{1/2} n)^{(2\alpha_0+1)/(4\alpha_0+1)}$, we obtain that $V^{\tiny{opt}}-V^{\pi^*}(\widehat{d})=O( n^{-\alpha_0/(1+4\alpha_0)}\log^{\alpha_0/(1+4\alpha_0)} n )$, with probability at least $1-O(n^{-2})$. The proof is hence completed. 

\subsection{Proof of Lemma 8}
For a given interval $\mathcal{I}'\in \widehat{\mathcal{P}}$, the set of intervals $\mathcal{I}$ considered in Lemma \hyperlink{lemma6}{8} can be classified into the following three categories. 

\noindent \textit{Category 1:} $\mathcal{I}=\mathcal{I}'$. Then it is immediate to see that $\|\theta_{0,\mathcal{I}}-\theta_{0,\mathcal{I}'}\|_2=0$ and the assertion automatically holds. 

\noindent \textit{Category 2:} There exists another interval $\mathcal{I}^*\in \mathfrak{I}(m)$ that satisfies $\mathcal{I}'=\mathcal{I}^*\cup \mathcal{I}$. Notice that the partition $\widehat{\mathcal{P}}^*=\widehat{\mathcal{P}} \cup \{\mathcal{I}^* \} \cup \mathcal{I}-\{\mathcal{I}' \}$ also belongs to $\mathcal{B}(m)$. By definition, we have
\begin{eqnarray*}
	&&\frac{1}{n}\sum_{i=1}^n\sum_{\mathcal{I}_0\in \widehat{\mathcal{P}}^*} \mathbb{I}(A_i\in \mathcal{I}_0) (Y_i-\overline{X}_i^\top \widehat{\theta}_{\mathcal{I}_0})^2+\lambda_n |\mathcal{I}_0|\|\widehat{\theta}_{\mathcal{I}_0}\|_2^2+\gamma_n |\widehat{\mathcal{P}}^*|\\
	&\ge&\frac{1}{n}\sum_{i=1}^n\sum_{\mathcal{I}_0\in \widehat{\mathcal{P}}} \mathbb{I}(A_i\in \mathcal{I}_0) (Y_i-\overline{X}_i^\top \widehat{\theta}_{\mathcal{I}_0})^2+\lambda_n|\mathcal{I}_0| \|\widehat{\theta}_{\mathcal{I}_0}\|_2^2+\gamma_n |\widehat{\mathcal{P}}|,
\end{eqnarray*}
and hence
\begin{eqnarray*}
	\frac{1}{n}\sum_{i=1}^n \mathbb{I}(A_i\in \mathcal{I}) (Y_i-\overline{X}_i^\top \widehat{\theta}_{\mathcal{I}})^2+\lambda_n|\mathcal{I}| \|\widehat{\theta}_{\mathcal{I}}\|_2^2+\frac{1}{n}\sum_{i=1}^n \mathbb{I}(A_i\in \mathcal{I}^*) (Y_i-\overline{X}_i^\top \widehat{\theta}_{\mathcal{I}^*})^2+\lambda_n|\mathcal{I}^*| \|\widehat{\theta}_{\mathcal{I}^*}\|_2^2\\ 
	\ge \frac{1}{n}\sum_{i=1}^n \mathbb{I}(A_i\in \mathcal{I}') (Y_i-\overline{X}_i^\top \widehat{\theta}_{\mathcal{I}'})^2+\lambda_n|\mathcal{I}'| \|\widehat{\theta}_{\mathcal{I}'}\|_2^2-\gamma_n.
\end{eqnarray*}
It follows from the definition of $\widehat{\theta}_{\mathcal{I}^*}$ that
\begin{eqnarray*}
	\frac{1}{n}\sum_{i=1}^n \mathbb{I}(A_i\in \mathcal{I}^*) (Y_i-\overline{X}_i^\top \widehat{\theta}_{\mathcal{I}^*})^2+\lambda_n|\mathcal{I}^*| \|\widehat{\theta}_{\mathcal{I}^*}\|_2^2\le \frac{1}{n}\sum_{i=1}^n \mathbb{I}(A_i\in \mathcal{I}^*) (Y_i-\overline{X}_i^\top \widehat{\theta}_{\mathcal{I}'})^2+\lambda_n|\mathcal{I}^*| \|\widehat{\theta}_{\mathcal{I}'}\|_2^2.
\end{eqnarray*}
Therefore, we obtain
\begin{eqnarray}\label{cat2eq10}
&&\frac{1}{n}\sum_{i=1}^n \mathbb{I}(A_i\in \mathcal{I}) (Y_i-\overline{X}_i^\top \widehat{\theta}_{\mathcal{I}})^2+\lambda_n|\mathcal{I}| \|\widehat{\theta}_{\mathcal{I}}\|_2^2 \\ &\ge& \nonumber
\frac{1}{n}\sum_{i=1}^n \mathbb{I}(A_i\in \mathcal{I}) (Y_i-\overline{X}_i^\top \widehat{\theta}_{\mathcal{I}'})^2+\lambda_n|\mathcal{I}| \|\widehat{\theta}_{\mathcal{I}'}\|_2^2-\gamma_n.
\end{eqnarray}

\noindent \textit{Category 3: }There exist two intervals $\mathcal{I}^*,\mathcal{I}^{**}\in \mathfrak{I}(m)$ that satisfy $\mathcal{I}'=\mathcal{I}^*\cup \mathcal{I} \cup \mathcal{I}^{**}$. Using similar arguments in proving \eqref{cat2eq10}, we can show that
\begin{eqnarray*}
	\frac{1}{n}\sum_{i=1}^n \mathbb{I}(A_i\in \mathcal{I}) (Y_i-\overline{X}_i^\top \widehat{\theta}_{\mathcal{I}})^2+\lambda_n|\mathcal{I}| \|\widehat{\theta}_{\mathcal{I}}\|_2^2 \ge 
	\frac{1}{n}\sum_{i=1}^n \mathbb{I}(A_i\in \mathcal{I}) (Y_i-\overline{X}_i^\top \widehat{\theta}_{\mathcal{I}'})^2+\lambda_n|\mathcal{I}| \|\widehat{\theta}_{\mathcal{I}'}\|_2^2-2\gamma_n.
\end{eqnarray*}
Hence, regardless of whether $\mathcal{I}$ belongs to Category 2, or it belongs to Category 3, we have
\begin{eqnarray}\nonumber
&&\frac{1}{n}\sum_{i=1}^n \mathbb{I}(A_i\in \mathcal{I}) (Y_i-\overline{X}_i^\top \widehat{\theta}_{\mathcal{I}})^2+\lambda_n|\mathcal{I}| \|\widehat{\theta}_{\mathcal{I}}\|_2^2 \\ \nonumber
&\ge& 
\frac{1}{n}\sum_{i=1}^n \mathbb{I}(A_i\in \mathcal{I}) (Y_i-\overline{X}_i^\top \widehat{\theta}_{\mathcal{I}'})^2+\lambda_n|\mathcal{I}| \|\widehat{\theta}_{\mathcal{I}'}\|_2^2-2\gamma_n\\\label{cat2eq20}
&\ge& \frac{1}{n}\sum_{i=1}^n \mathbb{I}(A_i\in \mathcal{I}) (Y_i-\overline{X}_i^\top \widehat{\theta}_{\mathcal{I}'})^2-2\gamma_n.
\end{eqnarray}
Notice that $|\mathcal{I}'|\ge |\mathcal{I}|$. Under the event defined in \eqref{event1}, we obtain that 
\begin{eqnarray*}
	\|\widehat{\theta}_{\mathcal{I}'}-\theta_{0,\mathcal{I}'}\|_2\le \frac{c_0\sqrt{\log n}}{\sqrt{|\mathcal{I}|n}}.
\end{eqnarray*}
Similar to \eqref{someinequality2}, we can show the following event occurs with probability at least $1-O(n^{-2})$, 
\begin{eqnarray}\label{cat2eq2.5}
\left|\frac{1}{n}\sum_{i=1}^n \mathbb{I}(A_i\in \mathcal{I})(Y_i-\overline{X}_i^\top \theta_{0,\mathcal{I}})\overline{X}_i^\top(\widehat{\theta}_{\mathcal{I}'}-\theta_{0,\mathcal{I}'})\right|\le O(1) n^{-1}\log n,
\end{eqnarray}
where $O(1)$ denotes some positive constant. Similarly, using Cauchy-Schwarz inequality, we can show with probability at least $1-O(n^{-2})$ that
\begin{eqnarray*}
	&&\left|\frac{1}{n}\sum_{i=1}^n \mathbb{I}(A_i\in \mathcal{I})(\overline{X}_i^\top \theta_{0,\mathcal{I}}-\overline{X}_i^\top \theta_{0,\mathcal{I}'})\overline{X}_i^\top(\widehat{\theta}_{\mathcal{I}'}-\theta_{0,\mathcal{I}'})\right|\\ &\le& \frac{1}{4n} \sum_{i=1}^n \mathbb{I}(A_i\in \mathcal{I}) (\overline{X}_i^\top \theta_{0,\mathcal{I}}-\overline{X}_i^\top \theta_{0,\mathcal{I}'})^2+\frac{1}{n}\sum_{i=1}^n \mathbb{I}(A_i\in \mathcal{I})\{\overline{X}_i^\top(\widehat{\theta}_{\mathcal{I}'}-\theta_{0,\mathcal{I}'})\}^2\\ &\le& \frac{1}{4n} \sum_{i=1}^n \mathbb{I}(A_i\in \mathcal{I}) (\overline{X}_i^\top \theta_{0,\mathcal{I}}-\overline{X}_i^\top \theta_{0,\mathcal{I}'})^2+O(1) n^{-1}\log n,
\end{eqnarray*}
where $O(1)$ denotes some positive constant. This together with \eqref{cat2eq2.5} yields
\begin{eqnarray*}
	&&\left|\frac{1}{n}\sum_{i=1}^n \mathbb{I}(A_i\in \mathcal{I})(Y_i-\overline{X}_i^\top \theta_{0,\mathcal{I}})\overline{X}_i^\top(\widehat{\theta}_{\mathcal{I}'}-\theta_{0,\mathcal{I}'})\right|\\
	&\le& \frac{1}{4n} \sum_{i=1}^n \mathbb{I}(A_i\in \mathcal{I}) (\overline{X}_i^\top \theta_{0,\mathcal{I}}-\overline{X}_i^\top \theta_{0,\mathcal{I}'})^2+ O(1) n^{-1}\log n,
\end{eqnarray*}
with probability at least $1-O(n^{-2})$, where $O(1)$ denote some positive constant. Using similar arguments in proving  \eqref{someinequality1}, we can show the following event occurs with probability at least $1-O(n^{-2})$, 
\begin{eqnarray}\label{cat2eq3}
\frac{1}{n}\sum_{i=1}^n \mathbb{I}(A_i\in \mathcal{I}) (Y_i-\overline{X}_i^\top \widehat{\theta}_{\mathcal{I}'})^2\ge \frac{1}{n}\sum_{i=1}^n \mathbb{I}(A_i\in \mathcal{I}) (Y_i-\overline{X}_i^\top \theta_{0,\mathcal{I}'})^2\\ \nonumber
-\frac{1}{2n} \sum_{i=1}^n \mathbb{I}(A_i\in \mathcal{I}) (\overline{X}_i^\top \theta_{0,\mathcal{I}}-\overline{X}_i^\top \theta_{0,\mathcal{I}'})^2-O(1) n^{-1}\log n,
\end{eqnarray}
where $O(1)$ denotes some positive constant. 

In addition, it follows from the definition of $\widehat{\theta}_{\mathcal{I}}$ that
\begin{eqnarray*}
	\frac{1}{n}\sum_{i=1}^n \mathbb{I}(A_i\in \mathcal{I}) (Y_i-\overline{X}_i^\top \widehat{\theta}_{\mathcal{I}})^2+\lambda_n|\mathcal{I}| \|\widehat{\theta}_{\mathcal{I}}\|_2^2\le \frac{1}{n}\sum_{i=1}^n \mathbb{I}(A_i\in \mathcal{I}) (Y_i-\overline{X}_i^\top \theta_{0,\mathcal{I}})^2+\lambda_n|\mathcal{I}| \|\theta_{0,\mathcal{I}}\|_2^2.
\end{eqnarray*}
By \eqref{boundbeta0I} and the condition that $\lambda_n=O(n^{-1}\log n)$, we obtain that
\begin{eqnarray*}
	\frac{1}{n}\sum_{i=1}^n \mathbb{I}(A_i\in \mathcal{I}) (Y_i-\overline{X}_i^\top \widehat{\theta}_{\mathcal{I}})^2+\lambda_n|\mathcal{I}| \|\widehat{\theta}_{\mathcal{I}}\|_2^2\le \frac{1}{n}\sum_{i=1}^n \mathbb{I}(A_i\in \mathcal{I}) (Y_i-\overline{X}_i^\top \theta_{0,\mathcal{I}})^2+O(1) n^{-1}\log n,
\end{eqnarray*}
where $O(1)$ denotes some positive constant. This together with \eqref{cat2eq20} and \eqref{cat2eq3} yields
\begin{eqnarray}\label{cat2eq4}
~~~~~~~~~~~~~&&\sum_{i=1}^n \mathbb{I}(A_i\in \mathcal{I}) (Y_i-\overline{X}_i^\top \theta_{0,\mathcal{I}})^2\\ \nonumber
\ge&& \sum_{i=1}^n \mathbb{I}(A_i\in \mathcal{I}) (Y_i-\overline{X}_i^\top \theta_{0,\mathcal{I}'})^2-2n\gamma_n -O(1) \log n\\ \nonumber
&&-\frac{1}{2} \sum_{i=1}^n \mathbb{I}(A_i\in \mathcal{I}) (\overline{X}_i^\top \theta_{0,\mathcal{I}}-\overline{X}_i^\top \theta_{0,\mathcal{I}'})^2,
\end{eqnarray}
with probability at least $1-O(n^{-2})$, where $O(1)$ denotes some positive constant. 

Notice that
\begin{eqnarray*}
	&&\sum_{i=1}^n \mathbb{I}(A_i\in \mathcal{I}) (Y_i-\overline{X}_i^\top \theta_{0,\mathcal{I}'})^2=\sum_{i=1}^n \mathbb{I}(A_i\in \mathcal{I}) (Y_i-\overline{X}_i^\top \theta_{0,\mathcal{I}}+\overline{X}_i^\top \theta_{0,\mathcal{I}}-\overline{X}_i^\top \theta_{0,\mathcal{I}'})^2\\
	&=&\sum_{i=1}^n \mathbb{I}(A_i\in \mathcal{I}) (Y_i-\overline{X}_i^\top \theta_{0,\mathcal{I}})^2+2\underbrace{\sum_{i=1}^n \mathbb{I}(A_i\in \mathcal{I}) (Y_i-\overline{X}_i^\top \theta_{0,\mathcal{I}})(\overline{X}_i^\top \theta_{0,\mathcal{I}}-\overline{X}_i^\top \theta_{0,\mathcal{I}'})}_{\chi_{10}}\\
	&+&\sum_{i=1}^n \mathbb{I}(A_i\in \mathcal{I}) (\overline{X}_i^\top \theta_{0,\mathcal{I}'}-\overline{X}_i^\top \theta_{0,\mathcal{I}})^2.
\end{eqnarray*}
Combining this with \eqref{cat2eq4} yields that 
\begin{eqnarray}\label{cat2eq5}
\sum_{i=1}^n \mathbb{I}(A_i\in \mathcal{I}) (\overline{X}_i^\top \theta_{0,\mathcal{I}'}-\overline{X}_i^\top \theta_{0,\mathcal{I}})^2\le 4n\gamma_n+O(1)\log n+4|\chi_{10}|.
\end{eqnarray}
Under the event defined in \eqref{someevent2}, we obtain that
\begin{eqnarray}\label{cat2eq6}
\sum_{i=1}^n \mathbb{I}(A_i\in \mathcal{I}) (\overline{X}_i^\top \theta_{0,\mathcal{I}'}-\overline{X}_i^\top \theta_{0,\mathcal{I}})^2\ge c_5 n |\mathcal{I}| \|\theta_{0,\mathcal{I}'}-\theta_{0,\mathcal{I}}\|_2^2.
\end{eqnarray}
By the definition of $\theta_{0,\mathcal{I}}$, we have $\Mean \mathbb{I}(A\in \mathcal{I})(Y-\overline{X}^\top \theta_{0,\mathcal{I}})\overline{X}=0$. Under the event defined in \eqref{event2}, it follows from Cauchy-Schwarz inequality that
\begin{eqnarray*}
	|\chi_{10}|\le \frac{2}{c_5 n |\mathcal{I}|}\left\|\sum_{i=1}^n \mathbb{I}(A_i\in \mathcal{I}) (Y_i-\overline{X}_i^\top \theta_{0,\mathcal{I}})\overline{X}_i\right\|_2^2+\frac{c_5 n}{8} |\mathcal{I}| \|\theta_{0,\mathcal{I}'}-\theta_{0,\mathcal{I}}\|_2^2\\
	\le \frac{2c_0^2 \log n}{c_5}+\frac{c_5 n}{8} |\mathcal{I}| \|\theta_{0,\mathcal{I}'}-\theta_{0,\mathcal{I}}\|_2^2.
\end{eqnarray*}
This together  with \eqref{cat2eq5} and \eqref{cat2eq6} yields that
\begin{eqnarray*}
	|\mathcal{I}|\|\theta_{0,\mathcal{I}}-\theta_{0,\mathcal{I}'}\|_2^2\le \frac{8\gamma_n}{c_5}+O(1)n^{-1}\log n,
\end{eqnarray*}
with probability at least $1-O(n^{-2})$, where $O(1)$ denotes some positive constant. Since $\gamma_n\gg n^{-1}\log n$, for sufficiently large $n$, we obtain with probability at least $1-O(n^{-2})$ that
\begin{eqnarray*}
	|\mathcal{I}|\|\theta_{0,\mathcal{I}}-\theta_{0,\mathcal{I}'}\|_2^2\le 9c_5^{-1}\gamma_n.
\end{eqnarray*}
The proof is hence completed. 
\change{\subsection{Proof of Theorem 7}  
The proof of Theorem \hyperlink{thm6}{7} relies on the following result that is proven in Lemma E.4 of \citet{cai2021deep}\footnote{See \url{https://openreview.net/attachment?id=rvKD3iqtBdk&name=supplementary_material}.}: 
For any interval $\mathcal{I}\in \mathfrak{I}(m)$ with $|\mathcal{I}|\gg \gamma_n$ and any interval $\mathcal{I}'\in \widehat{\mathcal{P}} $ with $\mathcal{I}\subseteq \mathcal{I}'$, we have  with probability approaching 1 (w.p.a.1.) that
	\begin{eqnarray}\label{mid_thm6}
	\Mean |q_{\mathcal{I},0}(X)-q_{\mathcal{I}',0}(X)|^2\le \bar{C} |\mathcal{I}|^{-1}\gamma_n,
	\end{eqnarray}
for some constant $\bar{C}>0$.  

The rest of the proof is divided into two parts. In the first part, we show assertion (i) in Theorem \hyperlink{thm6}{7} holds. In the second part, we present the proof for assertion (ii) in Theorem \hyperlink{thm6}{7}. It is worth mentioning that results in Lemmas \hyperlink{lemma1_thm5}{5} and \hyperlink{lemma2_thm5}{7} do not rely on the assumption that $Q(\cdot)$ is piecewise function. These lemmas hold under the conditions in Theorem \hyperlink{thm6}{7} as well. 
\smallskip

\noindent \textit{Proof of Part 1:}  Consider a sequence $\{d_n\}_n$ such that $d_n\to 0$ and $d_n\gg \gamma_n$. We aim to show 
$$\max_{\substack{a\in \mathcal{I}' \\ \mathcal{I}'\in \widehat{\mathcal{P}}  }} \Mean [|Q(X,a)-\widehat{q}_{\mathcal{I}'}(X)|^2]=O_p\big( \gamma_n^{\frac{2\alpha_0}{2\alpha_0+1}} \big)+O_p\big( (n\gamma_n)^{-\frac{2\beta}{2\beta+p}}\log^4n \big),$$ 
where the expectation is taken with respect to the marginal distribution of $X$. 

By Lemma \hyperlink{lemma1_thm5}{5}, it suffices to show
\begin{eqnarray}\label{eqn:keyequation0}
	\max_{\substack{a\in \mathcal{I}' \\ \mathcal{I}'\in \widehat{\mathcal{P}}  }} \Mean |Q(X,a)-q_{\mathcal{I}',0}(X)|^2=O_p\big( \gamma_n^{\frac{2\alpha_0}{2\alpha_0+1}} \big). 
\end{eqnarray}
Suppose $|\mathcal{I}'|\ge d_n$. Then according to \eqref{mid_thm6}, we can find some $\mathcal{I}$ such that $|\mathcal{I}|=d_n$ and $a\in\mathcal{I}\subseteq \mathcal{I}'$,
$$\Mean |q_{\mathcal{I},0}(X)-q_{\mathcal{I}',0}(X)|^2\le \bar{C} \frac{\gamma_n}{d_n}.$$
In addition, it follows from H{\"o}lder smoothness assumption that
\begin{eqnarray*}
 	\max_x\max_{\mathcal{I}}\max_{a\in \mathcal{I}} |Q(x,a)-q_{\mathcal{I}}(x)|\le \max_x\max_{\mathcal{I}}\max_{a_1,a_2\in \mathcal{I}} |Q(x,a_1)-Q(x,a_2)|=O(d_n^{\alpha_0}).
\end{eqnarray*} 
By setting $d_n$ to proportional to $\gamma_n^{1/(1+\alpha_0)}$, it is immediate to see that \eqref{eqn:keyequation0} holds. 

Next, suppose $|\mathcal{I}'|<\gamma_n^{1/(1+\alpha_0)}$. Then it follows from the H{\"o}lder smoothness condition that  \eqref{eqn:keyequation0} is satisfied as well. 
This completes the proof for the result (i). 

\smallskip

\noindent \textit{Proof of Part 2:}  
This part follows the second part of the proof of Theorem \hyperlink{thm4}{6}. Recall by the definition of the value function that
\begin{eqnarray}\label{valuefundefi_thm5}
V^{\tiny{opt}}-V^{\pi^*}(\widehat{d})=\Mean \left(\sup_{a\in [0,1]} Q(X,a) \right)-\Mean \left( \int_{\widehat{d}(X)}Q(X,a)\pi^*(a;X,\widehat{d}(X))da \right),
\end{eqnarray}
where the expectation is taken with respect to the marginal distribution of $X$.

Using similar arguments in \eqref{Holder4}, it follows from the result in Part 1 that
\begin{eqnarray}\label{chi13}
	\Mean \left( \sup_{a\in [0,1]} Q(X,a) \right)-\Mean \left( \sup_{\mathcal{I}\in \widehat{\mathcal{P}}} q_{\mathcal{I},0} (X)\right)=O_p\big(\gamma_n^{-\frac{\alpha_0}{2\alpha_0+1}}\big).
\end{eqnarray} 
Similarly, we can show that
\begin{eqnarray*}
	\Mean Q(X,\widehat{d}(X))-\Mean \left( \int_{\widehat{d}(X)} Q(X,a) \pi^*(a;X,\widehat{d}(X))da \right) =O_p\big(\gamma_n^{-\frac{\alpha_0}{2\alpha_0+1}}\big).
\end{eqnarray*}
This together with \eqref{valuefundefi_thm5} and \eqref{chi13} yields that,
\begin{eqnarray}\label{Holder3_thm5}
V^{\tiny{opt}}-V^{\pi^*}(\widehat{d})\le \Mean \left( \sup_{\mathcal{I}\in \widehat{\mathcal{P}}} q_{\mathcal{I},0} (X)\right)-\Mean Q(X,\widehat{d}(X)) +O_p\big(\gamma_n^{-\frac{\alpha_0}{2\alpha_0+1}}\big).
\end{eqnarray} 
	Using similar arguments in \eqref{Holder4}, we can obtain that
\begin{eqnarray*}
	\Mean \left( \sup_{\mathcal{I}\in \widehat{\mathcal{P}}} q_{\mathcal{I},0} (X)\right)-\Mean \left(  \sup_{\mathcal{I}\in \widehat{\mathcal{P}}} \widehat{q}_{\mathcal{I}} (X) \right)\le \bar{C}' \sup_{\mathcal{I}\in \widehat{\mathcal{P}}}  \sqrt{\Mean[ |q_{\mathcal{I},0}(X)-\widehat{q}_{\mathcal{I}} (X)|^2 ]},\\
	\Mean Q(X,\widehat{d}(X))-\Mean   \widehat{ Q}(X,\widehat{d}(X))\le \bar{C}'\sup_{\mathcal{I}\in \widehat{\mathcal{P}}}  \sqrt{\Mean[ |q_{\mathcal{I},0}(X)-\widehat{q}_{\mathcal{I}} (X)|^2 ]},
\end{eqnarray*}
for some constant $ \bar{C}'>0$. Since $\sup_{\mathcal{I}\in \widehat{\mathcal{P}}} \widehat{q}_{\mathcal{I}} (X) =\widehat{ Q}(X,\widehat{d}(X))$, it follows from Lemma \hyperlink{lemma1_thm5}{5} and \eqref{Holder3_thm5} that
\begin{eqnarray*}
V^{\tiny{opt}}-V^{\pi^*}(\widehat{d})= O_p\big(\gamma_n^{\frac{\alpha_0}{2\alpha_0+1}}\big)+O_p\big((n\gamma_n)^{-\frac{\beta}{2\beta+p}}\log^4 n\big).
\end{eqnarray*} 
The proof is completed by setting $\gamma_n$ to be proportional to $n^{-1/\big( 1+\frac{\beta(1+2\alpha_0)}{\alpha_0(p+2\beta)} \big)}$. }

\end{document}

%% file: math_commands.tex
\usepackage{amsmath,amsfonts,bm}

\def\eqref#1{equation~\ref{#1}}

\def\floor#1{\lfloor #1 \rfloor}
\def\1{\bm{1}}

\DeclareMathAlphabet{\mathsfit}{\encodingdefault}{\sfdefault}{m}{sl}
\SetMathAlphabet{\mathsfit}{bold}{\encodingdefault}{\sfdefault}{bx}{n}

\DeclareMathOperator*{\argmax}{arg\,max}
\DeclareMathOperator*{\argmin}{arg\,min}